\pdfoutput=1
\documentclass[11pt]{article}

\usepackage{jheppub}
\usepackage[dvipsnames]{xcolor}

\usepackage{epsfig}
\usepackage{amssymb}
\usepackage{amsmath}
\usepackage{tikz}
\usepackage{mathrsfs}
\usepackage{hyperref}
\usepackage{multirow}
\usepackage{scalerel}
\usepackage{mathtools}
\usepackage{textcomp}
\usepackage[all]{xy}
\usepackage{dsfont}
\usepackage{amsthm}
\usepackage{blkarray}
\usepackage{bm}
\usepackage{todonotes}
\usepackage{tikz-cd}
\usepackage{float}

\numberwithin{equation}{section}

\usepackage{caption}
\usepackage{subcaption}
\usepackage{framed}

\usepackage[mathscr]{eucal}

\usepackage{graphicx}
\usepackage{amsmath,amssymb,mathtools} % Math
\usepackage{xspace}
\usepackage{wrapfig}
\usepackage{slashed}
\usepackage[normalem]{ulem}
\usepackage[utf8]{inputenc}
\usepackage[T1]{fontenc}
\usepackage{mathtools}
\usepackage{stmaryrd}
\usepackage{appendix}

\usepackage{scrextend}

\usepackage{tabularx}
\usepackage{tensor}

\newcommand{\eq}[1]{Eq.~\eqref{eq:#1}}
\newcommand{\eqs}[2]{Eqs.~\eqref{eq:#1} and \eqref{eq:#2}}
\newcommand{\secn}[1]{Sec.~\ref{sec:#1}}

\newcommand{\app}[1]{App.~\ref{app:#1}}
\newcommand{\fig}[1]{Fig.~\ref{fig:#1}}
\newcommand{\figs}[2]{Figs.~\ref{fig:#1} and \ref{fig:#2}}

\newcommand{\df}{\mathrm{d}}

\newcommand{\tr}{\textrm{tr}}
\newcommand{\ra}{\rightarrow}

\newcommand{\as}{\alpha_s}

\newcommand{\eps}{\epsilon}

\newcommand{\cS}{\mathscr{S}}

\newcommand{\bn}{{\bar{n}}}

\newcommand{\nn}{\nonumber}

\newcommand{\cut}{\mathrm{cut}}
\renewcommand{\max}{\mathrm{max}}

\newcommand{\zcut}{z_{\rm cut}}
\newcommand{\qcut}{Q_{\rm cut}}
\newcommand{\tzcut}{\tilde z_{\rm cut}}

\newcommand{\Ok}{\Omega_{1\kappa}^{\figeight}}
\newcommand{\Uka}{\Upsilon_{1,0\kappa}^{\bndry}}
\newcommand{\Ukb}{\Upsilon_{1,1\kappa}^{\bndry}}

\newcommand{\Oq}{\Omega_{1q}^{\figeight}}
\newcommand{\Uqa}{\Upsilon_{1,0q}^{\bndry}}
\newcommand{\Uqb}{\Upsilon_{1,1q}^{\bndry}}

\newcommand{\Og}{\Omega_{1g}^{\figeight}}
\newcommand{\Uga}{\Upsilon_{1,0g}^{\bndry}}

\newcommand{\Pythia}{\texttt{Pythia}\xspace}

\def\ln{\textrm{ln}}
\def\df{\textrm{d}}
\def\nn{\nonumber}
\def\MS{\overline{\rm MS}}

\def\lambdaqcd{\Lambda_{\rm QCD}}
\def\bndry{\varocircle}
\def\figeight{\circ\!\!\circ}

\newcommand{\ee}{e^+e^-}

\def\inv{{-1}}

\def\veps{\varepsilon}
\allowdisplaybreaks[2]

\DeclareRobustCommand{\Sec}[1]{Sec.~\ref{#1}}

\DeclareRobustCommand{\Ref}[1]{Ref.~\cite{#1}}
\DeclareRobustCommand{\Refs}[1]{Refs.~\cite{#1}}
\DeclareRobustCommand{\eq}[1]{Eq.~(\ref{eq:#1})}
\DeclareRobustCommand{\eqs}[2]{Eqs.~(\ref{eq:#1}) and (\ref{eq:#2})}

\newcommand{\eqh}[1]{\begin{equation}\begin{aligned}#1\end{aligned}\end{equation}} 
\def \be {\begin{equation}}
\def \ee {\end{equation}}

\preprint{\vbox{\hbox{MIT--CTP 5461}\hbox{DESY--22--156}
\hbox{UWThPh 2022--15}
}}

\title{Prospects for strong coupling measurement at hadron colliders using soft-drop jet mass}

\author[a,b]{Holmfridur S. Hannesdottir,}%
\affiliation[a]{Institute for Advanced Study, Einstein Drive, Princeton, NJ 08540, USA}
\affiliation[b]{Department of Physics, Harvard University, Cambridge, MA 02138, USA}
\author[c,d,e]{Aditya Pathak,}%
\affiliation[c]{Deutsches Elektronen-Synchrotron DESY, Notkestr. 85, 22607 Hamburg, Germany}
\affiliation[d]{University of Manchester, School of Physics and Astronomy, Manchester, M13 9PL, United Kingdom}
\affiliation[e]{University of Vienna, Faculty of Physics, Boltzmanngasse 5, A-1090 Vienna, Austria}
\author[b]{Matthew D. Schwartz,}%
\author[f]{Iain W. Stewart}%
\affiliation[f]{Center for Theoretical Physics, Massachusetts Institute of Technology, Cambridge, MA 02139, USA}
\emailAdd{hofie@ias.edu}
\emailAdd{aditya.pathak@desy.de}
\emailAdd{schwartz@g.harvard.edu}
\emailAdd{iains@mit.edu}

\abstract{
We compute the soft-drop jet-mass distribution from $pp$ collisions to NNLL accuracy while including nonperturbative corrections through a field-theory based formalism.
Using these calculations, we assess the theoretical uncertainties on an $\as$ precision measurement due to higher order perturbative effects, nonperturbative corrections, and PDF uncertainty. We identify which soft-drop parameters are well-suited for measuring $\as$, 
and find that
higher-logarithmic resummation has a qualitatively important effect on the shape of the jet-mass distribution.
We find that quark jets and  gluon jets have similar sensitivity to $\alpha_s$, and 
emphasize that 
experimentally distinguishing  quark and gluon jets is not required for an $\as$ measurement. We conclude that measuring $\as$ to the 10\% level is feasible now, and with improvements in theory a 5\% level measurement is possible. Getting down to the 1\% level to be competitive with other state-of-the-art measurements will be challenging. 
}


\begin{document}
\maketitle

\section{Introduction}

The QCD strong coupling constant $\alpha_s$ is an essential ingredient in theoretical predictions for the Large Hadron Collider (LHC). Using a variety of high-energy event-shape measurements in scattering processes~\cite{L3:1992nwf,SLD:1994idb,DELPHI:1996oqw,ALEPH:2003obs,DELPHI:2004omy,OPAL:2004wof,Dissertori:2007xa,Becher:2008cf,Davison:2009wzs,Bethke:2009ehn,Dissertori:2009ik,Abbate:2010xh,Abbate:2012jh,Hoang:2015hka,Hoang:2014wka,CMS:2019oeb,ATLAS:2020mee,Aoki:2021kgd}, the value of $\alpha_s$ has been measured to a precision of $\lesssim 1$\%~\cite{ParticleDataGroup:2018ovx,ParticleDataGroup:2020ssz}. However, there is currently more than 3$\sigma$ tension between the most precise extractions, which are from low-energy extractions in lattice QCD~\cite{Aoki:2021kgd}, and from thrust and C-parameter measurements in $e^+e^-$ collisions at LEP~\cite{Abbate:2010xh,Abbate:2012jh,Hoang:2015hka}.
To resolve this discrepancy, it is important to consider alternative observables for determining $\alpha_s$. 
One such candidate is the soft-drop (SD)~\cite{Larkoski:2014wba} jet-mass cross section at hadron colliders.
Soft-drop jet mass is an important jet-substructure observable~\cite{Larkoski:2017cqq,Larkoski:2017iuy,Baron:2018nfz,Kang:2018vgn,Makris:2018npl,Kardos:2018kth,Chay:2018pvp,Napoletano:2018ohv,Lee:2019lge,Hoang:2019ceu,Gutierrez-Reyes:2019msa,Kardos:2019iwa,Marzani:2019evv,Mehtar-Tani:2019rrk,Kardos:2020ppl,Larkoski:2020wgx,Lifson:2020gua,Caucal:2021cfb} because the soft-drop grooming reduces pile-up contamination and sensitivity to hadronization while still allowing for precise theoretical calculations. The perturbative state-of-the-art of the soft-drop jet mass is next-to-next-to-leading logarithmic (NNLL) accuracy for the general case~\cite{Frye:2016aiz}, and results are even known to N$^3$LL for $e^+e^-$ collisions with no angular weight to the grooming~\cite{Kardos:2020gty}.
Thus a precise extraction of $\as$ through soft-drop observables is foreseeable.

The feasibility of using soft-drop thrust to measure $\as$ has already been investigated in detail in Ref.~\cite{Marzani:2019evv} for $e^+e^-$ colliders. Here our focus is on the use of soft-drop jet mass for $\alpha_s$ extraction in $pp$ colliders. 
Although extractions from $pp$ colliders are more challenging than those at $e^+e^-$ colliders, due to the more complicated environment, given the upcoming high-luminosity phase of the Large Hadron Collider (LHC), it is timely to probe the prospects of this observable for precision $\as$ determination.  
Points of reference for current precision of hadron collider $\alpha_s$ measurements are the 2019 PDG value of $\alpha_s(m_Z)=0.1159\pm 0.0034$~\cite{ParticleDataGroup:2020ssz} which has 3\% uncertainty, as well as the more recent TEEC measurement by ATLAS~\cite{ATLAS:2020mee}, $\alpha_s(m_Z)=0.1196\pm 0.004(\text{exp.}){}^{+0.0072}_{-0.0105}(\text{theo.})$ which has 9\% uncertainty.
In this work, we will consider the cross-section differential in the soft-drop jet mass for boosted jets with $p_T \geq 600$ GeV, with an eye towards the opportunity for future higher precision experimental measurements of soft-drop jet-mass cross sections. 

The soft-drop procedure traces the clustering history of the jet, grooming branches with soft radiation, until a sufficiently hard branch is found. More precisely, the jet is first re-clustered using the Cambridge/Aachen algorithm, and next, the groomer un-clusteres the jet, rejecting the softer subjet at each step, unless the pair satisfies the soft-drop criteria given by
\begin{align}\label{eq:SDpp}
    \frac{\min(p_{T_i},p_{T_j})}{p_{T_i}+p_{T_j}} > z_\cut \Big(\frac{\Delta  R_{ij}}{R_0}\Big)^\beta\,,
\end{align}
where $\Delta R_{ij}$ is the distance between the two branches and $\zcut$, $\beta$ and $R_0$ are parameters that control the strength of the groomer. The first time a branch is found that satisfies \eq{SDpp}, the groomer stops, and all the remaining emissions are kept.

Compared to \textit{ungroomed} observables, the soft-drop jet-mass spectrum 
is less sensitive to pile-up and hadronization corrections, for a  wide range of jet masses.
In addition, jet grooming allows for a more robust perturbative prediction: in the region soft drop is active, the non-global logarithms (NGL) arising from soft radiation crossing the jet boundary can only modify the normalization~\cite{Frye:2016aiz}.
The impact on the shape of the spectrum arises only due to differences in the NGL-effect on the normalization in quark and gluon-initiated jets, and hence is typically smaller compared to ungroomed jets.
Thus, the soft-drop jet mass is well-suited for hadron-collider predictions and may provide a powerful cross check on the existing $\as$ measurements based on ungroomed event shapes in $e^+e^-$ collisions.

\begin{figure}[t!]
    \centering
    \includegraphics[width=0.49\textwidth]{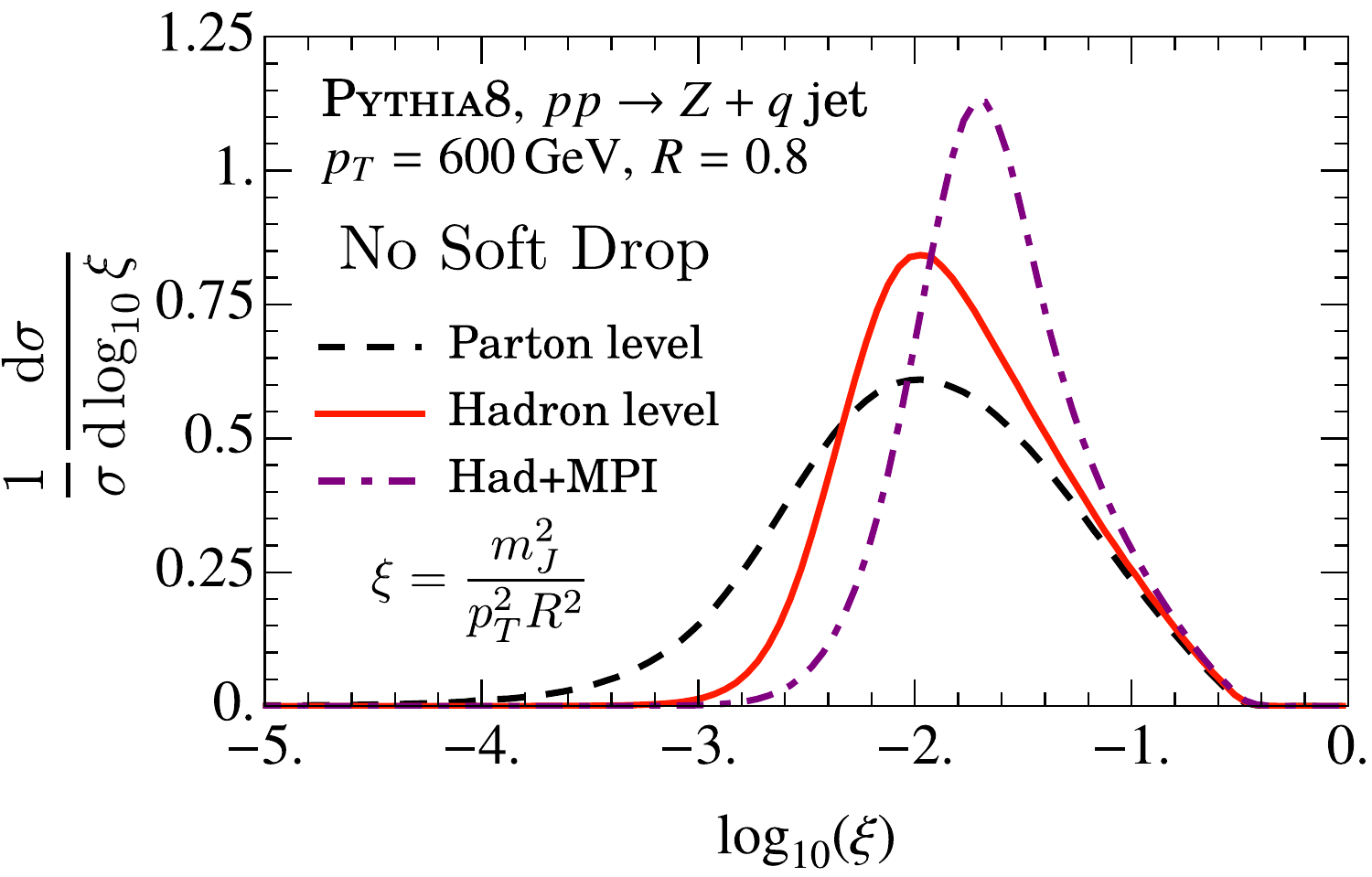}
    \includegraphics[width=0.49\textwidth]{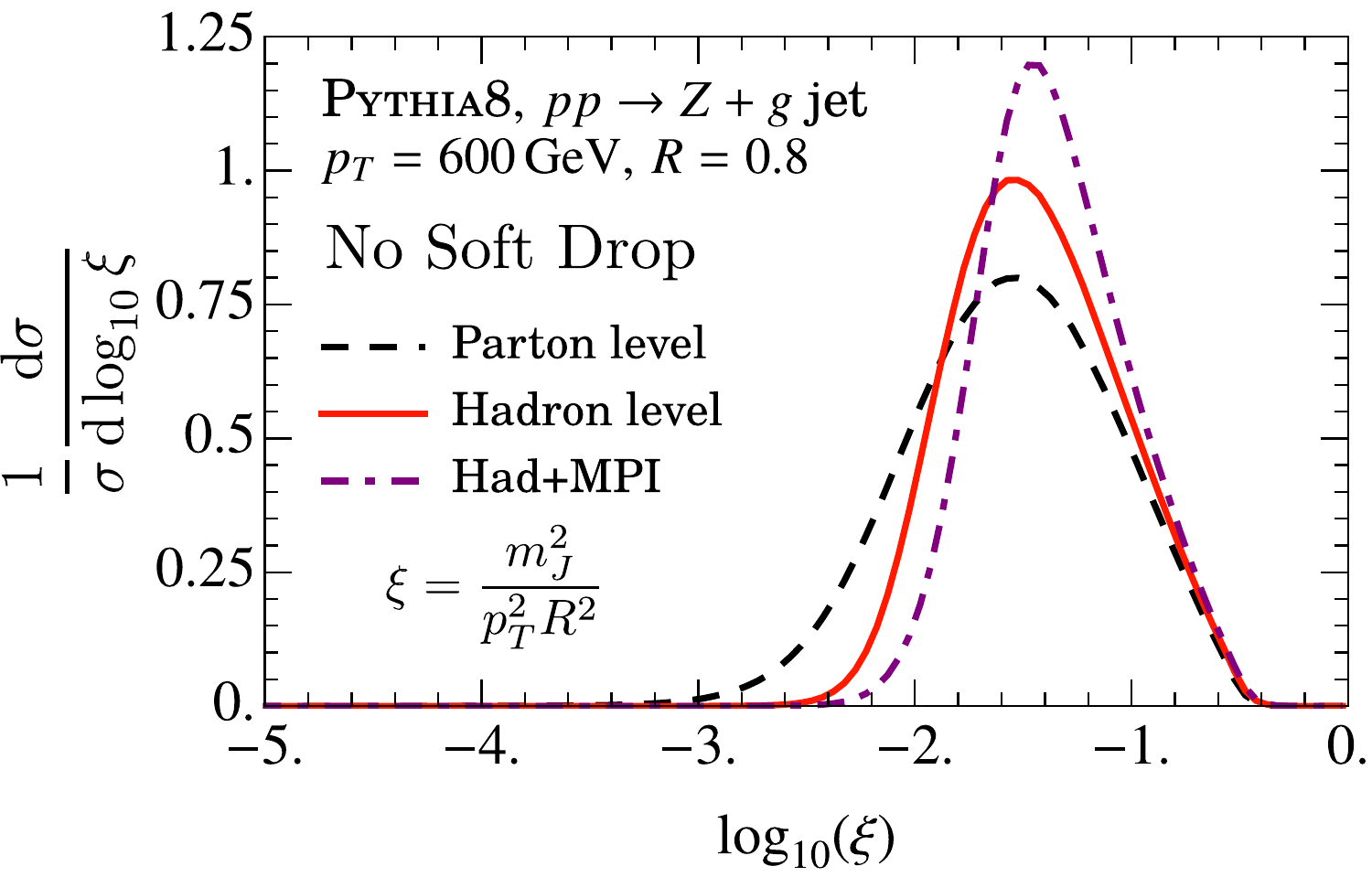}
    \includegraphics[width=0.49\textwidth]{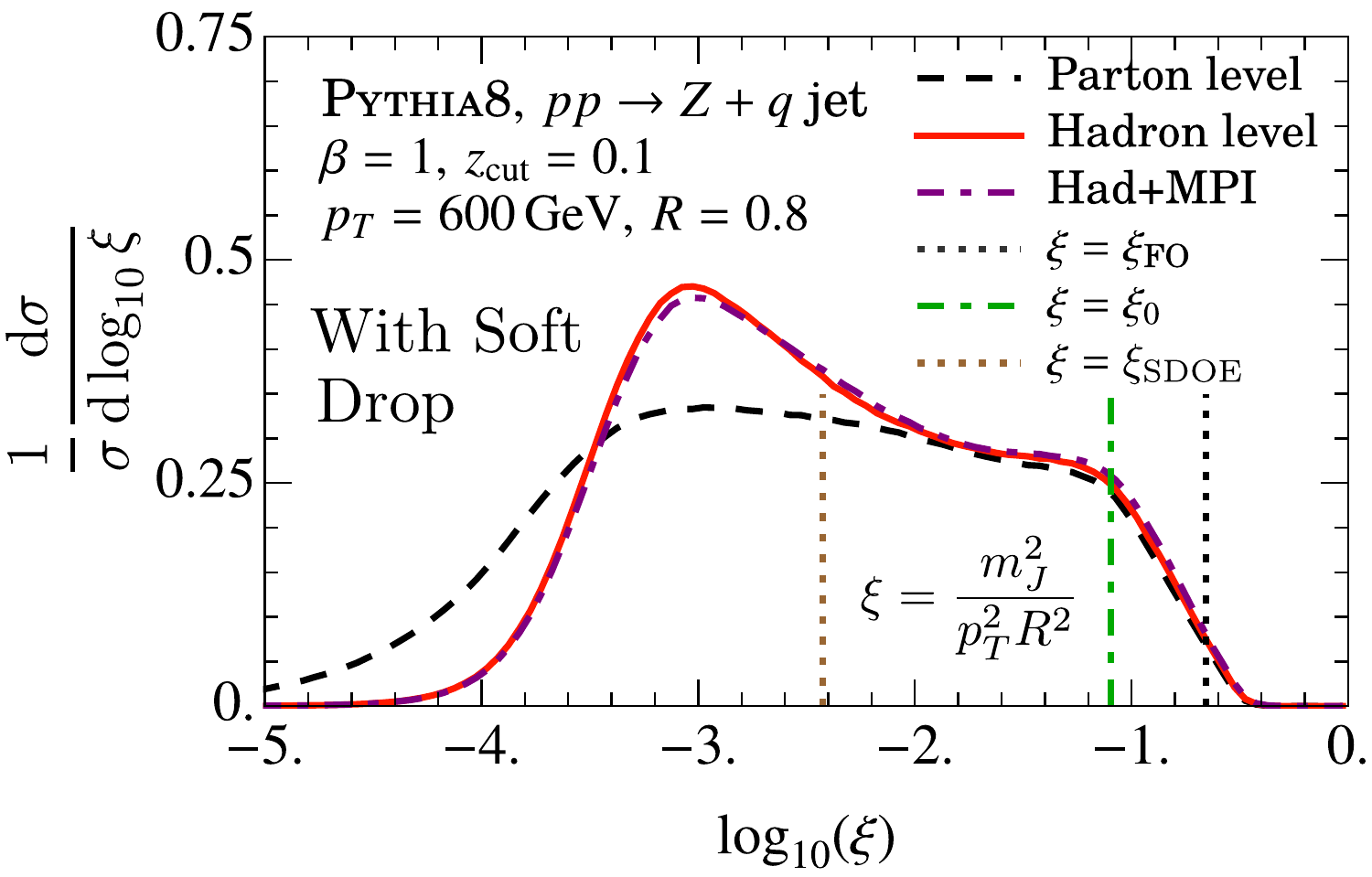}
    \includegraphics[width=0.49\textwidth]{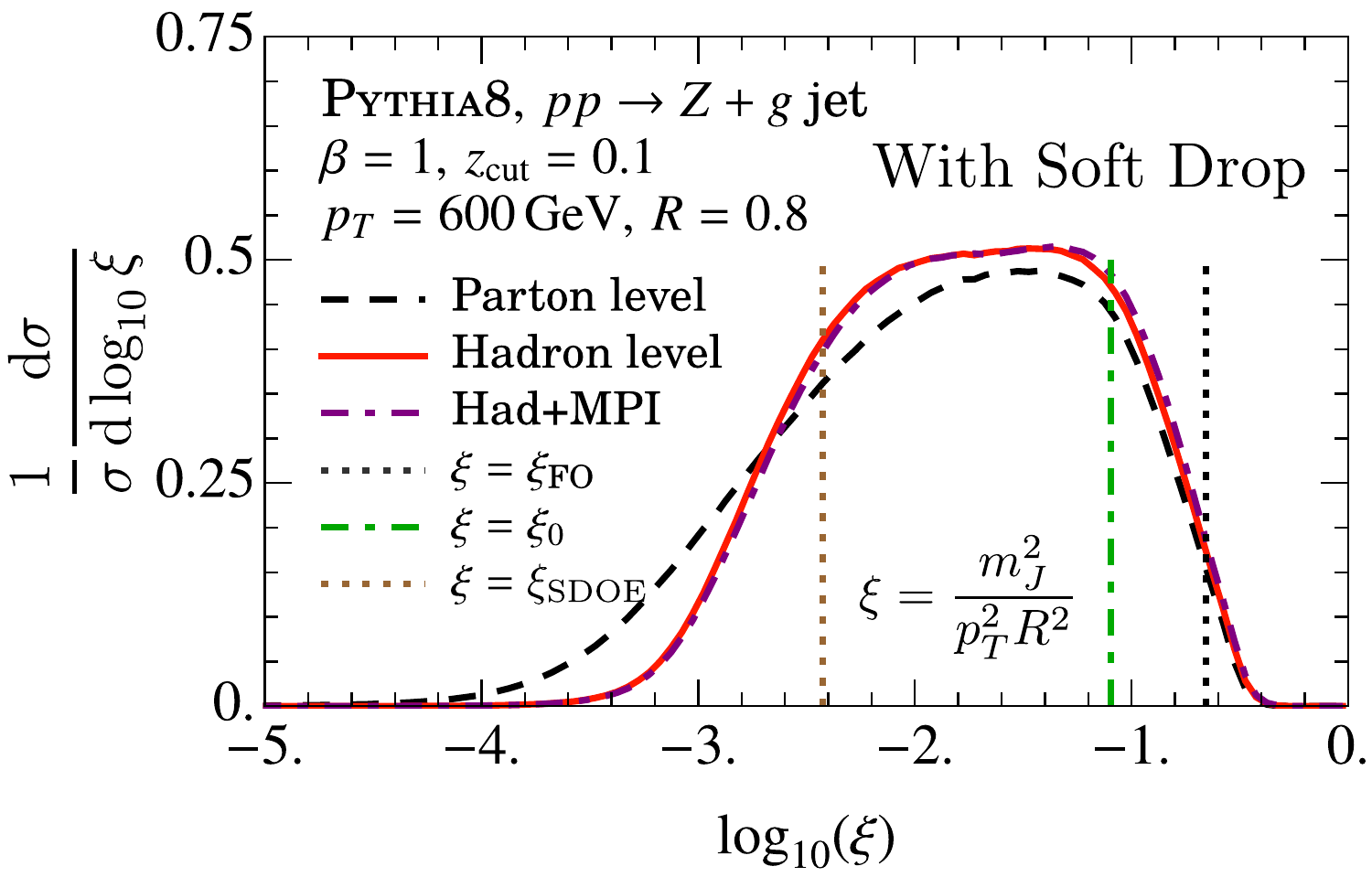}
    \caption{Jet-mass spectrum for quark and gluon jets in \textsc{Pythia8}, at parton and hadron levels and with the underlying event (modeled  as multi-parton-interaction (MPI)) with and without soft-drop grooming. We use the shorthand $\xi=m_J^2/(p_T R)^2$. The green vertical dot-dashed line indicates the jet masses below which soft-drop grooming becomes active and the groomed spectrum deviates significantly from the corresponding ungroomed spectrum. To the right of the black-dotted line, for $\xi \gtrsim \xi_{\rm FO}$ ,the fixed order non-singular corrections to the spectrum become important. For the soft-drop spectrum, the nonperturbative corrections become ${\cal O}(1)$ only to the left of the brown dotted line given by $\xi = \xi_{\rm SDOE}$.}
    \label{fig:MCPlot}
\end{figure}
Some of the characteristic features of the soft-drop jet-mass distribution are shown in \fig{MCPlot}. The figure shows the differential jet-mass cross section simulated in \textsc{Pythia8} for quark and gluon jets with $p_T \in [550,650]$ GeV and $R_{\rm jet} = 0.8$ without (upper panel) and with (lower panel) grooming,  choosing representative values of soft-drop parameters: $(\zcut, \beta) = (0.1, 1)$.
The figure demonstrates that in the
%`$m_J^2 \ra 0$ region'
low jet-mass region, the soft-drop jet mass has a much larger range that is dominated by perturbation theory, compared to its ungroomed counterparts, allowing for fits to $\as$ in a region with a larger cross section.

It is important to delineate the region where soft-drop grooming is active and predictions for the spectrum are dominated by perturbation theory, with nonperturbative effects giving small power-suppressed corrections.
The criterion that ensures that nonperturbative effects are power corrections is 
\begin{align}\label{eq:SDOE}
	\frac{Q \Lambda_{\rm QCD}}{m_J^2}
	\Bigl( \frac{m_J^2}{Q Q_{\rm cut}}\Bigr)^{\frac{1}{2+\beta}}  
	\ll 1 
	\qquad 
	\Rightarrow \,\text{soft-drop stopping emission is perturbative.}
\end{align}
For jet masses $m_J$ that satisfy this condition, 
the transverse momentum $k_t$ of the soft-drop stopping emission will be perturbative, i.e. $k_t\gg \lambdaqcd$, and consequently the nonperturbative effects are subleading. More precisely, we take $\lambdaqcd=1$ GeV and replace the inequality with  $\frac{Q \Lambda_{\rm QCD}}{m_J^2}
	\Bigl( \frac{m_J^2}{Q Q_{\rm cut}}\Bigr)^{\frac{1}{2+\beta}}=\frac{1}{5}$. This transition point demarcates the start of the \textit{soft drop operator expansion region} (SDOE)~\cite{Hoang:2019ceu} and is shown in \fig{MCPlot} 
by the brown dotted line at $\xi=\xi_{\rm SDOE}$, where $\xi=m_J^2/(p_T R)^2$. 
This region is also the one where resummation of logarithms of ratio of scales corresponding to soft drop passing and failing emissions is important, and a rigorous perturbative factorization theorem can be established. 
At even lower values of the jet mass, when  $k_t\sim \lambdaqcd$ for the emission that stops soft drop, then we are in the \textit{soft-drop nonperturbative region} (SDNP), i.e.\ when $\frac{Q \Lambda_{\rm QCD}}{m_J^2}
	\Bigl( \frac{m_J^2}{Q Q_{\rm cut}}\Bigr)^{\frac{1}{2+\beta}}=1$.
We define this region as $\xi < \xi_{\rm SDNP}$, with a precise definition given in Eq.~\eqref{eq:SDNPprecise} below.
Here nonperturbative effects are ${\cal O}(1)$ and we will not attempt to compute the cross section in this region.
As the jet mass increases above the green dot-dashed line at $\xi=\xi_{\rm 0}$ in \fig{MCPlot}, the soft drop grooming becomes less effective.
This occurs because for jet masses $\xi > \xi_0$ the very first subjet seen upon declustering the jet is already energetic enough to pass the groomer. The criterion for this to happen is
\begin{align}\label{eq:SDFO}
	\frac{m_J^2}{Q\qcut} \gtrsim 1
	\qquad 
	\Rightarrow \,\text{soft drop is less significant.}
\end{align}
Hence, we refer to this region as the \textit{plain jet-mass resummation} region. As can be see in from \fig{MCPlot} the parton, hadron and underlying event level spectra look almost identical for $\xi > \xi_0$ for both groomed and ungroomed cases. Finally, the region, for which $\xi > \xi_{\rm FO}$ is referred to as the \textit{fixed-order region}. In this analysis, we will not investigate fitting $\as$ for $\xi > \xi_0$, since obtaining the same precision as in the SDOE region would require a more dedicated study of the cusp region at $\xi\sim\xi_{\rm 0}$. 
We discuss the various regions in detail and provide precise definitions of these transition points in \secn{sdDef}.

In our analysis we compute NNLL perturbative predictions, in addition to employing the formalism of \Ref{Hoang:2019ceu} to include the leading nonperturbative power corrections. We also
include soft-drop non-singular fixed-order effects, such as the ones in Ref.~\cite{Marzani:2019evv}, but, as discussed above, we do not include resummation in the cusp region as investigated in Ref.~\cite{Benkendorfer:2021unv}. 
We also show in Sec.~\ref{sec:mJPC} below that the effects of NGLs at LL and large $N_c$ limit is within the NNLL perturbative uncertainty, 
and hence we do not directly study the effects of NGL normalization effects on the $\as$-sensitivity of the groomed jet mass.
A more detailed comparison with previous work is summarized below in \secn{compare}.

As we can see from \fig{MCPlot}, the hadronization corrections are still non-negligible in the perturbative region and must be taken into account in order to achieve a percent-level precision measurement on $\as$. 
The nonperturbative (NP) effects to the soft-drop jet mass were studied in~\Ref{Hoang:2019ceu} using a 
field-theoretic method that is independent of any hadronization model, 
and are given by
\begin{align}\label{eq:sigfullk1}
	\frac{1}{\sigma_\kappa}\frac{\df\sigma_\kappa}{\df m_J^2}
 =   \frac{1}{\hat \sigma_\kappa}
	\frac{\df\hat \sigma_\kappa}{\df m_J^2}   -Q \Ok\frac{\df}{\df m_J^2} \bigg( C_1^\kappa(m_J^2)\frac{1}{\hat \sigma_\kappa}\frac{\df\hat \sigma_\kappa}{\df m_J^2}\bigg) 
	+
	\frac{Q\Upsilon_{1\kappa}^{\bndry}(\beta)}{m_J^2} C_{2}^\kappa(m_J^2)\frac{1}{\hat \sigma_\kappa}\frac{\df \hat \sigma_\kappa}{\df m_J^2}  
	\, .
\end{align}
In this expression, $\df \hat \sigma_\kappa$ is the partonic cross section,  and the hadron-level cross section $\df \sigma_\kappa$ is obtained by including the NP power corrections represented by the two latter terms.
$\df\sigma_\kappa$ represents the normalized cross section for jets initiated by parton $\kappa$
and will be precisely defined below.
The functions $C_1^\kappa$ and $C_2^\kappa$ are parton-level objects that describe how the power corrections vary as a function of jet-kinematic and grooming parameters. 
\eq{sigfullk1} is valid at LL accuracy where a distinct two-pronged geometry of the groomed jet can be defined. At higher orders there can be subleading corrections that are distortions to this LL-approximation. Nevertheless, as shown in \Refs{Pathak:2020iue,AP}, a more accurate computation of the Wilson coefficients leads to improved description of the leading NP corrections. They have been computed to NNLL accuracy in Refs.~\cite{Pathak:2020iue,AP}, which we will employ in our analysis below. 
The power corrections for each $\kappa=q,g$ involve six unknown nonperturbative constants $\Ok$, $\Uka$ and $\Ukb$, with $\Upsilon_{1\kappa}^{\bndry}(\beta)=\Uka + \beta \Ukb$, where $\kappa$ indicates that the constants differ for quark and gluon jets. 
These constants are universal in the sense that they are independent of the jet kinematics and the grooming parameters.
The superscripts signify specific soft-drop related geometric constraints that we review in \secn{np}.

Since the impact of the power corrections is non-uniform across the SDOE region, as can be seen in \fig{MCPlot}, the nonperturbative constants must ideally be fit for along with $\as$. It will, however, be challenging to fit the theoretical prediction to experimental data for 7 parameters: $\as$ along with the 6 constants. Instead of analyzing the prospects for such a fit, we instead assess the uncertainty on a precision $\alpha_s$ measurement due to the lack of knowledge of these NP parameters. This allows us to estimate the ultimate precision on $\as$ that can be achieved in absence of any perturbative uncertainty, while not attempting to constrain the NP power corrections.
In this work, we use values of these 6 constants from studies of Monte Carlo hadronization models~\cite{Ferdinand:2022xxx} for the estimate,  
and show that the size of the power corrections turns out to be comparable to perturbative uncertainty at NNLL accuracy. 

Given the setup above, we will investigate the sensitivity of the cross section to $\as$ in the SDOE region, and to what extent perturbative variations and nonperturbative power corrections impact the uncertainty in $\as$ determination. The two main features of the jet-mass spectrum that can be exploited for a precision-$\as$ measurement are its shape and the integral of the jet-mass curve. We will state results for two ways of normalizing the cross section:
\begin{align}\label{eq:NormChoice}
	&\text{Normalize to inclusive cross section in the $p_T$-$\eta$ bin:} \quad
    \frac{1}{ \sigma_{\rm incl}}
	\frac{\df^3 \sigma}{\df p_T \df \eta\df \xi}
	\,,
	\\\label{eq:NormChoice2}
	&\text{Normalize to cross section in a fixed jet-mass range:}
 \quad
 \frac{1}{\sigma_{\rm fitrg}(\xi_{\rm SDOE},\xi_{\rm 0}')}
 \frac{\df^3 \sigma}{\df p_T \df \eta\df \xi}
	\,,
\end{align}
where 
\begin{align} \label{eq:sigma_norms}
 \sigma_{\rm incl} &=\frac{\df^2 \sigma}{\df p_T \df \eta} \,, 
 &\sigma_{\rm fitrg}(\xi_{\rm SDOE},\xi_{\rm 0}') 
    &= \int_{\xi_{\rm SDOE}}^{\xi_{0}'}\df \xi \: \frac{\df^3 \sigma}{\df p_T \df \eta\df \xi}
 \,.
\end{align}
Both normalization choices reduce sensitivity to the PDFs, non-global logarithms and the luminosity.
Here $\xi_0'$ is the transition point for the NNLL spectrum, that differs slightly from $\xi_0$, and is defined in \eq{xi0p}. Hence, the dominant region of interest for $\as$ fits is taken to be $\xi \in [\xi_{\rm SDOE}, \xi_0']$.
It is customary to normalize the spectrum by its integral in the SDOE region, i.e.\ the choice in \eq{NormChoice2}, as was done in the recent soft-drop jet-mass measurements by the ATLAS collaboration~\cite{ATLAS:2017zda,ATLAS:2019mgf,ATLAS:2021urs}. 
While normalizing the cross section in a fixed range reduces perturbative uncertainty, we will show below that it loses almost all sensitivity to $\as$, since the shape of the cross section in the SDOE region remains largely invariant upon variations in $\as$.
The sensitivity to $\as$ therefore lies in the integral of the cross section in the perturbative region (SDOE region and beyond). 
Thus, the fixed-range normalization choice greatly reduces sensitivity to $\as$ variations compared with the variations due to perturbative uncertainty and nonperturbative corrections, and hence increases uncertainties in any potential $\as$ measurements.
This insensitivity is particularly severe for quark jets where, at NNLL accuracy, the spectrum in the SDOE region flattens and variations in $\as$ have almost no effect on the cross section normalized to a fixed range.
Based on this study, we will justify our recommendation for the first choice in \eq{NormChoice}, and avoid self-normalizing the cross section in the SDOE region.

The outline of the paper is as follows: In \secn{theory} we review the details of resummation of the soft-drop jet-mass spectrum using soft-collinear effective field theory (SCET)~\cite{Bauer:2000ew,Bauer:2000yr,Bauer:2001ct,Bauer:2001yt}. The framework for the NP power corrections is discussed in \secn{np}. Finally, in \secn{analysis} we employ this setup in testing the sensitivity to $\as$ and impact of the NP power corrections via the procedure outlined above. We conclude in \secn{conclusion}. The appendices provide details of derivations and supplementary material on the numerical analysis.

\section{Perturbative predictions}

\label{sec:theory}

The various regions shown in \fig{MCPlot} are distinguished by specific logarithmic  or power corrections involving ratios of different energy scales  that become significant. We implement the resummation of these logarithms and the treatment of nonperturbative power corrections using SCET. 

\subsection{Kinematic features}
\label{sec:sdDef}
We first review the main kinematic features of the soft-drop spectrum. Although we are primarily concerned with $\as$ determination at the hadron colliders, our presentation of the formalism is done such that it is equally applicable to $e^+e^-$ collisions. 
In the case of $pp$ collisions, we will primarily focus on jets identified inclusively with $R/2 \ll 1$. The results for $e^+e^-$ remain valid for large $R$ jets and reduce to familiar expressions for hemisphere jets when $R=\pi/2$.
We refer the reader to \Refs{Frye:2016aiz,Hoang:2019ceu,Pathak:2020iue} for a more thorough discussion and derivation of some of the results stated here. 

The large momentum associated with the jet is given by
\begin{align}\label{eq:Q}
	&Q = p_T R& 
	 &(\textrm{$pp$ case})\,, &
	 &Q = 2 E_J\tan \frac{R}{2}&
	& (\textrm{$e^+e^-$ case})\,. & 
\end{align}
We will be concerned with the region of the jet-mass spectrum that is dominated by soft and collinear emissions. We define variables $z_i$, the $p_T$ or energy fraction of a soft subjet, given by 
\begin{align}\label{eq:Qdef}
	&z_i = \frac{p_{T_i}}{p_{T_J}} &
	 &(\textrm{$pp$ case})\,, &
	&z_i = \frac{E_i}{E_J}& 
	& (\textrm{$e^+e^-$ case})\,, & 
\end{align}
 and angle $\Delta R_{iJ}$ or $\theta_{iJ}$ that it makes with the jet axis. The soft-drop criteria in \eq{SDpp} for passing for soft collinear limit $z_i\ra 0$ and $\theta_i \ra 0$ can be compactly expressed as
\begin{align}
	&z_i > \zcut' \Delta R_{iJ}^\beta &
		 &(\textrm{$pp$ case})\,, &
    &z_i > \tzcut \Big(2 \sin \frac{\theta_{iJ}}{2}\Big)^\beta &
    & (\textrm{$e^+e^-$ case})\,, & 
\end{align}
where in defining $\zcut'$ and $\tzcut$ we have absorbed the parameter $R_0$:
\begin{align}\label{eq:tzcut}
    &\zcut' \equiv
    \zcut \big(R_0^{(pp)}\big)^{-\beta}  &
    &
    (\textrm{$pp$ case})\,, &
    &\tzcut \equiv \zcut\bigg[\sqrt{2}\sin\Big(\frac{R^{e^+e^-}_0}{2}\Big)\bigg]^{-\beta}&
    &
    (\textrm{$e^+e^-$ case})\,. &
\end{align}
With this definition, we can identify the energy scale $\qcut$ associated with soft drop:
\begin{align}\label{eq:qcut}
	&\qcut = p_T \zcut' R^{1+\beta}& 
	&\text{($pp$ case)} \, ,&
	&\qcut = \Big(2\tan \frac{R}{2}\Big)^{1+\beta}E_J \tzcut &
&(\textrm{$e^+e^-$ case}) \, .&
\end{align}
% the scale $\qcut = 2^\beta Q \tzcut$ can be used interchangeably for $pp$ and $e^+e^-$ collisions. 
% 
Finally, we will find it convenient to work with a dimensionless combination of the jet mass and the large momentum $Q$
\begin{align}\label{eq:xiDef}
	\xi \equiv \frac{m_J^2}{Q^2} \, .
\end{align}

As shown in \fig{MCPlot}, the soft-drop jet-mass spectrum can be divided into four distinct regions. 
The regions encountered as we reduce $\xi$ from its maximum value $\xi = 1$ are as follows: 
\begin{enumerate}
    \item \textit{The fixed-order region}: This region corresponds to groomed jet masses close to the end point of the spectrum for $\xi \lesssim 1$. In this region the logarithms $\ln (\xi)$ generated at each perturbative order are ${\cal O}(1)$ and the region  can be adequately described in fixed-order perturbation theory.  In practice, at a given order in $\as$ expansion, only a finite number of particles contribute to the jet mass, and hence the end point is seen at somewhat smaller values. For leading-order (LO) analysis, $\xi_{\rm LO} = 0.25$, and for next-to-leading-order (NLO) we have $\xi_{\rm NLO} \simeq 0.4$~\cite{Marzani:2017kqd}. In our analysis we will treat this region to ${\cal O}(\alpha_s)$, the LO accuracy.  We also retain all the finite $\zcut$ terms in this region. 
    \item \textit{Plain jet-mass resummation region}: This region is dominated by soft and collinear emissions and corresponds to $\xi_0< \xi \ll1$, and the logarithms, $\alpha_s^n\ln^m( \xi)$ with $m \leq 2n$ become large and must be resummed. Here
    \begin{align} \label{eq:xi0def}
        &\xi_0  =\zcut' R^\beta&
          &(\text{$pp$ case})\,,& 
         &\xi_0  = \tzcut \Big(2 \tan \frac{R}{2}\Big)^\beta& 
        &(\text{$e^+e^-$ case})\, .&
    \end{align}
    From \eqs{Q}{qcut}, we see that
    \begin{align}\label{eq:qcut2}
		\qcut = Q \xi_0 \, .
    \end{align}
    As discussed below in \secn{plain}, power corrections of ${\cal O}(\xi)$ are small and contributions from soft and collinear modes in this region factorize. However, soft drop does not significantly groom the jet in this region, as one of the first few de-clusterings pass the soft-drop condition. Consequently, the jet mass after grooming is only slightly smaller than its value prior to grooming. Thus, in this region the effect of soft drop alone can be treated in fixed-order perturbation theory. 
    Below the $\xi_0$ transition point this situation changes, and resummation must take into account a change in the logarithmic structure due to soft drop.
	Note, however, that close to the cusp, the soft emissions passing soft drop are at angles $\Delta R_{ij} \sim R$ and hence one must employ the version of soft drop without the collinear approximation $\Delta R_{ij}\ra 0$. Doing so leads to the ``soft wide-angle'' transition point $\xi_0'$  differing slightly from $\xi_0$~\cite{Baron:2018nfz}:
    \begin{align}\label{eq:xi0p}
    	\xi_0^\prime \equiv \frac{\xi_0}{(1 + \zeta^2)^{\frac{2+\beta}{2}}} \, .
    \end{align}
	where $\zeta$ for the two cases is given by:
	\begin{align}\label{eq:zetaDef}
		\zeta \equiv \tan\frac{R}{2}   \quad (\text{$e^+e^-$ case}) \,, \qquad 
		\zeta \equiv \frac{R}{2 \cosh \eta_J}   \quad (\text{$pp$ case}) \, .
	\end{align}
    The value of $\xi_0'$ corresponds to the location of the distinct \textit{soft-drop cusp}. Defining the parameter $\zeta$ will prove useful in universally keeping track of wide-angle soft effects. 
    
\item \textit{Soft-drop resummation region}: As we move past $\xi_0$ to  yet smaller jet masses $\xi < \xi_0$, the application of soft drop removes a significant amount of soft, wide-angle radiation. The hierarchy $\xi \ll \xi_0$ results in large logarithms $\ln (\xi/\xi_0)$, and a different resummation from the previous case is required. 
     At the same time, perturbative power corrections of ${\cal O}(\xi/\xi_0)$ are small and can be neglected. 
	We will also restrict ourselves to the case $\xi_0 \ll 1$, such that power suppressed terms in $\xi_0$, or equivalently $\zcut$, can be ignored and the leading power factorization derived in \Ref{Frye:2016aiz} can be employed. 
     This enables us to further factorize the soft function from the plain jet-mass region into contributions that fail and pass soft drop, such that the large logarithms $\alpha_s^n\ln^m( \xi/\xi_0)$ can be resummed. 
     (For the special case of $\beta = 0$ the leading logarithms are single logs with $m = n$.) 
     From the point of view of nonperturbative power corrections, as long as the first subjet that stops soft-drop is perturbative (with  $k_t \gg \Lambda_{\rm QCD}$), which amounts to satisfying the criterion in \eq{SDOE}, the perturbative prediction  dominates.  This is the SDOE region where nonperturbative power corrections can be incorporated through constant parameters that appear in a systematic expansion.  In terms of $\xi$, the condition  in \eq{SDOE} for being in the SDOE region can be expressed as
    \begin{align}\label{eq:xiSDOEdef}
        \Big(\frac{\xi_{\rm SDNP}}{\xi}\Big)^{\frac{1+\beta}{2+\beta}} \ll 1\,,\qquad \text{(SDOE region)} \, ,
    \end{align}
    where the parameter $\xi_{\rm SDNP}$, defined by \eq{xiSDNPdef} below,  characterizes the region below which nonperturbative corrections become ${\cal O}(1)$. 
    
    \item \textit{Soft-drop nonperturbative region}: 
    Here the nonperturbative effects are ${\cal O}(1)$. This happens for jet masses 
    \begin{align}\label{eq:xiSDNPdef}
		\xi \gtrsim \xi_{\rm SDNP} \equiv \xi_0 \bigg(\frac{1\,{\rm GeV}}{\qcut}\bigg)^{\frac{2+\beta}{1+\beta}}\,.
    \end{align}
     In the SDNP region, the perturbative resummation of large logarithms in SDOE region still carries over. However, the collinear-soft function describing dynamics of the emissions that pass the groomer becomes nonperturbative. This region is equivalent to the shape-function region in ungroomed jets, however the shape function that appears here is not normalized to one and involves convolution in fractional power $k_{\rm NP}^{\frac{2+\beta}{1+\beta}}$ of the nonperturbative momentum $k_{\rm NP}$~\cite{Hoang:2019ceu}.
     It remains to be explored how  the nonperturbative power corrections in the SDOE region (reviewed in \secn{np}) are related to the shape function in the SDNP region.
     
\end{enumerate}

Further below we will review
the perturbative resummation that is necessary in the plain jet-mass and the soft-drop resummation region.  We note that close to the soft-drop cusp, a more refined treatment along the lines of \Ref{Benkendorfer:2021unv} can be performed by identifying a resolved soft emission that stops the groomer. Our treatment nonetheless fully captures the dominant leading order effect at the soft-drop cusp including all the perturbative power corrections related to the jet mass and soft drop at ${\cal O}(\alpha_s)$, and we estimate that the impact from the lack of cusp resummation is smaller than all other uncertainties for the large values of $p_T$ considered for analysis of $\as$ in $pp$ collisions. We leave to future work a more detailed integration of the cusp region resummation with the setup presented here.

\subsection{Quark/gluon fraction} \label{sec:qgfraction}
The quark gluon fraction is a very useful {\it theoretical} concept, regardless of whether there is a direct experimental way to measure the number of quark jets and gluon jets. To compute the soft-drop jet mass we do not even need to know whether
the two classes of jets can be distinguished. What we compute is the total jet-mass distribution, inclusive over quarks and gluons. To do so, we compute the individual jet-mass distribution for quark jets and for gluon jets separately in SCET, add them together using perturbative predictions for
the quark and gluon jet cross sections or 
equivalently
the quark/gluon fraction, and then match the sum to full QCD results
to include additional fixed-order corrections.

At tree level, the quark/gluon cross sections are identical in SCET and QCD. Beyond leading order, the quark/gluon cross sections are ambiguous in QCD but not in SCET. 
In SCET, whether a jet is quark or gluon is determined by the quantum numbers of the collinear fields in the hard scattering operator. Thus, in SCET the quark/gluon fraction is well-defined and systematically calculable order-by-order in perturbation theory. Whatever ambiguity there is for the quark/gluon jet definition in QCD is immaterial, since after matching to the NLO inclusive (over quarks and gluons) jet-mass distribution we no longer need to discuss quark/gluon fractions. So at NLO and beyond the quark/gluon fraction is not meaningful in QCD, but it does not have to be. 
The quark/gluon fraction is nevertheless a useful concept because the leading-order prediction gets small corrections at NLO in SCET and provides valuable intuition for what the full matched distribution might look like. 

Note that the prediction for the cross section also depends on PDFs, as can be seen from \eqs{inclJ}{sigExcl}. It has been argued in~\Ref{Forte:2020pyp} that for observables that are sensitive to PDFs, a consistent determination of $\alpha_s$ cannot be carried out unless PDFs are simultaneously determined. The arguments in~\Ref{Forte:2020pyp} apply to inclusive cross-section measurements, while here we study jet substructure measurements. We remove the dependence on the inclusive cross section by appropriately normalizing the cross section as shown in \eq{NormChoice}, so that our observables become mostly insensitive to PDFs. The remaining small uncertainty is propagated into our result through the quark/gluon fractions, but we will show that the resulting uncertainties on an $\as$ measurement are a subdominant effect compared with the uncertainties due to scale-variations of the perturbative cross section, as well as those due to nonperturbative effects.

\subsubsection{Impact of PDF variations on quark/gluon fractions}

To compute the quark gluon fraction we begin at leading order by computing the rates for dijet production using the different partonic process $q \bar{q} \to g g$, $q q' \to q q'$, etc. The quark fraction at leading order is then
\begin{equation}
    x_{q} =  \frac{2\sigma (pp \to q q) + \sigma (pp \to gq)}{2\sigma (pp \to {\rm dijet})}
 \,,
\end{equation}
where $q$ heuristically refers to quarks or antiquarks, identical or distinct. The gluon fraction is $x_g=1-x_q$.

We show in Table~\ref{tab:qgpdfs} the tree-level quark/gluon fraction computed with various PDFs. These fractions were computed with Madgraph~\cite{Alwall:2014hca} for $pp$ collisions at 13 TeV with a $p_T$ cut of 600 GeV on the jets. 
We show results for a representative sample from among various available PDF sets.
We see that there is surprisingly little variation among the PDFs with a typical quark gluon fraction of $x_q= 0.51$ with about a 2\% uncertainty. 

\begin{table}[t]
    \centering
    \begin{tabular}{|c|c|c|c|}
    \hline
    \text{PDF} & $\alpha_s$~ \text{used} & $x_q$& \text{\%~change} \\
    \hline
 \text{NNPDF 23 LO} & 0.119 & 0.479 &  -6.0 \\
 \text{NNPDF 23 NLO} & 0.119 &  0.517 & 1.3 \\
 \text{NNPDF 23 NNLO} & 0.119 & 0.523 &  2.5\\
 \text{NNPDF 23 NNLO} & 0.120 &  0.514 &  0.84 \\
 \text{CT18NLO$\_$as$\_$0119} & 0.119 & 0.514& 0.87 \\
 \text{CT18NNLO$\_$as$\_$0119} & 0.119& 0.507 &  -0.49 \\
 \text{MSTW2008nlo68cl} & 0.120 & 0.510 &  0.063 \\
 \text{MSTW2008nlo68cl} & 0.117 &  0.514&  0.87 \\ 
 \hline
 \text{mean} & -- & 0.510 & 1.6 \\ 
 \hline
    \end{tabular}
    \caption{Quark/Gluon fraction for different parton distribution functions which average to $x_q = 0.510$ with a standard deviation of $0.013$. The \% changes are listed relative to this mean and the average \% change is the average of the absolute value of these \%'s.  }
    \label{tab:qgpdfs}
\end{table}

To systematically improve the computation for the quark gluon fraction, we include perturbative corrections using SCET factorization,
as discussed below in \Sec{sec:excl}.
In particular, we provide expressions for quark/gluon fractions for exclusive soft drop groomed dijets identified via a $p_T^{\rm veto}$ on additional jets.
Taking the result from there (in a schematic form) we have
\begin{align}
  x_q = \frac{1}{N} \sum_{\{\kappa_i\} } \int\!\! \df\xi_a \df\xi_b \:
    f_{\kappa_a}(\xi_a,\mu) f_{\kappa_b}(\xi_b,\mu)  
    \otimes 
    \big({\cal I}_{\kappa_a\kappa_{a'}} {\cal I}_{\kappa_b\kappa_{b'}}\big)
    \ {\rm tr}\Big[
   {\widehat H}_{\kappa_{a'}\kappa_{b'}}^{\kappa_1\kappa_2}(\Phi_{\rm born}) \otimes 
    {\widehat S}_{G,a'b'}^{\kappa_1\kappa_2}
    \Big] \,,
\end{align}
where here $\kappa_{a'}\kappa_{b'} \to \kappa_1\kappa_2$ is the partonic exclusive dijet hard scattering process, and $\sum_{\{\kappa_i\}}$ indicates a sum over all channels for each $\kappa_i$ variable, except that at least one of the final-state jet labels, $\kappa_1$ or $\kappa_2$, is fixed to be a quark.
Here $f_\kappa$ is a PDF, ${\cal I}_{\kappa\kappa'}$ accounts for initial-state radiation which depends on the exclusive jet veto and that may or may not change the identity of the parton participating in the hard scattering, $\widehat H$ encodes the hard scattering including virtual corrections, and ${\widehat S}_G$ are global soft functions. $N$ is a normalization factor.

We would like to determine the impact on $x_q$ from using PDFs obtained at different orders in perturbation theory, which also requires systematically including higher order terms in other parts of the factorization theorem.  
Since we are interested in an estimate for these effects in a context that is more general than exclusive jets, we choose to ignore the jet veto dependence as well as higher order corrections from the global soft functions, and hence work with tree level functions ${\cal I}_{\kappa\kappa'}$ and ${\widehat S}_{G}$. Thus, we focus on including the perturbative corrections to $H$ in a manner that is consistent with the use of higher order PDFs (in particular, for example, with the same choice of factorization scale $\mu$ in the $f$s and $H$ so that the $\mu$ dependence cancels to the order being considered). 
At this order the soft functions are just matrices, for example in the $qq\to qq$ channel
\begin{equation}
S_{IJ}^{0} = 
    \begin{pmatrix}
     N^2 & 0 \\
     0 & \frac{C_F N}{2}
    \end{pmatrix}
    \,.
\end{equation}
This and the other tree-level soft functions can be found in~\cite{Broggio:2014hoa}. The hard functions have been computed to NLO~\cite{Kelley:2010fn} and NNLO~\cite{Broggio:2014hoa}. 

Using the CTEQ NNLO PDFs with $\alpha_s = 0.119$ we find the LO, NLO and NNLO quark/gluon fractions to be
\begin{equation}
    x_q^{\text{LO}} = 0.507,\quad
    x_q^{\text{NLO}} = 0.530,\quad
    x_q^{\text{NNLO}} = 0.538,\quad
\end{equation}
corresponding to an 4.5\% increase in the fraction at NLO and a further 1.5\% increase at NNLO. Thus, we see that the impact of variations of PDF induces at most 2\% uncertainty on quark gluon fractions.
As we will see in Section~\ref{sec:fracalpha}, this translates to around a 1\% uncertainty on $\as$, which we will show is quite subdominant compared to other uncertainties in an extraction of $\alpha_s$ from soft-drop jet-mass spectra. 

\subsubsection{Measurements of inclusive jets}

We will be concerned with jets produced in hadron collisions with an underlying process such as $pp\ra \text{dijets} +X $ or $pp \ra Z + j_\kappa +X $ with the jet initiated by the parton $\kappa = q,g$.   We will derive the factorization in the context of inclusive jets measurement but our conclusions will also hold for the exclusive case.  We begin by considering the cross section, differential in $p_T$ and $\eta$, for inclusive jets~\cite{Ellis:1996mzs,Kang:2018jwa}:
\begin{align}\label{eq:inclJ}
	\frac{\df^3 \sigma}{\df p_T \df \eta \df \xi } 
	= \sum_{abc}\int \frac{\df x_a \df x_b \df z}{x_a x_b z}  f_a(x_a,\mu)  f_b (x_b, \mu)  H_{ab}^c \Big(x_a, x_b, \eta , \frac{p_T}{z} , \mu\Big) {\cal G}_c (z, \xi,  p_T ,R,  \mu) \, ,
\end{align}
where $f_{a,b}$ are the parton distribution functions, $H_{ab}^c$ is the hard function describing inclusive production of parton $c$ with transverse momentum ${\cal O}(p_T)$, which initiates a signal jet carrying momentum fraction $z$. The inclusive jet function~\cite{Kang:2016mcy} ${\cal G}_c$ describes the formation of jet of radius $R$ initiated by a parton $c$ and the jet-mass measurement $\xi$. 
The factorization is valid for $R/2 \ll 1$, which implies that the relevant modes describing the jet dynamics are the hard-collinear modes with transverse momentum $k_t \sim p_TR$ (but with energy $\sim p_T$), which is parametrically smaller than the scale $p_T$ characterizing the jet production. Numerically, however, it has been shown to work very well for $R/2 \lesssim 0.4$~\cite{Dasgupta:2014yra,Becher:2015hka,Chien:2015cka,Hornig:2016ahz,Dasgupta:2016bnd,Kolodrubetz:2016dzb}. 
The $\mu$ evolution in \eq{inclJ}  resums the large (single) logarithms $\alpha_s^n \ln^k (p_TR/p_T)$, with $k \leq n$:
\begin{align}\label{eq:DGLAP}
	\mu\frac{\df }{\df \mu} {\cal G}_i \big(z ,\xi, p_T R, \mu\big) = \frac{\alpha_s(\mu)}{\pi} \sum_j \int_z^1 \frac{\df z'}{z'}
	\: P_{ji}\bigg(\frac{z}{z'}, \mu\bigg) {\cal G}_{j}\big(z', \xi,  p_T, R, \mu\big) 
	\, ,
\end{align}
where $P_{ji}$ are the time-like splitting functions. As the DGLAP evolution between the hard and the hard collinear scale $p_T R$ involves branching of the original parton leading to formation of multiple jets. See also \Ref{Neill:2021std} for an extension of this formalism to leading jets. 

While the quark/gluon fraction is internal to the entire machinery for a theory prediction, 
	it is nevertheless useful to pull them out 
	such that we can define the intermediate objects corresponding to quark- and gluon-initiated jets, and analyze their properties separately.
The integral over the jet function in \eq{inclJ} defines the event averaged number of jets for quarks and gluons~\cite{Neill:2021std},
\begin{align}
    \int_0^1\! \df z\, \df \xi \: {\cal G}_\kappa (z, \xi, p_T ,R, \mu) = \langle N_{\kappa,\rm jets}\rangle \, , \qquad \kappa = q,g
  \,.
\end{align}
The number density is generated dynamically via QCD fragmentation process and depends on the center of mass energy and the jet algorithm. As such it is not straightforward to identify quark and gluon fractions associated with inclusive jet measurement. 

Our goal, however, is to analyze jets in the region of small jet masses, where, as we describe below, the hard collinear modes at the scale $ p_T R$ factorize from the collinear and soft dynamics governed by the jet-mass measurement. The additional constraint of the small jet-mass measurement enables us to speak of \textit{quark- and gluon-initiated jets}. When we deem a jet to be a quark or a gluon jet in perturbation theory, we are not referring to the color charge of all the partons inside but instead the color charge of the \textit{only collinear sector} that is singled out due to the small jet-mass constraint at leading power, i.e. with corrections suppressed at ${\cal O}(\xi)$. In general, there will be non-global emissions of soft gluons from outside of the jet but they do not modify this aspect. In the limit of small jet mass, the inclusive jet function in \eq{inclJ} factorizes as 
\begin{align}\label{eq:GcFact}
    {\cal G}_c\big(z, \xi , p_T , R, \mu \big) = \sum_i {\cal H}_{c \ra i} \big(z, p_T R, \mu\big) \: {\cal J}_i(\xi, p_T, \eta, R, \mu) \,, \qquad \xi \ll 1 \, .
\end{align}
The ${\cal H}_{c\ra i}$ function involves hard-collinear modes integrated out at scales $p_TR$, while ${\cal J}_i$ is still a multiscale function whose factorization depends on the jet-mass region, which will be treated in subsequent sections.
The ${\cal H}_{c\ra i}$ obeys the following evolution equation,
\begin{align}
    \mu\frac{\df}{\df \mu} {\cal H}_{i\ra j}  \big(z, p_T R, \mu \big) = \sum_k \int_z^1 \frac{\df z'}{z'} \: 
    \gamma_{ik} \bigg(\frac{z}{z'}, p_T R, \mu\bigg) {\cal H}_{k\ra j} \big(z', p_T R, \mu\big) \, ,
\end{align}
where the anomalous dimension $\gamma_{ij}$ is given by
\begin{align}
    \gamma_{ij} \big(z, p_T R ,\mu\big) = \delta_{ij} \delta (1-z) \Gamma_i \big(p_TR , \mu\big) + \frac{\alpha_s}{\pi} P_{ji}(z) \, .
\end{align}
We see that the same splitting function appears that is tied to the $z$ dependence in ${\cal H}_{i \ra j }$. The functions $\Gamma_i$ are responsible for Sudakov double logarithmic evolution between the hard-collinear, collinear and soft modes in the small jet-mass region. 
This diagonal evolution is consistent with the RG evolution of soft and collinear modes in ${\cal J}_i$ in \eq{GcFact}.
It is helpful to separate the purely diagonal piece at all orders by writing~\cite{Cal:2019hjc} 
\begin{align}\label{eq:splitHij}
    {\cal H}_{i \ra j } \big(z, p_T R , \mu\big) =  J_{ij}\big(z, p_T R ,\mu\big)N_{\rm incl}^j \big(p_T R , \mu\big) \, , 
\end{align}
where the functions $J_{ij}(z, p_T R , \mu)$ also track the flavor of the jet which the original parton fragments into and describes the dependence of their corresponding energy fraction.
They satisfy the same DGLAP evolution equation as the inclusive jet function in \eq{DGLAP}, whereas the functions $N_{\rm incl}^j$ satisfy multiplicative RG equations:
\begin{align}\label{eq:Nincl}
    \mu \frac{\df}{\df \mu}  \ln N^i_{\rm incl} \big(p_T R ,\mu \big) = \Gamma_i \big(p_T R, \mu \big)\,.
\end{align} 
The evolution equations for $N^i_{\rm incl}$ and $J_{ij}$ alone do not fix their all-orders expressions, and hence there is some freedom in separating the constant, non-logarithmic pieces of ${\cal H}_{i \ra j }$ between the two functions. At each order, we define $J_{ij}$ such that summing over the flavor index $j$ leads to the inclusive jet function:
\begin{align}\label{eq:sumJij}
	\sum_{j}J_{ij}\big(z, p_T R ,\mu\big) = J_{i}\big(z, p_T R ,\mu\big) \, .
\end{align}
This also fixes the $N_{\rm incl}^\kappa$.

This enables us to write an all-orders definition of quark and gluon fractions in this small jet-mass region:\footnote{We thank Kyle Lee for discussions on this.}
\begin{equation}\label{eq:xqgDef}
    x_\kappa\big(p_T R, \eta, \mu)  \equiv \frac{\sum_{a,b,c} f_a \otimes f_b \otimes H_{ab}^c \otimes J_{c\kappa}}{\sigma_{\rm incl}(p_T , \eta)} \, , 
\end{equation}
with
\begin{equation}
    \sigma_{\rm incl}(p_T, \eta)  \equiv \frac{\df^2 \sigma}{\df p_T \df \eta} = \sum_{a,b,c,d} f_a \otimes f_b \otimes H_{ab}^c \otimes J_{cd} \, ,\nn 
\end{equation}
where the convolutions denoted by $\otimes$ are written out as in \eq{inclJ}. In the denominator we have summed over all the indices resulting in the inclusive cross section. Thus, the normalized jet-mass cross section can be written as
\begin{align} \label{eq:sigdecomp}
    \frac{1}{\sigma_{\rm incl}(p_T, \eta)} \frac{\df^3 \sigma}{\df p_T \df \eta\df \xi} 
   = x_q\, \tilde{ \cal G}_q \big(\xi, p_TR, \mu \big) 
   + x_g\, \tilde{ \cal G}_g \big(\xi, p_TR, \mu \big) \, ,
\end{align}
where we have defined~\cite{Cal:2020flh}
\begin{align}\label{eq:tcalGDef}
    \tilde{ \cal G}_\kappa \big(\xi, p_TR, \mu \big)  
    \equiv \frac{1}{\sigma_\kappa^{\rm incl}}\frac{\df \sigma_\kappa}{\df \xi} (p_T, \eta)= N_{\rm incl}^\kappa (p_TR, \mu) {\cal J}_\kappa (\xi, p_T, \eta, R, \mu) \, .
\end{align}
In the expression above we have suppressed the effects of non-global logarithms for now, and we discuss them in detail below.

\subsection{The fixed-order region}

Using \eqs{sumJij}{tcalGDef}, at ${\cal O}(\alpha_s)$ we can write \eq{GcFact} as
\begin{align}\label{eq:GkappaExplicit}
    {\cal G}_\kappa (z, \xi,    p_TR,\mu)  
  &= \delta(1-z) \big[ \delta (\xi)  + \tilde {\cal G}^{[1]}_\kappa (\xi,  \alpha_s(\mu)) \big]+  \delta (\xi)  \sum_{j} J_{ \kappa j} (z, p_T R, \mu)  \nn \\[-2mm]
    &= \delta(1-z) \big[ \delta (\xi)  + \tilde {\cal G}^{[1]}_\kappa (\xi,  \alpha_s(\mu)) \big]+  \delta (\xi) J_{\kappa} (z, p_T R, \mu)  
    \, .
\end{align}
In the second line various pieces $ J_{ \kappa j} $ have been summed to yield the semi-inclusive jet function $J_\kappa$ with  $\tilde {\cal G}^{[1]}_\kappa$ being the ${\cal O}(\alpha_s)$ term. 
Furthermore, our extension of $\tilde {\cal G}_\kappa$ to include the large $\xi$ fixed-order region is defined in such a way that
\begin{align}
    \int_0^{1} \df \xi \: \tilde{\cal G}_\kappa(\xi, p_T R ,\mu) =  1 \, .
\end{align}
 
We first state the results for the ungroomed cross section which will prove useful for the groomed case.
We evaluated the one-loop, fixed-order results for $\tilde {\cal G}_\kappa$ in \eq{tcalGDef} for ungroomed jet-mass distribution  for  quark and gluon jets clustered with $k_T$-type algorithm:
\begin{align}\label{eq:calG}
	&\tilde {\cal G}^{\rm FO}_{q,{\rm No\,sd}} (\xi ,  \alpha_s(\mu)) = \delta (\xi) + \frac{\alpha_s(\mu) C_F}{2\pi} \Bigg\{ \delta (\xi) \bigg(-3 + \frac{\pi^2}{3}\bigg) \\
	&\qquad  + \Theta(1-4\xi) \Bigg[ - 2 {\cal L}_1 (\xi)  +  {\cal L}_0 (\xi) \bigg(4 \log \Big(\frac{1 + \sqrt{1-4 \xi}}{2}\Big) - \frac{3}{2}\sqrt{1- 4\xi}
	\bigg)
	\Bigg]
	\Bigg\}
	\nn \, , \\
	&\tilde {\cal G}^{\rm FO}_{g,{\rm No\, sd}}(\xi, \alpha_s (\mu)) = \delta (\xi) + \frac{\alpha_s(\mu)}{2\pi} \Biggl\{ \delta (\xi) \bigg(C_A \Big(\frac{1}{4} + \frac{\pi^2}{3}\Big) - \frac{13}{12}\beta_0\bigg) + \Theta(1-4\xi) \Bigg[ - 2 C_A {\cal L}_1 (\xi)\nn \\
	&\qquad  +  {\cal L}_0(\xi) \bigg(4 C_A  \log \Big(\frac{1 + \sqrt{1-4 \xi}}{2}\Big) -\frac{\beta_0}{2}\sqrt{1-4\xi} + \frac{C_A - 2 n_f T_F}{3} 4\xi \sqrt{1-4\xi}\bigg) 
	\Bigg]
	\Biggr\} \, , \nn 
\end{align}
with $\mathcal{L}_ n (x ) \equiv [\Theta(x)\ln^n x /x]_+$. 
We derived these results using one-loop calculations for quark and gluon collinear functions that appears in the soft-drop double-differential jet-mass and groomed jet-radius factorization~\cite{Pathak:2020iue}, discussed below in \secn{mJPC}. See \app{Gnosd} for more details.
We see that at LO the end point of the jet-mass distribution is at $\xi = 0.25$. It is straightforward to check that these expressions are normalized to 1, such that the ${\cal O}(\alpha_s)$  piece upon integration over $\xi$ vanishes:
\begin{align}\label{eq:xiNorm}
    \int_0^1 \df \xi \: \tilde {\cal G}^{\rm FO}_{\kappa, {\rm No\, sd}} (\xi, \alpha_s(\mu)) = 1\, .
\end{align}

As we show in \app{FOGr}, the one-loop calculation for the groomed jet-mass spectrum can be expressed in terms of the results for the ungroomed spectrum stated above, plus an additional correction:
\begin{align}\label{eq:calGSD}
    \tilde {\cal G}^{\rm FO}_{\kappa,\rm sd} (\xi ,\tzcut ,\beta, R,  \alpha_s(\mu))  &=  \tilde {\cal G}^{\rm FO}_{\kappa,{\rm No\,sd}} \big(\xi   ,\alpha_s(\mu)\big) + \Delta  \tilde {\cal G}^{\rm FO}_{\kappa, {\rm sd}} (\xi ,\tzcut ,\beta, R,  \alpha_s(\mu)) \, .
\end{align}
The groomed cross section $\tilde {\cal G}^{\rm FO}_{\kappa,\rm sd} $ is similarly normalized, such that all the effects of grooming in the cumulative jet-mass cross section vanish at the end point. The evaluation of the remainder $\Delta {\cal G}^{\rm FO}_{\kappa, \rm sd}$ is detailed in \app{FOGr}.

%========================================
\subsection{The plain jet-mass region}
%========================================

We now turn to plain jet-mass region with $\xi \ll1$ where the large logs of $\ln(\xi)$ in $\tilde {\cal G}_\kappa$ necessitate factorization and resummation. We first review  the results for jet-mass measurement without grooming and include the soft-drop related pieces in the subsequent section.  
%========================================
\subsubsection{The ungroomed jet-mass spectrum}
%========================================
\label{sec:ungroomed}
We first set up some useful notation. 
For a jet of radius $R$ emitted in the direction $\vec n$ at rapidity $\eta_J$, we define the light-like vectors
\begin{align}\label{eq:LCDef}
    n^\mu \equiv \zeta^{-1}  (1, \vec n) \,, \qquad \bn^\mu \equiv \zeta (1, -\vec n) \,, 
\end{align}
where we have performed a large boost in the direction of the jet using the parameter $\zeta$ defined above in \eq{zetaDef}.
Doing so allows us to work with hemisphere-jets like coordinates and all the intermediate factors of jet radius appearing in the factorization formulae are easily accounted for.  In terms of these light-like vectors, we define the light-cone decomposition of a momentum $q^\mu$ as
\begin{align}\label{eq:LC}
    q^\mu =q^+  \frac{\bn^\mu}{2} +  q^-  \frac{n^\mu}{2} + q_\perp^\mu \, . 
\end{align}

The factorization of the ${\cal G}_\kappa$ function in \eq{inclJ} for ungroomed jet-mass measurement on inclusive jets involves double logarithms and NGLs due to soft emissions:
\begin{align}\label{eq:factPlainNGL}
  {\cal G}_{\kappa,\rm No\,sd}^{\rm plain} (z, \xi , Q,\mu ) &=\frac{\df}{\df\xi}\Bigg[ \sum_{\kappa'} \sum_j \widehat {\cal H}_{\kappa'\ra \kappa}^{j}(z, Q, \mu) \otimes_\Omega \widehat  \Sigma_\kappa^{j} (\xi,Q, \mu) \Bigg]  \, .
\end{align}
The hard collinear scale $Q$ was defined above in \eq{Q}.
This factorization formula is valid to all orders in $\alpha_s$ and at leading power in the power counting set by $\xi$, such that its relation to the full theory result is given by
\begin{align}
    {\cal G}_{\kappa,\rm No\,sd} (z, \xi , Q, \alpha_s(\mu))  = {\cal G}_{\kappa,\rm No\,sd}^{\rm plain} (z, \xi , Q,\mu )  \big(1 + {\cal O}(\xi) \big) \, .
\end{align}
In \eq{factPlainNGL} $\widehat \Sigma_\kappa^j$ is the cumulative jet-mass cross section that accounts for the contributions from the soft and collinear modes. 
The index $j$ runs over the number of hard partons outside of the jet that contribute to soft radiation everywhere, part of which is clustered with the jet.
Such configurations lead to non-global logarithms (NGL) and the $\otimes_\Omega$ denotes angular integrals between hard collinear and soft radiation. 
 The cumulative jet-mass cross section is given by convolution in the $+$ component between the soft and collinear contributions:
\begin{align}\label{eq:Sigmaj}
        \widehat  {\Sigma}^{j}_\kappa(\xi,Q, \mu) &=
    \int_0^\infty \df s  \int_0^\infty \df \ell_c^+\: J_\kappa \big( s , \mu \big)\:
\widehat \cS_{c_m}^{\kappa,j} \big(\ell_c^+,   \mu \big)  \: \delta \bigg(\xi - \frac{s}{Q^2} - \frac{\ell_c^+}{Q}\bigg)
\,.
\end{align}
Here $J_\kappa(s,\mu)$ is the standard inclusive jet function discussed below in \app{jetFunc} and $\widehat \cS_{c_m}^{\kappa,j}$ is the cumulative ungroomed soft function discussed below.  
 For simplicity, we will treat the NGLs in the convolution in \eq{factPlainNGL} at NLL accuracy (and at LL and large-$N_c$ in our numerical studies below) where they can be included  multiplicatively~\cite{Kang:2019prh}:
\begin{align}\label{eq:SimplifyNGLPlain}
    &\sum_j \widehat {\cal H}_{\kappa'\ra \kappa}^j (z, Q, \mu) \otimes_\Omega \widehat  \Sigma_\kappa^{j} (\xi,Q, \mu)\ra\langle \widehat {\cal H}_{\kappa'\ra \kappa} (z, Q, \mu)  \rangle  \,  {\cal S}_{\rm NGL}^\kappa \big(t [\xi Q,  Q]\big) \, \langle \widehat \Sigma_\kappa (\xi,Q, \mu)\rangle \,, 
\end{align}
where defining $\cS_{c_m}^{\kappa}\big  (    \ell_c^+,   \mu \big )=\bigl\langle \widehat \cS_{c_m}^{\kappa}\big  (    \ell_c^+,   \mu \big ) \bigr\rangle$ we have
\begin{align}
   \langle \widehat \Sigma_\kappa(\xi,Q, \mu) \rangle 
   =     \int_0^\infty \df s  \int_0^\infty \df \ell_c^+\: J_\kappa \big( s , \mu \big)\:  \cS_{c_m}^{\kappa}\big  (    \ell_c^+,   \mu \big ) \: \delta \bigg(\xi - \frac{s}{Q^2} - \frac{\ell_c^+}{Q}\bigg)
    \, .
\end{align}
Here $\langle \ldots\rangle$ denotes angular averaging after solid angle integration and 
\begin{align}
    t[\mu_0,\mu_1] \equiv \frac{1}{2\pi} \int_{\mu_0}^{\mu_1} \frac{d \mu'}{\mu'} \: \alpha_s (\mu') \,.
\end{align}
Similarly ${\cal H}_{\kappa'\ra \kappa} (z, Q, \mu) =\langle \widehat {\cal H}_{\kappa'\ra \kappa} (z, Q, \mu)  \rangle$, where ${\cal H}_{\kappa'\ra \kappa} (z, Q, \mu)$ is the same function appearing above in \eq{GcFact}. 
At the lowest order for jets clustered with anti-$k_T$ algorithm, the non-global piece ${\cal S}_{\rm NGL}^{\kappa}$ is given by
\begin{align}
    {\cal S}_{\rm NGL}^\kappa \big(t [\mu_0, \mu_1]\big)  = 1 - C_\kappa C_A \Big(\frac{\alpha_s}{2\pi}\Big)^2 \frac{\pi^2}{3} \log^2 \Big(\frac{\mu_1}{\mu_0}\Big) + \ldots \, .
\end{align}
The function $\cS^\kappa_{c_m}$ now no longer depends on the angles of hard partons, and
can be expressed as a cumulative integral of the differential soft function:
\begin{align}
    \cS^\kappa_{c_m} \big( \ell_c^+, \mu\big)=  \int_0^{\ell_c^+} d  \ell^+ S^\kappa_{c_m} \big(\ell^+,  \mu\big) 
  \, .
\end{align}
The one-loop result is given in \eq{Scm1L}. 

From \eqs{tcalGDef}{factPlainNGL} the differential cross section stripped of its DGLAP evolution at ${\cal O}(\alpha_s)$ can be written as
\begin{align}\label{eq:FactPlain}
    \tilde {\cal G}^{\rm plain}_{\kappa,{\rm No\,sd}} (\xi , Q, \mu )  =  N^\kappa_{\rm incl}( Q, \mu) {\cal J}^{\rm plain}_{\kappa,{\rm No\,sd}}(\xi, Q, \mu) \bigg(1 + {\cal O}(\xi)\bigg) \, ,
\end{align}
The ${\cal O}(\alpha_s)$ expressions of $N_{\rm incl}^\kappa$ are given below in \eq{Nincl1loop}.

From \eq{SimplifyNGLPlain} we have
\begin{align}\label{eq:FactPlainKappa}
    {\cal J}^{\rm plain}_{\kappa,{\rm No\,sd}}(\xi,Q, \mu)  =   \frac{\df}{\df\xi} \Big( {\cal S}_{\rm NGL}^\kappa \big(t_S \big) \,
    \Sigma_\kappa(\xi,Q,\mu) 
    \Big) \, , \qquad t_S \equiv t [Q \xi,  Q] \, .
\end{align}
With the RG evolution made explicit, \eq{FactPlainKappa} reads
\begin{align}\label{eq:FactPlainResum}
    {\cal J}^{\rm plain}_{\kappa,{\rm No\,sd}}(\xi, Q, \mu)  &=\Bigg( {\cal S}_{\rm NGL}^\kappa \big( t_S \big) {\cal J}^\kappa_{\rm plain} [\partial_\Omega ; \xi, Q, \mu] \frac{e^{\gamma_E\Omega}}{\Gamma(-\Omega)} \\
    &\quad + \Big(\frac{\df}{\df \ln \xi} {\cal S}_{\rm NGL}^\kappa \big( t_S \big) \Big)  {\cal J}^\kappa_{\rm plain} [\partial_\Omega ; \xi, Q, \mu] \frac{e^{\gamma_E\Omega}}{\Gamma(1-\Omega)}\Bigg)
    \Bigg|_{\Omega = \tilde \omega (\mu_s, \mu_J)} \, , \nn 
\end{align}
where the function of the derivative operator ${\cal J}^\kappa_{\rm plain} [\partial_\Omega ; \xi, Q, \mu]$ is defined in terms of Laplace transforms of the jet and the soft functions and the explicit expression is provided in \eq{PlainOperator}.

Next, we define the non-singular piece that is needed to recover the full plain jet-mass one-loop cross section in \eq{calG} including power suppressed terms. It is obtained by subtracting the plain contribution from
the full theory result,
\begin{align}\label{eq:nsNoSD}
    {\cal G}_{\kappa, \rm No\,sd}^{\rm n.s.} (z, \xi, Q, \alpha_s(\mu)) =   {\cal G}^{\rm FO}_{\kappa,\rm No\,sd} (z, \xi ,Q,  \alpha_s(\mu))  - {\cal G}_{\kappa,\rm No\,sd}^{\rm plain} (z, \xi , Q,\mu ) \, ,
\end{align}
which at ${\cal O}(\alpha_s)$ yields
\begin{align}
    \tilde {\cal G}_{\kappa,{\rm No\,sd}}^{\rm n.s.[1]} (\xi, \alpha_s(\mu))&=  \tilde {\cal G}^{\rm FO[1]}_{\kappa,{\rm No\,sd}} (\xi , \alpha_s(\mu)) \\ 
    &\quad 
    -  \Theta(1-4\xi)\bigg(\delta(\xi) N_{\rm incl}^{\kappa[1]} (Q ,\mu)+ Q^2 J_{\kappa}^{[1]} (\xi Q^2,\mu) + Q S_{c_m}^{\kappa[1]} (Q \xi, \mu)\bigg) \nn \, .
\end{align}
Hence we have
\begin{align}\label{eq:NoSDNS}
    \tilde {\cal G}_{q,{\rm No\,sd}}^{\rm n.s.[1]} (\xi, \alpha_s(\mu))&= \frac{\alpha_s C_F}{2\pi} \Theta(1-4\xi ) {\cal L}_0(\xi)
    \Bigg(4 \log\bigg(\frac{1+\sqrt{1-4\xi}}{2}\bigg) - \frac{3}{2}\big(\sqrt{1-4\xi} - 1\big)\Bigg) \, , \nn \\
    \tilde {\cal G}_{g,{\rm No\, sd}}^{\rm n.s.[1]} (\xi,\alpha_s(\mu)) &= \frac{\alpha_s }{2\pi} \Theta(1-4\xi)    {\cal L}_0(\xi) \bigg(4 C_A  \log \Big(\frac{1 + \sqrt{1-4 \xi}}{2}\Big) -\frac{\beta_0}{2}\big( \sqrt{1-4\xi} - 1\big)  \nn \\
    &\qquad \qquad + \frac{C_A - 2 n_f T_F}{3} 4\xi \sqrt{1-4\xi}\bigg) 
\, .
\end{align}
Thus the total resummed ungroomed jet-mass cross section including ${\cal O}(\as)$ power suppressed terms reads
\begin{align}\label{eq:FactPlainFull}
    \frac{1}{\sigma_\kappa}\frac{\df \sigma^{\rm No\,sd}_\kappa}{\df \xi} =   N^\kappa_{\rm incl}( Q, \mu) {\cal J}^{\rm plain}_{\kappa,{\rm No\,sd}}(\xi, Q, \mu)  + \tilde {\cal G}^{\rm n.s.[1]}_{\kappa, {\rm No\,sd} }(\xi , \alpha_s)  \, .
\end{align}
%

%========================================
\subsubsection{The groomed jet-mass spectrum}
%========================================
\label{sec:plain}
In the plain jet-mass region the modifications due to soft drop can be treated in fixed-order perturbation theory. Because up to ${\cal O}(\zcut)$ the jet grooming primarily only affects the  measurement on the soft radiation,  
this amounts to making in  \eq{SimplifyNGLPlain} the replacement
\begin{align}\label{eq:Ssd}
    \cS_{c_m}^{\kappa}\big( \ell_c^+,   \mu \big)
\ra 
\cS_{\rm sd}^{\kappa}\big( \ell_c^+,  Q \xi_0, \beta, \zeta ,\mu \big) \equiv 
\cS_{c_m}^{\kappa}\big( \ell_c^+,   \mu \big) +   \int_0^{\ell_c^+} d \ell^+ \: \Delta  S_{\rm sd}^\kappa  \big( 
\ell^+,  Q \xi_0' ,\beta  , \zeta, \alpha_s(\mu) \big) \, ,
\end{align}
where $\Delta  S_{\rm sd}^\kappa$ is the fixed-order soft-drop related correction. At ${\cal O}(\alpha_s)$ we have
\begin{align}\label{eq:Splain1}
    &\Delta S^{\kappa[1]}_{\rm sd}  \big(\ell^+  , Q \xi_0 ', \beta, \zeta , \alpha_s(\mu) \big)   \\
    &\qquad = 
    -\frac{\alpha_s (\mu)C_\kappa}{\pi} \Biggl[\frac{\Theta(\ell^+)}{\ell^+} \Theta\bigg(\Big(\frac{Q\xi_0'}{\ell^+}\Big)^\frac{2}{2+\beta}  - 1\bigg)\ln \bigg(\Big(\frac{Q\xi_0}{\ell^+}\Big)^\frac{2}{2+\beta} - \zeta^2\bigg) \Biggr]^{[Q\xi_0']}_+  \, , \nn
\end{align}
where $\zeta$ was defined above in \eq{zetaDef}. 
Here, the plus function is defined as
\begin{align}
	\big[\Theta(x)q(x) \big]_+^{[x_0]}
	\equiv \lim_{\eps\ra 0}
	\Big( \Theta(x-\eps) q(x) 
	- \delta (x- \eps) \int_{\eps}^{x_0} \df x'\: q(x')
	\Big)  \, .
\end{align}
Note that because the soft modes defining this function are at wide angles, the end-point of the plus function appears at the soft wide-angle transition point $\xi_0'$ given in \eq{xi0p}.
We derive this one-loop expression in \app{FOGr}. 

This replacement allows us to write down the factorized formula for soft-drop jet-mass cross section in the plain jet-mass region:
\begin{align}\label{eq:factPlainSD}
    {\cal G}_{\kappa,\rm sd}^{\rm plain} (z, \xi , Q, \mu) &=\frac{\df}{\df\xi}\Bigg[ \sum_{\kappa'}   {\cal H}_{\kappa'\ra \kappa} (z, Q, \mu) \, \Sigma_{\kappa,\rm sd} (\xi,Q, \mu) \Bigg]  \, ,
\end{align}
where $\Sigma_{\kappa,\rm sd}$ is defined analogously to \eq{Sigmaj} with additional soft-drop related fixed-order pieces in the soft function in \eq{Ssd}. Its relation to the full theory cross section is given by
\begin{align}
        {\cal G}_{\kappa,\rm sd}  \big(z, \xi , Q,\alpha_s (\mu)\big) &=     {\cal G}_{\kappa,\rm sd}^{\rm plain} (z, \xi , Q, \mu)\Big(1 + {\cal O}(\xi, \xi_0) \Big)
\end{align}
Thus, analogous to \eq{FactPlain} the resummed cross section (to NNLL accuracy) is given by
\begin{align}\label{eq:xsecResummedPlain0}
     \tilde {\cal G}_{\kappa,\rm sd}^{\rm plain} \big(\xi, \zcut, \beta, R, \mu\big) = N_{\rm incl}^{\kappa} (Q , \mu) {\cal J}_{\kappa,\rm sd}^{\rm plain}(\xi,Q,\mu) \,,
\end{align}
where
\begin{align}\label{eq:xsecResummedPlain}
   &{\cal J}_{\kappa,\rm sd}^{\rm plain}(\xi,Q,\mu)
    =    {\cal S}_{\rm NGL}^\kappa \big( t_S \big) {\cal J}^\kappa_{\rm plain} [\partial_\Omega ; \xi, Q, \mu] \bigg(
    \frac{e^{\gamma_E\Omega}}{\Gamma(- \Omega)}
    + 
    {\cal Q}_{\rm sd}^{\kappa} \bigg[\Omega,  \frac{\xi}{\xi_0}, \beta,R,  \mu \bigg] \nn
    \bigg)   \Bigg|_{\Omega = \tilde \omega(\mu_{s}, \mu_J)} \\
    & \quad+ \Big(\frac{\df}{\df \ln \xi} {\cal S}_{\rm NGL}^\kappa \big( t_S \big) \Big)  {\cal J}^\kappa_{\rm plain} [\partial_\Omega ; \xi, Q, \mu] \bigg(\frac{e^{\gamma_E\Omega}}{\Gamma(1-\Omega)} + {\cal Q}_{\rm sd}^{\kappa} \bigg[\Omega-1,  \frac{\xi}{\xi_0}, \beta,R,  \mu \bigg] \bigg) \Bigg|_{\Omega = \tilde \omega(\mu_{s}, \mu_J)} 
    \,.
\end{align}
The kernel ${\cal Q}_{\rm sd}^\kappa$ accounts for the effects of soft drop, including resummation between the jet and soft scales. The explicit expression is given in \eq{Qsd}.
 
The additional ${\cal O}(\xi_0)$ power corrections dropped in \eq{xsecResummedPlain0} include finite $\zcut$ terms and are captured by the following non-singular term:
\begin{align}\label{eq:GSDNS}
    \Delta \tilde {\cal G}^{\rm n.s.[1]}_{\kappa, \rm sd} = \Delta \tilde {\cal G}_{\kappa, \rm sd}^{\rm FO[1]} (\xi, \tzcut, \beta, R, \alpha_s(\mu))- Q \Delta S_{\rm sd}^{[1]}\big(Q \xi, Q \xi_0', \beta, \zeta, \alpha_s(\mu) \big)\, .
\end{align}
where $ \Delta \tilde {\cal G}_{\kappa, \rm sd}^{\rm FO} $ appeared in \eq{calGSD}. At one-loop the factor of $Q$ cancels against a $1/(Q\xi)$ from $\Delta S_{\rm sd}^{[1]}$ such that $\Delta \tilde {\cal G}^{\rm n.s.[1]}_{\kappa, \rm sd}$ is independent of $Q$.
The $\Delta S_{\rm sd}^{[1]}$ accounts for all the soft singularities which allows us to evaluate this non-singular piece via simple numerical integration. 
Including the non-singular corrections in \eqs{NoSDNS}{xsecResummedPlain0} accounts for all perturbative power corrections at ${\cal O}(\alpha_s)$ in the plain jet-mass region:
\begin{align}\label{eq:SDNS}
     \tilde {\cal G}_{\kappa,\rm sd} = N_{\rm incl}^{\kappa} (Q , \mu) {\cal J}_{\kappa,\rm sd}^{\rm plain}(\xi,\mu) + 
     {\cal G}_{\kappa, \rm No\,sd}^{\rm n.s.[1]} \big(\xi, \alpha_s(\mu)\big) + 
        \Delta \tilde {\cal G}^{\rm n.s.[1]}_{\kappa, \rm sd}(\xi, \tzcut, \beta, R, \alpha_s(\mu)) \,,
\end{align}
where we have dropped some of the function arguments for simplicity.
In \app{FOGr} we will show that $\Delta \tilde {\cal G}^{\rm n.s.[1]}_{\kappa, \rm sd}$ results in a numerically small contribution which can be ignored in comparison to perturbative uncertainty from scale variation at NNLL. Hence, we will drop $\Delta \tilde {\cal G}^{\rm n.s.[1]}_{\kappa, \rm sd}$ in our $\as$-sensitivity analysis.

%========================================
\subsection{The soft-drop resummation region}
%========================================
 
In the soft-drop resummation region for $\xi \ll \xi_0$, the (differential) soft function $S_{\rm sd}^\kappa$ in \eq{Ssd} can be factorized into pieces that correspond to soft-drop passing and failing soft radiation, such that
\begin{align}\label{eq:SplainFact}
    & S^\kappa_{c_m} \big(\ell^+,  \mu\big) + \Delta S^\kappa_{\rm sd} \big( 
    \ell^+,  Q \xi_0'  ,\beta  , \zeta, \mu \big)  \\
    &\qquad =  S_G^\kappa  \big(\qcut, \beta, \zeta , \mu\big) \qcut^{\frac{1}{1+\beta}}S_c^\kappa  \big(\ell^+ \qcut^{\frac{1}{1+\beta}} , \beta , \mu\big) 
    \Bigg[1 + {\cal O}   \bigg( \Big(\frac{\ell^+}{\qcut}\Big)^{\frac{2}{2+\beta}}\bigg) \Bigg] \, , \nn 
\end{align}
where the term in the square brackets indicates the size of the power corrections. 
The $S_G^\kappa$ and $S_c^\kappa$ are the global soft and collinear soft functions~\cite{Frye:2016aiz} that respectively describe physics of soft modes that fail and pass the soft-drop test. 
$\qcut$, defined in \eqs{qcut}{qcut2}, defines the scale of the global soft function. 
The one-loop expressions are provided in \eqs{SG1loop}{sc1loop}. 
Here we have made use of a dimension $\big(\frac{2+\beta}{1+\beta}\big)$ variable in the argument of $S_c^\kappa$ following~\cite{Hoang:2019ceu}. 
We show in \app{csoft} that the collinear-soft function is independent of $\qcut$, as $\qcut$ is a high scale from the perspective of collinear-soft emissions~\cite{Frye:2016aiz}.
It is also independent of the jet radius $R$, since collinear-soft modes are at parametrically smaller angles than the jet radius.
Thus, from \eq{SplainFact} we have the following factorization formula in this region:
\begin{align}\label{eq:SDFact}
    {\cal G}^{\rm resum}_{\kappa, \rm sd} &\big(z, \xi, Q ,\zcut,\beta ,\mu \big) = \sum_{\kappa'} {\cal H}_{\kappa' \ra \kappa} (z, Q, \mu) S_{G}^\kappa  \big(\qcut, \beta, \zeta,\mu\big) {\cal S}_{\rm NGL}^\kappa \big( t_{gs} \big) \\
    &\qquad \times \int \df \tilde k \int \df s \: J_\kappa \big(s, \mu\big) S_c^\kappa \big(\tilde k, \beta, \mu \big)
    \delta \bigg(\xi - \frac{s}{Q^2} -\frac{\tilde k(\qcut)^{\frac{-1}{1+\beta}}}{Q}  \bigg) 
    \nn \, .
\end{align}
The relation between this factorized cross section and the full theory result is given by
\begin{align}\label{eq:SDFactPC}
    {\cal G}_{\kappa, \rm sd} \big(z, \xi, Q ,\zcut,\beta ,\alpha_s(\mu) \big) = {\cal G}^{\rm resum}_{\kappa, \rm sd} \big(z, \xi, Q ,\zcut,\beta ,\mu \big) \Bigg[
    1  +  {\cal O}   \bigg( \xi, \xi_0, \Big(\frac{\xi}{ \xi_0}\Big)^{\frac{2}{2+\beta}}\bigg) \Bigg] \, , 
\end{align}
where we have displayed all the power corrections that are small in the soft-drop resummation region. We discuss these corrections in detail in the next section. 
Next, we note that since the wide-angle modes are groomed away in this region, the NGLs give a contribution that is independent of the jet mass. The argument of the NGL piece is given by
\begin{align}
    t_{gs} \equiv t \big[\qcut , Q\big] \, .
\end{align}
Comparing with \eq{FactPlainKappa} we see that the evolution of the NGLs is frozen to the fixed scale $\qcut$.

Finally, at one-loop we can remove the DGLAP piece and write
\begin{align}\label{eq:SDResum}
    \tilde {\cal G}^{\rm resum}_{\kappa, \rm sd} (\xi, Q, \qcut, \beta, R, \mu)= N_{\rm incl}^{\kappa} (Q, \mu) 
    {\cal J}_\kappa^{\rm sd\, resum}(\xi, Q, \qcut, \beta, R, \mu)
    \, , \qquad (\xi < \xi_0) \, ,
\end{align}
where 
\begin{align}\label{eq:GsdFactResum}
    {\cal J}_\kappa^{\rm sd\, resum}  &=    
      S_G^\kappa  \big(\qcut, \beta, \zeta , \mu\big) 
     {\cal S}_{\rm NGL}^\kappa \big( t_{gs} \big) {\cal J}^\kappa_{\rm sd} [\partial_\Omega ; \xi, Q, \mu]
     \frac{e^{\gamma_E\Omega}}{\Gamma(- \Omega)}\Bigg|_{\Omega = \tilde \omega(\mu_{cs}, \mu_J)} 
     \, .
\end{align}
The function ${\cal J}^\kappa_{\rm sd} [\partial_\Omega ; \xi, Q, \mu]$ now involves Laplace transforms of the jet and the collinear-soft functions, and is given in \eq{SDOperator}. Because the NGLs are independent of $\xi$,  the  terms in the second line in \eq{xsecResummedPlain} are not present in the soft-drop resummation region. 

%========================================
\subsection{Soft drop on exclusive jets}
%========================================
 \label{sec:excl}
 
Another way of isolating jets is via an exclusive measurement which involves vetoing additional jets. For example, for $pp \ra $ dijets with a veto on transverse momentum of additional jets $p_{T_i} < p_T^{\rm veto}$, one can derive a factorization formula with a $p_T$ jet veto~\cite{Stewart:2010tn,Berger:2010xi,Banfi:2012yh,Becher:2012qa,Dasgupta:2012hg,Banfi:2012jm,Chien:2012ur,Jouttenus:2013hs,Becher:2013xia,Stewart:2013faa,Stewart:2014nna,Stewart:2015waa,Banfi:2015pju} for a soft drop groomed jet-mass measurement
\begin{align}\label{eq:sigExcl}
	\frac{\df \sigma(p_{T}^{\rm veto})}{\df \xi_1 \df \xi_2 \df \Phi_{\rm born}} 
	&= \sum_{\{\kappa_1,\kappa_2,\kappa_a,\kappa_b\}} {\cal B}_n^{\kappa_a}\big(x_a, Q, p_T^{\rm veto}, R^2 ,\mu ,\nu \big)
	{\cal B}_\bn^{\kappa_b}\big(x_b, Q, p_T^{\rm veto}, R^2 ,\mu ,\nu \big) \\
	&\qquad \times \tr \Bigg[\widehat H_{ab}^{\kappa_1\kappa_2} \big(Q_{i}, \Phi_{\rm born}, \mu \big) 
	\otimes_{\Omega}
	\widehat S_{G,ab}^{\kappa_1 \kappa_2} \big( p_T^{\rm veto}, R  ,Q_{1\rm cut}, Q_{2\rm cut} , \beta , \mu ,\nu  \big)	\Bigg]	\nn 
	\\
	&\qquad  \times 
	\int \df s_1\,  \df \tilde k_1  \:  J_{\kappa_1} \big(s_1, \mu \big)\, 
	S_c^{\kappa_1}(\tilde k_1, \beta, \mu )\delta \bigg(\xi_1 - \frac{s_1+Q_1 Q_{1\rm cut}^{-\frac{1}{1+\beta}} \tilde k_1}{Q_1^2}  \bigg)
	\nn \\
	&\qquad\times 
	\int \df s_2\, \df \tilde k_2 \:  J_{\kappa_2} \big(s_2, \mu \big)
	S_c^{\kappa_2} \big(\tilde k_2, \beta, \mu \big)
	\delta \bigg(\xi_2-  \frac{s_2+Q_2 Q_{2\rm cut}^{-\frac{1}{1+\beta}} \tilde k_2}{Q_2^2}   \bigg)
	\nn
	\, .
\end{align}
Here $Q_1$, $Q_2$ and $Q$ are appropriate large momenta associated with the two jets and the beam functions respectively, and $Q_{i\,\rm cut}$ are the global soft scales:
\begin{align}
	Q_i = p_{T_i} \cosh \eta_{J_i} \,, \qquad Q_{i\rm cut} = (2\cosh \eta_{J_i})^{1 + \beta} p_{T_i} \zcut \, .
\end{align}
The momentum fractions $x_{a,b}$ are constrained by the born kinematics.
The born kinematics phase space $\Phi_{\rm born}$ is captured in the hard function $\widehat H_{ab}^{\kappa_1\kappa_2}$. 
The sum runs over various partonic channels $\kappa_a \kappa_b \ra \kappa_1 \kappa_2$ which lead to the production of two jets.
Color is traced over, and there are two color channels for $\kappa_a$-$\kappa_b=$ quark-quark, three for quark-gluon, and nine for gluon-gluon. The parton distribution function dependence appears through the beam functions ${\cal B}_n^{\kappa_a}$ and ${\cal B}_{\bar n}^{\kappa_b}$, via a further factorization between perturbative (${\cal I}$) and nonperturbative ($f$) contributions in the direction of the beams:
\begin{align}
  {\cal B}^\kappa\big(x, Q, p_T^{\rm veto}, R^2 ,\mu ,\nu \big)
   &= \sum_{\kappa'} \int_x^1 \frac{\df\xi}{\xi} \: 
   {\cal I}_{\kappa\kappa'}(x/\xi,Q, p_T^{\rm veto}, R^2 ,\mu ,\nu)
   f_{\kappa'}(\xi,\mu) \,.
\end{align}
The global soft function $\widehat S_{G,a b}^{\kappa_1 \kappa_2}$ describes the dynamics of soft particles emitted off the initial state partons as well as those  that are clustered in the  final state jets but groomed away. 
As a consequence of jet grooming the jet-mass spectra of the two jets are decoupled. 

It is useful to separate \eq{sigExcl} into quark-gluon fractions that are independent of the jet-mass measurement, times a function that depends on the jet mass. To make this meaningful we must carry out the separation in a $\mu$-independent manner. To this end we define
\begin{align}\label{eq:xExcl}
	x^{\rm excl}_{\kappa_1   \kappa_2} (p_T^{\rm veto}, \Phi_{\rm born})
	&=  \sum_{\{\kappa_a,\kappa_b\}} {\cal B}_n^{\kappa_a}\big(x_a, Q, p_T^{\rm veto}, R^2 ,\mu ,\nu \big) 
	{\cal B}_\bn^{\kappa_b}\big(x_b, Q, p_T^{\rm veto}, R^2 ,\mu ,\nu \big) 
 \\
	&\quad \times
	\frac{ \tr \Big[\widehat H_{ab}^{\kappa_1 \kappa_2} \big(Q_{i}, \Phi_{\rm born}, \mu \big) 
	\otimes_{\Omega}
	\widehat S_{G,a b}^{\kappa_1 \kappa_2} \big( p_T^{\rm veto}, R  ,Q_{1\rm cut}, Q_{2\rm cut} , \beta , \mu ,\nu  \big)	\Big]}
    { N_{\rm incl}^{\kappa_1}(Q_1, \mu) S_G^{\kappa_1}(Q_{1\rm cut}, \beta ,R, \mu )  \ N_{\rm incl}^{\kappa_2}(Q_2, \mu) S_G^{\kappa_2}(Q_{2\rm cut}, \beta ,R, \mu ) }
	\nn 
	\,,
\end{align}
such that the exclusive jet cross section in the soft drop resummation region is given by
\begin{align}
 &\frac{\df \sigma(p_{T}^{\rm veto})}{\df \xi_1 \df \xi_2 \df \Phi_{\rm born}} 
  \\
  &\quad = \sum_{\{\kappa_1 , \kappa_2\}} x^{\rm excl}_{\kappa_1   \kappa_2} (p_T^{\rm veto}, \Phi_{\rm born}) \:
  \tilde {\cal G}^{\rm resum}_{\kappa_1, \rm sd} \big(\xi_1, Q_1, Q_{1\rm cut} ,\beta, R, \mu\big) \,
  \tilde {\cal G}^{\rm resum}_{\kappa_2, \rm sd} \big(\xi_2, Q_2, Q_{2\rm cut} ,\beta, R, \mu\big) \nn \,.
\end{align}
This equation is analogous to \eq{sigdecomp} for the inclusive case, except that here we have fractions $x_{\kappa_1\kappa_2}$ that depend on the quark-gluon identities $\kappa_i$ of each of the dijets. The jet-mass shape dependence is described by the same partonic cross section $\tilde {\cal G}_{\kappa, \rm sd}$ for flavor $\kappa$ defined in \eq{tcalGDef},
whose resummed version $\tilde {\cal G}_{\kappa, \rm sd}^{\rm resum}$ was defined above in \eq{SDResum}. 
The definition of $x_{\kappa_1\kappa_2}^{\rm excl}$ in \eq{xExcl} is RG invariant since the $\mu$-dependence arising from the $J_{\kappa_i}$ and $S_c^{\kappa_i}$ functions associated with the final state jets in \eq{sigExcl}, is equal but opposite to that of the factors in the denominator of \eq{xExcl}.

In \secn{qgfraction} we made use of the exclusive dijet factorization to determine the impact that variations in the PDFs cause to the quark-gluon fractions. Since there the fraction for a single quark jet was examined, we 
can make the connection with the fractions defined here using
\begin{align}
    x_q^{\rm excl} &= \frac{2 x_{qq}^{\rm excl} +x^{\rm excl}_{qg}}{2 ( x_{qq}^{\rm excl} + x_{gg}^{\rm excl} + x_{qg}^{\rm excl})} 
  \,,
  & x_g^{\rm excl} &= \frac{x^{\rm excl}_{qg}+2 x^{\rm excl}_{gg}}{2 ( x_{qq}^{\rm excl} + x_{gg}^{\rm excl} + x_{qg}^{\rm excl})} 
  \,.
\end{align}
Here a subscript $q$ indicates a sum over all quark and antiquark flavor channels. 
Note that by construction $x_q^{\rm excl}+x_g^{\rm excl}=1$.

%========================================
\subsection{Power corrections in the soft-drop resummation region}
%========================================
\label{sec:pc}
We notice from \eq{SDFactPC} that there are three kinds of power corrections appearing on top of the soft-drop resummed result. The ${\cal O}(\xi)$ is related to jet-mass resummation, the  ${\cal O}(\xi_0)$ is related to finite $\zcut$ effects that are ignored in the soft approximation, and lastly the ${\cal O}((\xi/\xi_0)^{\frac{2}{2+\beta}})$ results from collinear-soft approximation of the soft-drop passing subjet. We investigate the ${\cal O}(\xi_0)$ power correction  in \eq{GSDNS} in resummation region at ${\cal O}(\alpha_s)$. Thus, we can refine \eq{SDFactPC} at ${\cal O}(\alpha_s)$ as 
\begin{align}\label{eq:SDFactPC2}
    \tilde {\cal G}_{\kappa, \rm sd} \big( \xi, Q ,\zcut,\beta ,\alpha_s(\mu) \big) &=  \tilde {\cal G}^{\rm resum}_{\kappa, \rm sd} \big( \xi, Q ,\zcut,\beta ,\mu \big) \Bigg[
    1  +  {\cal O}   \bigg( \xi, \Big(\frac{\xi}{ \xi_0}\Big)^{\frac{2}{2+\beta}}\bigg) \Bigg] \nn \\
    &\quad +  \Delta \tilde {\cal G}^{\rm n.s.[1]}_{\kappa, \rm sd}(\xi, \zcut, \beta, R, \alpha_s(\mu))    \, .
\end{align}
As remarked above, a fixed-order analysis of $\Delta \tilde {\cal G}^{\rm n.s.[1]}_{\kappa, \rm sd}$ in \app{FOGr} shows that this is a small correction. Thus, we will suppress the corrections from $\Delta \tilde {\cal G}^{\rm n.s.[1]}_{\kappa, \rm sd}$ in the analysis below.

The remaining two power corrections, however, require a careful treatment. Unlike \eqs{NoSDNS}{xsecResummedPlain0} where the power corrections were of purely fixed-order origin, we cannot simply set all the scales in \eq{SDFactPC} to be equal as these power corrections are related to turning off one, \textit{but not every}, resummation. We now describe our strategy to include these power corrections using profile functions. 

%-----------------
\subsubsection{Soft-drop power corrections}
%-----------------

Let us first consider the ${\cal O}\big((\xi/\xi_0)^{\frac{2}{2+\beta}}\big)$  power correction. \eq{SplainFact} prescribes calculating this power correction by setting scales of the soft pieces in the difference of two factorization formulae in \eqs{xsecResummedPlain0}{SDResum}. On the other hand, the jet scale must be left where it should be in the two cases\footnote{The canonical jet scale is the same in the two cases but can differ in the precise implementation through profile functions as we show below.}, as we would still like to retain the jet-mass resummation in the two cases. This can be straightforwardly implemented via the following set of $\xi$-dependent scales
\begin{align}\label{eq:plainSDProf}
    &\text{Plain jet-mass profiles}: & 
    &\mu_{\rm plain}&&  \equiv \{\mu_J(\xi) , \mu_s(\xi) , \mu_N \} \, ,&
    \\
    &\text{soft-drop jet-mass profiles}: &
    &\mu_{\rm sd}&   &\equiv \{\mu_J(\xi) , \mu_{gs}, \mu_{cs}(\xi) , \mu_N \}  \, ,& \nn 
    \\
    &\text{soft-drop to plain jet-mass transition}:&
    &\mu_{\rm sd \ra plain} &&\equiv \{\mu_J(\xi), \mu_{gs} , \mu_{cs \ra s} (\xi), \mu_N\}
    \, . &\nn
\end{align}
They are chosen to individually minimize logarithms in the momentum space expressions of the factorization functions. We describe their implementation in detail in \secn{prof}, but for now it suffices to note that in the $\mu_{\rm plain}$ profile  the $\mu_{cs}$ and $\mu_{gs}$ scales merge into a single soft scale $\mu_{s}$~\cite{Chien:2019osu}, turning off any soft-drop related resummation. This, however, continues to retain resummation between the jet and the soft scales. 
We have also defined a profile $\mu_{\rm sd\ra plain}$ that merges the collinear and global soft scales for $\xi \geq \xi_0$.  

Using these three profiles, the soft-drop power corrections can be captured by considering the combination:
\begin{align}\label{eq:Gmatched0}
    \tilde {\cal G}^{\rm matched}_{\kappa, \rm sd} (\xi) \equiv  \tilde {\cal G}^{\rm resum}_{\kappa,\rm sd} \big(\xi,  \mu_{\rm sd \ra plain} \big) + \tilde  {\cal G}^{\rm plain}_{\kappa, \rm sd} \big(\xi , \mu_{\rm plain} \big) -  \tilde {\cal G}_{\rm sd}^{\rm resum} \big(\xi,  \mu_{\rm plain} \big) + \ldots \, ,
\end{align}
where the arguments $\mu_{\rm plain}$ etc. in these expressions are understood as inserting each scale from the set in \eq{plainSDProf} into the respective factorization function. For simplicity we have suppressed other arguments. The behavior of $\mu_{\rm sd \ra plain}$ ensures that in the plain jet-mass region the first and the third term cancel, leaving the fixed-order soft-drop cross section. Note that the choice $\mu_{cs \ra s}(\xi)$ in $\mu_{\rm sd \ra plain}$ is arbitrary as long as the same profiles are used in the first and the third term in the $\xi > \xi_0$ region. This matching also ensures  that the NGL pieces cancel with each other in precisely the same fashion.
Additionally, in the implementation below, the $\mu_{\rm plain}$ profile is designed to saturate to $\mu_{\rm N}$ for $ \xi > \xi_0$, but ahead of the LO jet-mass end point $\xi = 1/4$, such that close to the endpoint all the resummations are turned off. 
Finally, the terms `$\ldots$' in \eq{Gmatched0} correspond to jet-mass dependent power corrections which we now address.

%-----------------
\subsubsection{Jet-mass power corrections}
%-----------------
\label{sec:mJPC}
The jet-mass power corrections need careful treatment. We cannot simply extrapolate the formula for the non-singular term in \eq{nsNoSD} into the groomed region as it lacks resummation related to any soft-drop factorization that we would like to retain. \eq{nsNoSD} is actually connected to a soft-collinear factorization where the soft modes live at the boundary of the jet. This is perfectly fine for large jet masses beyond the soft-drop cusp and \eq{nsNoSD} remains legitimate in the ungroomed region. However, in the region soft drop is active, the soft modes cannot lie at angles close to the jet radius, but must be confined within the maximum kinematically allowed groomed jet radius. We define
\begin{align}\label{eq:rgDef}
	r_g \equiv \frac{R_g}{R} \,, \qquad (\text{$pp$ case}) \,, \qquad 
	r_g \equiv \frac{\tan \frac{R_g}{2}}{\tan\frac{R}{2}} \,, \qquad (\text{$e^+e^-$ case}) \, ,
\end{align}
such that the groomed jet radius is constrained by the jet mass as~\cite{Pathak:2020iue}
\begin{align}\label{eq:psiMaxDef}
   r_g  <r_g^{\rm max} (\xi) &=  \min \Bigg\{\bigg(\frac{m_J^2}{Q \qcut}\bigg)^{\frac{1}{2+\beta}}, 1\Bigg \}    \\
   &=   \min \bigg\{\Big(\frac{\xi}{\xi_0}\Big)^{\frac{1}{2+\beta}}, 1\bigg\} \nn 
\, .
\end{align}
Because of this restriction, there is a non-trivial resummation between the global soft and collinear soft scales. The second argument of min is valid when $\xi > \xi_0$, the ungroomed region.

To properly evaluate these mass-dependent power corrections, we instead need to consider the cross section, differential in groomed jet mass, and simultaneously cumulative in the groomed jet radius, as worked out in \Ref{Pathak:2020iue}. The jet-mass measurement constrains the range of groomed jet measurement, with the maximum allowed value given by \eq{psiMaxDef}. When $r_g \lesssim r_g^{\rm max}$, one has a distinct collinear-soft subjet that stops the soft drop. \eq{SDFact} is in fact a special case when the (cumulative) $r_g = r_g^{\rm max}$, such that all groomed jet radius values are allowed. 
On the other hand, for a given $\xi$ measurement, the groomed jet radius can be as small as 
\begin{align}
	r_g\gtrsim r_g^{\rm min}(\xi) = \sqrt{\xi} \, .
\end{align}
This corresponds  to the situation where there is a single hard collinear mode that fills up the entire jet, and the soft drop is stopped by a haze of soft radiation at the angles $R_g$ to the jet. 
In this  region, the jet-mass measurement is not factorized, but groomed jet radius is, with the corresponding factorization formula given by
\begin{align}\label{eq:FactSmallRg}
	{\cal G}^{\rm min}_{\kappa, \rm sd}\big(z, \xi, r_g ,\zcut,\beta ,\mu \big)&= \sum_{\kappa'} {\cal H}_{\kappa' \ra \kappa} (z, Q, \mu) S_{G}^\kappa  \big(\qcut, \beta, \zeta,\mu\big) {\cal S}_{\rm NGL}^\kappa \big( t_{gs} \big) \\
    &\quad \times S_{c_g}^{\kappa} \big(\qcut r_g^{1+\beta}, \beta, \mu\big) {\cal S}_{\rm NGL}^\kappa \big( t_{c} \big)\frac{1}{r_g^2}  {\cal C}^\kappa \bigg(\frac{\xi}{r_g^2}, Qr_g, \mu \bigg) 
    \nn \, .
\end{align}
The soft function $ S_{c_g}^{\kappa}$ describes dynamics of soft modes at the groomed jet boundary. The soft-drop resummation is included through an RG evolution between the global soft and $S_{c_g}$ scales. 
Here ${\cal C}^\kappa$ is the collinear function which, as can be seen from \eqs{Cq}{Cg}, is precisely the fixed-order result for ungroomed cross section in \eq{calG} with a jet radius $r_g R$, differing only in the $\delta (\xi)$ pieces at one-loop to satisfy RG consistency (which are moved into the DGLAP evolution piece). 
Next, in addition to the non-global logarithms present at the jet boundary, probing the groomed jet radius results in additional NGL at the $R_g$ boundary~\cite{Kang:2019prh}, 
with the argument of this NGL piece being
\begin{align}\label{eq:tc}
    t_c \equiv t \big[\qcut r_g^{1+\beta}, Q r_g \big] \, .
\end{align}
The two arguments are canonical scales of the $S_{c_g}^\kappa$ and ${\cal C}^\kappa$ functions. Since we are only interested in capturing the ${\cal O}(\as)$ jet-mass related power suppressed terms we will turn off this NGL piece in the numerical analysis. 

For our purpose, it is important to note that the jet-mass dependence in collinear function factorizes as in the case of an ungroomed jet with the jet radius $R\ra r_gR$:
\begin{align}\label{eq:FactCalC}
    \frac{1}{r_g^2}{\cal C}^\kappa \Big(\frac{\xi}{r_g^2} \Big) &= \int \df s \int \df \ell^+ \: J_\kappa \big(s, \mu\big) \, S^\kappa_{c_m}(\ell^+, \mu) \, \delta \bigg(\frac{\xi}{r_g^2} - \frac{s}{Q^2r_g^2} -\frac{\ell^+}{Qr_g} \bigg)\bigg(1 + {\cal O}\Big(\frac{\xi}{r_g^2}\Big)\bigg) \nn \\ 
    &= \Sigma_\kappa \bigg(\frac{\xi}{r_g^2}, Q r_g,\mu \bigg) \bigg(1 + {\cal O}\Big(\frac{\xi}{r_g^2}\Big)\bigg)\,,
\end{align}
where $\Sigma_\kappa$ was defined above in \eq{Sigmaj}.
However, the collinear function is not RG invariant alone, but only in combination with the other pieces in \eq{FactSmallRg}. The cross section resulting from \eq{FactCalC} defines the ``intermediate-$R_g$'' factorization:
\begin{align}\label{eq:FactIntRg}
    {\cal G}^{\rm int}_{\kappa,\rm sd} (z, \xi , r_g ,\zcut,\beta ,\mu)  &= \frac{\df }{\df \xi} \Bigg[
  \sum_{\kappa'} {\cal H}_{\kappa' \ra \kappa} (z, Q, \mu) S_{G}^\kappa  \big(\qcut, \beta, \zeta,\mu\big)
    {\cal S}_{\rm NGL}^\kappa(t_{gs}) \nn \\
    & \quad\times  S_{c_g}^{\kappa} \big(\qcut r_g^{1+\beta}, \beta, \mu\big) {\cal S}_{\rm NGL}^\kappa(t_{cs}) \Sigma_{\kappa}\bigg(\frac{\xi}{r_g^2}, Qr_g, \mu\bigg)   \Bigg] \, ,
\end{align}
where, as before, we have NGLs associated with the groomed jet boundary, and this time the arguments are 
\begin{align}\label{eq:tcs}
    t_{cs} \equiv t\bigg[\qcut r_g^{1+\beta}, \frac{Q \xi}{r_g} \bigg]\, .
\end{align}

Since we are not constraining the groomed jet radius, we set it to its maximum allowed value in \eq{psiMaxDef} in these expressions. Thus, from \eq{FactCalC} we see that all the jet-mass dependent power corrections in the soft-drop resummation region are given by the difference of two factorized cross sections in \eqs{FactSmallRg}{FactIntRg} at $r_g = r_g^{\rm max}(\xi)$ by setting the scale of jet and collinear soft functions to be the same. The scale of the global soft function must be left to its canonical value to retain soft-drop resummation. To this end, we define the profile:
\begin{align}\label{eq:profMin}
    &\text{Min-$R_g$ factorization profile:}& 
    &\mu_{\rm min} \equiv \{ \mu_{\cal C}(\xi), \mu_{cs}(\xi), \mu_{gs} , \mu_{N} \}\,,&
\end{align}
such that the jet-mass dependent power corrections in the soft-drop resummation region are given by
\begin{align}
    &\text{${\cal O}(\xi)$ p.c. for $\xi < \xi_0$}:&
     &\tilde {\cal G}^{\rm min}_{\kappa, \rm sd} \big(\xi ,r_g^{\rm max}(\xi) ,  \mu_{\rm min} \big) 
     - \tilde {\cal G}_{\rm sd}^{\rm int} \big(\xi,  r_g^{\rm max}(\xi), \mu_{\rm min} \big)  \, ,&
\end{align}
where we  defined the maximum groomed jet radius in \eq{psiMaxDef}.

As a consequence of setting $r_g = r_g^{\rm max}(\xi)$, the same collinear soft scale $\mu_{cs}(\xi)$ appears in the $S_{c_g}^\kappa$ function in \eqs{FactSmallRg}{FactIntRg}. The new scale $\mu_{\cal C}(\xi)$ minimizes logarithms in the collinear function and is defined below in \secn{prof}. Following \eq{FactCalC}, we ought to use this same scale in the $J_\kappa$ and $S_{c_m}$ functions in the intermediate $R_g$ cross section. We give explicit formulae for evaluating these power corrections in \app{MassSub}. 

We would like to connect  this formula with the jet-mass power corrections  in the ungroomed region in \eq{nsNoSD}. This is quite straightforward as the profile $\mu_{\cal C}(\xi)$ automatically becomes $\mu_N$ when $r_g^{\rm max} = 1$ in the ungroomed region. We only need to ensure that $\mu_{cs}(\xi)$ is switched to $\mu_{gs}$ for $\xi > \xi_0$ to turn off any soft-drop resummation in this region. Thus, we will define  a $\mu_{\rm min \ra plain}$ profile, which includes this behavior of the collinear soft profile, 
\begin{align}
     &\text{Min-$R_g$ to plain jet-mass transition:}& 
    &\mu_{\rm min\ra plain } \equiv \{ \mu_{\cal C}(\xi), \mu_{cs \ra gs}(\xi), \mu_{gs} , \mu_{N} \}\,.&
\end{align}
Using this result we arrive at the complete expression for the matched cross section in \eq{Gmatched0}:
\begin{align}\label{eq:Match}
    \tilde {\cal G}^{\rm matched}_{\kappa, \rm sd} (\xi) &\equiv  
    \textcolor{Orange}{\tilde {\cal G}^{\rm resum}_{\kappa,\rm sd} \big(\xi,  \mu_{\rm sd \ra plain} \big)} 
    + 
    \textcolor{Green}{\tilde {\cal G}^{\rm plain}_{\kappa, \rm sd} \big(\xi , \mu_{\rm plain} \big)} 
    - 
    \textcolor{Red}{\tilde {\cal G}_{\rm sd}^{\rm resum} \big(\xi,  \mu_{\rm plain} \big)} \\ 
    &\quad +  
    \textcolor{Blue}{\tilde {\cal G}^{\rm min}_{\kappa, \rm sd} \big(\xi ,r_g^{\rm max}(\xi) ,  \mu_{\rm min\ra plain} \big)} 
    - 
    \textcolor{Magenta}{\tilde {\cal G}_{\rm sd}^{\rm int} \big(\xi,  r_g^{\rm max}(\xi), \mu_{\rm min \ra plain} \big)} 
    \, . \nn 
\end{align}
\begin{figure}[t]
	\centering
	\includegraphics[width=0.49\textwidth]{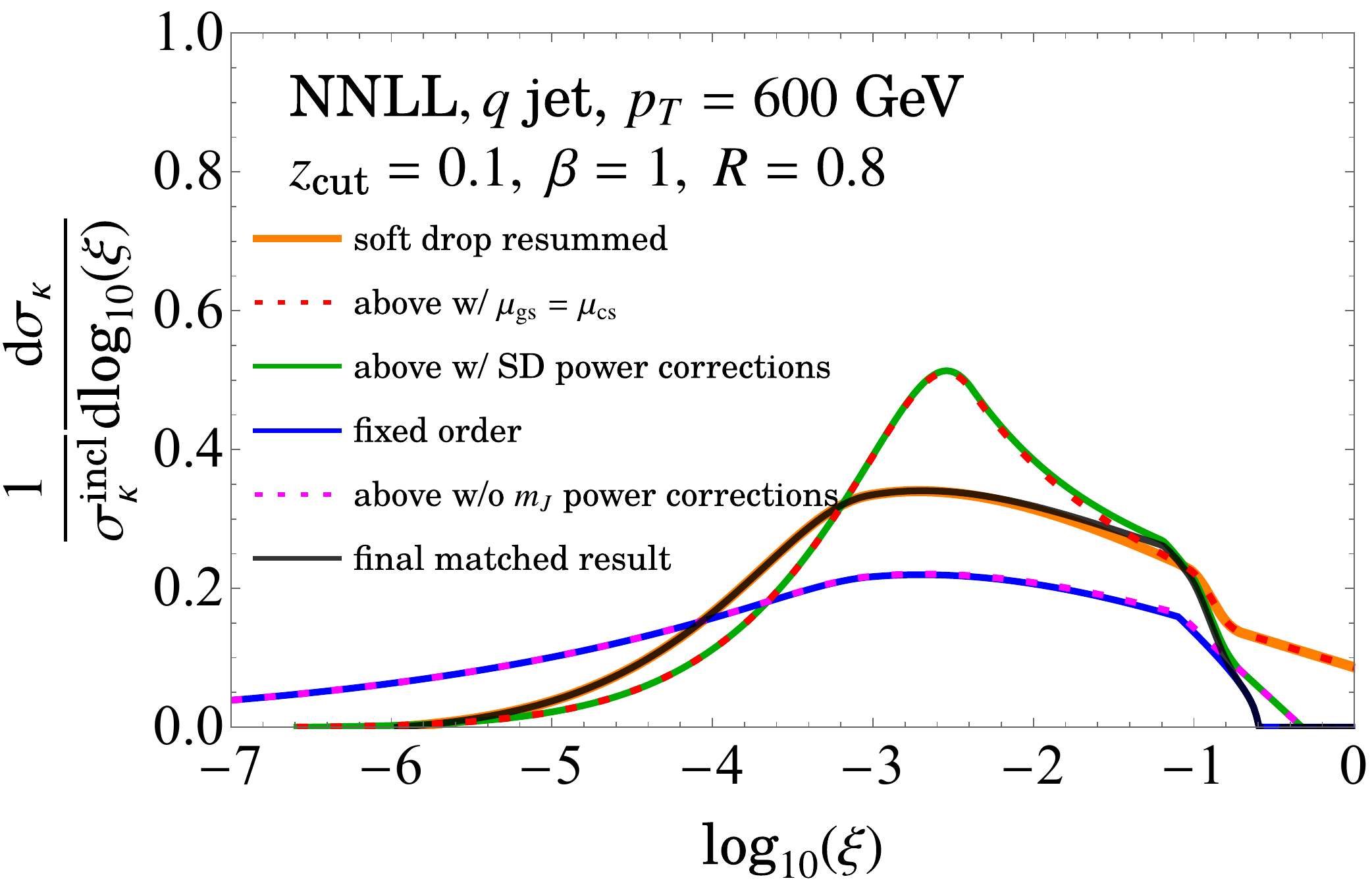}
	\includegraphics[width=0.49\textwidth]{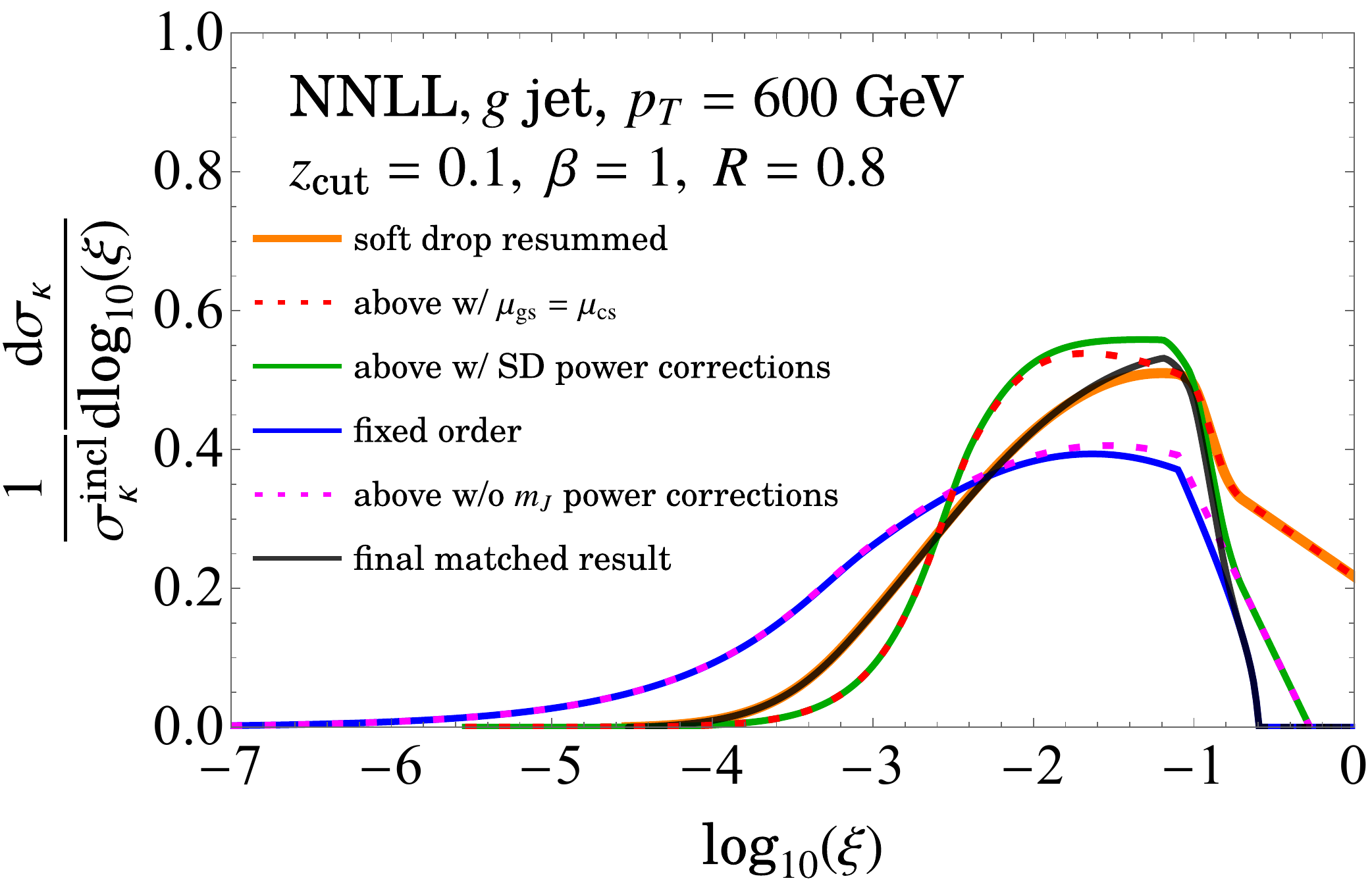}
	\caption{The matched curves at NNLL for quark jets (left) and gluon jets (right). The dotted curves correspond to overlap terms that are subtracted in the matched cross section.}
	\label{fig:matching}
\end{figure}
The second line now captures all the jet-mass dependent power corrections. We display in \fig{matching}  how this works in practice. Recall from \eq{tcalGDef} that $\tilde{\cal G}_\kappa$ is defined as the normalized cross section $\frac{1}{\sigma_\kappa^{\text{incl}}} \frac{\df \sigma_\kappa}{\df \xi}$, which is what we plot in \fig{matching}. For now we have turned off all the NGL terms, which we discuss separately below. We have colored each term to match the corresponding curve in \fig{matching}. We see from \fig{matching} that the two power corrections have opposite signs and their size can depend on the kinematic and grooming parameters, as well as the flavor of the jet.%
\footnote{In the figure we follow a slightly different implementation than written in \eqs{plainSDProf}{Match} by letting the collinear soft scale merge with global soft scale for both $\tilde {\cal G}^{\rm resum}_{\kappa,\rm sd}$ pieces. This still has the same effect of the two pieces canceling out each other entirely in the ungroomed resummation region, as can be seen in \fig{matching} for $\log_{10}\xi >\log_{10 }\xi_0 \sim -1$.
} 
Next, we see that the pieces $\tilde {\cal G}^{\rm plain}_{\kappa,\rm sd}$ and $\tilde {\cal G}^{\rm int}_{\rm sd}$ pieces cancel each other in the same region. This is straightforward to see from \eq{FactIntRg} where $r_g = 1$ results in complete (internal) cancellation of $S_{c_g}^\kappa$ and $S_{G}^\kappa$ functions as well as canceling of $\Sigma_\kappa(\xi)$ between the two cross sections. 
\begin{figure}[t]
	\centering
	\includegraphics[width=\textwidth]{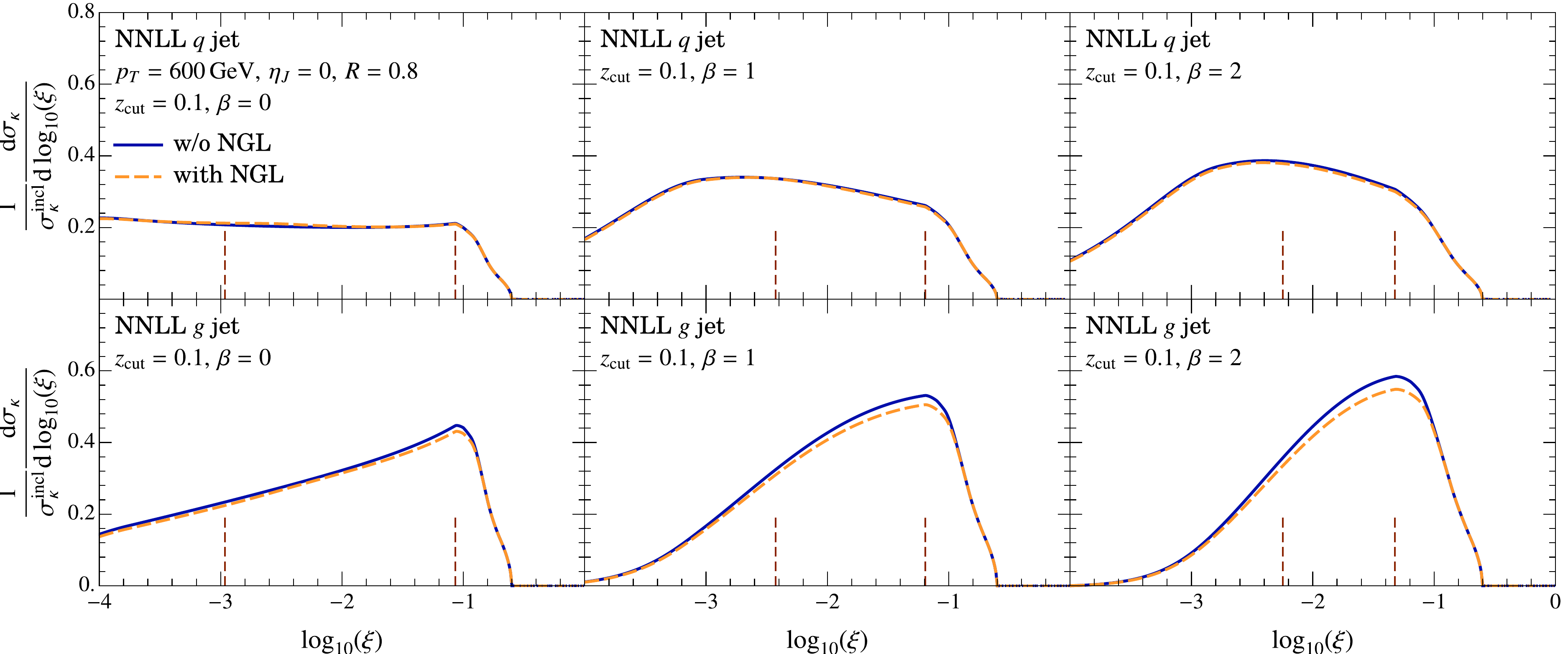}
	
	\caption{Impact of non-global logarithms on the matched groomed jet-mass spectrum. Here we treat NGLs at LL and large $N_c$ limit. The differences between the solid and dashed curves are within NNLL perturbative uncertainty.}
	\label{fig:ngl}
\end{figure}

We now turn to the NGL pieces.  In the implementation of the NGL, we make use of the canonical scales of the set of profiles used above. This allows us to employ the same matching scheme in \eq{Match} to merge NGLs that differ in each of the terms. 
In this work we will limit ourselves to analyzing the impact of NGLs at LL in the large $N_c$ limit using the parameterization provided in \Ref{Dasgupta:2001sh}. In \fig{ngl} we show the NNLL spectrum for quark and gluon jets for $\beta \in \{0,1,2\}$.  
We find that the impact of NGL on the quark jets is very small, while the difference for gluon jets is somewhat larger.
As mentioned earlier, despite the fact that the NGLs appear only in the normalization in the SDOE region, these differences in their relative impact on quark and gluon jets can lead to modification of the shape. 
However, we will see that the differences in \fig{ngl} are well within the NNLL perturbative uncertainty due to scale variation. 
(The only borderline case is $\beta = 2$ gluon jets.) 
Thus for simplicity we not include $\as$ variations in these NGL pieces in our analysis. We leave a more detailed exploration of NGL effects on groomed jet mass to future work 
when higher precision results become available for $pp$ collisions, such ${\cal O}(\alpha_s^2)$ terms or the full N$^3$LL resummation.

%=======================================================================
\subsection{Profile functions}
%=======================================================================

\label{sec:prof}

In this section we describe the implementation of profile functions~\cite{Ligeti:2008ac,Abbate:2010xh} as well as the scheme for varying them to estimate perturbative uncertainty. 
We will employ the  profile functions  described in \Ref{Pathak:2020iue} and review the essential ingredients here.
We start with the two scales $\mu_N$ and $\mu_{gs}$, from which other profile functions will be derived:
\begin{align}\label{eq:muNmuGS}
    \mu_N = e_N Q \,, \qquad 
    \mu_{gs} =  e_{gs} \qcut 
    \, ,
\end{align}
where the parameters $e_N$ and $e_{gs}$ give a handle for varying these scales, as described in the next subsection. For central profiles, we take $e_N=e_{gs}=1$.

%-------------------
\subsubsection{Soft-drop resummation region}
%-------------------
We first consider the soft-drop profiles valid in the region $\xi < \xi_0$.
The value $\xi_0$, defined in \eq{xi0def}, is the point at which the collinear-soft and global-soft scales merge.
To define the jet-mass dependent soft and collinear scales we first introduce the auxiliary function $ f_{\rm run}(r_g) $ defined by
\begin{align}
	\label{eq:frunRg}
	f_{\rm run}(r_g)
	&\equiv \left\{
	\begin{array}{ll}
		y_0 \Big(1+ \frac{r_g^2}{4y_0^2}\Big)	& ~~~~~~~~~~r_g \leq  2 y_0\\[4pt]
		r_g  & ~~~~~~~~~~  2 y_0 < r_g \leq 1
	\end{array}
	\right.\;.
\end{align}
Here
\begin{align}\label{eq:y0}
	y_0  \equiv \frac{n_0}{(\mu_{gs}/1\, {\rm GeV})}   
	>\frac{\Lambda_{\rm QCD}}{\qcut}
	\, ,
\end{align}
and $n_0$ is an ${\cal O}(1)$ number. We take $n_0 = 1$ GeV as the central value for our numerical analysis. This number governs the point $\xi_{\rm SDNP}$ at which the jet-mass dependent scales will freeze,
\begin{align}\label{eq:xiSDNPDef}
    \xi_{\rm SDNP} \equiv \xi_0 \,  y_0^{\frac{2+\beta}{1+\beta}}
    = \frac{n_0}{\mu_N}\Big(\frac{n_0}{\mu_{gs}}\Big)^{\frac{1}{1+\beta}}
  \,.
\end{align}
Using the auxiliary function, the central collinear-soft profile $\mu_{cs}(\xi)$ is given by
\begin{align}\label{eq:muCS}
	\mu_{cs} (\xi )  \equiv 
	\mu_{gs}  \, f_{\rm run} \bigg ( \Big(\frac{\xi}{\xi_0}\Big)^{\frac{1+\beta}{2+\beta}}\bigg) \,, 
	\qquad\quad \xi < \xi_0\,.
\end{align}
To derive the jet function scale $\mu_J$ from $\mu_{cs}$, we first derive an auxiliary ungroomed soft scale
\begin{align}\label{eq:tmuS}
    \tilde \mu_s (\xi)
    \equiv \mu_{cs}(\xi) \: \bigg(\frac{\mu_{cs}(\xi)}{\mu_{gs}}\bigg)^{\frac{1}{1+\beta}}
    \,,
    \qquad\quad \xi < \xi_0
    \, ,
\end{align}
such that the jet function scale is given by
\begin{align}\label{eq:muJ}
	\mu_J (\xi) &\equiv \Big [\mu_N \tilde \mu_s(\xi) 
	\Big]^{\frac{1}{2}}
	= \Bigg[ 
	\mu_N\: \mu_{cs}(\xi) \: \bigg(\frac{\mu_{cs}(\xi)}{\mu_{gs}}\bigg)^{\frac{1}{1+\beta}}
	\Bigg]^\frac{1}{2}
	\,,
	\qquad\quad \xi < \xi_0
	\, .
\end{align}
Finally the collinear function scale $\mu_{\cal C}$ is defined as
\begin{align}
    \mu_{\cal C}(\xi) = \mu_N \Big(\frac{\mu_{cs}(\xi)}{\mu_{gs}}\Big)^{\frac{1}{1+\beta}} \, .
\end{align}
Thus $\mu_{\cal C}$ inherits its variations from $\mu_{cs}$ and $\mu_{gs}$.
%-------------------
\subsubsection{Ungroomed region}
%-------------------
At the value $\xi_0$ the global-soft and collinear-soft scales merge, i.e.\ $\mu_{cs} \vert_{\xi_0}= \mu_{gs}\vert_{\xi_0}$. Beyond this value, we enter the ungroomed resummation region. We first derive the  profile for the ungroomed soft function, $\mu_s$.
Following \Ref{Lustermans:2019plv} its profile can be written as:
\begin{align}
	\label{eq:muS}
	\mu_s(\xi)
	\equiv \mu_N \, f_{\rm run}^{\rm plain}(\xi) \, ,
\end{align}
where $ f_{\rm run}^{m_J}(\xi) $ is given by
\begin{align}
	\label{eq:frunmJ}
	f_{\rm run}^{\rm plain}(\xi)
	&\equiv \left\{
	\begin{array}{ll}
		x_0 \big(1+ \frac{\xi^2}{4 x_0^2}\big)	& ~~~~~~~~~~\xi \leq  2 x_0\\[2pt]
		\xi  & ~~~~~~~~~~  2 x_0 < \xi \leq x_1\\[2pt]
		\xi + \frac{(2 - x_2 - x_3)(\xi - x_1)^2}{2(x_2 - x_1)(x_3 - x_1)} & ~~~~~~~~~~ x_1 < \xi \leq x_2\\[2pt]
		1- \frac{(2 - x_1- x_2)(\xi - x_3)^2}{2(x_3 - x_1)(x_3 - x_2)}  & ~~~~~~~~~~  x_2 < \xi \leq x_3\\[2pt]
		1 & ~~~~~~~~~~x_3 < \xi \leq 1\\[2pt]
	\end{array}
	\right.\;.
\end{align}
Here $x_0$ corresponds to the point where the scale is frozen in the nonperturbative region. To be consistent with \eq{y0}, we set it to
\begin{align}\label{eq:x0}
	x_0 = \frac{n_0}{(\mu_N/1 {\rm GeV})} \, .
\end{align}
The purpose of this function is to give a smooth transition between the linear dependence on $\xi$ for $2 x_0< \xi<x_1$, and a constant value for $\xi>x_3$. To this end, we add and subtract a quadradic function in $\xi$ in the regions $[x_1,x_2]$, and $[x_2,x_3]$, respectively. When choosing values for $x_1$, $x_2$ and $x_3$, we must ensure that the transition to the fixed-order region, whenever possible, starts only after the soft drop turns off, but in addition to finishing before the jet-mass endpoint at $\xi = 0.25$. To this end, we take $x_1 = 1.15 \xi_0$, $x_3 = 0.2$, and $x_2 = (x_1 + x_3)/2$.\footnote{We deal with cases with aggressive grooming, such as $\zcut = 0.2$ and $\beta = 0$, where the soft-drop transition point can appear in the fixed-order region with $0.2 < \xi_0 < 0.25$ by setting $x_3  = (0.25 + \xi_0)/2$. For yet higher $\zcut$, when $\xi_0 > 0.25$, we set $x_3 = \min\{\xi_0 + 0.1, 1\}$. In each of these two cases we set $x_1 = 1.15\xi_0 < x_3$ or $x_1 = \xi_0 < x_3$, and $x_2 = (x_1+x_3)/2$.}

Analogously to  Eq.~\eqref{eq:muJ}, the jet scale for ungroomed profiles is  derived from $\mu_s(\xi)$ profile
\begin{align}\label{eq:muJPlain}
	\mu_J^{\rm plain} (\xi)\equiv \Big[\mu_N\: \mu_s \Big]^{\frac{1}{2}}
	\, ,  
\end{align}
such that the jet scale also has the plain jet-mass transition points. In the implementation of the set of profiles $\mu_{\rm sd\ra plain}$ the jet scale in \eq{muJ} will be replaced by the profile in \eq{muJPlain} once $\xi > \xi_0$ and the global and collinear-soft scales by the ungroomed soft scale in \eq{muS}. 

\begin{figure}[t]
    \centering
    \includegraphics[width=0.49\linewidth]{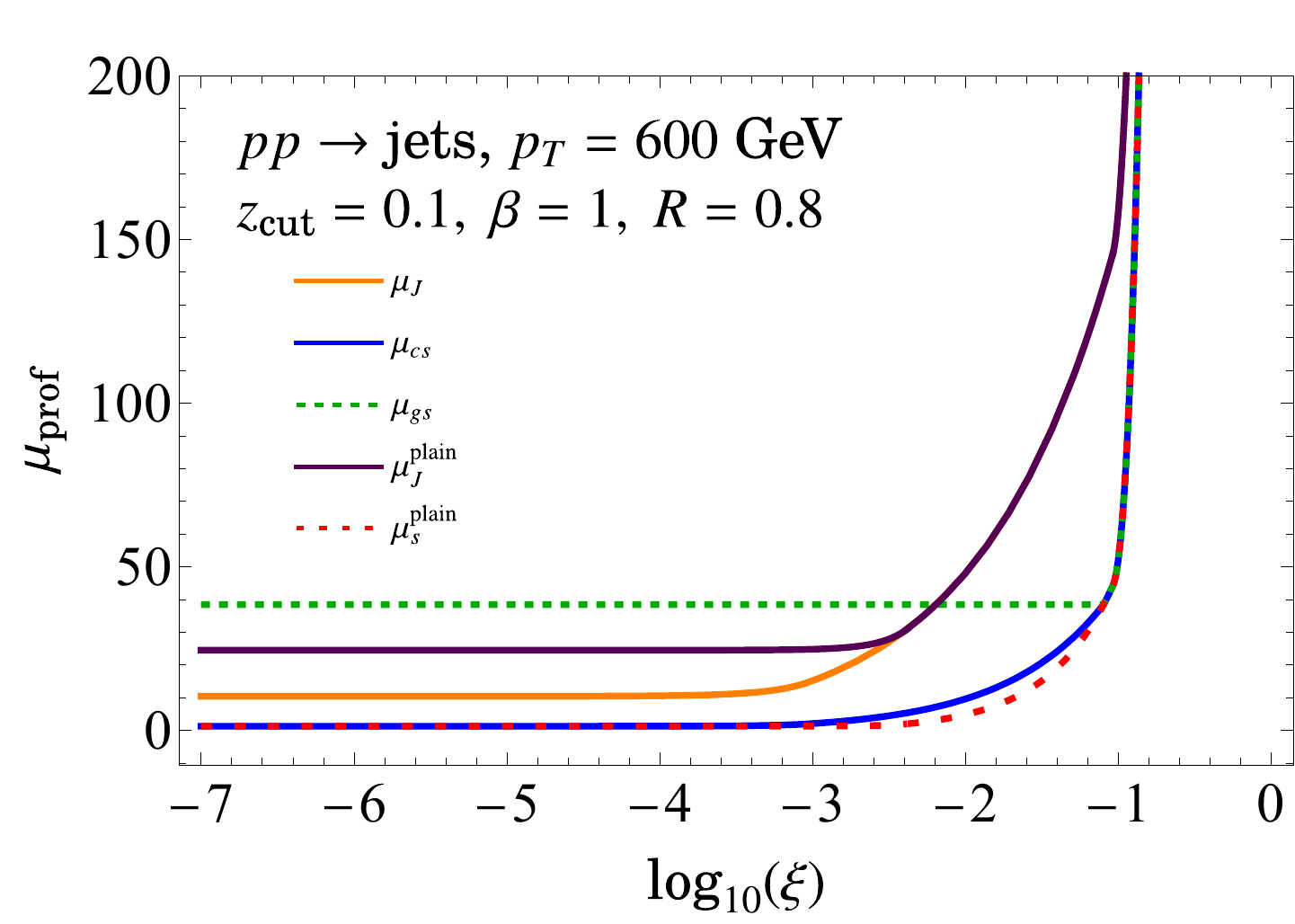} \\
    \includegraphics[width=0.49\linewidth]{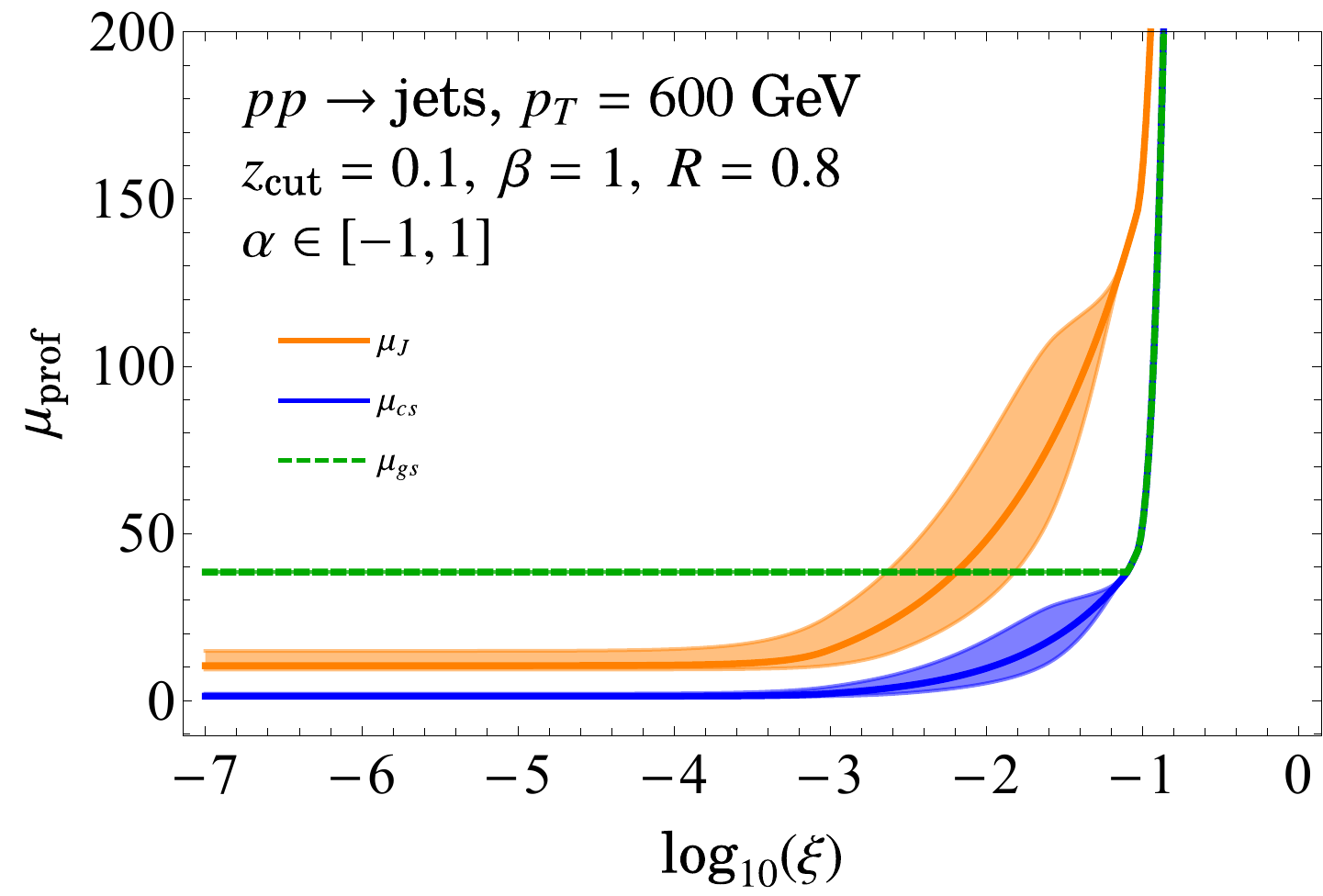}
    \includegraphics[width=0.49\linewidth]{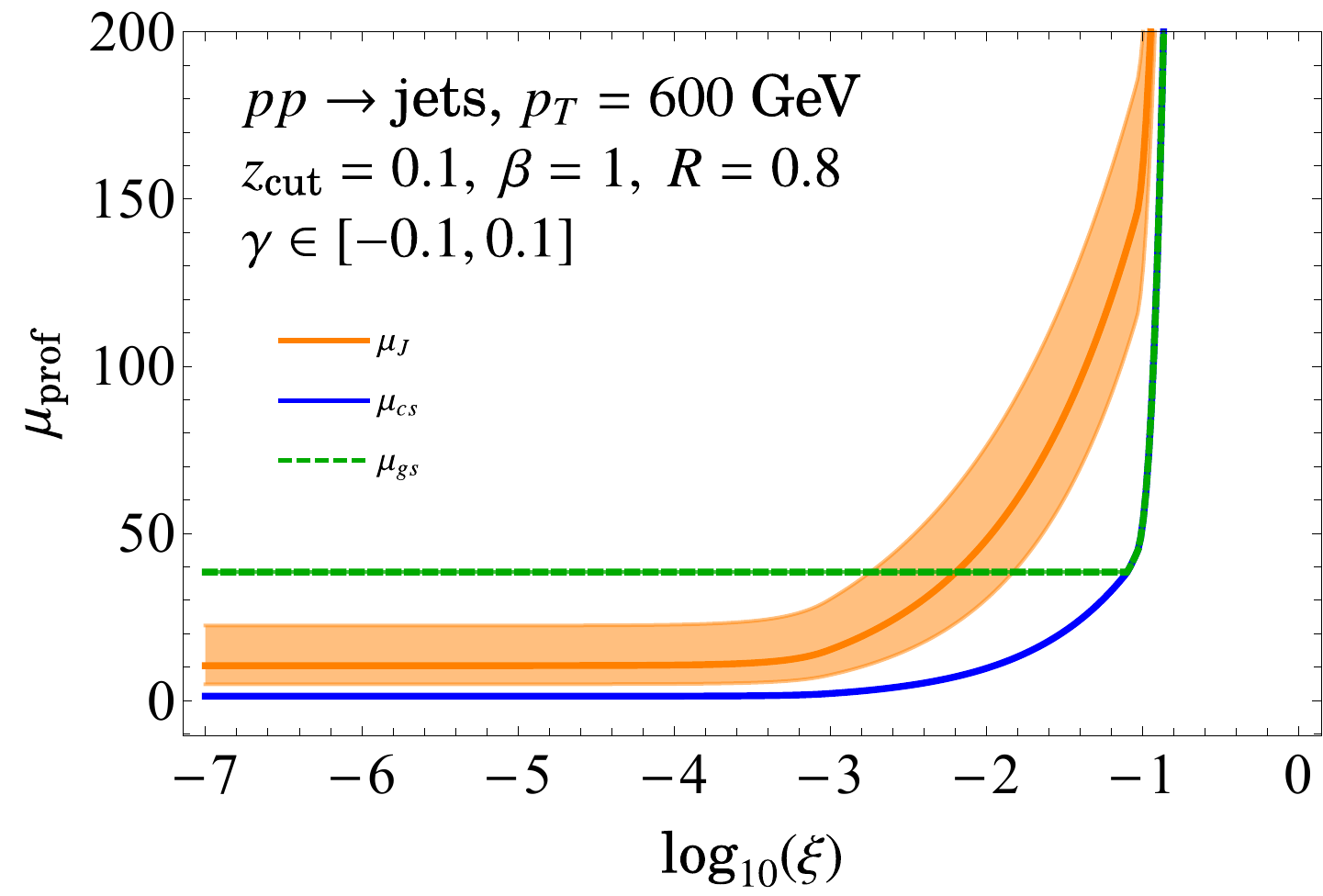}
    \caption{Profile functions for $pp$ collisions at $p_T=600$ GeV. Top panel shows profiles used for the soft-drop resummed and plain jet-mass cross sections. The bottom panel shows the profile variations with $\alpha$ (left panel) and with $\gamma$ (right panel) in the soft-drop resummation region.}
    \label{fig:profiles}
\end{figure}
%
%-------------------
\subsubsection{Profile variation}
%-------------------
\label{sec:profvar}

We now describe the profile variations used to assess perturbative uncertainties. They are implemented through a set of parameters introduced above,
\begin{equation}
\{e_i,\alpha,\lambda,\gamma,\rho,n_0\} \,.
\end{equation}
The order in which the parameters are introduced in the section also corresponds with how they are implemented in the code: the variation of parameters introduced later are impacted by simultaneous variation of parameters introduced earlier, but not vice versa. We now turn to describing the effects of varying each parameter.

First, the hard and global soft scales are varied by varying the $e_i\equiv e_{N}=e_{gs}$, where we have set $e_N = e_{gs}$ to prevent the soft-drop cusp point $\xi_0$ from varying. These variations affect all the other scales derived from these scales, as shown in the previous section, and below.

Second, we implement an independent variation of the collinear-soft scale, by making use of a trumpet function $f_{\rm vary}(r_g)$:
\begin{equation}\label{eq:fVary}
    f_{\rm vary}(r_g)
    = \left\{ \begin{array}{l r}
        2(1 - r_g^2)\,,                                 & \qquad    r_g < 0.5 \\ 
        1 + 2(1- r_g)^2\,,                            &  \qquad 0.5 \leq r_g \leq 1
    \end{array}
    \right. ,
\end{equation}
which implements a variation for $r_g<1$.
Following \eq{muCS}, the variation of $\mu_{cs}$ scale is given by
\begin{align} \label{eq:alphadef}
    \tilde \mu_{cs}^{\rm vary}(\xi, \alpha)  &\equiv \mu_{gs} \Bigg[ f_{\rm vary}\bigg ( \Big(\frac{\xi}{\xi_0}\Big)^{\frac{1+\beta}{2+\beta}}\bigg) \Bigg]^{\alpha}
    \, f_{\rm run} \bigg ( \Big(\frac{\xi}{\xi_0}\Big)^{\frac{1+\beta}{2+\beta}}\bigg) \, , \qquad \xi < \xi_0
    \, .
\end{align}
The choice $\alpha = 0$ returns the default profile, and varying $\alpha = \pm 0.5$ allows for variation in the resummation region up to a factor of $\sqrt{2}$. 
The tilde indicates that this is not quite the final scale. This is because, as written, the $\alpha$ variation will drive this scale below the nonperturbative scale $n_0$ in \eq{y0}. To prevent that, we re-freeze this scale in the nonperturbative region:
\begin{align}
     \mu_{cs}^{\rm vary}(\xi, \alpha)  \equiv f_{\rm freeze}\big[   \tilde \mu_{cs}^{\rm vary}(\xi, \alpha)\big]
\end{align}
where
\begin{equation}\label{eq:ffreeze}
    f_{\rm freeze}[\mu]
    \equiv \left\{ \begin{array}{l r}
        \mu\,,                                 & \qquad   \mu \geq 2n_0 \\ 
       n_0 \Big(1 +\frac{\mu^2}{4n_0^2}\Big)\,,                            &  \qquad  \mu < 2n_0 
    \end{array}
    \right. .
\end{equation}
Having defined $\alpha$-variation of $\mu_{cs}$, the corresponding variations of $\mu_{J}$ is derived using 	\eq{muJ} by replacing $\mu_{cs} \ra \mu_{cs}^{\rm vary}$. As mentioned above, these two scales inherit the $e_{gs}$ variation through the $\mu_{gs}$ scale. 

Third, we consider the variation that breaks the following canonical relation between the soft scales: $\mu_{cs}^{2+\beta} = \mu_s^{1+\beta} \mu_{gs}$.
It is implemented via a parameter $\rho$ as follows:
\begin{align}\label{eq:rhodef}
   \tilde  \mu_{cs}^{\rm vary}(\xi, \alpha, \rho) &\equiv \Bigg[ f_{\rm vary}\bigg ( \Big(\frac{\xi}{\xi_0}\Big)^{\frac{1+\beta}{2+\beta}}\bigg) \Bigg]^{\alpha} \big(\tilde \mu_s(\xi)\big)^{\frac{1+\beta + \rho}{2+\beta}} \big(\mu_{gs}\big)^{\frac{1-\rho}{2+\beta}} \, , \nn 
    \\ 
      \mu_{cs}^{\rm vary}(\xi, \alpha, \rho)  &\equiv f_{\rm freeze}\big[ \tilde  \mu_{cs}^{\rm vary}(\xi, \alpha, \rho) \big] \, . 
\end{align}
Here the auxiliary scale $\tilde \mu_s$ is defined in \eq{tmuS}, i.e.\ without any $\alpha$ variation.

As a final variation in the resummation region, we break the canonical relation between the jet and the hard scale in \eq{muJ}, via parameter $\gamma$:
\begin{align}\label{eq:muJVary}
    \mu_{J}^{\rm vary}(\xi, \alpha,\rho,\gamma) = 
    \big(\mu_{N}^{\rm vary} \big)^{\frac{1}{2}+\gamma}
     \big(\tilde \mu^{\rm vary}_s(\xi, \alpha,\rho)\big)^{\frac{1}{2}-\gamma}
     \, ,
\end{align}
where, analogously to before, the auxiliary scale $\tilde \mu^{\rm vary}_s(\xi, \alpha,\rho)$ inherits the $\alpha$ and $\rho$ variation from being defined by substituting $\mu_{cs}(\xi)$ in \eq{tmuS} by $\mu_{cs}^{\rm vary}(\xi,\alpha,\rho)$ defined in \eq{rhodef}:
\begin{align}
    \tilde \mu_{s}^{\rm vary}(\xi, \alpha, \rho)
    \equiv \mu_{cs}^{\rm vary} \Big(\frac{\mu_{cs}^{\rm vary}}{\mu_{gs}^{\rm vary}}\Big)^{\frac{1}{1+\beta}} \, .
\end{align}

Next, we consider profile variation in the ungroomed region. The hard scale is varied as before using the parameter $e_i$. The trumpet variation of the $\mu_s(\xi)$ scale, defined in \eq{muS}, is implemented only in the ungroomed resummation region $\xi_0 \leq \xi \leq x_3$, where the parameter $x_3$ appears in $f_{\rm run}^{\rm plain}$ defined in \eq{frunmJ}. To this end, we first define
\begin{align}
    f^{\rm plain}_{\rm vary}(\xi)
    \equiv \left\{ \begin{array}{l l}
        1 &\qquad 0 < \xi \leq \xi_0 \\
         \zeta\big(\xi, \xi_0, x_{\rm mid}; 1 , 2\big)\,,                                 & \qquad    \xi_0 \leq \xi < x_{\rm mid} \\ 
         \zeta \big(\xi, x_{\rm mid}, x_3; 2, 1\big)\,,                            &  \qquad x_{\rm mid} \leq \xi \leq x_3 \\
         1 & \qquad x_3 < \xi \leq 1 
    \end{array}
    \right. .
\end{align}
where
\begin{align}
    x_{\rm mid} \equiv \frac{\xi_0 + x_3}{2} \, , 
\end{align}
and
\begin{align}
    \zeta \big(\xi, x_{\rm start}, x_{\rm end}; a_1, a_2\big) \equiv
    \left\{ \begin{array}{l l}
        a_1 &\qquad \xi \leq x_{\rm start} \\
         a_1 + \frac{2(a_2 - a_1)(\xi - x_{\rm start})^2}{(x_{\rm end} - x_{\rm start})^2}
         \,,  
         & \qquad    x_{\rm start} \leq \xi < \frac{x_{\rm start} + x_{\rm end}}{2} \\ 
         a_2 - \frac{2(a_2 - a_1)(x_{\rm end} - \xi)^2}{(x_{\rm end} - x_{\rm start})^2}
        \,,  
         &  \qquad \frac{x_{\rm start} + x_{\rm end}}{2} \leq \xi \leq x_{\rm end} \\
         a_2 & \qquad x_{\rm end} < \xi  
    \end{array}
    \right. .
\end{align}
Thus, the profile $\mu_s(\xi)$ is varied as
\begin{align}
    \tilde{\mu}^{\rm vary}_s(\xi , \lambda) \equiv \mu_N \Big[f_{\rm vary}^{\rm plain}(\xi)\Big]^\lambda
    f_{\rm run}^{\rm plain}(\xi) \, ,
\end{align}
but since the variation with $\lambda$ may spoil the freezing in the nonperturbative region, we must re-freeze the soft scale by taking 
\begin{equation}
    \mu^{\rm vary}_s(\xi , \lambda) = f_{\rm freeze}[\tilde{\mu}^{\rm vary}_s(\xi , \lambda)] \,.
\end{equation}
Finally, analogously to \eq{muJVary}, the variation of the scale $\mu_{J}^{\rm plain}$ is given by
\begin{align}\label{eq:muJPlainVary}
    \mu_J^{\rm plain, vary}(\xi, \lambda, \gamma) \equiv 
    \mu_N^{\frac{1}{2}+\gamma} \big(\mu^{\rm vary}_s(\xi, \lambda)\big)^{\frac{1}{2} -\gamma} \, .
\end{align}

In our analysis, we choose the following profile variation ranges for the six different parameters:
\begin{align} \label{eq:profrange}
    &e_i \in [0.5, 2] \,,&
    &\alpha \in [-1,1] \,,&
    &\lambda \in [-1, 0.3] \,,&
    \\
    &\gamma \in [-0.1,0.1] \,,&
    &\rho \in [-0.1,0.1] \,,&
    &n_0 \in [0.75, 1.25] \,.&\nn
\end{align}
We have adopted the same range of variations as in \Ref{Pathak:2020iue}, with the exception of $\lambda$ parameter which did not appear there. A key criterion in deciding the range of variation is to ensure that the hierarchies among various scales is maintained. We also ensure that variation does not interrupt the monotonic behavior of the scales or introduce unphysical kinks. These profiles with the variations are displayed in \fig{profiles}. 
Since it is \textit{a priori} not clear how the different variations might affect each other, we vary the profiles randomly in all six parameters, and take an envelope for our uncertainty.

\begin{figure}[t]
    \centering
    \includegraphics[width=\textwidth]{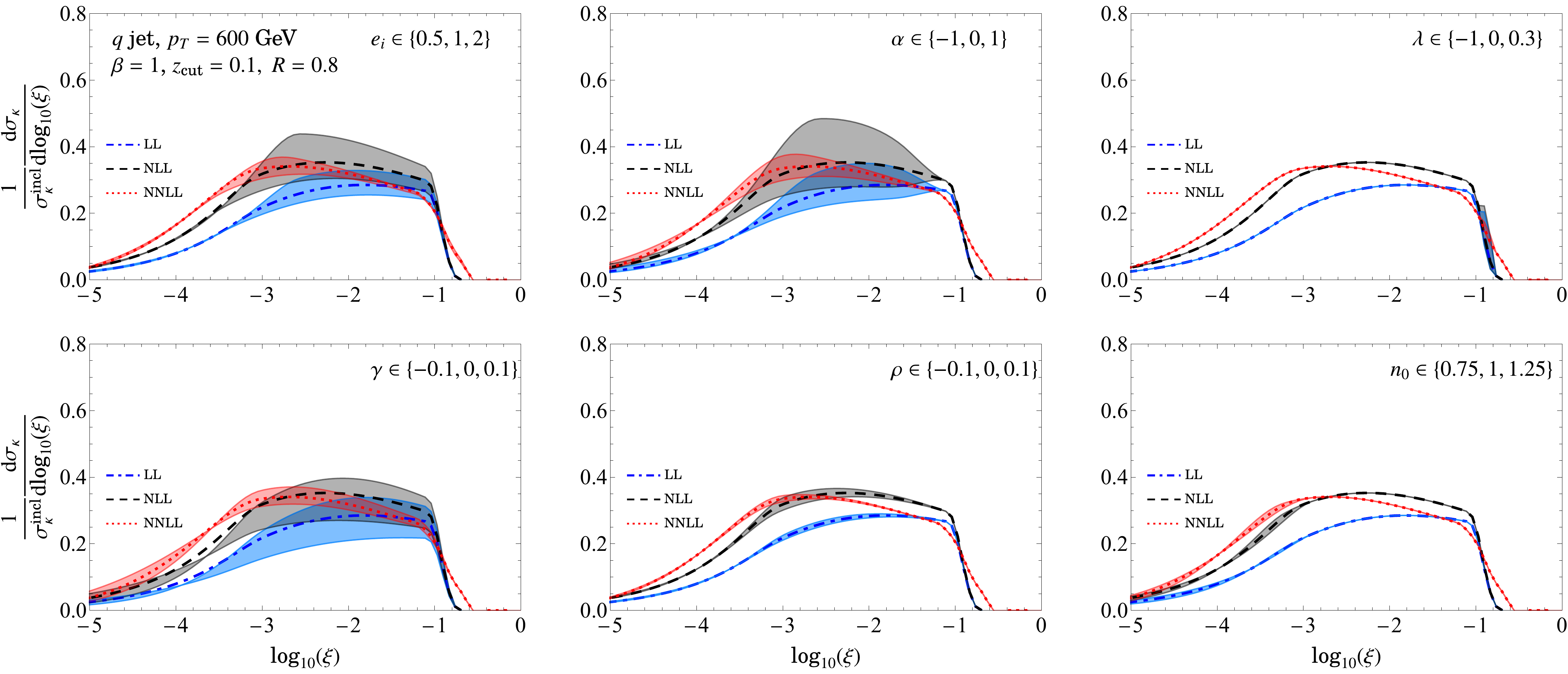}
    \\[20pt]
    \includegraphics[width=\textwidth]{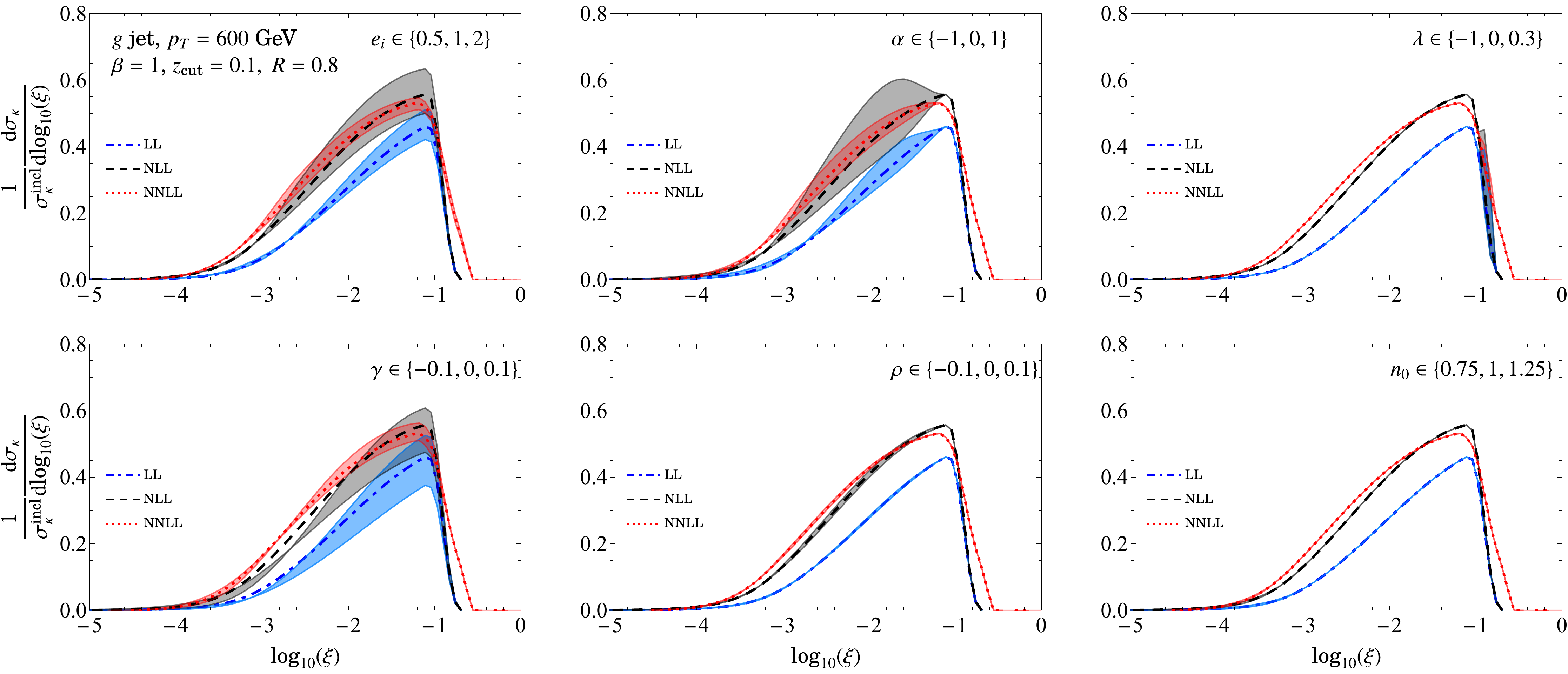}
    \caption{Variations for each of the six different profile-variation parameters individually, for quark and gluon jets at $p_T=600$ GeV, with $\beta=1$. Other values of $\beta$ are shown in App.~\ref{app:profvar}.}
    \label{fig:profvar}
\end{figure}

\fig{profvar} shows impact of profile variations for each of the parameter for the LL, NLL and NNLL jet-mass spectrum of quarks and gluons.
Note that at LL and NLL, the shape and the transition points are governed solely by profile functions entering the RG evolution kernels. The soft drop cusp for LL and NLL cross section is thus located at $\xi_0$, whereas it appears at $\xi_0'$ for NNLL once $\mathcal{O}(\as)$ terms are included. The end-point of the jet-mass spectrum for LL and NLL is located at $\xi = x_3$, where $x_3$ was defined above in \eq{frunmJ}. The NNLL curve, being matched to LO non-singular cross section, has the end point at $\xi = 0.25$. The transition to the nonperturbative region, however, happens for the same jet mass for all of the three orders. Next, since we are normalizing to the inclusive cross section according to Eq.~\eqref{eq:Match}, and indeed the integral over the curve for the NLL and NNLL spectra is close to 1, although the integral of the LL curve is around a factor of 20\% smaller.
We find that the dominant variation is induced by the overall scale variation due to $e_i$, the trumpet variation via $\alpha$, and the variation due to breaking of jet, soft and hard collinear canonical relation in \eqs{muJVary}{muJPlainVary} through $\gamma$. On the other hand, the variation of the plain soft function scale in the ungroomed region via $\lambda$, the breaking of the soft drop canonical relation via $\rho$ in \eq{rhodef}, and the nonperturbative scale variation via $n_0$ lead to subdominant effects. We also notice here a more significant dependence on the perturbative order for the slope of the spectrum in the soft drop resummation region for quark jets compared to gluon jets. We will discuss the impact of this effect on $\as$ below. The analogous profile variation figures for $\beta = 0$ and $\beta = 2$ are provided in \app{profvar}.

\subsection{Comparison with previous work}
\label{sec:compare}

We now make a comparison with earlier calculations of the groomed jet-mass spectrum and list the various improvements achieved in this paper. 
\begin{enumerate}
	\item \underline{Resummation and matching}: 
	The leading power factorization in the small $\zcut$ limit was formulated in SCET in \Ref{Frye:2016aiz}. 
	In \Refs{Marzani:2017mva,Marzani:2017kqd} the analysis was extended by including finite $\zcut$ terms in framework of diagrammatic QCD for $\beta = 0$ soft drop at NLL (single-logarithmic) accuracy and matched to NLO fixed-order results. 
	The finite $\zcut$ pieces were incorporated both at the level of fixed-order perturbation theory as well as by modifying the resummation kernels in the Sudakov double log exponent. 
	Using the $1\ra3$ splitting functions in the collinear limit, the calculation in this framework was improved to NNLL accuracy in \Ref{Anderle:2020mxj} providing a cross check for results in \Ref{Frye:2016aiz}. 
	In \Ref{Kang:2018jwa} the groomed and ungroomed jet-mass spectrum for inclusive jets in $pp$ collisions was calculated at NLL accuracy, and matched to LO fixed-order results. 
	Formulating the problem in terms of inclusive jets enabled determination of the absolute normalization up to NGLs. 
	
	In our analysis we have performed the state-of-the-art NNLL resummation with the ingredients available to-date. In order to extend the analysis to N$^3$LL accuracy, we need two loop constant pieces of the collinear-soft $S_c^\kappa$ and global soft $S_G^\kappa$ functions for quarks and gluon jets with generic jet radius $R < \pi/2$, as well as the corresponding 3-loop non-cusp anomalous dimension.
	This is partially achieved through the calculation for hemisphere quark jets for $\beta = 0$ in \Ref{Kardos:2020ppl}. 
	However, the 2-loop global soft constant term calculation for non-hemisphere jets is still lacking.
	
	Next, while our matching to fixed-order results is performed at LO, we have done so by computing fully analytic expressions for the jet-mass dependent pieces $\tilde {\cal G}_{\kappa, \rm No\,sd}$ and $\tilde {\cal G}_{\kappa, \rm sd}$ of the inclusive jet function, as defined in \eq{tcalGDef}. 
	In the groomed region, we have analytically captured the jet-mass related power corrections using the doubly differential groomed jet mass and groomed jet radius result in the $r_g \gtrsim r_g^{\rm min}(\xi)$ regime. 
	We have also checked that the finite $\zcut$ power corrections in the $\Delta \tilde {\cal G}^{\rm n.s.[1]}_{\kappa, \rm sd}$ term defined in \eq{GSDNS} are numerically insignificant for our analysis. This is consistent with the observation in \Ref{Larkoski:2020wgx}.
	
	\item \underline{Perturbative uncertainty}:
	Typically, the perturbative uncertainty has been estimated by varying resummation scales up and down by a factor of two around their canonical values. In \Ref{Kang:2018jwa} the jet scale and the global soft scales were defined as a function of the collinear soft (or ungroomed soft) scales, and hence their variations were analogously constrained. 
	In contrast, we have parameterized our variation in terms of parameters that each probe a different physical effect associated with resummation.
	Unlike \Ref{Kang:2018jwa} we have also included profile variations that probe breaking of the canonical relations related to the jet-mass and soft-drop measurements. In fact, breaking the jet, hard collinear and soft scale relation via parameter $\gamma$ leads to a variation comparable to other standard variations, and is the main determinant in the overall perturbative uncertainty in $\beta = 0$ soft drop (see \fig{profvarb0}).  
	Furthermore, to avoid being biased by correlations when varying scales one at a time, we consider a more comprehensive variation by randomly choosing points from the specified range of profile parameters.
	We have, however, chosen not to consider variations in the cusp location as in \Ref{Marzani:2019evv}, since this is a kinematically calculable endpoint.
	
	\item \underline{Transition into the ungroomed region}:
	The problem of consistently combining the results in groomed and ungroomed regions around the soft-drop cusp has been considered in the previous literature. 
	In \Refs{Frye:2016aiz} this matching was performed using numerical fixed-order code, and thus lacked resummation in the ungroomed region. 
	\Refs{Marzani:2017mva,Marzani:2017kqd} did include resummation in the ungroomed region and performed the matching to NLL ungroomed resummed result. 
	Note that no subtlety in the analysis of the transition point arises at this order since the entire shape of jet-mass distribution is described via resummation kernels and the transition can be straightforwardly implemented using collinear-soft transition point $\xi_0$. 
	The same applies to the NLL analysis of \Ref{Kang:2018jwa,Marzani:2019evv}. \Ref{Marzani:2019evv} however, did take into account the uncertainty induced due to differences between the collinear-soft transition point $\xi_0$ and the soft wide-angle transition point $\xi_0'$.  
	Finally, in \Ref{Baron:2018nfz}, the transition region was more closely analyzed by computing the cumulative of the fixed-order soft-drop correction $\Delta S_{\rm sd}^{\kappa[1]}$ in \eq{Splain1}. 
	Their implementation of matching between groomed and ungroomed region involved directly using the (cumulative of) $\Delta S_{\rm sd}^{\kappa[1]}$ correction augmented with running coupling corrections. 
	However, this partially resummed result could not remove in discontinuity in the matched groomed result at ${\cal O} (\as^2)$. 
	
	As shown above, our treatment of the cusp using profile functions results in a seamless transition between the groomed and ungroomed regions,
	such that the transition happens consistently at the soft wide-angle transition point $\xi_0'$ defined in \eq{xi0p} in the matched result. 
	This was possible only after including the resummation between the soft and the jet scales which was not accounted for in \Ref{Baron:2018nfz} (which can also be phrased as neglecting the multiple emission contributions).
	Furthermore, our results include complete NNLL resummation of Sudakov logarithms across the entire spectrum.
	 
	\item \underline{Resummation at the soft-drop cusp}:
	The resummation of logarithms of $\xi$ and $\xi_0$ in the cusp region becomes subtle because, similar to the consideration of NGL in ungroomed jet mass, there can be arbitrary number of widely separated soft subjets whose angular location must be tracked to determine the outcome of grooming. 
	In \Ref{Benkendorfer:2021unv} an all-orders resummation was performed for the case of single resolved soft subjet at ``NLL accuracy''. 
	Note that because the groomed jet-mass measurement is inclusive over the kinematics and the number of resolved soft subjets, 
	the calculation with a single resolved soft subjet alone does not capture the entire NLL tower of $\alpha_s^n \ln^n \xi$ terms. 
	Nevertheless, this calculation showed that the effect of resummation is to shift the location of the cusp and cause small variations in the normalization. 
	
	Because \Ref{Benkendorfer:2021unv} did not match the cusp-resummation results to the cross sections in the ungroomed jet-mass region above and the groomed jet-mass  resummation region with collinear-soft modes below, it is not clear how much the impact of resummation on the location of the cusp really is. For jets of $p_T > 600$ GeV that we consider in this analysis,  Fig.~6 of \Ref{Benkendorfer:2021unv}  suggests a positive shift in $\xi_0$ of about 20\%. A proper matching to results for larger and smaller jet masses may render the shift smaller. Furthermore, a 20\% shift may already be ruled out by the ATLAS data when compared to NNLL theory. Nevertheless, we find that a 20\% shift to $\xi_0$ is only about a 5\% modification to the range that we use for analyzing sensitivity to the $\as$. Thus our conclusions derived below while ignoring resummation of the cusp are expected to continue to hold, though more investigation of this is warranted.
	
\end{enumerate}

%============================================
\section{Nonperturbative corrections}
%============================================
\label{sec:np}

It was shown in  \Ref{Hoang:2019ceu} that nonperturbative power corrections in the soft drop resummation region take the form in \eq{sigfullk1}. It is helpful to rewrite the NP power corrections as~\cite{Pathak:2020iue}
\begin{align}\label{eq:sigHadxi}
	\frac{1}{ \sigma_\kappa} \frac{\df  \sigma^{\rm had}_\kappa}{\df \xi}  &= 
	\frac{1}{\hat \sigma_\kappa} \frac{\df \hat \sigma_\kappa}{\df \xi} 
	-  \frac{\Ok}{Q}
	\frac{\df}{\df\xi} \bigg(  
	\int_{r_g^{\rm min}(\xi)}^{r_g^\max(\xi)} \df r_g \:r_g \, \frac{1}{\hat \sigma_\kappa} \frac{\df \hat \sigma_\kappa }{\df \xi \df r_g}
	\bigg) \\
	&\quad 
	+ \frac{\Uka+ \beta\Ukb }{\qcut}
	\int_{r_g^{\rm min}(\xi)}^{r_g^\max(\xi)} \df r_g \:\frac{w_{\rm max}(\xi, r_g)}{r_g} \, \frac{\df}{\df \veps}\bigg(\frac{1}{\hat \sigma_\kappa (\veps)} \frac{\df \hat \sigma_\kappa(\veps) }{\df \xi \df r_g}\bigg) \bigg|_{\veps \ra 0 }
	+ \ldots 
	\, ,\nn 
\end{align}
where $\df  \sigma^{\rm had}_\kappa$ is the hadron-level jet-mass cross section for flavor $\kappa$ as defined above in \eq{tcalGDef} and  $\{\Ok, \Uka, \Ukb\}$ are ${\cal O}(\Lambda_{\rm{QCD}})$ constants that multiply $r_g$ moments of doubly differential, parton level cross sections $\df \hat \sigma_\kappa/(\df \xi \df r_g)$. 
Here $r_g = R_g/R$ defined above in \eq{rgDef}.
The meaning of the parameter $\veps$ and the function $w_{\rm max}$ is explained below.
Comparing with \eq{sigfullk1} we see how functions $C_{1,2}^\kappa$ are defined in terms of these moments.
The result in \eq{sigHadxi} is a \textit{model independent} statement derived from an effective field theory analysis, and a places strong constraint on any hadronization model. 
The leading hadronization corrections are described by the three ${\cal O}(\Lambda_{\rm{QCD}})$ constants that only depend on the quark or gluon flavor of the jet, but are independent of $\zcut$, $\beta$, $R$, $p_T$ and $\eta$. 
This implies that one can in principle constrain the parameters $\{\Ok, \Uka, \Ukb\}$  by comparing experimental data on groomed jet mass and theoretical predictions.

The statement of \eq{sigHadxi} is that all the hadronization power corrections can be descried in terms of these constants, whereas the $r_g$-weighted cross sections multiplying these numbers are perturbatively calculable and account for the dependence of these power corrections on grooming and kinematic parameters.
The nature of power corrections is clearly more involved than compared to the ungroomed jets. This mainly results from the dynamical nature of the groomed jet radius $R_g$ in contrast to a fixed jet radius $R$ in the ungroomed jets, and the requirement to pass the soft drop test at each stage of de-clustering. As a consequence, the entire power correction has both perturbative and nonperturbative components. The perturbative component, involving doubly differential cross sections, is responsible for describing the dynamics of the $R_g$ distribution, whereas the nonperturbative component is parameterized by three unknown ${\cal O}(\Lambda_{\rm{QCD}})$ constants at leading order for each of $\kappa=q,g$.
 The two terms that are added to the parton level cross section in \eq{sigHadxi} represent the ``shift'' and ``boundary'' nonperturbative corrections resulting from modification in the jet-mass measurement as well as change in the outcome of the soft drop test due to hadronization. 
 
The above factorization is valid in the LL strong ordering limit of perturbative collinear-soft (c-soft) emissions. This captures the dominant ``two-pronged'' configuration of the collinear and c-soft subjet that stop the soft drop and determine the groomed jet radius $R_g$.
The `$\ldots$' in \eq{sigHadxi} represent further subleading power corrections which include effects that are suppressed by higher powers in $\Lambda_{\rm{QCD}}$, and corrections at ${\cal O}(\Lambda_{\rm{QCD}})$ where strong ordering is not obeyed and higher-pronged configurations of c-soft subjets and collinear subjet contribute. The latter is thus an NLL effect. Nevertheless, similar to the expansion in number of resolved and ordered soft subjets in resummation of non-global logarithms~\cite{Larkoski:2015zka}, the two-pronged configuration of the ``LL factorization'' captures the leading nonperturbative power corrections, and is sufficient for the precision studies here. 
As we will show below, the sizes of these hadronization corrections are comparable to NNLL perturbative uncertainty, which impacts the precision to which these nonperturbative effects can be constrained. Hence, any further subleading effects not accounted in \eq{sigHadxi} are well beyond our current sensitivity. We also note that typically in analytical models of hadronization corrections, such as in \Ref{Dasgupta:2007wa}, effects of hadron masses are ignored and any ``universal nonperturbative parameters'' are understood with additional ${\cal O}(1)$ fudge factors. In our case, however, the factorization in \eq{sigHadxi} is more general, and the form of the NP corrections shown above remains the same even if hadron masses are explicitly taken into consideration because scaling related to $R_g$ is governed solely by perturbative dynamics.\footnote{However, the hadron mass effects do complicate the relation between the NP constants that appear in \eq{sigHadxi} from those in other groomed observables, for example, angularities. See  Ref.~\cite{Mateu:2012nk} for how hadron mass effects are incorporated in operator-based analyses.}

The two terms in \eq{sigHadxi} capture dependence on the groomed jet radius analogous to how the jet radius $R$ appears in the expressions for nonperturbative power corrections to (ungroomed) jet mass and jet $p_T$ respectively~\cite{Dasgupta:2007wa,Stewart:2014nna}. 
For ungroomed jets, the nonperturbative contribution to the jet mass $\Omega_{1\kappa}(R)$ is proportional to $R$, which for groomed jets is now replaced by $r_g$ times $\Ok$ in \eq{sigHadxi}.
The reason a term analogous to jet $p_T$ power correction is present is because for groomed jets, the recoil of the collinear soft subjet due to hadronization can change the outcome of the soft drop test relative to a reference parton level configuration, and is of the same order as the effect of shift in the jet mass. The recoil includes both shift in the c-soft subjet $p_{T,cs}$ (ignoring the subleading effect on the jet $p_T$) as well as a change in its direction relative to the collinear subjet, the parton level $r_g = R_g/R$. These two aspects of c-soft subjet recoil are accounted by the parameters $\Uka$ and $\Ukb$ respectively. 
Thus, the boundary correction includes a $1/r_g$ dependence familiar from the $1/R$ correction appearing in the jet $p_T$ shift due to hadronization.

Moreover, the boundary effect only appears for c-soft subjets that \textit{barely} pass or fail soft drop. As a result, the change in the parton level $r_g$ due to c-soft subjet recoil captured by $\Ukb$ is relevant only for $\beta > 0$.
The boundary effect can be probed by replacing the inequality in the soft drop condition by equality. To implement this, we evaluate the double differential cross section using a shifted version of the soft drop condition (for $pp$ collisions):
\begin{align}
	\overline \Theta_{\rm sd}(\veps) \equiv \Theta\bigg( \min\{z_i, z_j\} - \xi_0 \Big(\frac{\Delta R_{ij}}{R}\Big)^\beta + \xi_0 \veps \bigg)
   \,.
\end{align}
Taking the $\veps$ derivative then turns this into a $\delta$-function constraint. The \textit{weight function} $w_{\rm max}(\xi, r_g) = 1$  in the region where the $r_g \lesssim r_g^{\rm max}(\xi)$ defined in \eq{psiMaxDef}.  This condition ensures that there is a distinct collinear-soft subjet that stops the groomer. The contribution from the  region where $r_g \ll r_{g}^{\rm max}(\xi)$ is a subleading NLL effect, which must be, however, suppressed by hand using this weight function because of the $1/r_g$ enhancement. 

\begin{figure}[t!]
	\centering
	\includegraphics[width=0.49\textwidth]{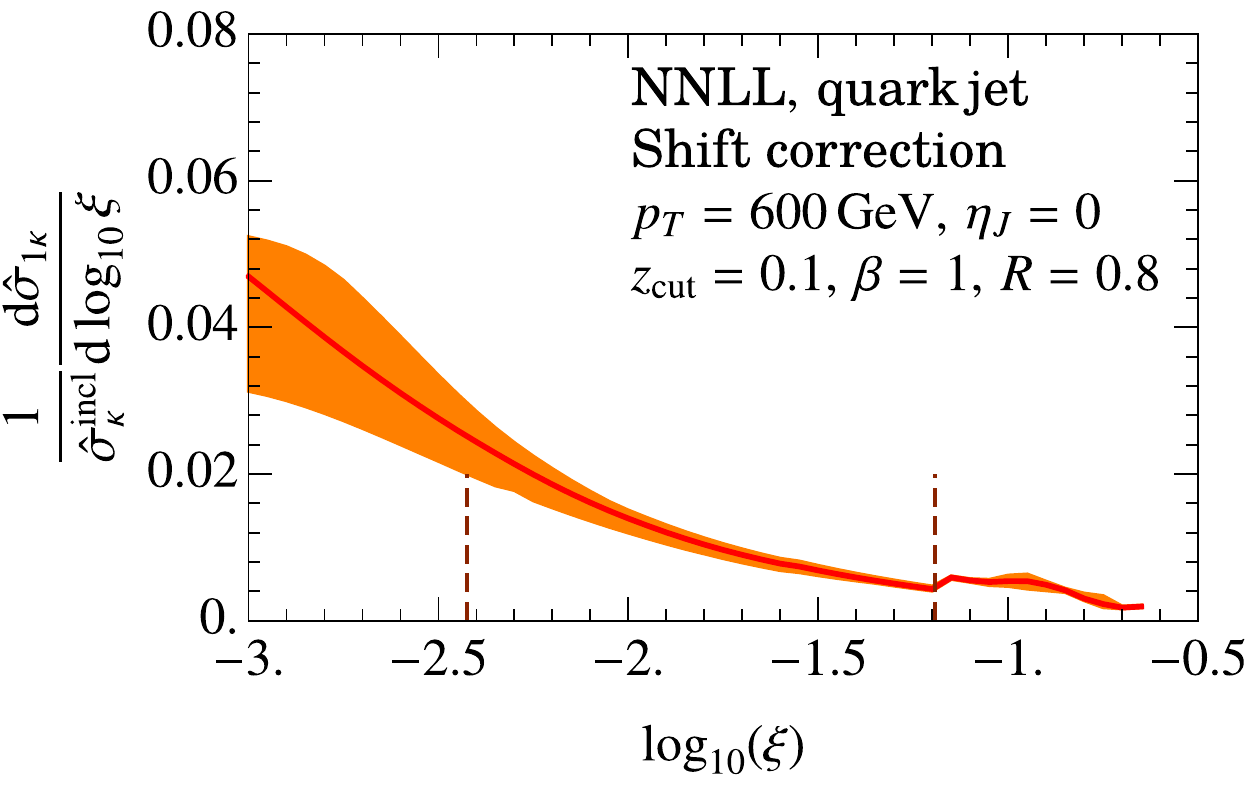}
	\includegraphics[width=0.49\textwidth]{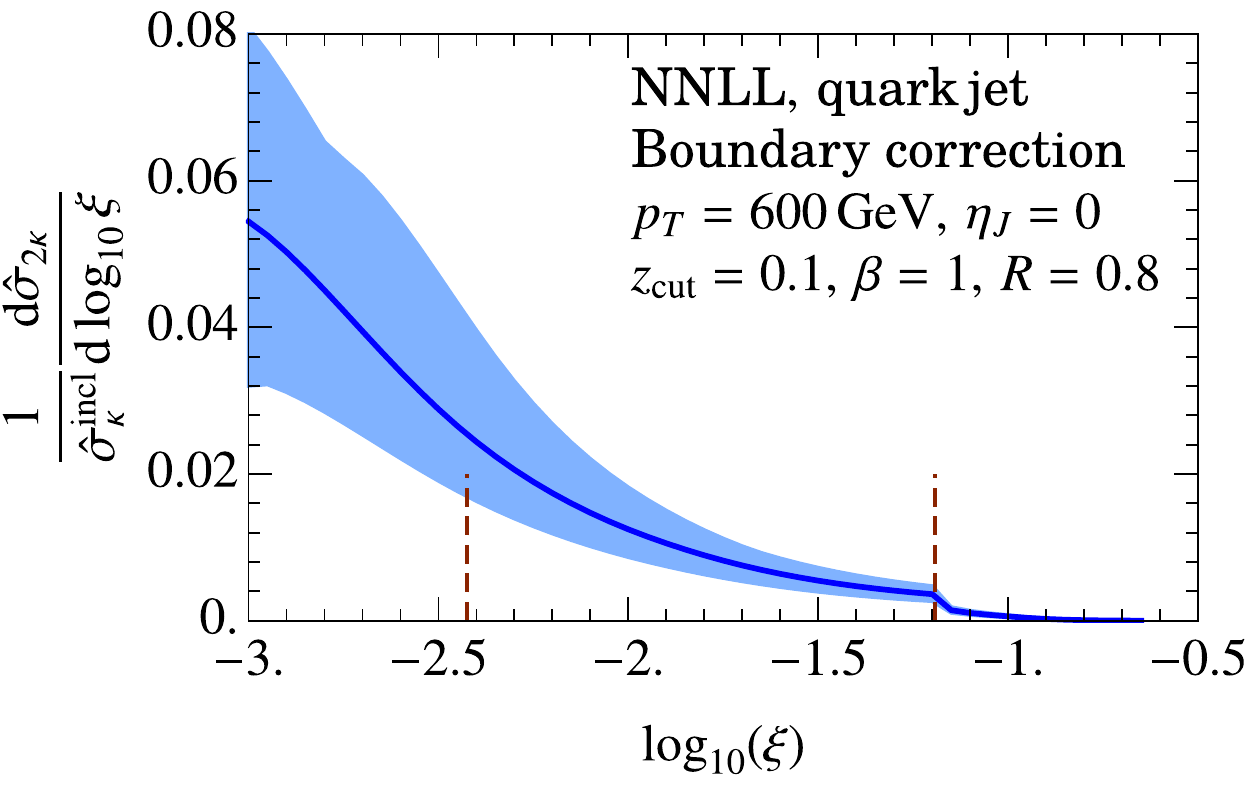}
	\includegraphics[width=0.49\textwidth]{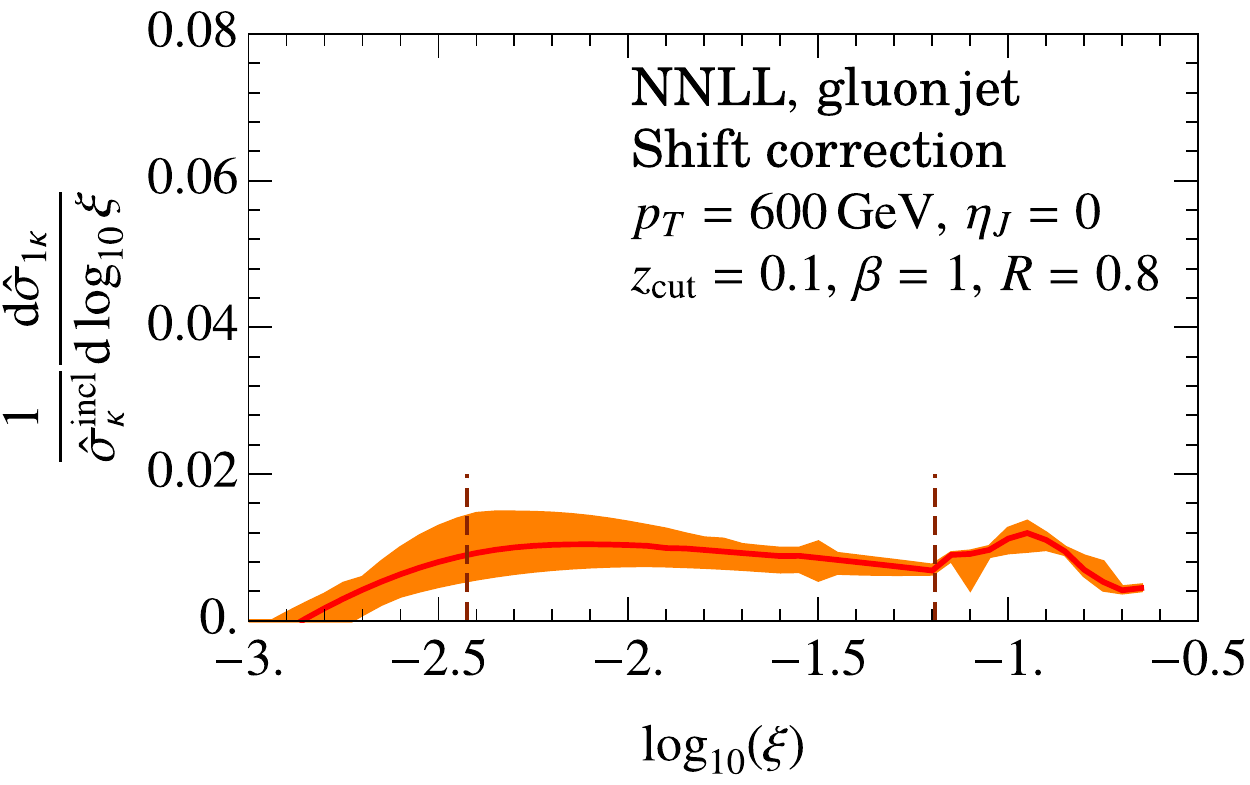}
	\includegraphics[width=0.49\textwidth]{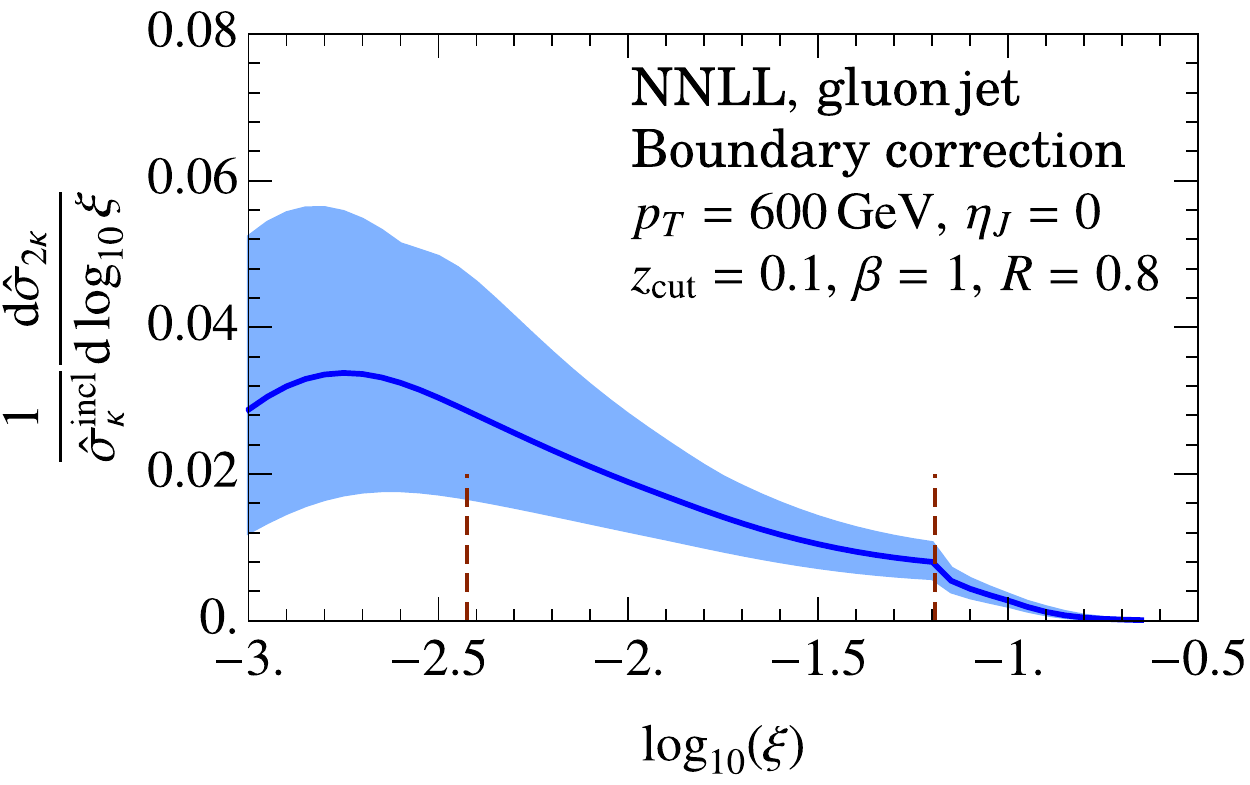}
	\caption{Weighted cross sections for shift and boundary power corrections. Vertical lines denote the extent of the SDOE region.}
	\label{fig:ciweighted}
\end{figure}
These moments being perturbative quantities can be in principle computed to arbitrarily high accuracy. 
Despite the fact that beyond LL factorization in \eq{sigHadxi} there are subleading corrections arising from higher pronged configurations, a more precise computation of these weighted cross section allows one to accurately capture the dominant geometrical effect of soft drop, and account for some of the higher order terms. These moments were first computed in \Ref{Hoang:2019ceu} at LL in the coherent branching framework.
In \Ref{Pathak:2020iue} the problem of computing the $r_g$ moments as a function of jet mass was cast into a computation of the double differential cross section. \Ref{Pathak:2020iue} developed the effective field theory for doubly differential cross section and used it to compute these moments at NLL$'$ accuracy. We have already reviewed and utilized some aspects of this EFT setup in \secn{mJPC}.
The calculation was further improved in \Ref{AP} 
by matching the doubly differential cross section to the ungroomed region, including an improved treatment of the cusp region along the lines of the analysis presented in \secn{pc}, and extending the resummation to NNLL accuracy.
In what follows we will make use of results from \Ref{AP}.

To assess the jet-mass dependence and the size of the NP power corrections in \eq{sigHadxi}, we write
\begin{align}\label{eq:sigLog10x}
	\frac{1}{\sigma_{\kappa}} \frac{\df \sigma_\kappa}{\df \xi} = 
	\frac{1}{\hat \sigma_{\kappa}}
	\frac{\df \hat{\sigma}_\kappa}{\df \xi} +
	\frac{\Ok}{1\,{\rm GeV}}	 \frac{1}{\hat \sigma_{\kappa}}
	\frac{\df \hat{\sigma}_{1\kappa}}{\df \xi}
	+ \frac{ \Uka +  \beta\Ukb }{1\,{\rm GeV}}	\frac{1}{\hat \sigma_{\kappa}}
	\frac{\df \hat{\sigma}_{2\kappa}}{\df \xi} \, ,
\end{align}
and the coefficients of the NP parameters, $(1/\hat \sigma_{\kappa})
	{\df \hat{\sigma}_{i\kappa}}/{\df \xi} $ for $i=1,2$, are defined by the functions that appeared in \eq{sigHadxi}.
The weighted cross sections for the two effects  are shown in \fig{ciweighted} for quark and gluon jets. The orange and blue bands denote the perturbative uncertainty associated with the NNLL calculation, ignoring subdominant NGL effects. The vertical lines mark the extent of the SDOE region given in \eq{SDNPprecise}.  Comparing with the results in \fig{profvar} we see that these weighted cross sections in the SDOE region are about factor of 10 smaller than the leading parton level cross section. Thus, as expected, for ${\cal O}(1\, {\rm GeV})$ values of these parameters, these corrections are small in the SDOE region. 

\section{Sensitivity of matched jet-mass cross section to \texorpdfstring{$\bm{\alpha_s}$}{L}}
\label{sec:analysis}

In this section, we consider how the various theoretical uncertainties translate into uncertainty on the extraction of $\as$. 
We will state results for two ways of normalizing the cross section, as shown in \eq{NormChoice}, namely either using the inclusive cross section $\sigma_{\rm incl}(p_T, \eta)$ or the cross section obtained by integrating over the fit range of jet-mass values $\sigma_{\rm fitrg}(p_T, \eta,\xi_{\rm SDOE},\xi_{\rm 0}')$. The fit range that we consider for $\as$ analysis is given by
\begin{align}\label{eq:SDNPprecise}
	&\xi \in \big[\xi_{\rm SDOE}, \xi_0'\big] \, ,&
	&\xi_{\rm SDOE} = \xi_{\rm SDNP} \, 5^{\frac{2+\beta}{1+\beta}} \, , & 
\end{align}
where
$\xi_{\rm SDNP}$ was defined in \eq{xiSDNPDef} and $\xi_0'$ in \eq{xi0p}. 
Here the factor of 5 has replaced the strong inequality in \eq{xiSDOEdef}. The choice of 5 is motivated by analyzing  the impact of perturbative uncertainty in the shift and boundary corrections shown in \fig{ciweighted}. For $\xi < \xi_{\rm SDOE}$ defined in \eq{SDNPprecise}, the perturbative uncertainty grows and leads to poor description of these two effects in the SDOE region.

Since it will be useful to consider the individual quark and gluon components of the cross section for these two choices of normalization, we also utilize \eq{sigdecomp} and write
\begin{align}
    \frac{1}{\sigma_{\rm incl}}\: \frac{\df^3 \sigma}{\df p_T \df \eta\df \xi} 
  &=  \frac{x_q}{\sigma_{q}^{\rm incl}} \frac{\df \sigma_q}{\df \xi} +  \frac{x_g}{\sigma_{g}^{\rm incl}} \frac{\df \sigma_g}{\df \xi} 
  \,, \\
 \frac{1}{\sigma_{\rm fitrg}(\xi_{\rm SDOE},\xi_{\rm 0}')}\: \frac{\df^3 \sigma}{\df p_T \df \eta\df \xi} 
  &=  \frac{x_q}{\sigma_{q}^{\rm fitrg}} \frac{\df \sigma_q}{\df \xi} +  \frac{x_g}{\sigma_{g}^{\rm fitrg}} \frac{\df \sigma_g}{\df \xi} 
  \,,
\end{align}
where $\sigma_{\rm incl}$ and $\sigma_{\rm fitrg}$ were defined in Eq.\eqref{eq:sigma_norms},
the $\sigma_\kappa^{\rm incl}$ are defined such that
\begin{equation}
    \int d\xi \frac{1}{\sigma_\kappa^{\rm incl}} \frac{\df\sigma_\kappa}{\df\xi} = 1\,, \qquad \qquad \int_{\xi_{\rm SDOE}}^{\xi_{\rm 0}'} d\xi \frac{1}{\sigma_\kappa^{\rm fitrg}} \frac{\df\sigma_\kappa}{\df\xi}=1 \,.
\end{equation}

To summarize the different scales, we have, in order from smallest to largest
\begin{itemize}
    \item $\xi_{\rm SDNP}$ defined in Eq.~\eqref{eq:xiSDNPdef} is where the nonperturbative effects become order 1. 
    \item $\xi_{\rm SDOE}$ defined in Eq.~\eqref{eq:SDNPprecise} is the lower end of the fit range, where nonperturbative corrections  can be treated in a systematic power expansion.
    \item $\xi_0'$ defined in Eq.~\eqref{eq:xi0p} is the location of the cusp, where soft wide-angle emissions become relevant. It is the upper end of our fit range.
    \item $\xi_0$ defined in Eq.~\eqref{eq:xi0def} is the transition point where the collinear-soft and global-soft functions merge into a single soft function.
    \item $\xi_{\rm FO}=x_3$, with $x_3$  defined in \eq{frunmJ}, is the point where fixed order perturbation theory is accurate.  For $\xi\ge \xi_{\rm FO}$ there are no large logarithms.
\end{itemize}

\subsection{Quark- and gluon-jet dependence on \texorpdfstring{$\alpha_s$}{L}}
\label{sec:aSdep}

\begin{figure}[t!]
	\centering
    \includegraphics[width=\textwidth]{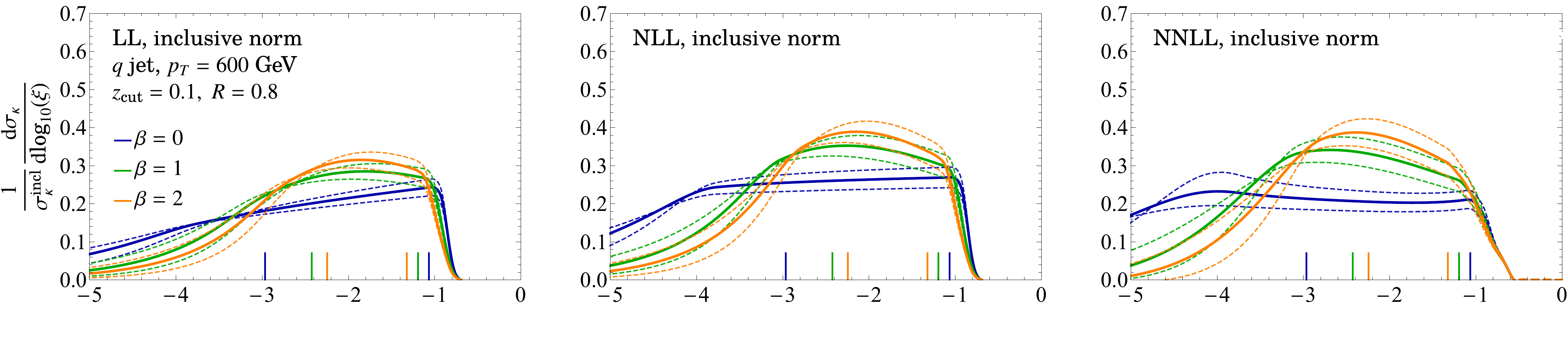}
    \\[0pt]
\includegraphics[width=\textwidth]{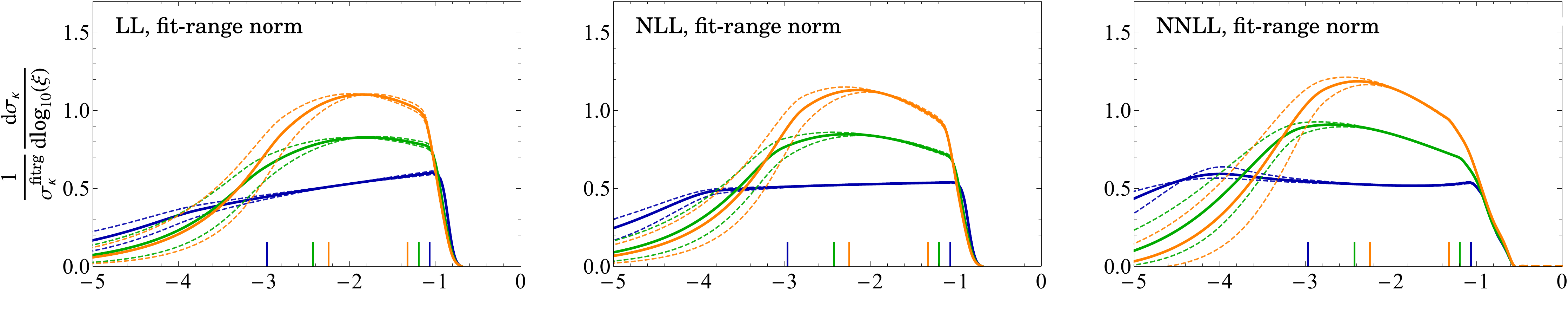}
    \\[10pt]
    \includegraphics[width=\textwidth]{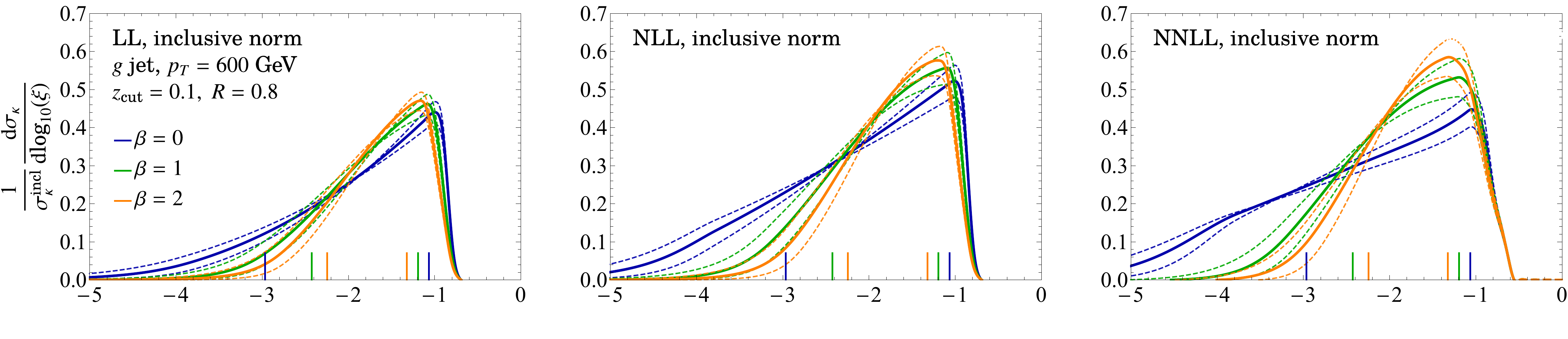}
    \\[0pt]
        \includegraphics[width=\textwidth]{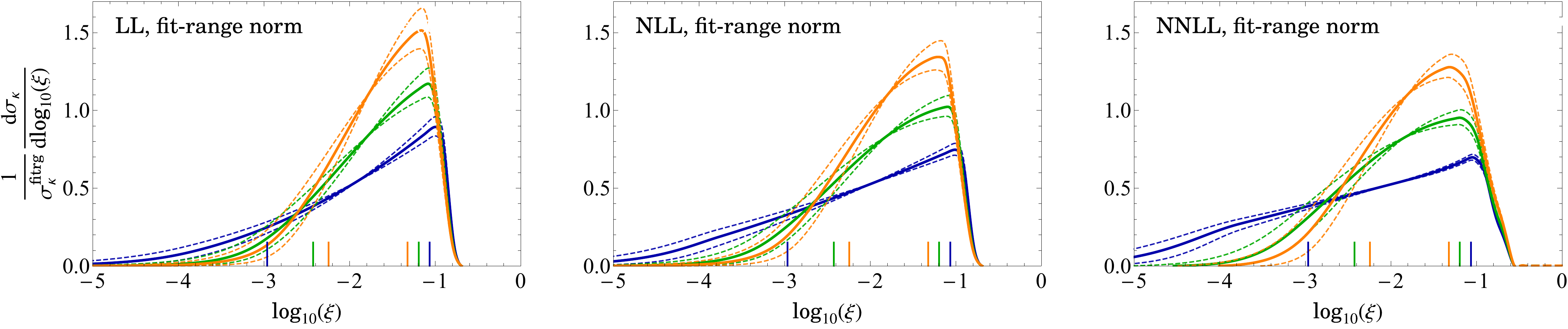}
	\caption{The $\alpha_s$ dependence of the normalized jet-mass spectrum of quark jets (top two rows) and gluon jets (bottom two rows), for various logarithmic orders (columns) and with different normalizations (rows 1 and 3 use the inclusive cross section normalization, while rows 2 and 4 are  normalized in the fit range). 
	Solid curves show matched jet-mass spectrum for $\alpha_s(m_Z)=0.118$, while dashed curves correspond those with a $\pm10$\% variation in $\alpha_s(m_Z)$. The start and end points of the fit range in which the curves are normalized is indicated with vertical lines, for each value of $\beta$. 
    }
	\label{fig:alpha_slope}
\end{figure}

Before analyzing the theoretical uncertainty on $\alpha_s$ measurement using soft-drop jet mass, let us pause to discuss how the jet-mass spectrum provides sensitivity to $\alpha_s$. For example, a \textit{leading logarithm} approximation for $\beta=0$ gives (cf.~\eq{SDOperator}),
\be 
    \frac{d\sigma^\kappa_{\rm sd}}{d \log(\xi)}
    \propto \text{exp} \Big[-\frac{\alpha_s(\mu) C_\kappa}{\pi} a_\kappa \log \xi \Big]
    =
    \xi^{-\frac{\alpha_s(\mu) C_\kappa}{\pi} a_\kappa}
    \label{eq:aSestimate}
\ee 
where $a_\kappa$ is a $\xi$-independent constant. 
This estimate shows that the LL formula predicts the slope of the jet-mass spectrum to be proportional to $\alpha_s$, as was pointed out in Ref.~\cite{Proceedings:2018jsb}. One can wonder whether this estimate carries over to NLL and NNLL predictions.

\begin{figure}[t!]
    \centering
    \includegraphics[width=0.49\textwidth]{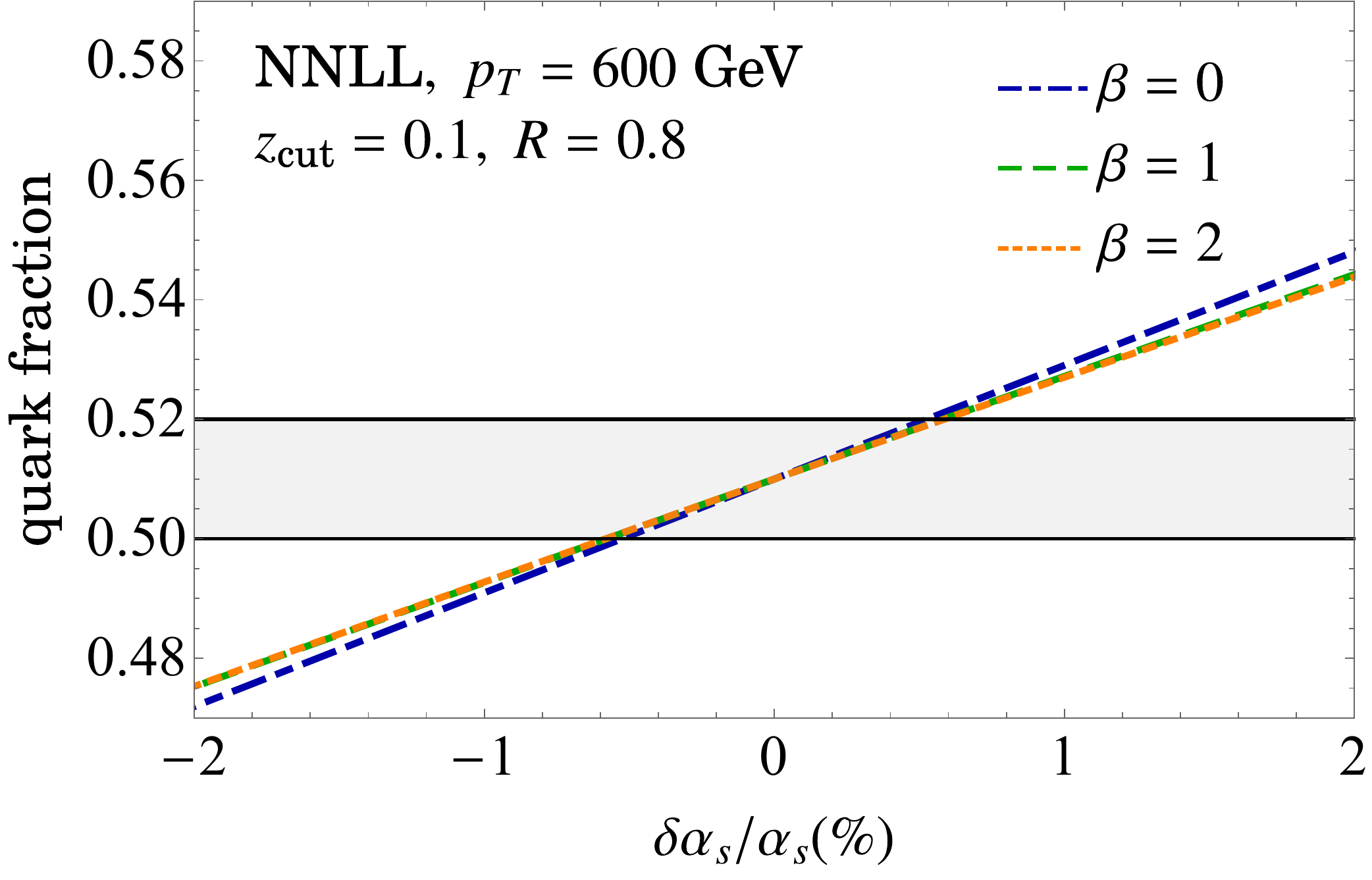}
    \caption{Variation of the quark fraction that would give a similar change as a certain variation in $\alpha_s$. We take the central values to be a 51\% quark fraction and 49\% gluon fractions, and the grey lines represent a one percentage-point variation in the fraction.
    }
    \label{fig:alpha_chi}
\end{figure}

To analyze how the slope of the jet-mass spectrum changes from LL to NLL to NNLL we give results for spectra at these orders in \fig{alpha_slope}. To get a visual idea of $\as$ sensitivity, we also show variations of $\alpha_s(m_Z)$ by $\pm10$\% by the dashed curves around each central solid curve.
For quark jets, we notice a striking feature: while the estimate from Eq.~\eqref{eq:aSestimate} can be argued to roughly hold for the LL curves with $\beta=0$ (varying the slope of the spectrum in the resummation region by $\pm10$\% gives a similar change as varying $\alpha_s(m_Z)$), the approximation is not useful at NLL and beyond, since the jet-mass spectrum flattens out. Thus, normalizing cross sections in the fit region washes out almost all dependence of the quark-jet spectrum on $\as$, and the only sensitivity comes from the area under the curve in this range.  For gluon jets, on the other hand, a linear dependence on $\alpha_s$ persists at higher logarithmic orders.

\subsection{Variation with quark/gluon fraction \label{sec:fracalpha}}

To assess how $\as$ sensitivity changes with variations to the quark and gluon fractions, we calculate the values of $\chi^2$ that are obtained by varying the fraction, and compare with the $\chi^2$ values obtained by varying $\as$, as shown in Figure~\ref{fig:alpha_chi}. The dependence is roughly linear, with the quark fraction satisfying $x_q \approx 0.51 + 2 \frac{\delta \alpha_s}{\alpha_s}$, which shows that a 1\% uncertainty in $\as$ would result from a variation in the quark fraction of roughly 4\%. This value is consistent with the previous observation that gluon jets are more sensitive to the variation in $\as$, from which we arrive at the estimate that any given percentage uncertainty on the quark/gluon fraction propagates to less than half of that value for the corresponding percent uncertainty in $\as$.  
Given the small uncertainty on the quark/gluon fraction discussed in Sec.~\ref{sec:qgfraction}, we find that this is a subdominant uncertainty for the extraction of $\alpha_s$.

\begin{figure}[t]
    \centering
    \includegraphics[width=0.49\textwidth]{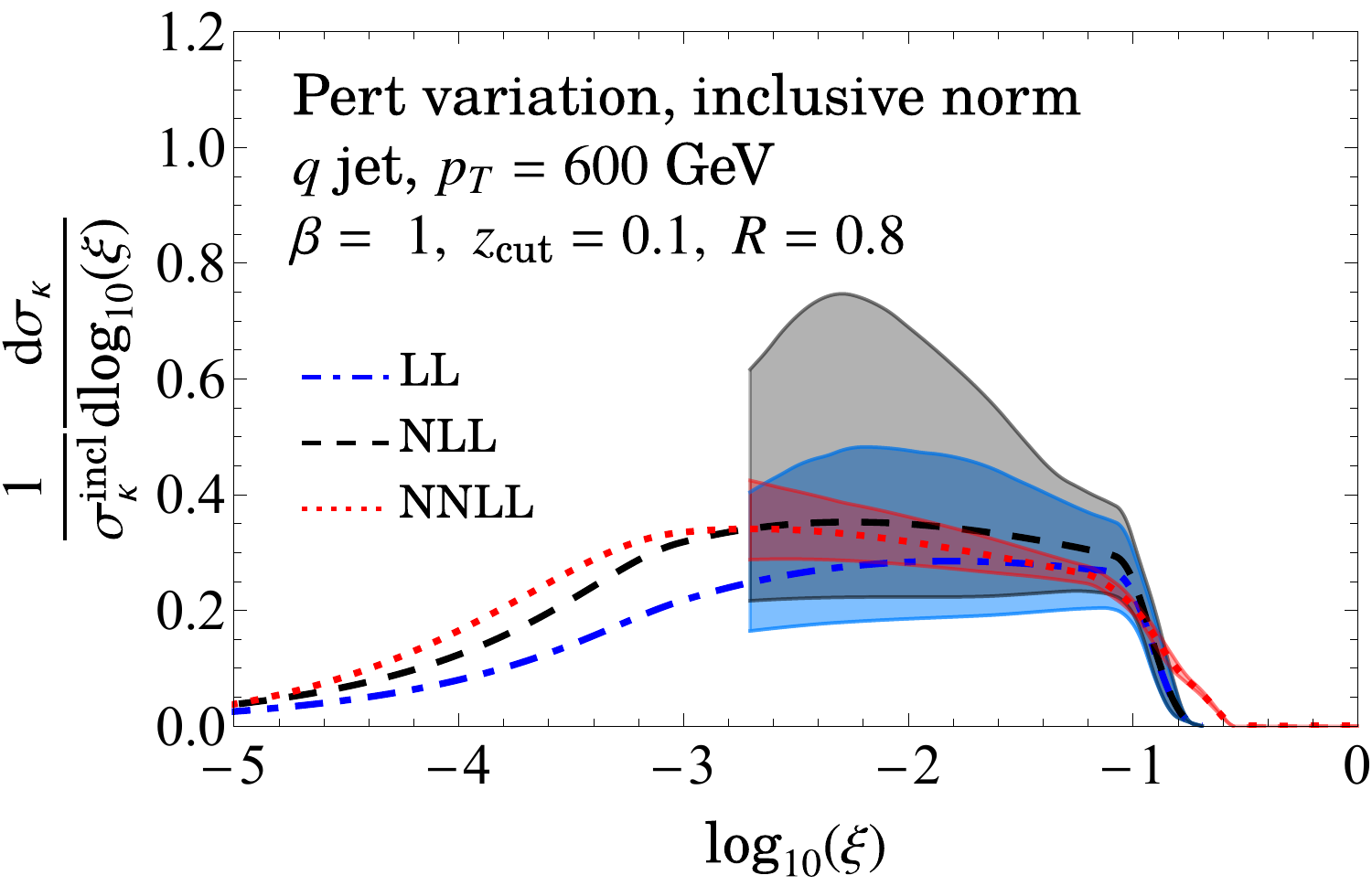}
    \includegraphics[width=0.49\textwidth]{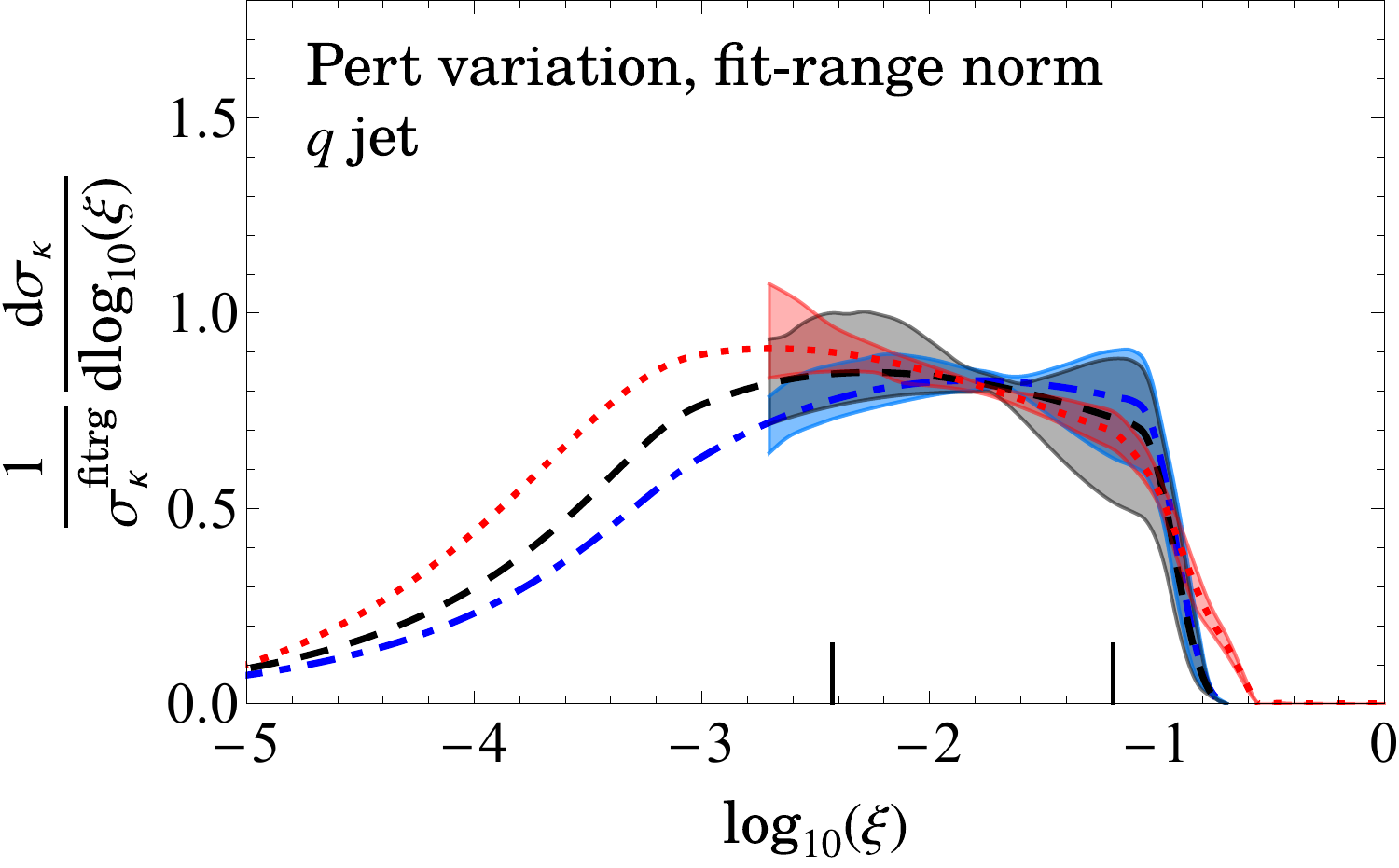}
    \includegraphics[width=0.49\textwidth]{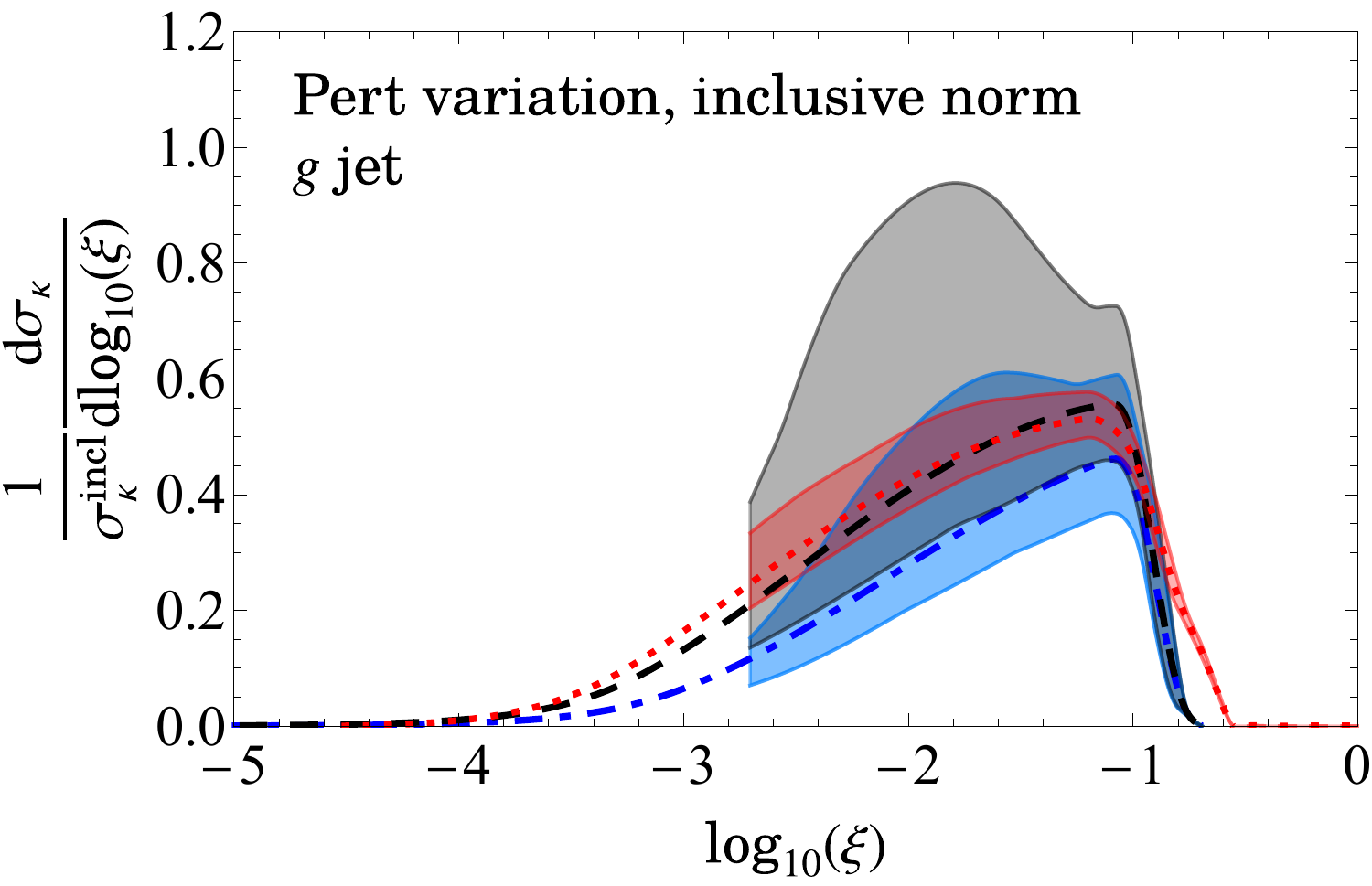}
    \includegraphics[width=0.49\textwidth]{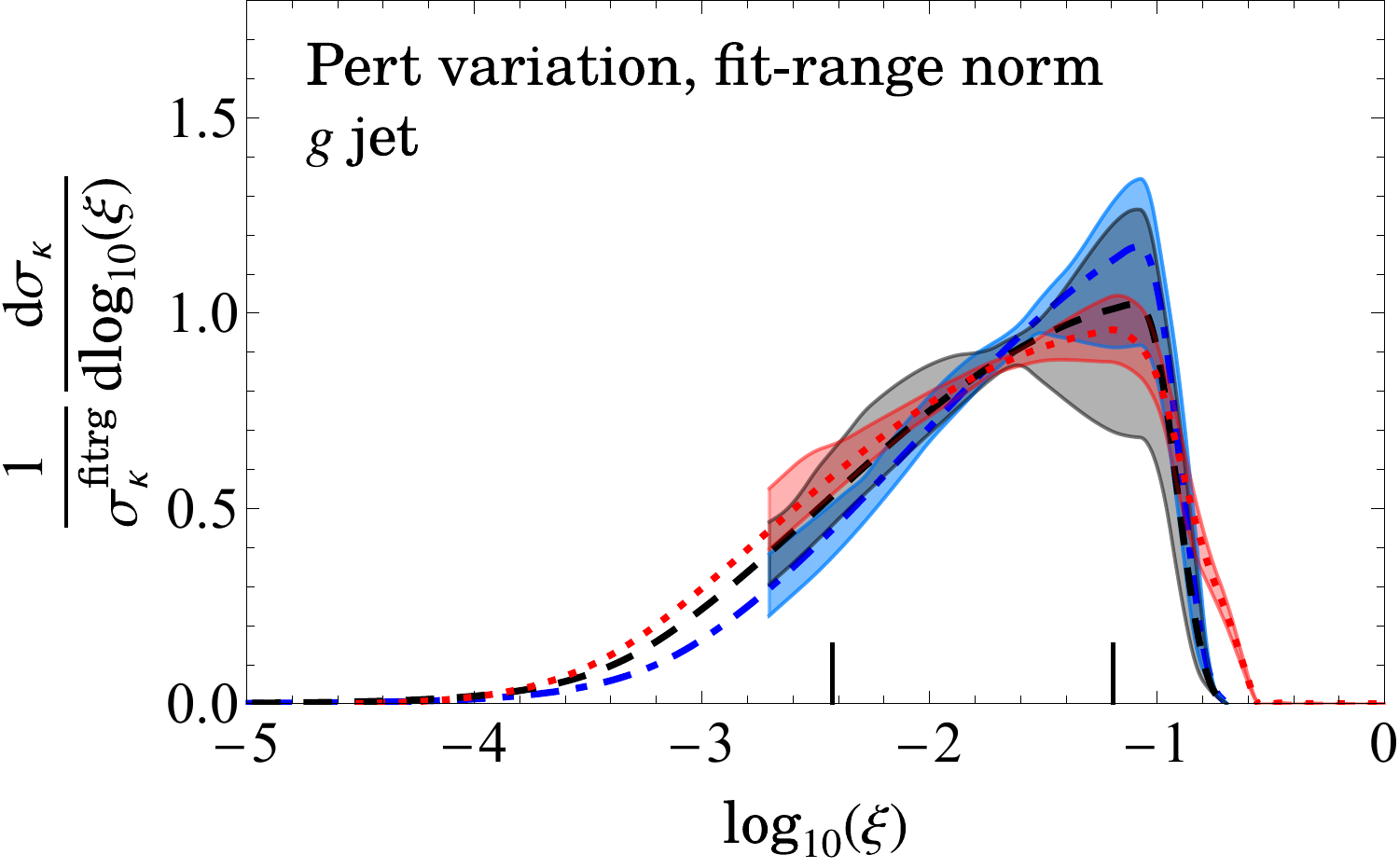}
    \caption{An envelope of the profile variations for quark and gluon jets respectively, at LL, NLL and NNLL. The perturbative uncertainties are only shown just beyond the fit range considered for $\as$-sensitivity analysis. The pinching in the uncertainties in the plots in the right column results from the fit range normalization.}
    \label{fig:profile_vars_all}
\end{figure}

\subsection{Variation with profile parameters}

To estimate the uncertainty in the theoretical calculation, we vary scales using profiles as described in Sec.~\ref{sec:profvar}. In \fig{profile_vars_all}  we show the combined perturbative uncertainty, obtained by taking the envelope of the curves obtained from random profile parameter choices from within the ranges given in \eq{profrange}.
We observe that, as expected, the uncertainties decrease as the order increases, and that the central NNLL curves are within the NLL uncertainty bands in the jet-mass region to be used for the $\alpha_s$ fits.

\begin{figure}[t]
    \centering
    \includegraphics[width=0.49\textwidth]{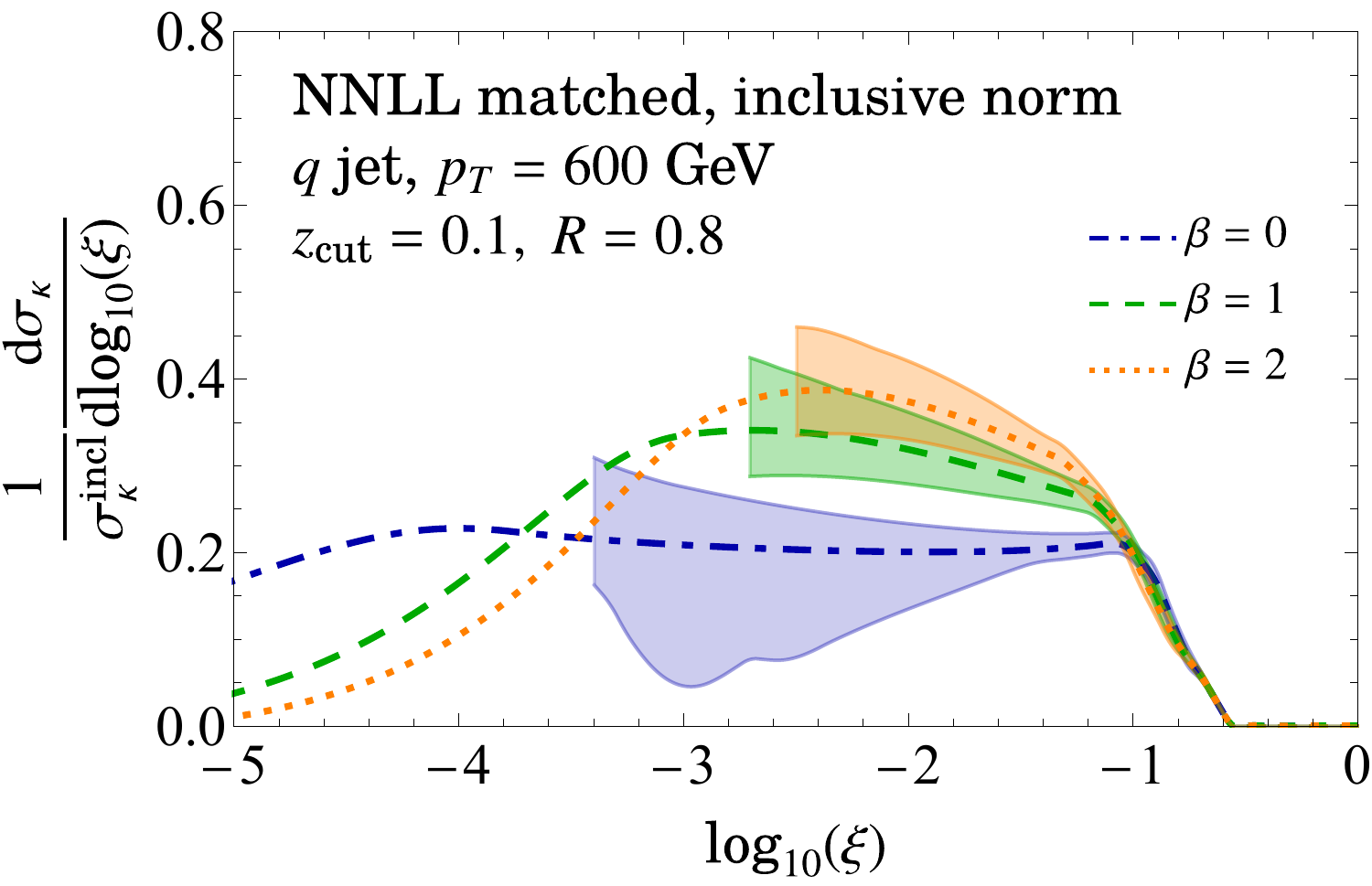}
    \includegraphics[width=0.49\textwidth]{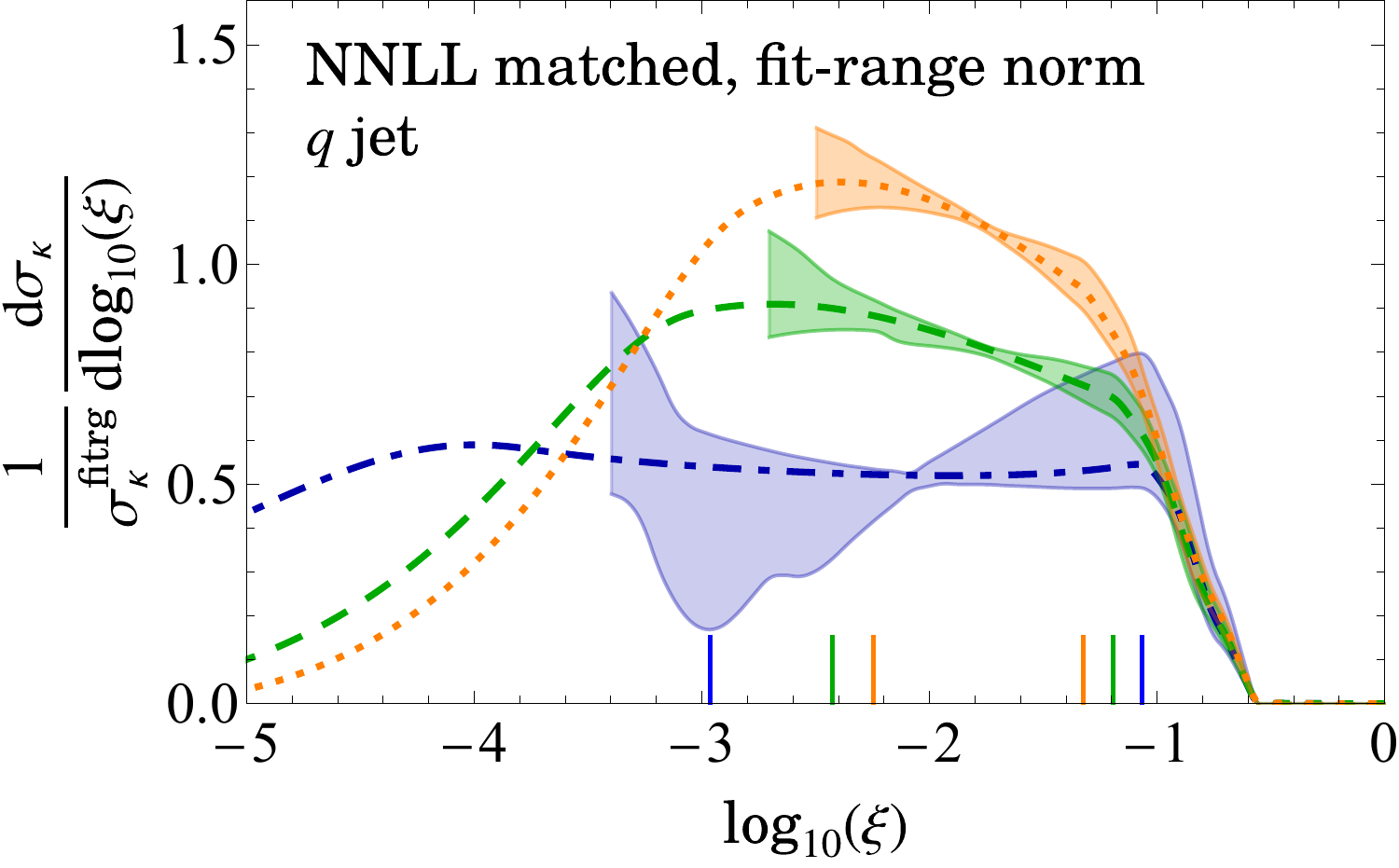}
    \includegraphics[width=0.49\textwidth]{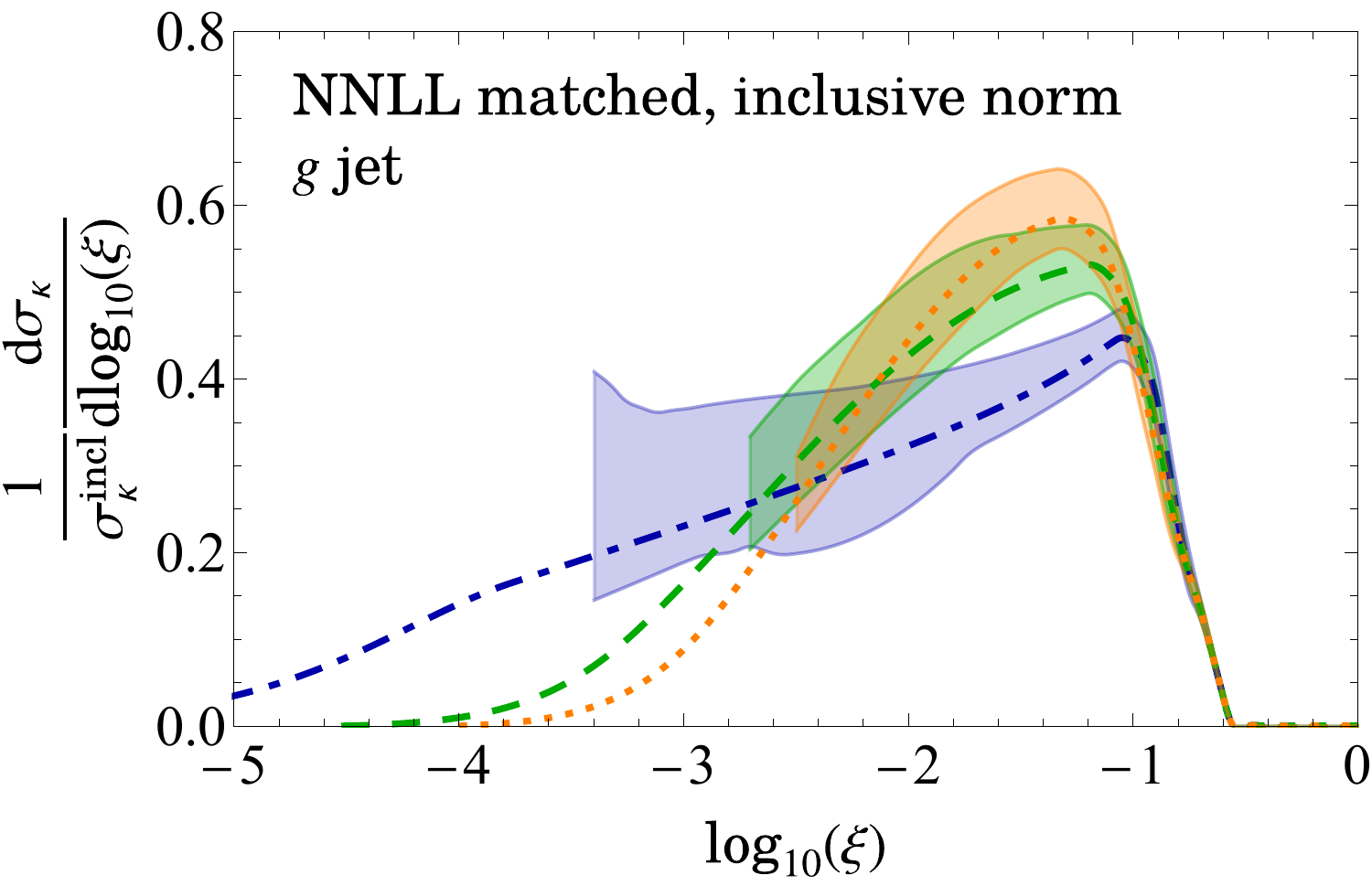}
    \includegraphics[width=0.49\textwidth]{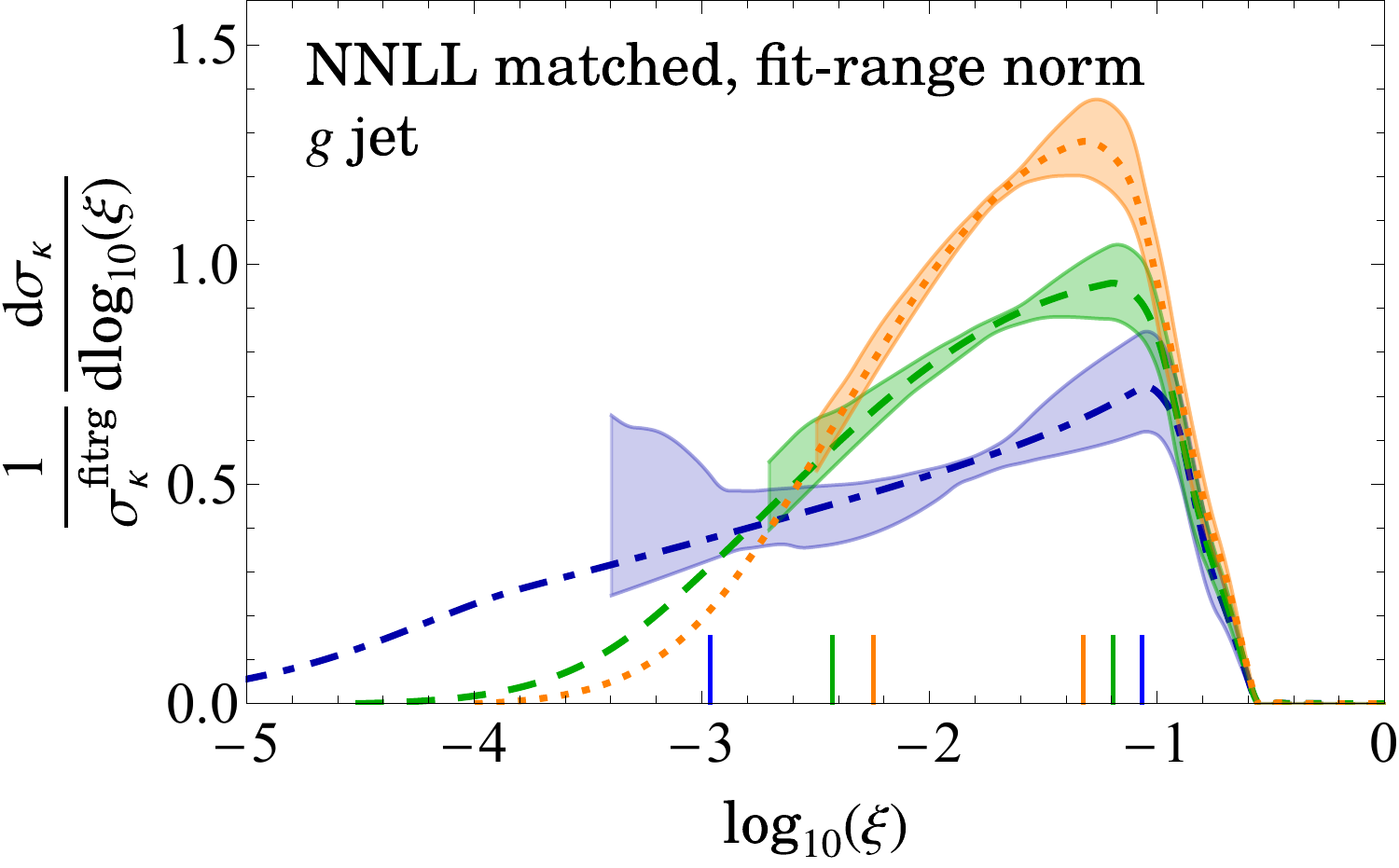}
    \caption{An envelope of the profile variations for quark and gluon jets, for different values of $\beta$.}
    \label{fig:profile_vars_NNLL}
\end{figure}

In Figure~\ref{fig:profile_vars_NNLL}, we show the NNLL jet-mass cross section for $\beta = 0,1,2$, with the bands corresponding to the envelope of results from the random scan of profile parameters, as specified in Sec.~\ref{sec:profvar}. As in \fig{profile_vars_all}, in the right column we show spectrum normalized in the fit range (shown as small vertical lines). As a result of fit-range normalization, the perturbative uncertainties are significantly reduced. 
However, we will see below that it also dramatically decreases the sensitivity to $\as$, especially for quark jets where the NNLL spectrum is much flatter in the fit range. 
We notice that as we move to smaller jet masses towards the nonperturbative region, the bands grow in size. This is due to small scales of $\as$ probed in the region.

\subsection{Variation with the nonperturbative parameters}

\label{sec:varyNP}
We note that the NP factorization is valid only in the SDOE region. At even smaller jet-mass values,
the nonperturbative contributions become large, since
corrections in higher powers of $\Lambda_{\rm{QCD}}$ are no longer suppressed,
and both perturbative and nonperturbative components must be treated on equal footing.
As described in \secn{sdDef}, this happens for jet masses $\xi < \xi_{\rm SDNP}$ where the collinear-soft subjet itself becomes nonperturbative. To stay well away from the SDNP region, we require the condition given in \eq{xiSDOEdef}.
More specifically, we implement NP corrections only in $\xi \in [\xi_{\rm SDOE}, \xi_0']$. We take the factor of 5 in \eq{xiSDOEdef} as a substitute for the strong inequality.  

Ideally, we would like to fit the cross section to the nonperturbative parameters $\Ok$, $\Uka$ and $\Ukb$. This fit would, however, involve seven different parameters;  $\alpha_s$ along with three nonperturbative constants for quark jets and three for gluon jets. Such a fitting of the jet-mass spectrum to seven parameters, using available collider data, would be challenging. 
Given the uncertainty on the experimental data, it does not seem viable to expect an extraction of all seven parameters simultaneously from the SDOE region alone. A possible solution to this problem can be to make use of the SDNP region. It will be very interesting to work out the relation between the NP parameters in the SDOE region and the shape function that appears in the SDNP region~\cite{Frye:2016aiz,Hoang:2019ceu}. Such an exercise will prove immensely useful in being able to better constrain the NP parameters, as the hadron and parton level curves differ significantly in the SDNP region. 
Note that this cannot be achieved by simply extending the SDNP shape function to the SDOE region as has been considered earlier~\cite{Frye:2016aiz,Kang:2018jwa}, see Ref.~\cite{Hoang:2019ceu}.
We leave this exploration to future studies. 

One might consider varying the nonperturbative parameters in the GeV range, thereby covering all possible values of these parameters on the order of magnitude of $\Lambda_{\rm{QCD}}$, as a probe for the uncertainty induced by not including the nonperturbative corrections. However, we found that such variations bury all predictions in an immense uncertainty. This fact can be understood from Eq.~\eqref{eq:sigHadxi}: the cross section is linear in the nonperturbative parameters and hence an uncertainty band from a variation of $~2$ GeV will be roughly 10 times larger than a variation of $~200$ MeV. 

Furthermore, a blind variation of these parameters is somewhat naive. In \Ref{Hoang:2019ceu}, by considering  various combinations of parton showers and MC hadronization models, it was observed that these parameters exhibited strong correlations. For example, while the shift correction is expected to be positive, the sign of $\Uka$ parameter turns out to be negative.  While the sign of this constant is not constrained from the derivation of the NP factorization, this observation is consistent with the negative shift observed in the $p_T$  of the ungroomed jets~\cite{Dasgupta:2007wa}.\footnote{Roughly speaking, the negative sign in the jet $p_T$ results from a recoil effect and a consequence of accommodating an additional NP particle for a  fixed energy budget of the hard scattering.}  
Additionally, as can be seen from the shift and boundary corrections in \fig{ciweighted}, the weights have approximately similar size, and the fits to MC hadronization models in \Ref{Hoang:2019ceu} for quark jets also resulted in similar absolute values of $\Oq$ and $\Uqa$ parameters. This cancellation is why the $\beta = 0$ for quark jets in MC simulations has almost negligible hadronization corrections in the SDOE region. This shows that mMDT ($\beta = 0$ SD) does not intrinsically make the groomed jet mass perturbative, but rather that (in MC) the very tiny hadronization corrections result from (an almost accidental) cancellation of two distinct physical effects of shift and boundary corrections. This is also confirmed from a more recent updated  analysis in \Ref{Ferdinand:2022xxx} of these NP corrections using NNLL calculations of the weighted cross sections multiplying these parameters in \eq{sigHadxi}. 

\begin{figure}[t]
    \centering
    \includegraphics[width=0.49\textwidth]{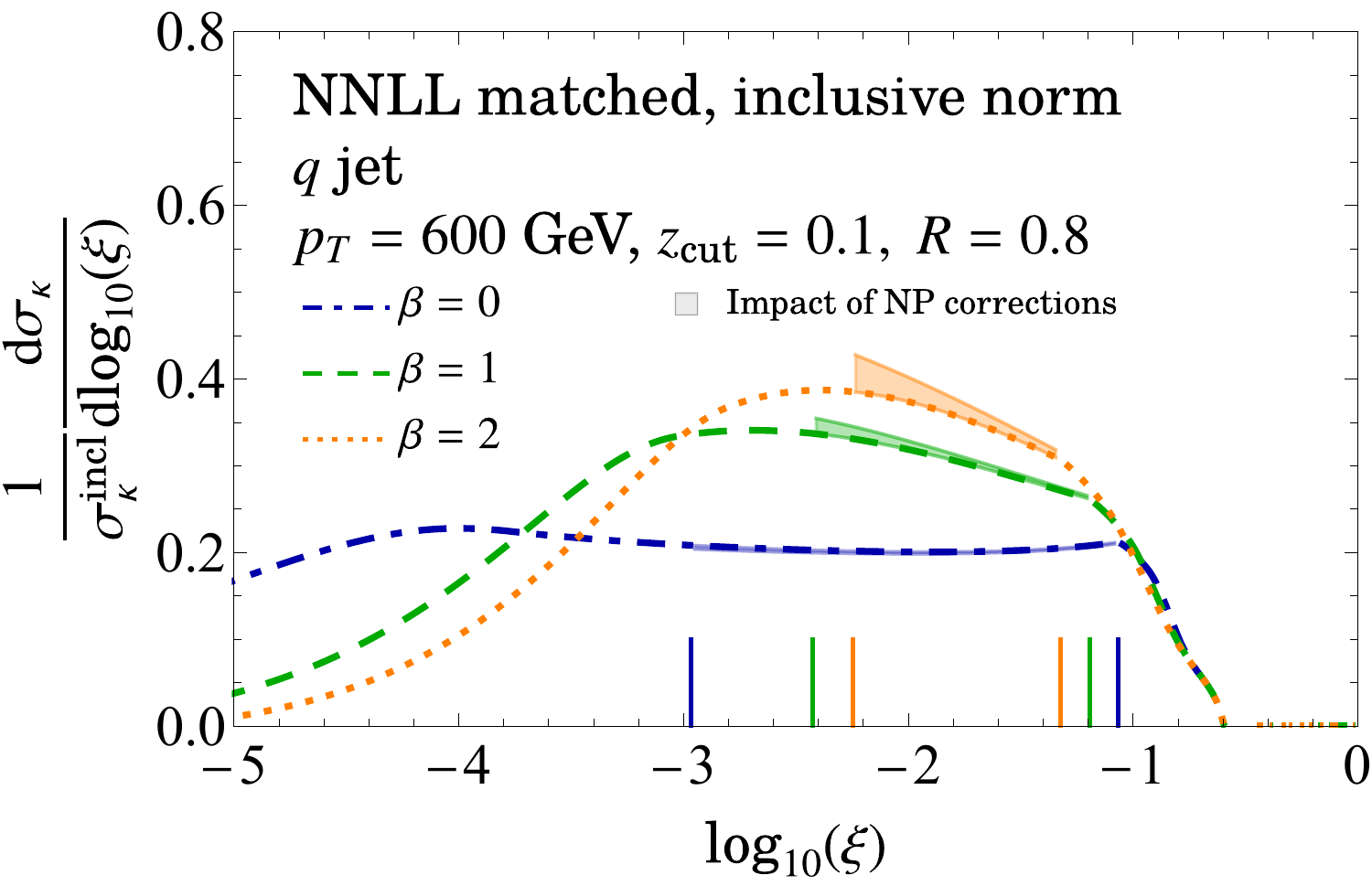}
    \includegraphics[width=0.49\textwidth]{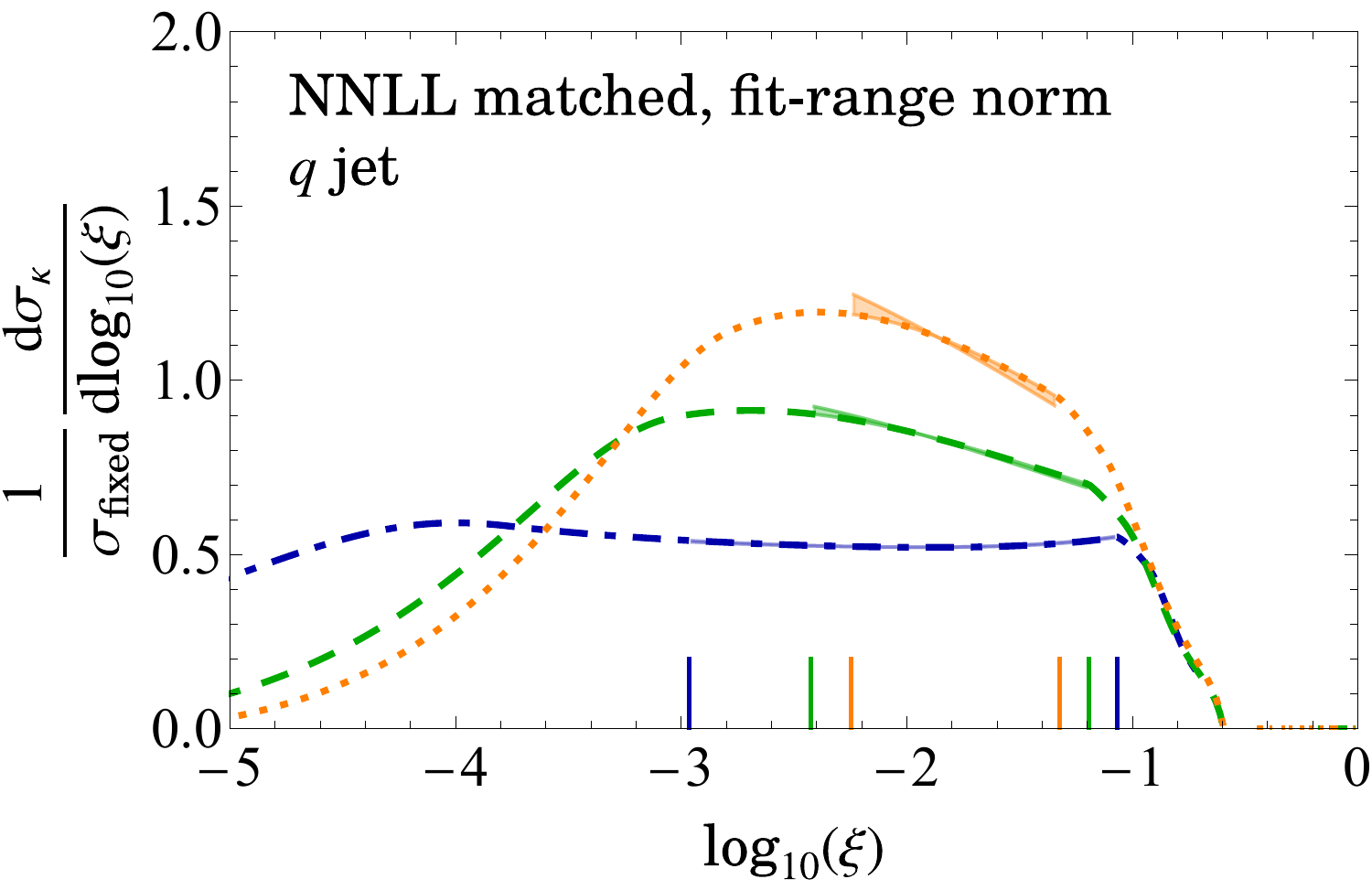}
    \includegraphics[width=0.49\textwidth]{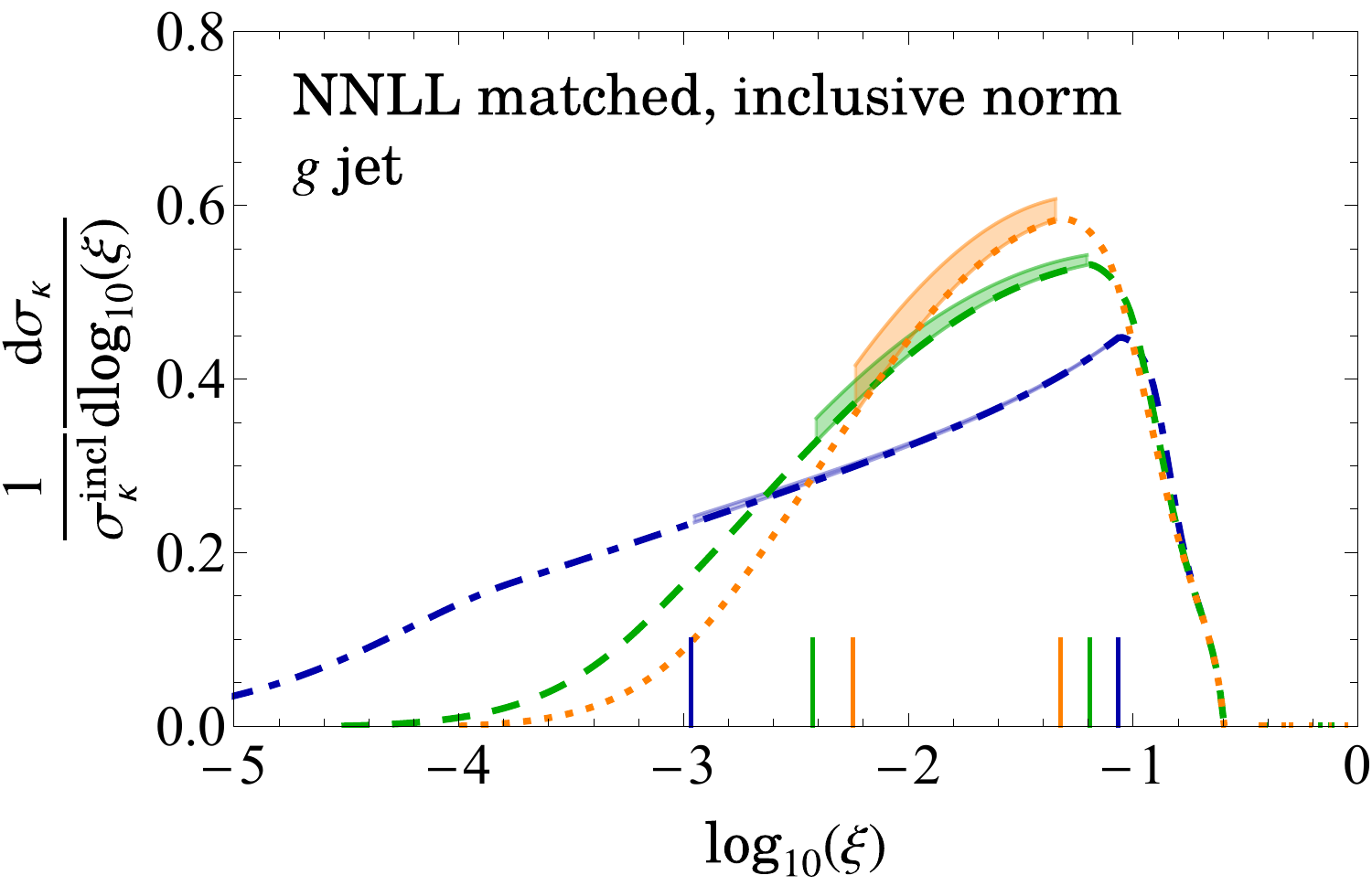}
    \includegraphics[width=0.49\textwidth]{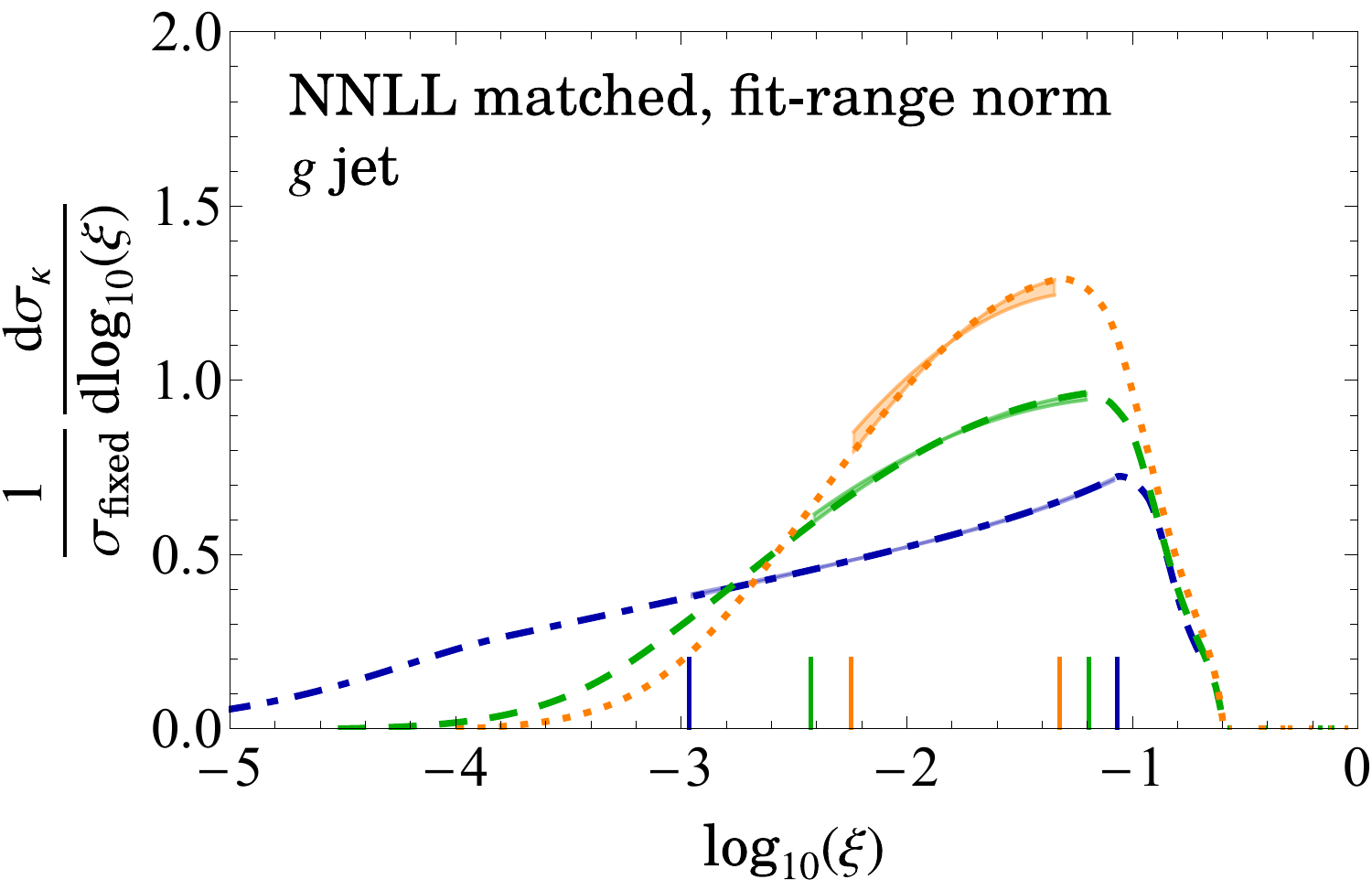}
    \caption{The dependence of the cross section on nonperturbative parameters. The start and end points of the fit range are indicated with vertical lines.}
    \label{fig:NPvar}
\end{figure}

Hence, instead of a blind variation we will incorporate these observations from the above-mentioned studies with hadronization models. Our approach will therefore be to estimate the uncertainty due to the nonperturbative corrections by comparing results without power corrections to results using values for the power correction parameters from
Monte Carlo simulations.
Explicitly, we use the default tune of  \Pythia\texttt{8.3} to fit for the parameters $\Ok$, $\Uka$ and $\Ukb$ following the analysis in \Ref{Ferdinand:2022xxx}:
\\
\begin{align}
\label{pythiavalues1}
    &\Omega_{1,q}^{\circ\!\!\circ}  = 0.55 \,{\rm GeV}\, ,& 
    &\Uqa= -0.73 \,{\rm GeV}\, ,& 
    &\Uqb= 0.90 \,{\rm GeV}  \,, &
    &\text{for quarks,}&
    \\
    &\Og  = 0.91\,{\rm GeV} \, ,&
    &\Uga  = -0.24 \,{\rm GeV}\, ,&
    &\Uga = 0.90 \,{\rm GeV} \, ,&
    &\text{for gluons.}&
\label{pythiavalues2}
\end{align}
As remarked above, the signs of $\Omega_{1\kappa}^{\circ\!\!\circ}$ and $\Uka$ are opposite. 
This fact turns out to work in favor of reduced uncertainties, since the two of the nonperturbative correction terms in Eq.~\eqref{eq:sigHadxi} always partially cancel. Such correlations between the parameters also explain why these values, despite being different from the ones in \Ref{Hoang:2019ceu}, give similar corrections to the partonic spectrum for quarks. We also note that the differences between these values and from those in \Ref{Hoang:2019ceu} are compatible within perturbative uncertainties, see \Ref{Ferdinand:2022xxx}. 
The values quoted in Eqs.~\eqref{pythiavalues1} and \eqref{pythiavalues2} make use weighted cross sections in \eq{sigLog10x}, shown in \fig{ciweighted} computed to NNLL accuracy, instead of just LL accurate coefficients used in \Ref{Hoang:2019ceu}.

\fig{NPvar} shows the effect of including these values of the nonperturbative parameters. We can see that these parameters have a larger effect at smaller values of jet mass, as expected, but also that the overall size of the effect is in general smaller than the scale variations shown in \fig{profile_vars_all}.
We estimate the nonperturbative uncertainty on the measured value of $\as$ from the difference between our reference parton-level curve and a ``hadron level curve'' obtained using the values in Eqs.~\eqref{pythiavalues1} and \eqref{pythiavalues2}. 
Such a procedure may overestimate the uncertainty, as the parton-level curve, without any nonperturbative corrections, cannot be correct. But it may also underestimate the uncertainty, as values larger than these may be more accurate. Note that when comparing the $\beta = 1$ results in \fig{NPvar} to those showed for \Pythia\texttt{8.3} in \fig{MCPlot}, both the partonic and the hadronic curves are different. The fit to \Pythia are used here only to get the nonperturbative constants $\Ok$ and $\Uka$ and $\Ukb$.

Note that our treatment here is not the same as using \Pythia to account for NP power corrections, as has been done previously, for example in \Ref{Marzani:2017mva}. We only use the values above as a useful guide for the size of these effects, while still employing the NNLL results for the weighted cross sections that multiply these constants. The same conclusions will follow if we use results of fits from other combinations of hadronization models and parton showers. Ultimately, our goal here is to show the typical size of these NP corrections, and the irreducible uncertainty on $\as$ that follows, should we not choose to fit for these parameters. Our choice for \Pythia is partially motivated by the observation in \Ref{Ferdinand:2022xxx} that fits performed with are found to be consistent with the universality predictions of \eq{sigHadxi} that the NP constants be independent of $\zcut$, there be a linear $\beta$ dependence of the boundary corrections, and that they be independent of production process (such as quark jets in $e^+e^- \ra q\bar q$ vs. $pp \ra Z + q$ jet).

\subsection{Statistical analysis}
\label{sec:stat}

\begin{figure}[t]
	\centering
	\includegraphics[width=0.49\textwidth]{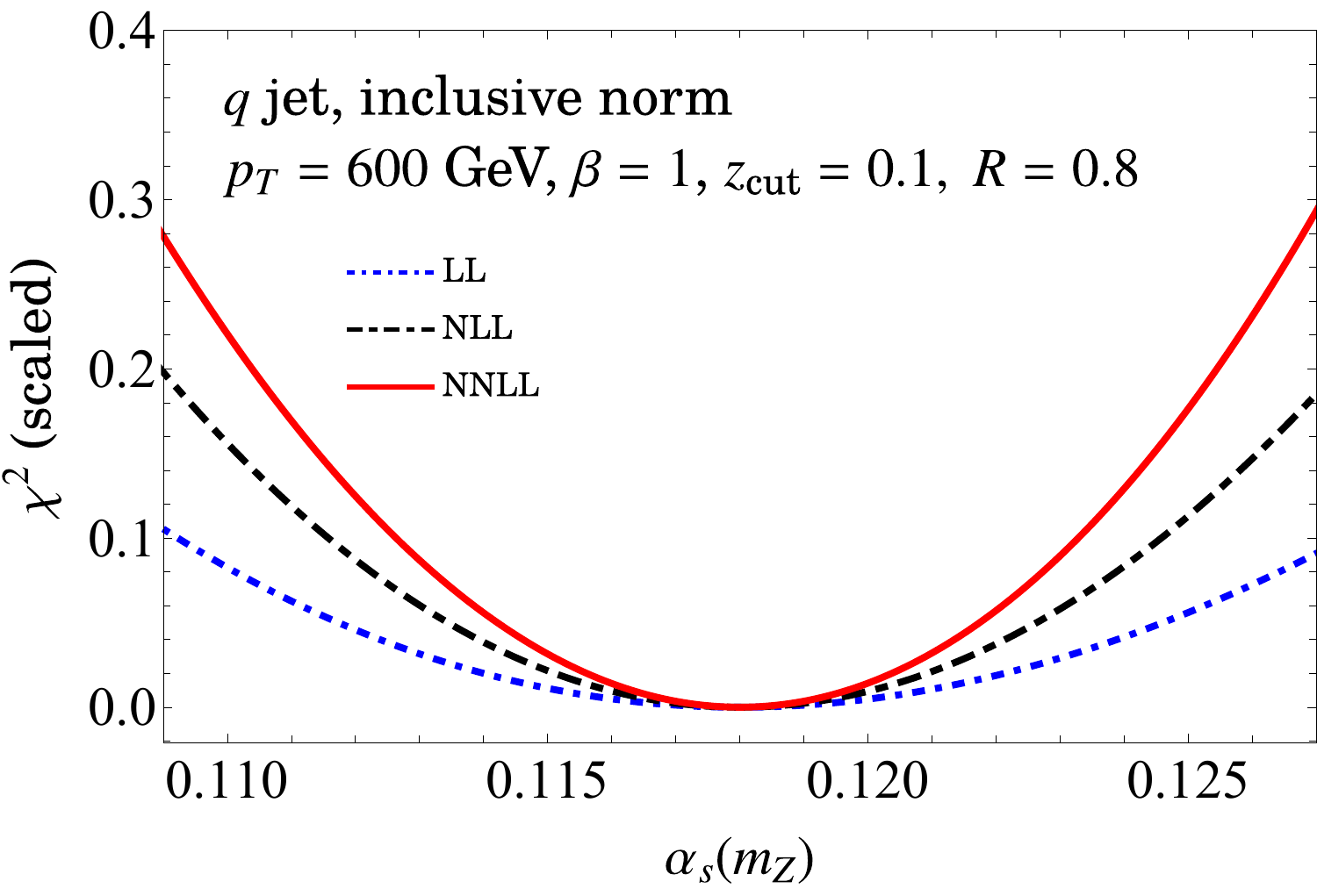}
	\includegraphics[width=0.49\textwidth]{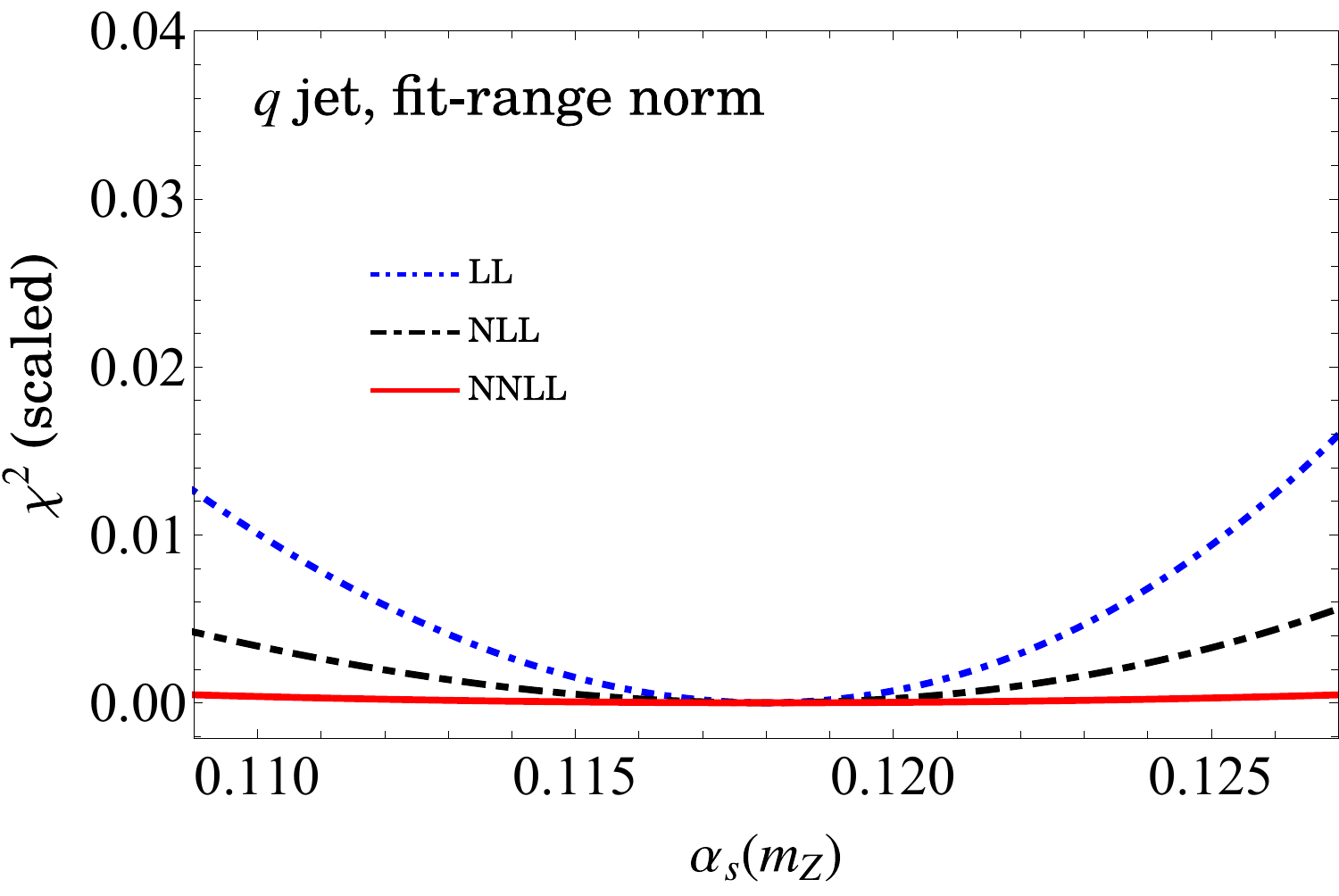}
	\includegraphics[width=0.49\textwidth]{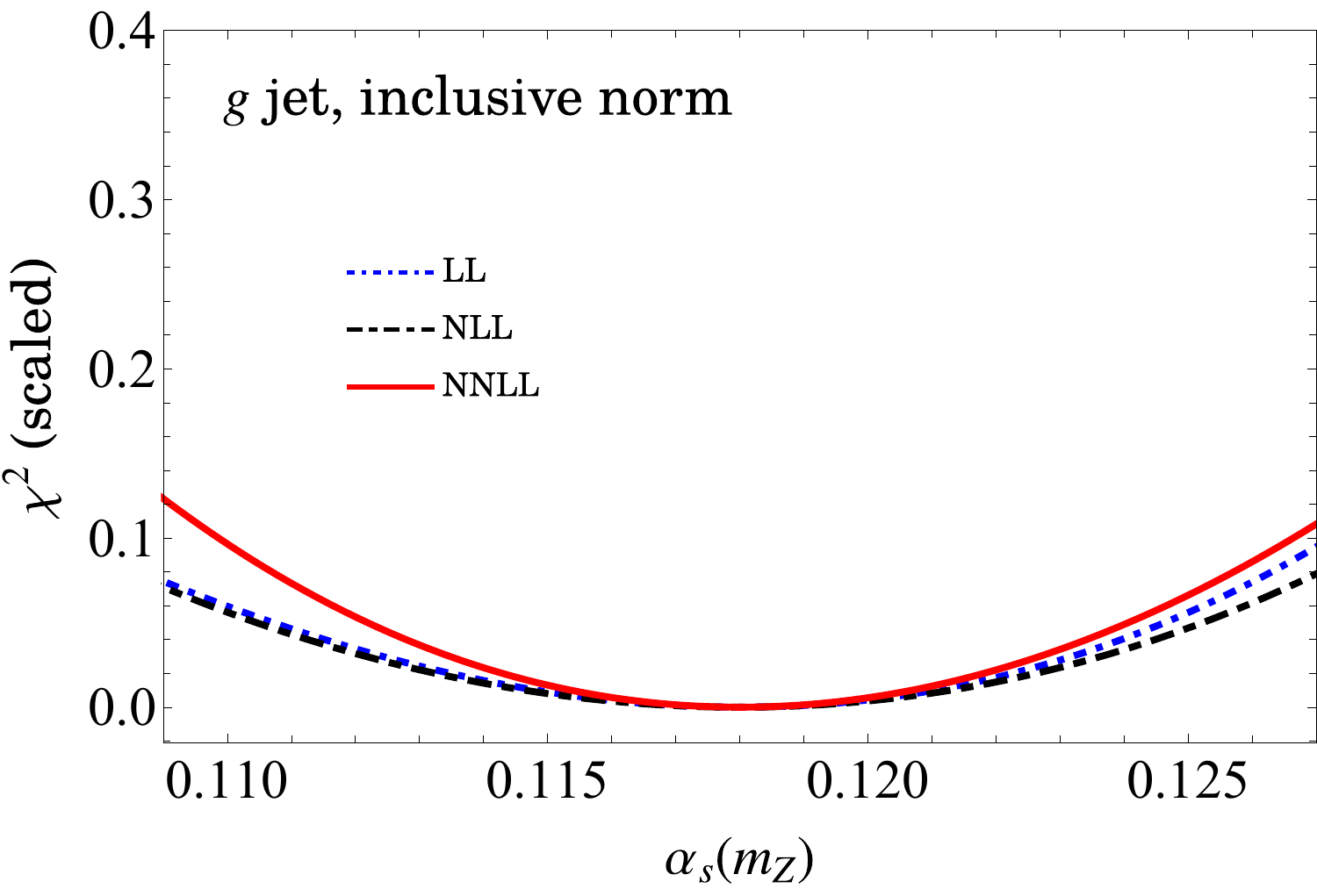}
	\includegraphics[width=0.49\textwidth]{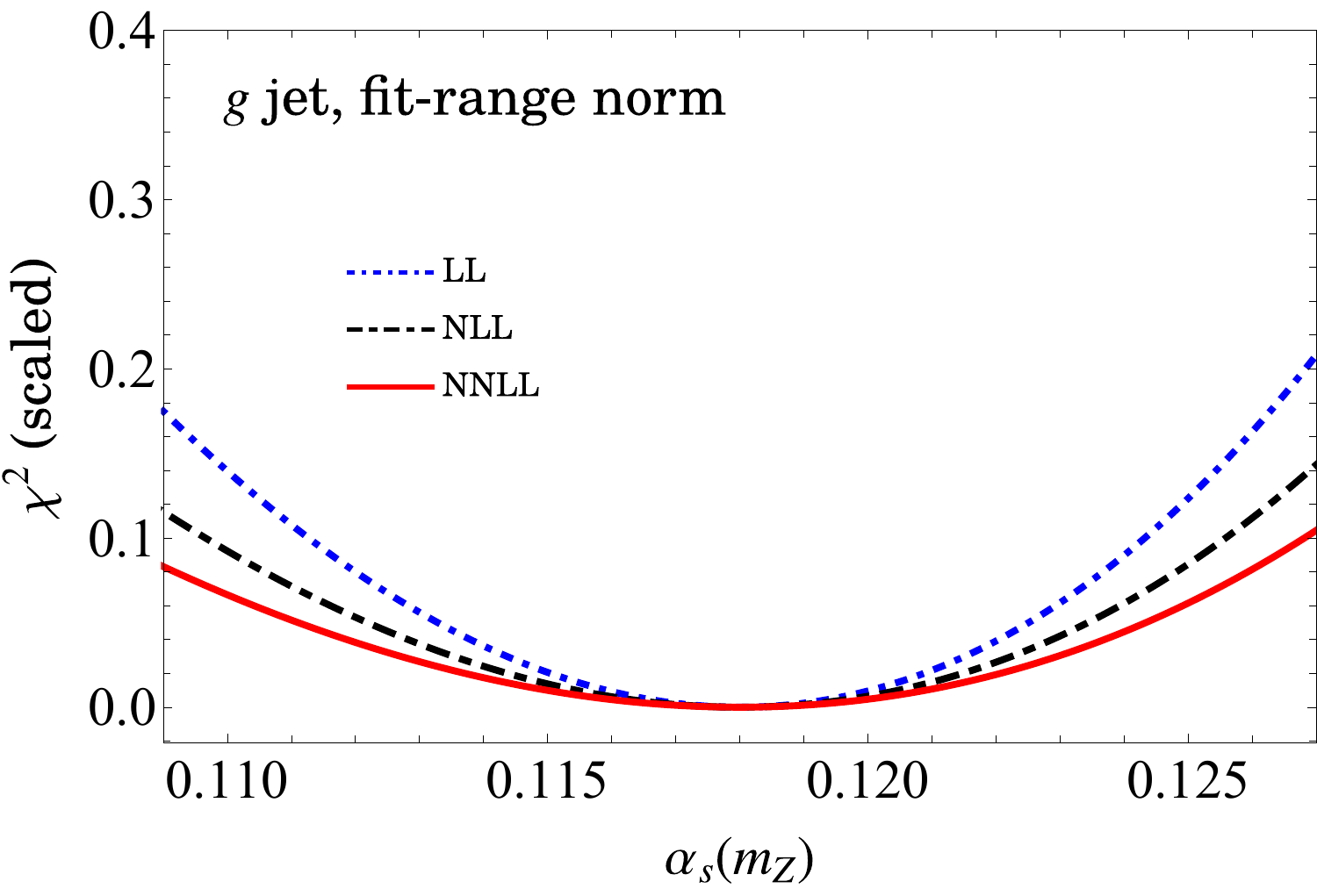}
	\caption{Values of $\chi^2$ obtained by varying $\alpha_s(m_Z)$. Left panels show quark jets and right panels show gluon jets. Upper panels normalize to the inclusive cross-section, while lower panels normalize to a fixed jet-mass fit range in the SDOE region.  
    When normalized to the fit range,  the quark jets loose sensitivity to $\as$ as seen from the different y-axis scales in the left panels.  The bottom panels also show that both quark and gluon jets loose sensitivity to $\as$ when cross sections are normalized in the fit range, whereas this is not the case when the inclusive cross section is used for the normalization.}
	\label{fig:alpha_chi0}
\end{figure}

In order to determine the sensitivity to $\alpha_s$, we first quantify the deviation of different variations from a central curve. Our central curve is calculated using the default profiles in Sec.~\ref{sec:profvar}.
To imitate a potential comparison with collider data, we split the range $\xi_{\rm{SDOE}}$ to $\xi_{0}'$ into a number of bins $n_{\rm{bins}}=10$.
After integrating the cross section over each bin and dividing by the bin size, we obtain the $n_{\rm{bins}}$-dimensional vectors $\Delta \vec{ \sigma}_{\rm{c}}$ and $\Delta \vec{ \sigma}_{\rm{v}}$  of cross-section values in each bin, for the central and variational values respectively. The $\chi^2$ value is calculated as
\begin{equation}
    \chi^2 \left(\Delta \vec{ \sigma}_{\rm{c}}, \Delta \vec{ \sigma}_{\rm{v}}\right)
    =
    \sum_{i}
    \frac{\left[ (\Delta \vec{ \sigma}_{\rm{c}})_i-(\Delta \vec{ \sigma}_{\rm{v}})_i\right]^2}{(\Delta \vec{ \sigma}_{\rm{c}})_i} \,.
\end{equation}
For the profile variations, where we perform a scan, probing a random variation of all parameters simultaneously, each point in the scan will give a certain $\chi^2$ value. In that case, we determine the $\chi^2$ to compare with other variations as the value for which 90\% of the $\chi^2$ values are contained within. We then compare the $\chi^2$ values with that obtained by a variation of the full cross section with $\alpha_s$, and estimate the uncertainty in the theoretical extraction of $\alpha_s$ as the variation which gives the same $\chi^2$ value. 

We show a scaled $\chi^2$ value corresponding to $\alpha_s(m_Z)$ variations for quark jets and gluon jets in the left and right panels of \fig{alpha_chi0} respectively, for the two different choices for  normalizing the cross section (upper versus lower panels). For quark jets the sensitivity to $\alpha_s$ is dramatically larger when normalized to the inclusive cross section, as compared to normalizing in the fit range. Quark jets normalized in the fit range also loose sensitivity to $\as$ as the order is increased from LL $\to$ NLL $\to$ NNLL. For both quark and gluon jets the sensitivity is largest at NNLL when normalizing to the inclusive cross section.
We will elaborate on these points in more detail below, as well as considering other values of $\beta$.

\subsection{Final results for \texorpdfstring{$\alpha_s$}{aS} uncertainties}

\begin{figure}[t]
    \centering
    \includegraphics[width=0.49\textwidth]{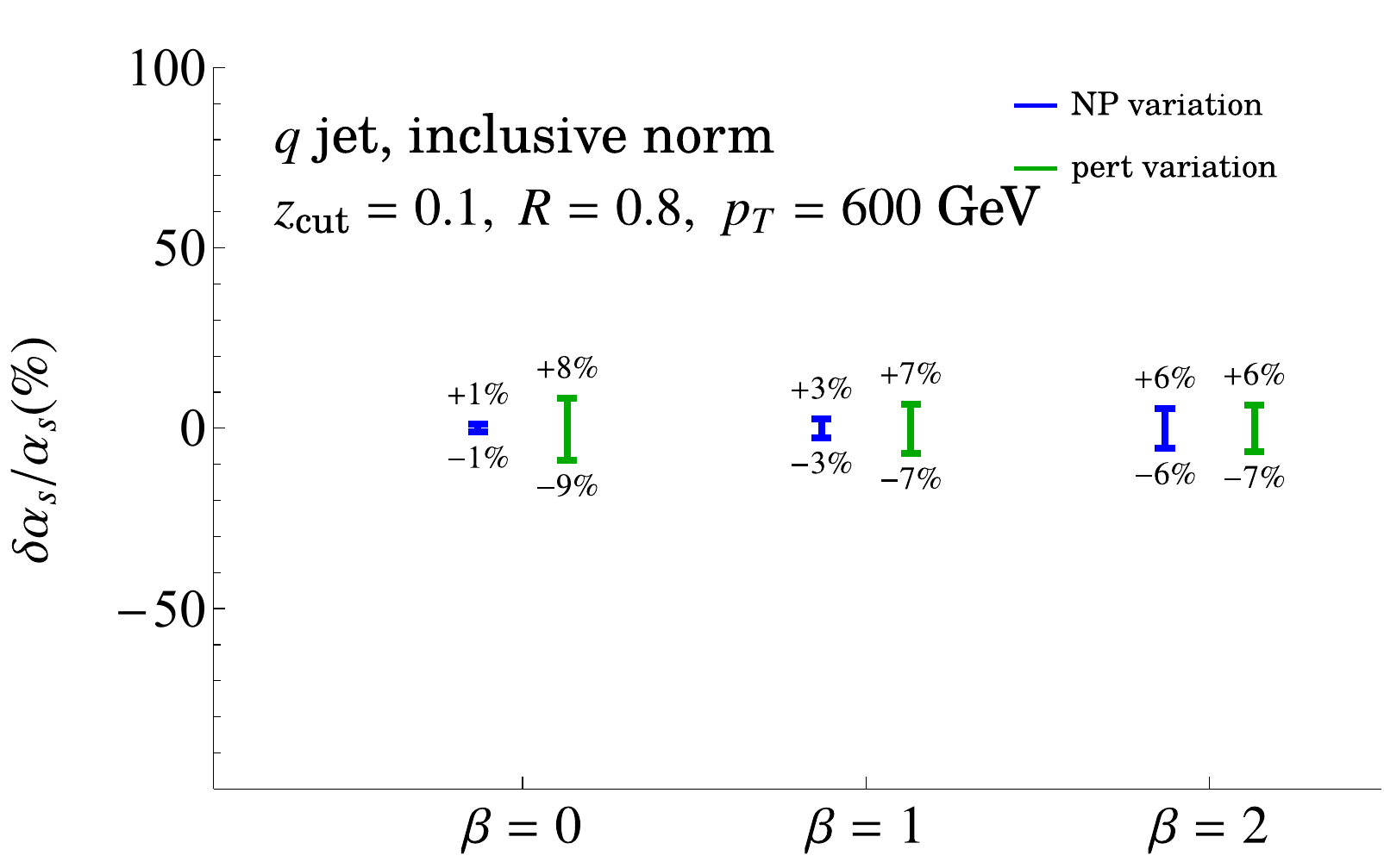}
    \includegraphics[width=0.49\textwidth]{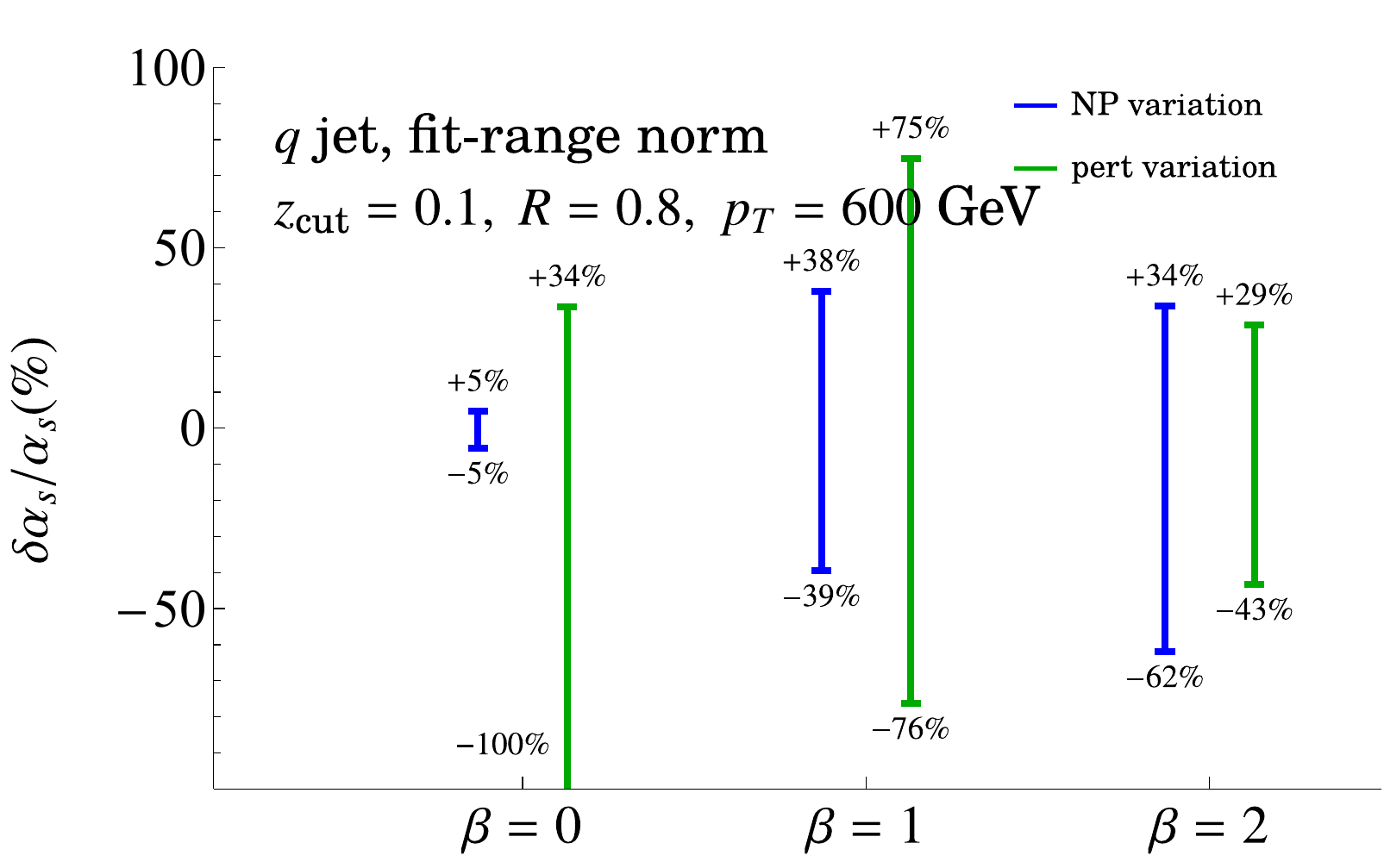}
    \includegraphics[width=0.49\textwidth]{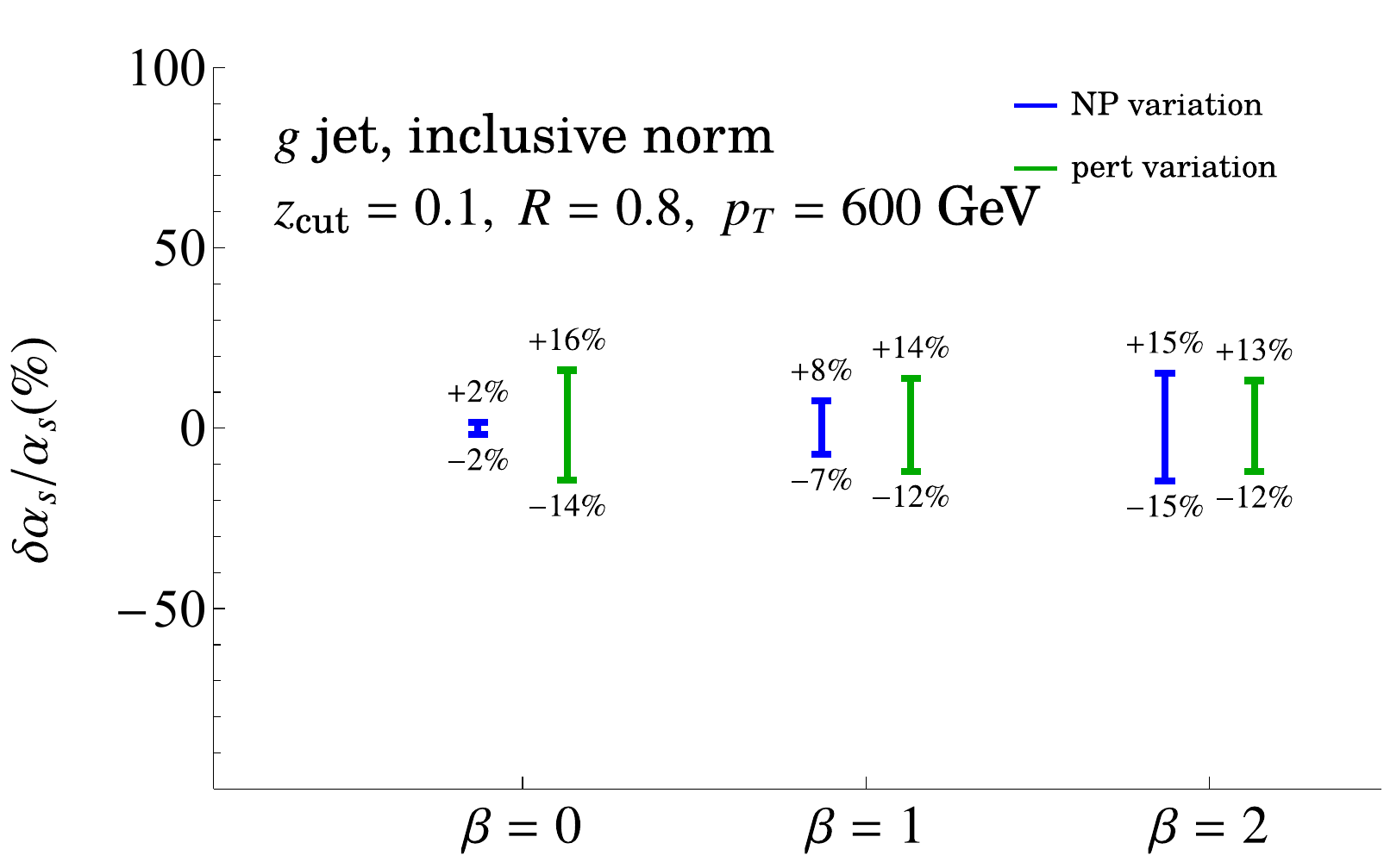}
    \includegraphics[width=0.49\textwidth]{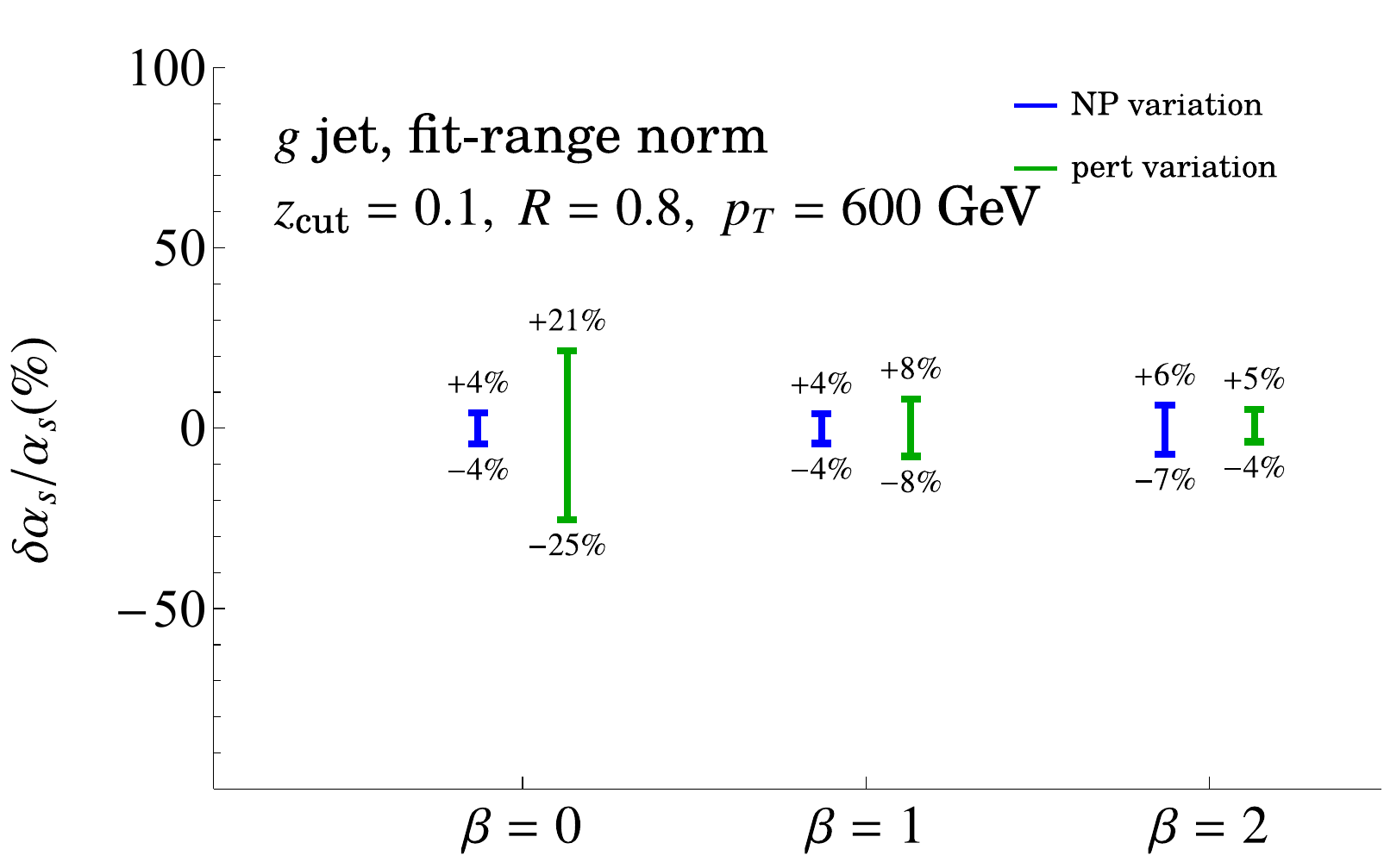}
    \caption{Uncertainty in $\alpha_s(m_Z)$ due to profile variations and nonperturbative-parameter uncertainty, for different values of $\beta$, for the normalized soft-drop jet-mass cross section.}
    \label{fig:Moneyplot1}
\end{figure} 

\begin{figure}[t]
    \centering
    \includegraphics[width=0.49\textwidth]{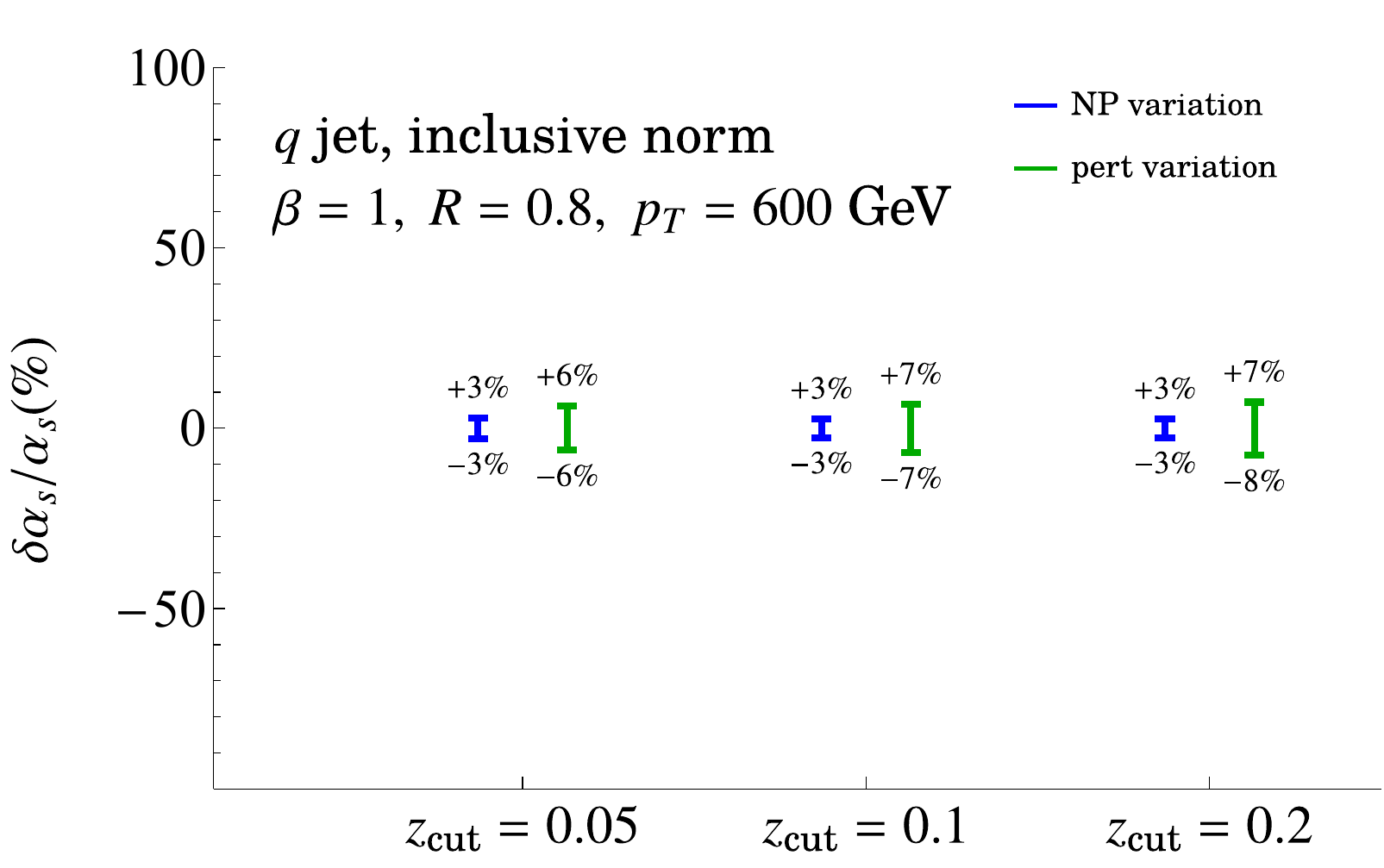}
    \includegraphics[width=0.49\textwidth]{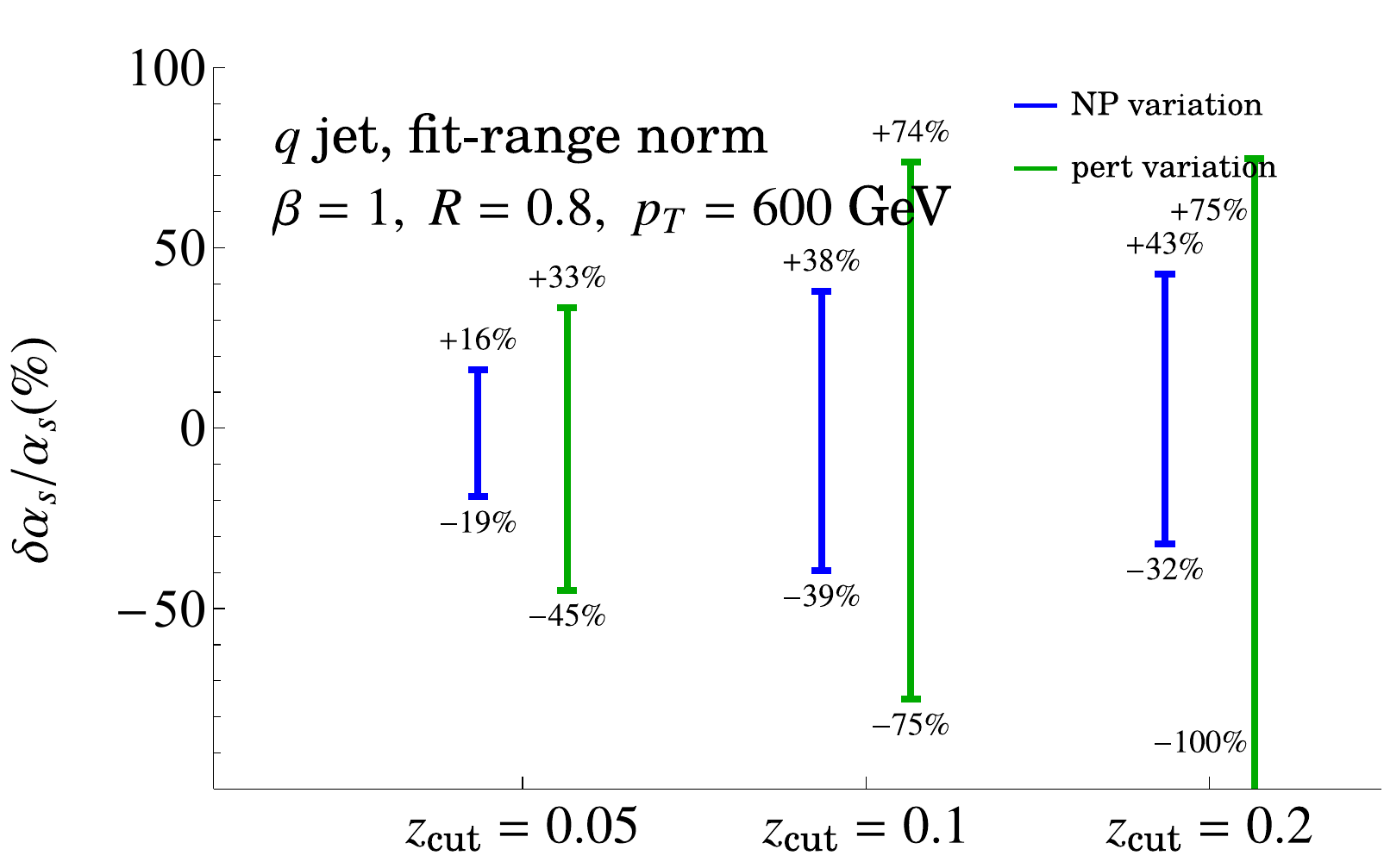}
    \includegraphics[width=0.49\textwidth]{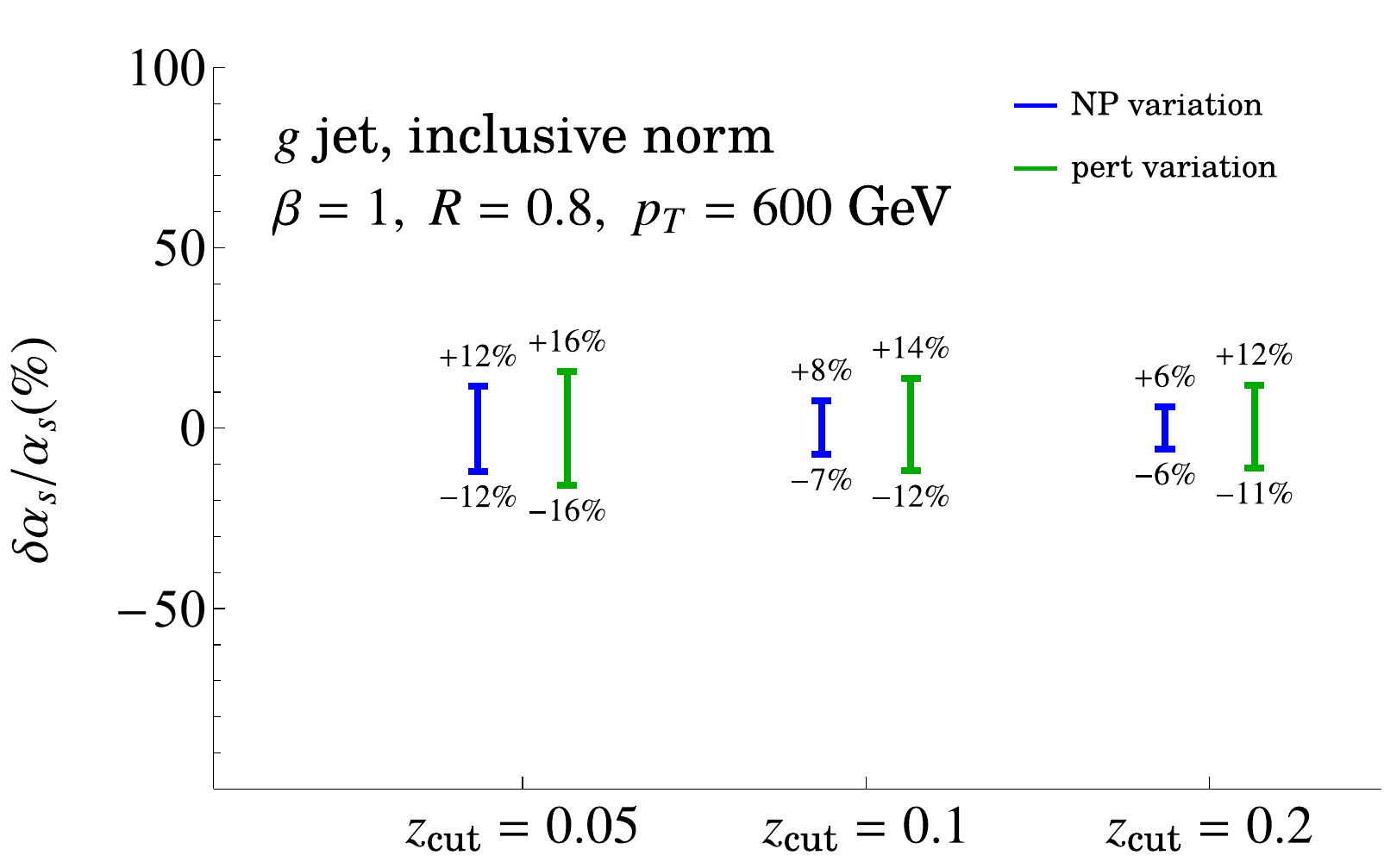}
    \includegraphics[width=0.49\textwidth]{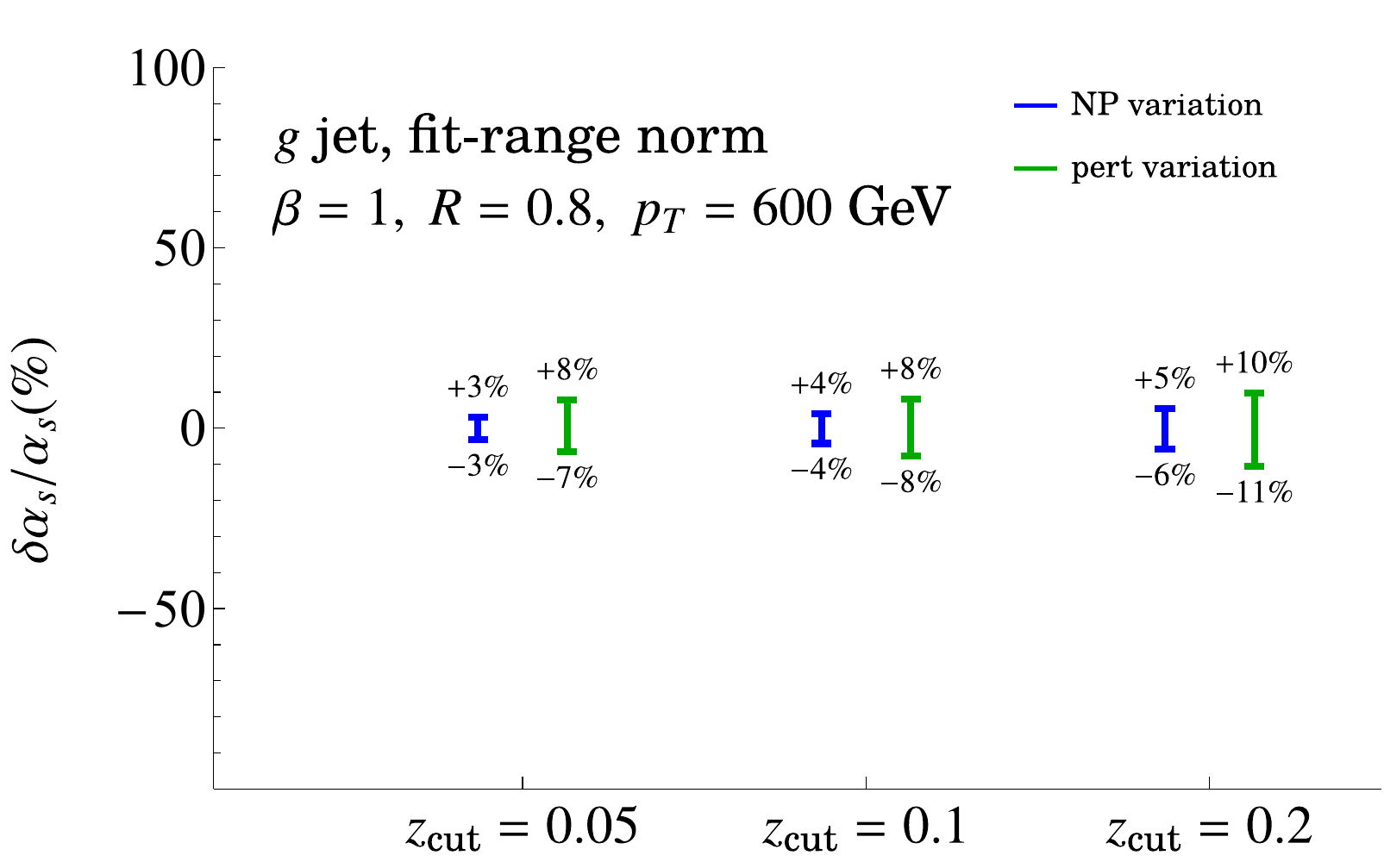}
    \caption{Uncertainty in $\alpha_s(m_Z)$ due to profile variations and nonperturbative-parameter uncertainty, for different values of $\zcut$, for the normalized soft-drop jet-mass cross section.}
    \label{fig:Moneyplot2}
\end{figure}

\begin{figure}[t]
    \centering
    \includegraphics[width=0.49\textwidth]{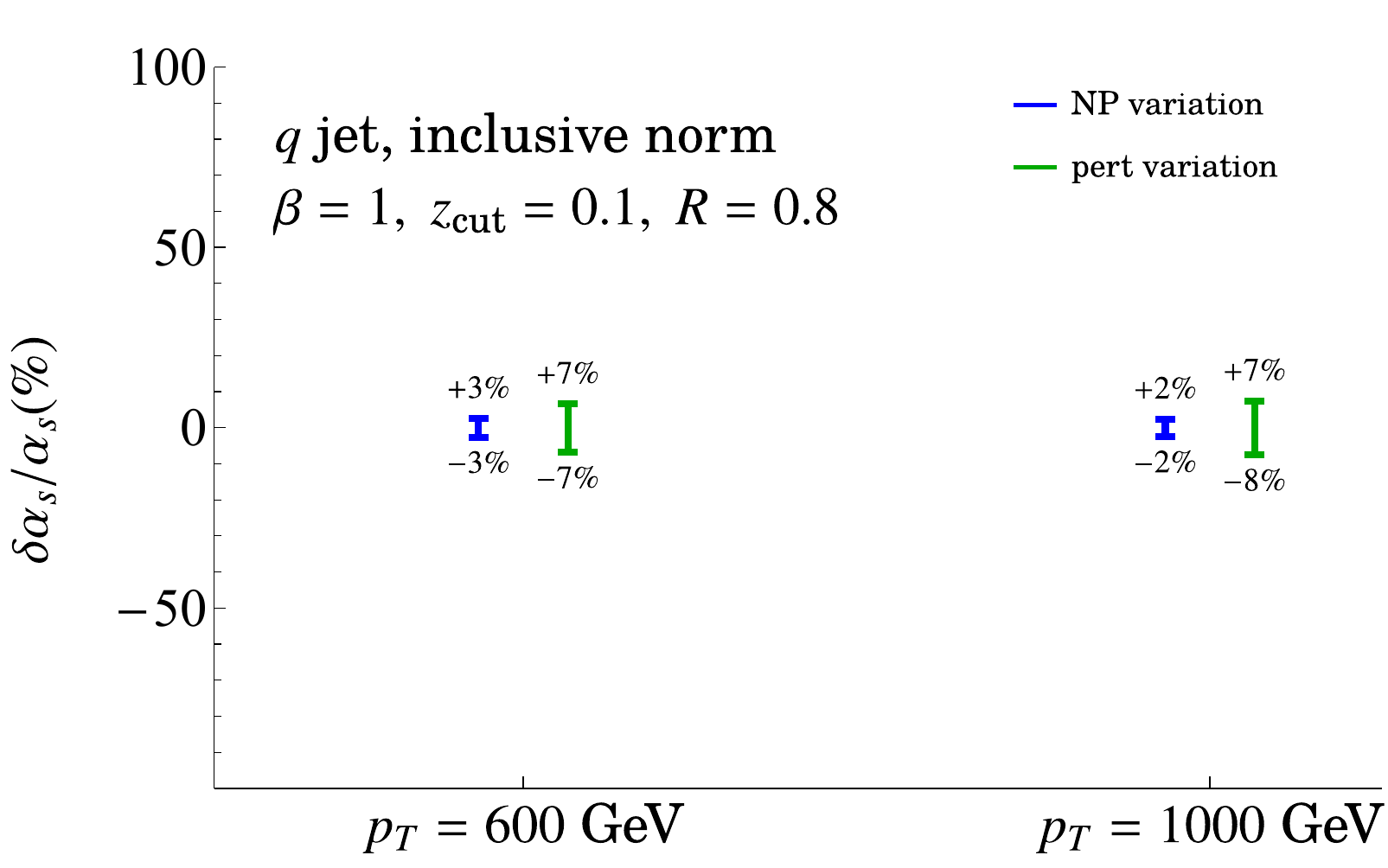}
    \includegraphics[width=0.49\textwidth]{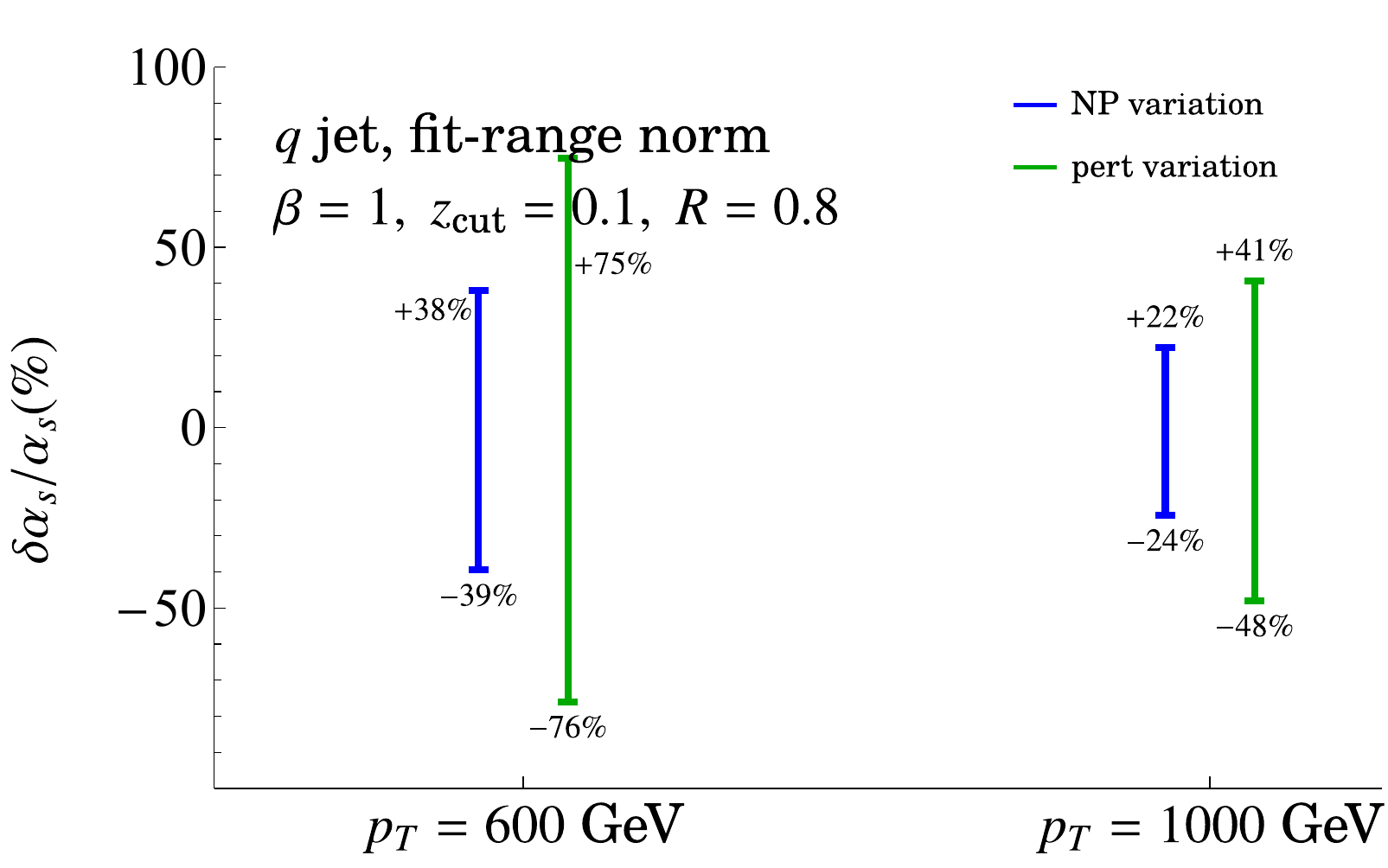}
    \includegraphics[width=0.49\textwidth]{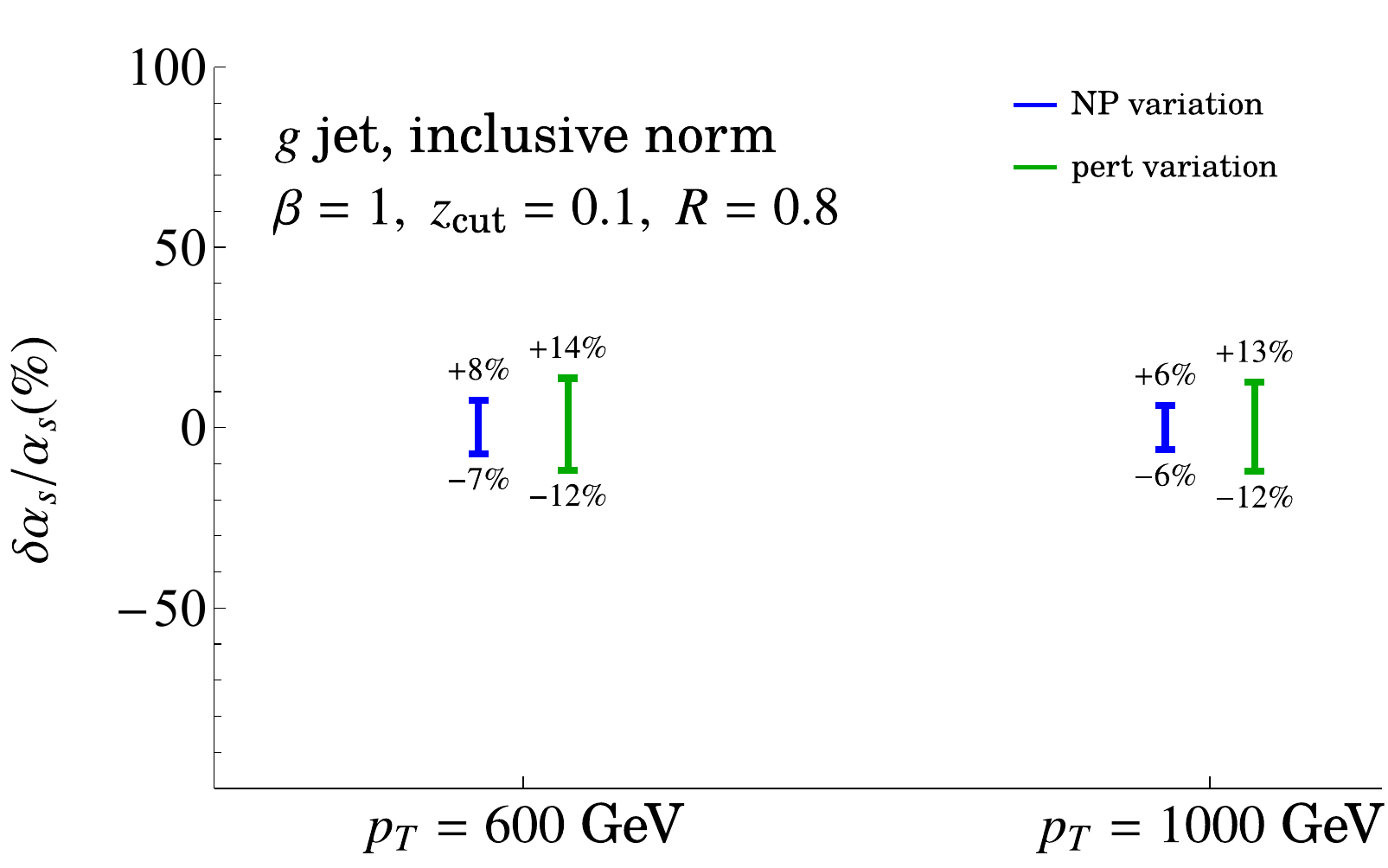}
    \includegraphics[width=0.49\textwidth]{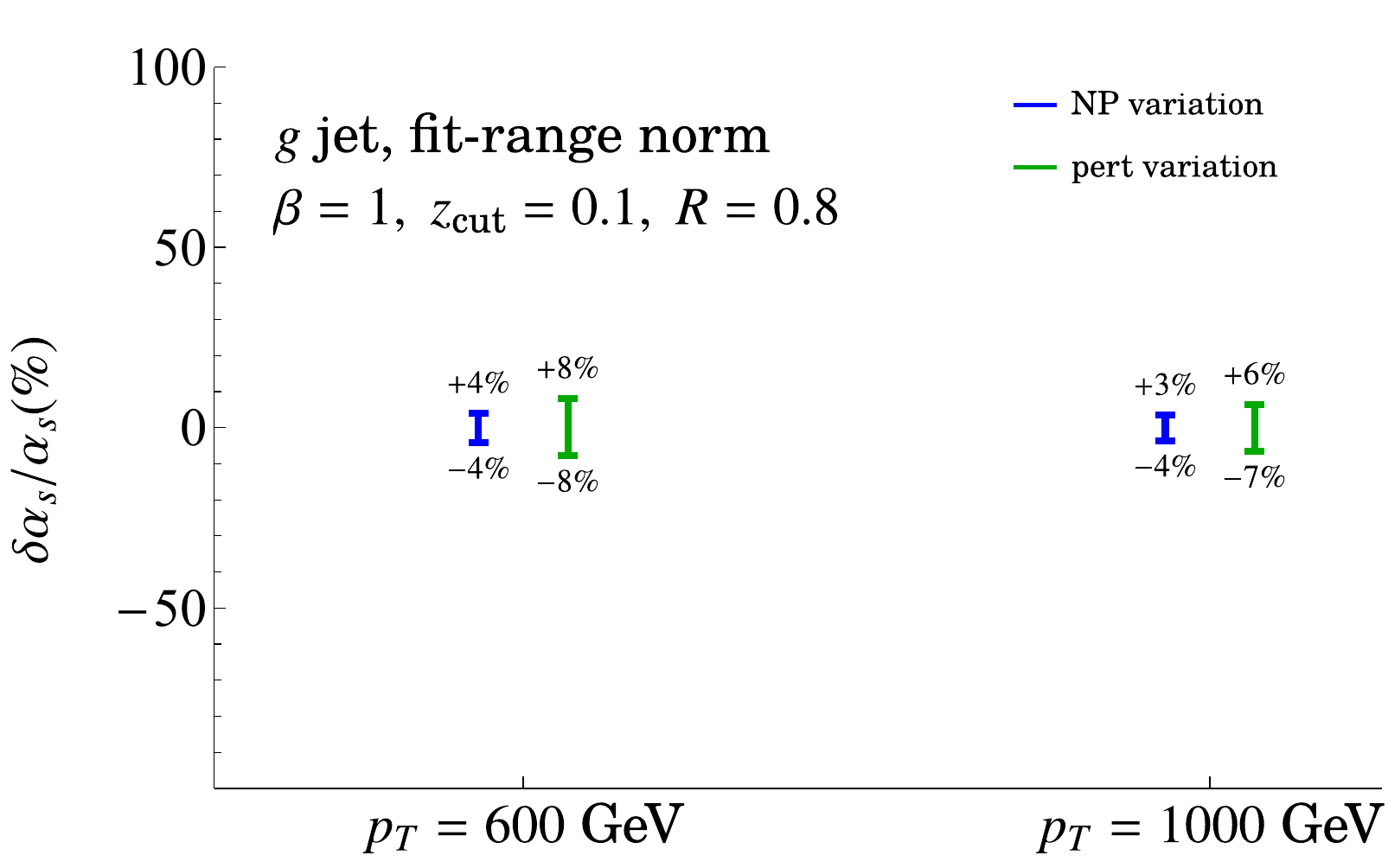}
    \caption{Uncertainty in $\alpha_s(m_Z)$ due to profile variations and nonperturbative-parameter uncertainty, for different values of $p_T$, for the normalized soft-drop jet-mass cross section.}
    \label{fig:Moneyplot3}
\end{figure}

Summaries of the aggregate uncertainties on $\alpha_s(m_Z)$ as we vary $\beta$, $z_\cut$ and $p_T$ are shown in Figs.~\ref{fig:Moneyplot1}$-$\ref{fig:Moneyplot3}, respectively. Both the perturbative and nonperturbative
uncertainties are obtained by comparing the $\chi^2$ values obtained by the method described in Sec.~\ref{sec:stat} to the ones obtained by varying $\alpha_s(m_Z)$. Each figure shows the result for quarks and gluon jets separately, and with two different choices of normalization as presented in Eq.~\ref{eq:sigma_norms}.

As discussed before, the quark jets lose sensitivity to variations in $\alpha_s$ when the spectrum is normalized in the fit range, i.e.\ in the range $\xi\in [\xi_{\text{SDOE}},\xi_0']$ as defined in \ref{eq:SDNPprecise}. Since the spectrum is close to being flat at NNLL in this range, and the shape is largely independent of variations in $\as$, both the perturbative and nonperturbative corrections 
are large when using the fit-range normalization, and much smaller for the inclusive normalization.
This feature is not present for gluon jets since the spectrum is not flat, as was seen in \fig{alpha_slope}. For gluon jets, the uncertainties obtained using inclusive normalization and fit-range normalization are of the same order of magnitude. 
Since data includes contributions from both quark and gluon jets, there is a clear preference for using the inclusive norm in future $\alpha_s(m_Z)$ analyses.  
The size of the perturbative uncertainty that we find on $\alpha_s(m_Z)$ is also of the expected size for a NNLL level calculation. For example, it can be compared to the $e^+e^-$ thrust fits of Ref.~\cite{Abbate:2010xh} which were done at different orders, where an 8\% uncertainty on $\alpha_s(m_Z)$ was found for a fit at NNLL.\footnote{This 8\% uncertainty on $\alpha_s(m_Z)$ at NNLL is reduced to 4\% at NNLL$^\prime$ and 2\% at N$^3$LL$^\prime$, and finally to 1\% after removing renormalons~\cite{Abbate:2010xh}.} This value is of the same size as the uncertainty we find for quark jets with the inclusive norm, which is the relevant comparison.

We can also ask how the uncertainties vary with different soft-drop parameters $\beta$ and $\zcut$, to identify the ideal parameters for a potential measurement. These results are shown in \figs{Moneyplot1}{Moneyplot2}.  We do not see any particularly compelling trends of note for the variation with $z_\cut$,
so simply fixing $z_{\rm cut}=0.1$ likely suffices for $\alpha_s$ analyses.
For $\beta=0$, it looks like the dominant uncertainty is perturbative, while for higher $\beta$ the perturbative and nonperturbative uncertainties are more comparable. 
In particular, for the favored inclusive norm choice we find that nonperturbative relative to perturbative uncertainties are: comparable for $\beta=2$, roughly half as large for $\beta=1$, and as much as a factor of eight smaller for $\beta=0$.
Note, however, that for $\beta=0$ there is a strong cancellation between 
$\Oq$ and $\Uqa$ which have similar magnitudes and opposite signs, making the precise nonperturbative uncertainties quoted for $\beta=0$ more sensitive to the numbers used for these parameters.
Our numbers for the nonperturbative corrections should be interpreted as estimates only, since their values are obtained using fits to Monte-Carlo data from \Pythia\texttt{8.3}. 
To improve the estimate of uncertainty, one can, for example, try to fit some or all of the nonperturbative constants $\Ok$, $\Uka$ and $\Ukb$.
At present, the values obtained in \Pythia\texttt{8.3} as given in Eq.~\eqref{pythiavalues1} and~\eqref{pythiavalues2}, give a sense of the order-of-magnitude of the expected uncertainties due to hadronization corrections. The values were fit using multiple $\beta$, $\zcut$ and $p_T$ values.
Our results should be taken as motivation to independently analyze data obtained with different values of $\beta$, in order to compare the consistency between the resulting values of $\alpha_s(m_Z)$.

A similar pattern for the uncertainties is seen at other values of $p_T$, as can be seen in \fig{Moneyplot3}.  In general, it will be important to simultaneously include data from multiple different bins in $p_T$ when fitting for $\alpha_s(m_Z)$, in particular if one attempts to reduce the nonperturbative uncertainty by simultaneoulsly fitting for those parameters.

\section{Implications and Conclusions}
\label{sec:conclusion}

We studied the prospects for a precision measurement of the strong-coupling constant $\as$ in $pp$ collisions by computing the soft-drop jet-mass distribution to NNLL order. Since this observable greatly reduces sensitivity to hadronization corrections compared with its non soft-drop counterpart, one can hope to use a wider range of jet-mass values $m_J^2$ for extraction from data. We estimated the theoretical uncertainties on an $\as$ measurement at the LHC, using a field-theoretic factorization-based framework for both perturbative and nonperturbative effects. We rely on Monte-Carlo simulation only to give a ballpark measure of the nonperturbative parameters, 
exploiting factorization to predict how nonperturbative effects are modified by the jet kinematics and jet grooming.
We use scale variations for the perturbative uncertainty, and estimate of the quark/gluon fraction uncertainty with variation among PDF sets.

To study the possibility of getting a precision $\as$ measurement, we computed the resummed jet-mass cross section to the current state-of-the-art of NNLL for $pp$ collisions, as outlined in \secn{analysis}.
Uncertainties on $\as$ can be seen in Figs.~\ref{fig:Moneyplot1}$-$\ref{fig:Moneyplot3}. The perturbative uncertainties were estimated using profile functions for the scale variations, as explained in Sec.~\ref{sec:prof}. At this logarithmic order for $\zcut=0.1$ and $p_T=600$ GeV, we find the perturbative uncertainties to translate into anywhere between $6\%$ and $16\%$
uncertainty on $\as$ in potential measurements, depending on which value of $\beta \in \{0,1,2\}$ is chosen. 
If the cross section were known to higher logarithmic order, the perturbative uncertainties would be correspondingly smaller, perhaps by as much as a factor of 2.
As discussed in Sec.~\ref{sec:compare}, some ingredients for a soft-drop N$^3$LL cross section are still lacking: the functions $S_c^\kappa$ and $S_G^\kappa$ that appear in the factorization formula in Eq.~\eqref{eq:SDFact}, are missing the two-loop-constant contributions, and the three-loop non-cusp anomalous dimension. These pieces have been computed in \Ref{Kardos:2020ppl} for $\beta=0$ with hemisphere quark jets in $e^+e^-$, while for $pp$ colliders one would need the analogous results for gluon jets at small $R$, as well as the two-loop global-soft function constant for quark jets at small $R$. As noted above in \secn{pc}, at this order the non-global logarithms may also become relevant and will need to be accounted for. Finally, for a complete prediction for soft-drop jet mass measured on inclusive jets, the results above will need to be combined with the production hard function, PDFs and the DGLAP evolution of the inclusive jet function.

An important finding of our work is how normalizing the jet-mass spectrum affects the uncertainties. If we normalize to the range in which $\as$ is to be fit as in Eq.~\eqref{eq:NormChoice2}, the sensitivity to $\as$ is greatly reduced for quark jets, compared to an inclusive normalization as in Eq.~\eqref{eq:NormChoice}. The reason is that the quark spectrum flattens out at NNLL order, so even though variations in $\as$ affect the slope and the height of the jet-mass spectrum, changing $\as$ has a very little effect on this fit-range normalized flat spectrum. However, when we use the inclusive normalization, additional information is included: the total number of jets in the sample, which by itself does not have a strong dependence on $\alpha_s$. This choice therefore retains sensitivity to $\as$, as the height of the spectrum in the fit region changes with variations in $\as$. One might be concerned that using the inclusive normalization makes the prediction sensitive to a region outside of the fit range, where the theory uncertainty is large. However, since soft drop never vetoes a jet entirely, one does not actually need a prediction of the shape outside the fit region at all; one can simply count the total number of jets falling in the respective $p_T$-$\eta_J$ bin. Since the $\as$ sensitivity is superior with the inclusive normalization than with the fit-range normalization this suggests that utilizing the number of jets in the fit range compared to the total is critical to measuring $\as$. This partially explains why gluon jets are not better than quark jets for measuring $\as$ despite their more complicated shape. 

A novelty of this work is how the nonperturbative effects are incorporated using model-independent power-corrections, as constructed in Ref.~\cite{Hoang:2019ceu} and presented in Eq.~\eqref{eq:sigHadxi}. The framework relies on six constants, $\Ok, \Uka, \Ukb$ (with $\kappa=q,g$ representing values for quarks and gluons, respectively), which cannot be computed using first-principle methods. 
For our analysis we used the values obtained from Monte-Carlo extractions in \Pythia\texttt{8.3} from Ref.~\cite{Ferdinand:2022xxx}. 
These values were only used to estimate the nonperturbative uncertainty on an $\as$ measurement due to lacking information about these nonperturbative parameters, as shown in Figs.~\ref{fig:Moneyplot1}$-$\ref{fig:Moneyplot3}.  
For example, for $\beta=1$ we find around $3\%$ nonperturbative uncertainty for quark jets and $8\%$ uncertainty for gluon jets. Other results can be found in the figures, and we note that they do depend on $\beta$, but are fairly independent of $z_{\rm cut}$ and $p_T$.
While these give us rough
estimates of the expected uncertainty due to the lack of knowledge of nonperturbative corrections, the specific values should not be taken too literally, as they rely on input from a Monte-Carlo hadronization model. 
These uncertainties are improvable as they can, in principle be measured. If the data is good enough, one could fit some or all of the nonperturbative parameters which could dramatically help improve the $\as$ measurement. The best prospect is perhaps obtained if the data can be acquired with very fine binning so that one can remove biases from the nonperturbative parameters while retaining sensitivity to the perturbative $\alpha_s$ sensitive cross section. 

Our results for uncertainties on $\alpha_s$ can be used as motivation for higher order perturbative computations that would enable fits to be extended to N$^3$LL. In particular, from \fig{Moneyplot1} we note that $\beta=0$ has the smallest nonperturbative uncertainty, being dominated by the perturbative uncertainty by a large margin.  This conclusion relies on the abnormally small nonperturbative effects for $\beta=0$, which arise due to a cancellation between positive and negative nonperturbative parameters. The negative sign is predicted by fits of the hadronization parameters to various Monte Carlos, and the exact magnitude of the cancellation may not be robust. However, even if one triples the size of the nonperturbative uncertainties we find for the $\beta=0$ case, one still concludes that extending $pp$ results for $\beta=0$ to N$^3$LL is the best target for carrying out higher order perturbative computations.

One strategy to perform the extraction on $pp$ soft drop groomed jet-mass data would be to carry out independent fits with different values of $\beta$ and fixed $z_{\rm cut}=0.1$, while simultaneously including multiple bins in $p_T$ and different rapidities $\eta$. The use of rapidity bins gives different quark/gluon fractions and possibly enable the study of a quark rich region~\cite{Gallicchio:2011xc}.

Summarizing we find a combined uncertainty of around $9\%$ for quark jets and $16\%$ for gluon jets with the inclusive normalization. 
Combining these findings, we therefore conclude that a current theory computation would result in around $\sim 10\%$ uncertainty on $\as$, which is roughly an order of magnitude higher than the state-of-the-art extractions from lattice QCD and from thrust and C-parameter data in $e^+e^-$ collisions. However, while a precision-$\as$ measurement from soft drop groomed jet mass in $pp$ colliders does not compete well with other methods at present; more experimental data and higher order perturbative calculations will make an $\as$ extraction using soft-drop jet mass more compelling in the future. 
In particular, the data could be used to reduce the nonperturbative uncertainties by fitting for the model-independent nonperturbative constants from Eq.~\eqref{eq:sigHadxi}, and the computation of the missing two- and three-loop ingredients to achieve N$^3$LL accuracy could bring the theory uncertainties down as well,
thus obtaining high precision data and carrying out a sustained theory effort is well motivated. 

\section*{Acknowledgements}
\enlargethispage{20pt}
We are grateful to Kyle Lee for numerous discussions on various parts of the manuscript. We thank Andrew Larkoski, Simone Marzani, Johannes Michel and Robert Szafron for helpful discussions, Anna Ferdinand for providing us with the results of fits to MC hadronization models, and Lorenzo Zoppi for having performed the calculation of the collinear function for gluon jets used in the results above.
We thank Johannes Michel for providing support with numerical implementation in \texttt{SCETlib}~\cite{scetlib}.
A numerical implementation in \texttt{C++} of the hadron level NNLL cross section built on the core functionalities of the \texttt{SCETlib} library will be made available as a part of \texttt{scetlib::sd} module~\cite{scetlibSD}.
HSH gratefully acknowledges support from the Simons Foundation (816048, HSH). AP acknowledges support from DESY (Hamburg, Germany), a member of the Helmholtz Association HGF. AP was a member of the Lancaster-Manchester-Sheffield Consortium for Fundamental Physics, which is supported by the UK Science and Technology Facilities Council (STFC) under grant number ST/T001038/1, and at University of Vienna was supported by FWF Austrian Science Fund under the Project No.~P28535-N27.
IWS  was supported by the U.S.~Department of Energy, Office of Science, Office of Nuclear Physics, from DE-SC0011090 and by the Simons Foundation through the Investigator grant 327942.
MS was supported by the U.S. Department of Energy under contract DE-SC0013607.

\appendix
%=======================================================================
%=======================================================================
\section{Fixed order results}
\label{app:Oneloop}
%=======================================================================
%=======================================================================

Here we derive one-loop results for the groomed and ungroomed jet mass in the fixed order region in \eqs{calG}{calGSD}. 
%----------------------------------------
\subsection{Matrix elements and phase space}
%----------------------------------------

In the following we will limit to ${\cal O} (\alpha_s)$ and hence consider splittings $i^* \ra jk$, where $i^*$ is the jet initiating parton and $j$ is the softer of the two final state partons. We identify two logarithmic variables, the jet mass and the energy fraction:
\begin{align}\label{eq:xyDef}
    s = p_{i^*}^2 = (p_i + p_j)^2  \, , \qquad x = \frac{q_j^-}{Q} \, , \qquad     y \equiv \frac{s}{Q^2} \, ,
\end{align}
where $Q$ was defined above in \eq{Q}, and we follow the light cone decomposition defined in \eqs{LCDef}{LC}. 
We will find it convenient to work with the dimensionless variable $y$. 
 In the limit when $j$ becomes soft, the jet mass can be expressed in terms of the plus component of the softer parton:
\begin{align}\label{eq:yDef2}
    s \ra Q q_j^+ \, ,\qquad y = \frac{q_j^+}{Q} \, , \qquad  (\text{soft limit})
\end{align}
with $x$ being the same as in \eq{xyDef}.

Next, the squared matrix element in the $\MS$ scheme for the $i^* \ra jk$ in collinear limit is given by
\begin{align}
    \sigma_2^c \big(x,s\big) = \mu^{2\eps}\iota^\eps \frac{2g^2}{s}\sum_j \hat P_{ji} (x) \,,  \qquad \iota^\eps = \Big(\frac{e^{\gamma_E}}{4\pi}\Big)^\eps \, ,
\end{align}
where
\begin{align}
    \hat P_{gq}(x) &= C_F \bigg[\frac{1 + (1-x)^2}{x} - \eps x \bigg] \, , \nn \\
    \hat P_{gg} (x) &=C_A \bigg[\frac{x}{1-x} + \frac{1-x}{x} + x(1-x)\bigg]\, , \nn \\
    \hat P_{qg} (x) &= n_f T_F \bigg[x^2 + (1-x)^2 - 2 \eps x(1-x)\bigg] \, .
\end{align}
The two-particle phase space for collinear radiation is given by
\begin{align}
\frac{1}{Q^2}2g^2 \mu^{2\eps}   \iota^\eps d \Phi_2^{c} (x,y) =\frac{\alpha_s}{2\pi}\frac{ e^{\eps \gamma_E}}{ \Gamma(1-\eps)}  \Big(\frac{\mu^2}{Q^2y}\Big)^\eps \frac{ dx\,dy}{\big[x(1-x)\big]^\eps} \, , 
\end{align}
where the pre-factors are included for later convenience.
We will also need the expression for one-particle soft emission phase space:
\begin{align}
    \frac{1}{Q^2}2 g^2 \iota^\eps \mu^{2\eps} d \Phi_1^s(x,  y) &\equiv \frac{1}{Q^2} 2 g^2 \iota^\eps \mu^{2\eps} \int \frac{d^d q}{(2\pi)^d}  2\pi \delta (q^2)\\
    &= 
    \frac{\alpha_s}{2\pi}
    \frac{e^{\eps\gamma_E } }{\Gamma(1-\eps)} 
    \Big(\frac{\mu^2}{Q}\Big)^\eps  \frac{dx\, dy}{y^{1+\eps}x^{\eps}}
    \, , \nn
\end{align}
The soft limit of the splitting functions that survive at leading power are given by:
\begin{align}
    &\hat P^{(0)}_{qg}(x) = \frac{2 C_F}{x} \, ,& 
    &\hat P^{(0)}_{gg}(x) = \frac{C_A}{x}  \, .&
\end{align}
For $g^*\ra gg$  splitting we also have another soft limit from $x\ra 1$ which compensates for the factor of 2.
%=======================================================================
\subsection{One-loop result for the ungroomed jet mass}
%=======================================================================
\label{app:Gnosd}
We evaluate here the inclusive jet function for a jet clustered by $k_T$-type clustering and yielding the jet-mass dependent variable $\xi$. We have cases where one of the partons after the $1\ra2$ splitting leaves the jet, in which case the energy fraction retained by the jet $z <1$ and when both the partons remain in the jet. We first consider the case where both the partons are present in the jet. In terms of the 2-particle phase space variables defined above, we have
\begin{align}\label{eq:GInteg}
    {\cal G}^{\rm [both\,in]}_{\kappa, \rm No\,sd}\big(z, Q , \xi,\mu\big)   
    &=\delta(1-z)\mu^{2\eps} \iota^\eps (2g^2) \int   \frac{\df \Phi_2^{c} (x,y)}{yQ^2} \Big(\sum_j \hat P_{ j\kappa} (x) \Big)  \overline \Theta_{k_T}(x, y) \big [ \delta (\xi  -y)   -\delta (\xi)\big]  \, ,
\end{align}
The energy fraction $z$ of constituents clustered with the jet is set to one for this case. In the second line we have the measurement function for an ungroomed jet. This involves the jet clustering constraint given by
\begin{align}
    &\overline \Theta_{k_T}(x,y) =  \Theta \big(x(1-x) -y\big) \, .
\end{align}
The $-\delta(\xi)$ term corresponds to virtual contributions. This calculation, in fact turns out to be precisely the same as that of the collinear function introduced above in \eq{FactSmallRg}, such that
\begin{align}
    {\cal G}^{\rm [both\,in]}_{\kappa,\rm No\,sd} (z, \xi, Q , \mu ) = \delta(1-z) \,  {\cal C}^\kappa \big(\xi, Q, \mu \big)
\end{align}
The result for quark jets was calculated in \Ref{Pathak:2020iue}, giving
\begin{align}\label{eq:Cq}
    {\cal C}^q \big[\xi,Q, \mu \big]  &= \delta(\xi) + \frac{\alpha_s (\mu) C_F}{2\pi} \Bigg\{
    \delta (\xi) \bigg(
    \frac{1}{2}  L_Q^2+ \frac{3}{2}  L_Q+ \frac{7}{2} - \frac{5\pi^2}{12} \bigg)  \\
    &\qquad  
    + \Theta \big(1-4 \xi\big) \Bigg[   - 2 {\cal L}_1 ( \xi) + {\cal L}_0(\xi) \bigg(
    4 \ln \Big(\frac{1 + \sqrt{1 - 4 \xi}}{2}\Big)  -\frac{3}{2} \sqrt{1-4\xi}  \bigg)
    \Bigg]
    \Bigg\}  \, ,\nn
\end{align}
where
\begin{align}\label{eq:LmuQRdef}
    L_Q = \ln \frac{\mu^2}{Q^2} \, .
\end{align}
The result for $\kappa = g$ is given by\footnote{We are grateful to Lorenzo Zoppi for this result.}
\begin{align}\label{eq:Cg}
    {\cal C}^g \big[\xi,Q, \mu \big]  &=
    \frac{1}{\big(r_g^{\rm max}(\xi)\big)^2} \Bigg[
    \delta (\xi) + \frac{\alpha_s(\mu)}{2\pi} \Biggl\{ \delta (\xi) \bigg(
    \frac{C_A}{2}  L_Q^2+ \frac{\beta_0}{2}  L_Q+ 
    C_A \Big(\frac{67}{18} - \frac{5\pi^2}{12}\Big) - \frac{10}{9}n_f T_F\bigg) \nn  \\
    &\qquad+ \Theta(1-4\xi) \Bigg[ - 2 C_A {\cal L}_1 (\xi) +  {\cal L}_0(\xi) \bigg(4 C_A  \log \Big(\frac{1 + \sqrt{1-4  \xi}}{2}\Big) \nn \\
    &\qquad \qquad  -\frac{\beta_0}{2}\sqrt{1-4\xi} + \frac{C_A - 2 n_f T_F}{3} 4\xi \sqrt{1-4\xi}\bigg) 
    \Bigg]
    \Biggr\} \Bigg] \, ,
\end{align}
Including the other cases where one of the partons is outside the jet, we have
\begin{align}\label{eq:GqgBare}
    {\cal G}^{[1]}_{q,\rm No\,sd} \big(z, Q , \xi,\mu \big)    &= \delta (\xi)  \Big[ \delta (1-z) + J_{q}^{[1]}  \big(z, Q , \mu \big) \Big]  \\
    &+ \frac{\alpha_sC_F}{2\pi} 
    \delta (1-z) \Bigg\{\delta (\xi)\Big(
    -3 +  \frac{\pi^2}{3} \Big) \nn   \\
    &   
    +  \Theta \big(1-4 \xi\big) \Bigg[   - 2 {\cal L}_1 (\xi) + {\cal L}_0(\xi) \bigg(
    4 \ln \Big(\frac{1 + \sqrt{1 - 4 \xi}}{2}\Big)  -\frac{3}{2} \sqrt{1-4\xi}  \bigg)
    \Bigg]
    \Bigg\}  \, , \nn\\
    {\cal G}^{[1]}_{g,\rm No\,sd} \big(z, Q , \xi ,\mu\big)     &= \delta (\xi)  \Big[ \delta (1-z) + J_{g}^{[1]}  \big(z, Q , \mu \big) \Big]  \nn \\
    &+ \frac{\alpha_sC_F}{2\pi} 
    \delta (1-z) \Biggl\{\delta (\xi) \bigg(C_A \Big(\frac{1}{4} + \frac{\pi^2}{3}\Big) - \frac{13}{12}\beta_0\bigg) + \Theta(1-4\xi) \Bigg[ - 2 C_A {\cal L}_1 (\xi) \nn\\
    &\hspace{-30pt} 
    +  {\cal L}_0(\xi) \bigg(4 C_A  \log \Big(\frac{1 + \sqrt{1-4 \xi}}{2}\Big) -\frac{\beta_0}{2}\sqrt{1-4\xi} + \frac{C_A - 2 n_f T_F}{3} 4\xi \sqrt{1-4\xi}\bigg) 
    \Bigg]
    \Biggr\} \, . \nn 
    \nn
\end{align}
Here $J_\kappa$ is the inclusive jet function and the one-loop results can be found in \Ref{Kang:2016mcy}.
Thus, we can read off the parts that correspond to $\tilde{\cal G}_\kappa$ function defined in \eq{inclJ} yielding the results given in \eq{calG}.

%=======================================================================
\subsection{One-loop result for the groomed jet mass}
%=======================================================================

\label{app:FOGr}
We now include the soft drop grooming condition in \eq{GInteg}: 
\begin{align}\label{eq:GIntegSD}
    {\cal G}^{\rm [both\,in]}_{\kappa, \rm sd}\big(z, Q , \xi, \alpha_s\big)   
    &=\delta(1-z)\mu^{2\eps} \iota^\eps (2g^2) \int d \Phi_2^{c} (x,y)  \frac{1}{y Q^2} \Big(\sum_j \hat P_{ j\kappa} (x) \Big) 
    \\
    &\quad\times  \overline \Theta_{k_T}(x,y) \big [\overline \Theta_{\rm sd}(x,y) \delta (\xi  - y) +  \Theta_{\rm sd}(x,y)\delta (\xi)  -\delta (\xi)\big] \nn \, ,
\end{align}
The soft drop constraint for either $pp$ or $e^+e^-$ collisions can be written as 
\begin{align}
    \overline \Theta_{\rm sd} (x,y) &= \Theta \bigg(\frac{\min \{x_1 ,x_2\}}{x_1 + x_2} - \xi_0 \Big(\frac{y}{4x_1x_2}\Big)^\frac{\beta}{2} \bigg) \, ,
    \qquad \Theta_{\rm sd} (x,y)\equiv  1 - \overline \Theta_{\rm sd}(x,y) \, ,
\end{align}
where the variables $x_{1,2}$ are given by
\begin{align}
    x_1(x,y) =  \frac{xy \zeta^2}{2} + \frac{1-x}{2} \, , \qquad 
    x_2(x,y) = \frac{(1-x) y \zeta^2}{2} + \frac{x}{2} \, .
\end{align}
$\zeta$ in was defined in \eq{zetaDef} and $\xi_0$ in \eq{xi0def}.

Including the other case when one of the partons leaves the jet, we re-write \eq{GIntegSD} as
\begin{align}
    {\cal G}_{\kappa, \rm sd} \big(z, Q , \xi, \alpha_s \big)   
    = {\cal G}_{\kappa, \rm No\,sd} \big(z, Q , \xi, \alpha_s\big)    
    +\delta(1-z) \Delta   {\cal G}_{\kappa,\rm sd}\big(Q , \xi, \alpha_s(\mu)\big)    \, ,
\end{align}
where the first term is the result for ungroomed jet-mass measurement given in \eq{GqgBare}, and the second, finite piece being the remainder that depends on the soft drop condition:
\begin{align}\label{eq:calGSDInteg}
     \Delta   {\cal G}_{\kappa,\rm sd}\big(Q , \xi, \alpha_s(\mu)\big)
    &=  -\frac{\alpha_s}{2\pi}\frac{ e^{\eps \gamma_E}}{ \Gamma(1-\eps)}  \Big(\frac{\mu}{Q}\Big)^{2\eps}
    \int_0^{1} \frac{dx\,dy   }{y^{1+\eps} \big[x(1-x)\big]^\eps}\Big(\sum_j \hat P_{ j\kappa} (x) \Big)    \nn \\
    &\qquad \qquad \times \overline \Theta_{k_T} (x,y)
    \Theta_{\rm sd} (x,y)
    \big[\delta (\xi - y) - \delta (\xi)\big] \, .
\end{align}
To proceed further, we consider the result for \eq{GIntegSD} in the soft limit, which will also be useful for other results for the soft function in the ungroomed region. Since soft-drop to ${\cal O} (\zcut)$ affects only the soft radiation, the soft limit of the above result will properly capture all the divergent and logarithmic parts of the cross section, and the remaining parts will be finite. The soft limit is found by letting $x \ll 1$ in \eq{GIntegSD} and retaining the pieces that survive in the leading power,
\begin{align}\label{eq:CalGBothInSD}
    {\cal G}^{(0)}_{\kappa, \rm sd}\big(z, Q , \xi ,\alpha_s\big)   
    &=\delta(1-z)\mu^{2\eps} \iota^\eps (2g^2) \int \frac{ d \Phi_1^{s} (x,y) }{yQ^2} \Big(\sum_j \hat P^{(0)}_{ j\kappa} (x) \Big) \\
    &\quad\times  \overline \Theta^{(0)}_{k_T}(x,y) \big [\overline \Theta^{(0)}_{\rm sd}(x,y) \delta (\xi  - y) +  \Theta^{(0)}_{\rm sd}(x,y)\delta (\xi)  -\delta (\xi)\big] \nn \, ,
\end{align}
where the constraints are now given by
\begin{align}\label{eq:Constr0}
   \overline \Theta^{(0)}_{k_T} (x, y) = \Theta (x -y ) \, , \qquad 
   \overline \Theta_{\rm sd}^{(0)} (x,y) = \Theta \Big(y \zeta^2 +  x - \xi_0^{\frac{2}{2+\beta}} y^{\frac{\beta}{2+\beta}}\Big) \, .
\end{align} 
We now re-write \eq{CalGBothInSD} as
\begin{align}
    {\cal G}^{(0)\rm [both\,in]}_{\kappa, \rm sd}\big(z, Q , \xi, \alpha_s\big)   &
    =  \delta(1-z)\Big [ {\cal G}^{(0)}_{\kappa, \rm No\,sd}\big(\xi, Q,\alpha_s  \big)    
    +
  \Delta   {\cal G}_{\kappa,\rm sd}^{(0)}\big(\xi, Q ,  \alpha_s(\mu) \big)  \Big] \, ,
\end{align}
where the first term on the r.h.s. corresponds to the soft limit of the ungroomed jet-mass measurement and the splitting functions in \eq{GInteg} (for the case when both the partons are in the jet). Similarly the second term is  the soft limit of the result in \eq{calGSDInteg}. The first piece is simply the ungroomed soft function written in terms of $\xi$:
\begin{align}\label{eq:G0nosd}
    {\cal G}^{(0)}_{\kappa, \rm No\,sd}\big( \xi, Q ,\alpha_s\big)    &= 
    \frac{\alpha_s C_\kappa}{2\pi} \Biggl\{\delta (\xi) \bigg(-\frac{1}{\eps^2} - \frac{L_Q}{\eps} - L_Q^2 + \frac{\pi^2}{12}\bigg) + {\cal L}_0(\xi) \bigg(\frac{2}{\eps} + 2L_Q \bigg) - 4 {\cal L}_1 (\xi)
    \Biggr\} \, ,
\end{align}
with $L_Q$ defined above in \eq{LmuQRdef},
and the finite remainder being
\begin{align}\label{eq:DeltaG0sd}
    \Delta   {\cal G}^{(0)}_{\kappa, \rm sd} \big(\xi, \alpha_s(\mu)\big)  
     &=  -\frac{\alpha_sC_\kappa}{\pi}\frac{ e^{\eps \gamma_E}}{ \Gamma(1-\eps)}  \Big(\frac{\mu}{Q}\Big)^{2\eps}
    \int_0^{\infty} \frac{dx\,dy   }{(xy)^{1+\eps} }  \overline \Theta^{(0)}_{k_T} (x,y)
    \Theta^{(0)}_{\rm sd} (x,y)
    \big[\delta (\xi - y) - \delta (\xi)\big] \nn \\ 
        &=-\frac{\alpha_s C_\kappa}{\pi} \Biggl[\frac{\Theta(\xi_0^\prime - \xi)\Theta(\xi)}{\xi} \ln \bigg((1+\zeta^2)\Big[\Big(\frac{\xi_0'}{\xi}\Big)^\frac{2}{2+\beta} -1 \Big] + 1\bigg) \Biggr]^{[\xi_0']}_+ \, .
\end{align}
The soft drop and jet clustering conditions ensure that $x$ is never allowed to be more than one, hence the upper limit can be sent to $\infty$. 
As a result, just as \eq{G0nosd} is simply the ungroomed soft function, so is this term nothing but the result for ${\cal O}(\alpha_s)$ soft drop correction in \eq{Splain1} with the replacement $\xi = \ell^+/Q$. This term thus will be correctly reproduced by the resummed cross section in \eq{xsecResummedPlain} in the fixed order region.

\begin{figure}[t]
	\centering
	\includegraphics[width=0.49\textwidth]{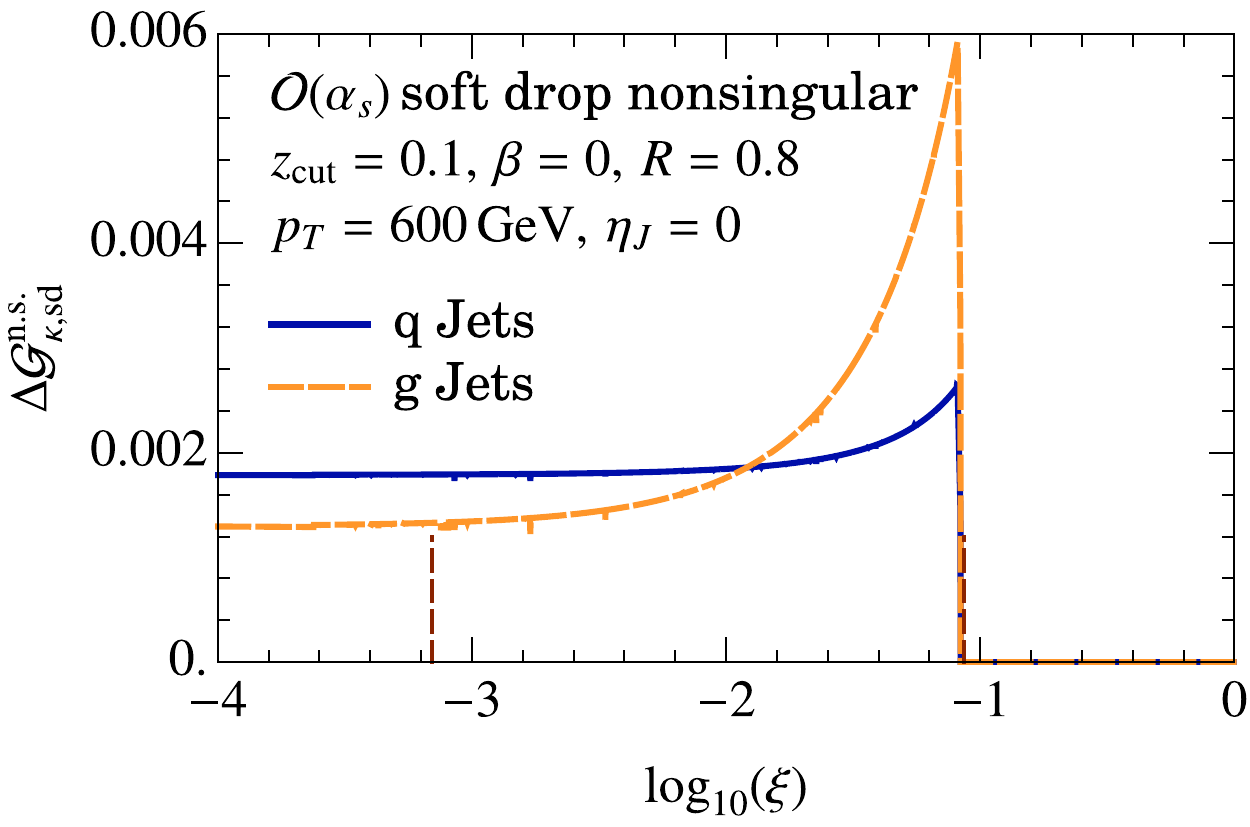}
	\includegraphics[width=0.49\textwidth]{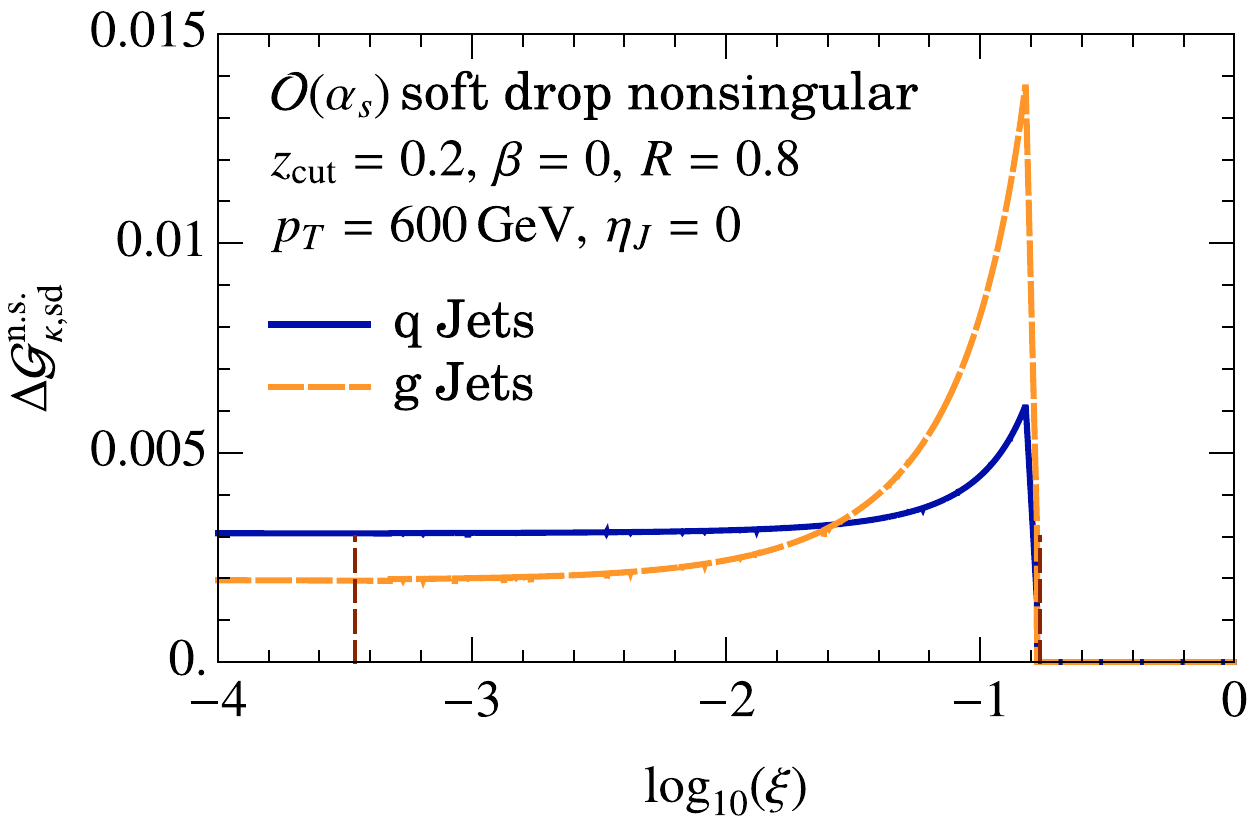}
	\includegraphics[width=0.49\textwidth]{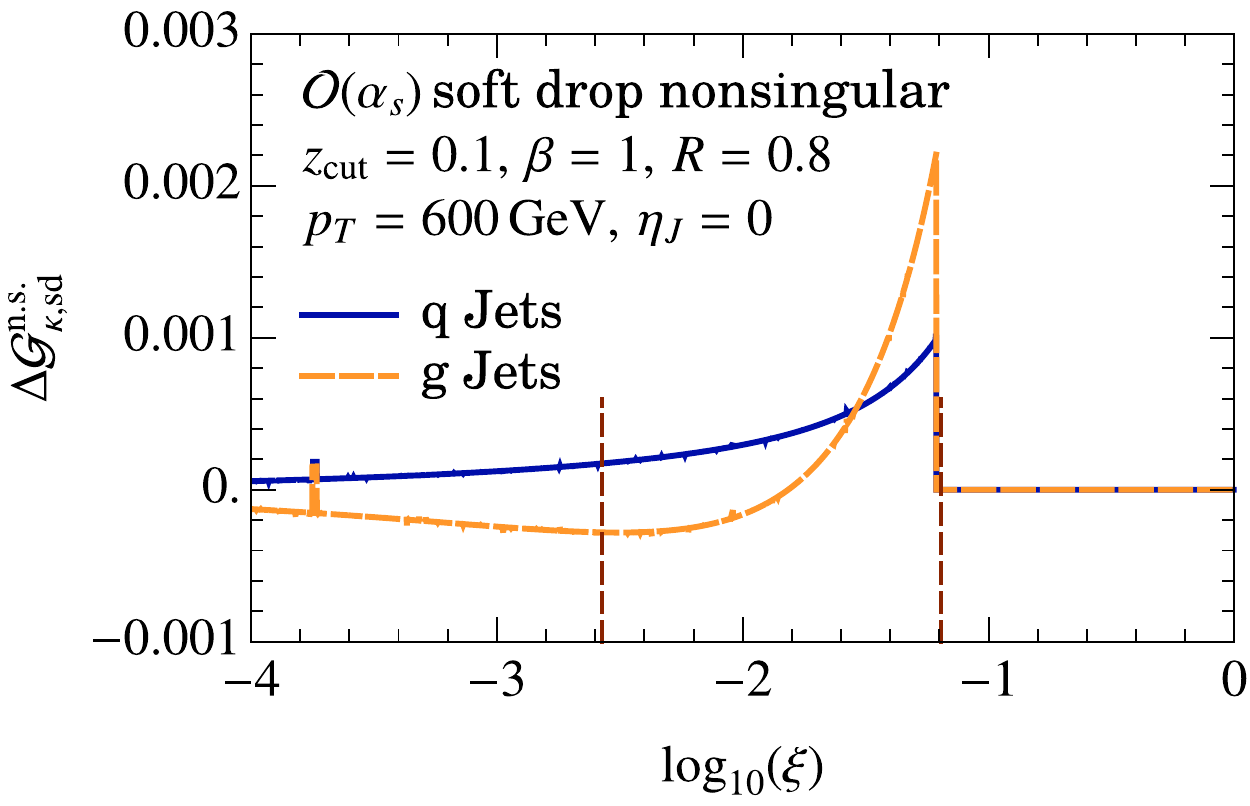}
	\includegraphics[width=0.49\textwidth]{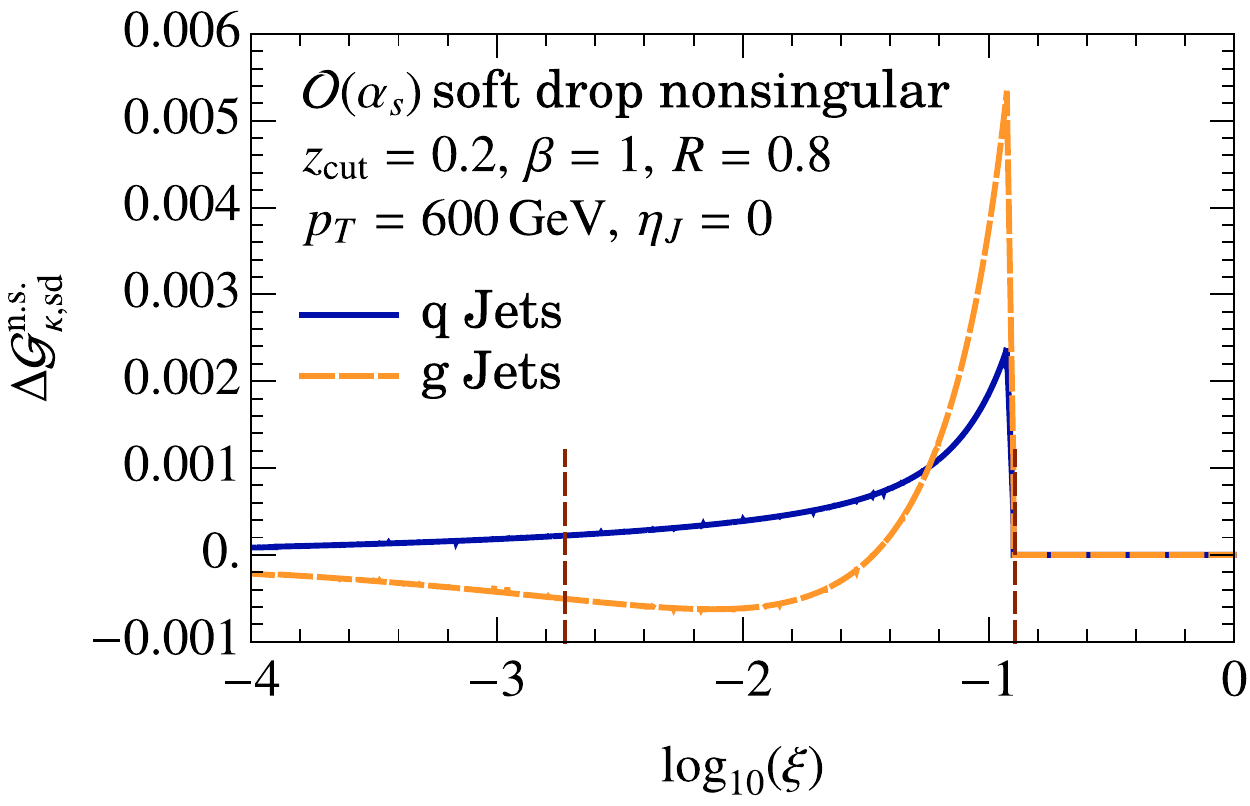}
	\includegraphics[width=0.49\textwidth]{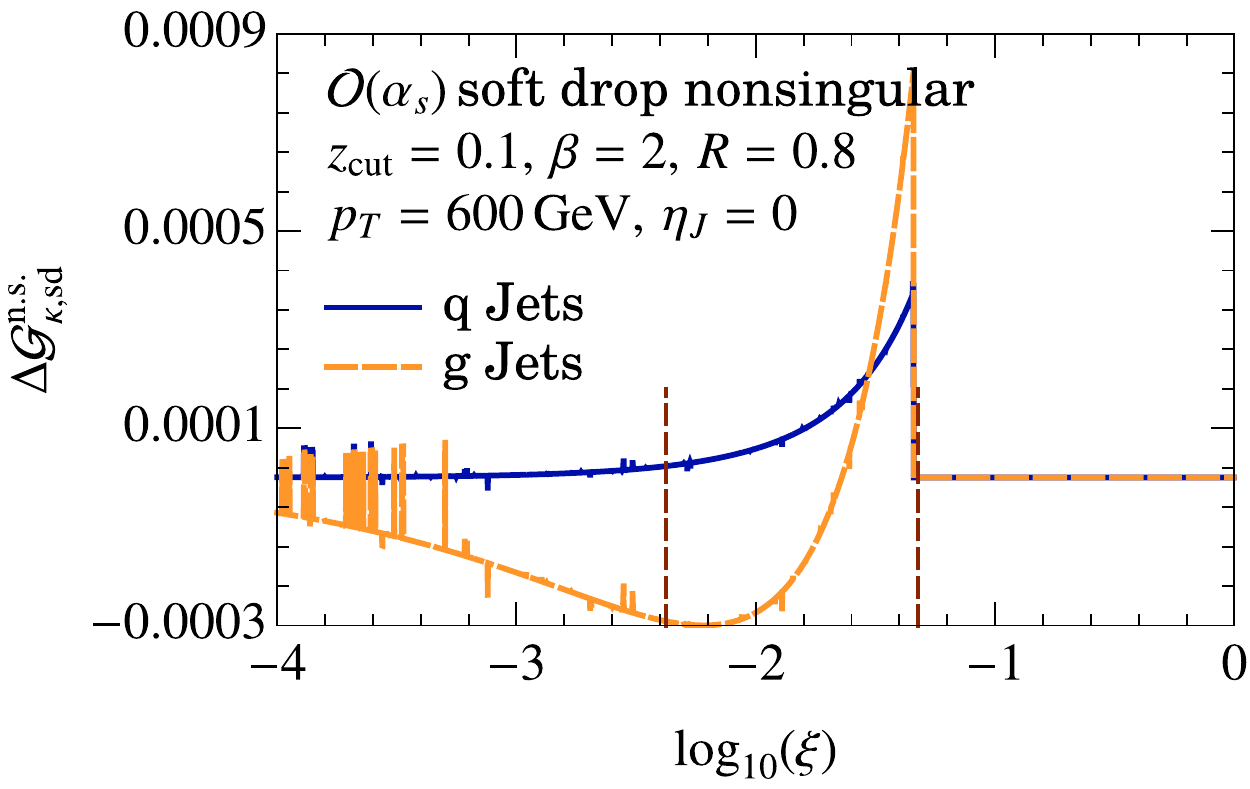}
	\includegraphics[width=0.49\textwidth]{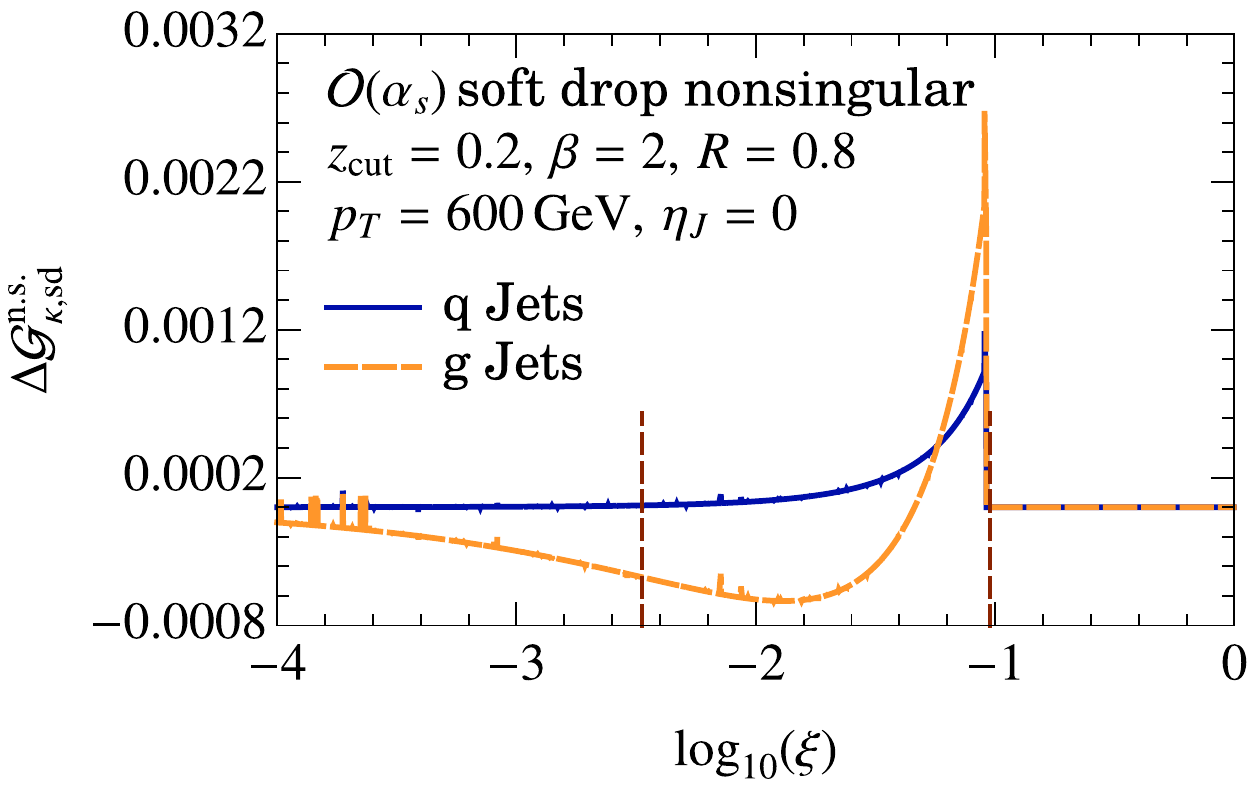}
	
	\caption{Soft drop non-singular corrections shown for various combinations of soft drop parameters and quark and gluon jets. The jaggedness is related to numerical integration. The dashed vertical lines denote the SDOE region given in \eq{SDNPprecise}, which is also the range chosen for the $\as$-sensitivity analysis.}
	\label{fig:sdns}
\end{figure}

Thus, we can evaluate the non-singular correction in \eq{GSDNS} by subtracting \eq{DeltaG0sd} from \eq{calGSDInteg}. Doing so regulates the soft singularity at $x\ra 0$, and thus we are able to numerically integrate  the remainder to obtain a residual fixed order correction that vanishes as $\xi \ra 0$:
\begin{align}\label{eq:NSInteg}
     \Delta \tilde {\cal G}^{\rm n.s.[1]}_{\kappa, \rm sd} &=  \Delta   {\cal G}_{\kappa, \rm sd} \big(\xi,Q, \alpha_s(\mu)\big)  
    -  \Delta   {\cal G}^{(0)}_{\kappa, \rm sd} \big(\xi,Q, \alpha_s(\mu)\big)  
    \\
    & = \frac{\alpha_s}{2\pi}  \Biggl[\int_0^1  \frac{dx}{\xi}  \sum_j  \big(\Theta_{\cal R}(x,\xi )   \hat P_{ j\kappa} (x) -\Theta^{(0)}_{\cal R}(x,\xi)   \hat P^{(0)}_{ j\kappa} (x)\big) \Biggr]_+   \, ,\nn
\end{align}
where the integration region is specified by the following constraints involving the soft drop (failing) and jet clustering conditions:
\begin{align}\label{eq:sdns}
    \Theta_{\cal R}(x,y) &\equiv \Theta\big(x(1-x) -y\big) 
    \Theta_{\rm sd} (x,y)
    \, , \\
    \Theta^{(0)}_{\cal R}(x,y)  &\equiv \Theta (x- y)   \Theta \big(\xi_0^{\frac{2}{2+\beta}} y^{\frac{\beta}{2+\beta}}  -x- y \zeta^2 \big) \, .\nn
\end{align}
In \eq{NSInteg} we set the subleading power cases for $P_{ j\kappa}^{(0)}$ to zero.

We show in \fig{sdns} the result of numerical integration of \eq{sdns} for $\zcut \in \{0.1, 0.2\}$ (columns) and $\beta \in \{0, 1, 2\}$ (rows) for quark and gluon jets. Comparing with the cross section in the first row in \fig{profile_vars_all} we see that these non-singular corrections are smaller by a factor of 100, and well within the perturbative uncertainty. This justifies our reasoning for dropping these terms for simplicity in our $\as$-sensitivity analysis. 
Furthermore, we see that these corrections are discontinuous at $\xi_0'$ as they are only ${\cal O}(\as)$ pieces. Including multiple emission contributions will be necessary to consistently describe ${\cal O}(\xi_0)$ power suppressed terms in the groomed jet-mass cross section.

\section{Results for factorization functions}
We follow precisely the same convention and notation for anomalous dimensions and Laplace transform as detailed in App.~A of \Ref{Pathak:2020iue}.

%=====================
\subsection{Hard collinear normalization}
%=====================

The contribution of the hard collinear mode to the normalization is given by
\begin{align}\label{eq:Nincl1loop}
	N_{\rm incl}^q(Q, \mu) &= 1 + \frac{\alpha_s(\mu) C_F}{2\pi}  \Bigg\{
	-\frac{1}{2} \ln^2 \frac{\mu^2}{Q^2} - \frac{3}{2}\ln\frac{\mu^2}{Q^2}
	-\frac{13}{2} + \frac{3\pi^2}{4} 	\Bigg\} \, , \\
	N_{\rm incl}^g (Q, \mu) &= 1 +  \frac{\alpha_s(\mu)}{2\pi} 
	\Bigg\{
	-\frac{C_A}{2} \ln^2 \frac{\mu^2}{Q^2} - \frac{\beta_0}{2}\ln\frac{\mu^2}{Q^2}
	+ C_A \Big(-\frac{5}{12} + \frac{3\pi^2}{4}\Big)	 - \frac{23}{12} \beta_0\Bigg\} \nn \, .
\end{align}
The cusp anomalous dimension is given by
\begin{align}
    \Gamma_{N^\kappa} \big[\alpha_s\big] = - 2 C_\kappa \Gamma^{\rm cusp} [\alpha_s]
\end{align}
and the non-cusp anomalous dimension being simply
\begin{align}
    \gamma^{N_\kappa}[\alpha_s] = 2 \gamma_\kappa \big[\alpha_s\big]
\end{align}
where $\gamma_\kappa$ is the anomalous dimension for flavor $\kappa$, such that
\begin{align}
    \gamma_0^{N^q} = - 6 C_F \,, \qquad 
    \gamma_0^{N^g} = -2 \beta_0 \, ,
\end{align}
\begin{align}
    \gamma_1^{N^q} &= \big(-3 + 4\pi^2 - 48 \zeta_3) C_F^2 
    + \bigg(-\frac{961}{27} - \frac{11\pi^2}{3} + 52 \zeta_3 \bigg) C_F C_A + \bigg(\frac{260}{27} + \frac{4\pi^2}{3}\bigg) C_F n_f T_F \, ,  \nn \\
    \gamma_1^{N_g} &= \bigg(-\frac{118}{9} + 4 \zeta_3\bigg) C_A^2 + \bigg( -\frac{38}{9} + \frac{\pi^2}{3}
    \bigg) C_A \beta_0 - 4 \beta_1 \, .
\end{align}
%=====================
\subsection{Jet function}
%=====================
\label{app:jetFunc}
The jet function for a quark jet collinear to the direction $n$ is given by
\begin{align}
    J_q (m_J^2,\mu) = \frac{\left(2\pi\right)^3}{N_C}
    {\rm tr} \bigl\langle 0 \big|
    \frac{\slashed{n}}{2} \chi_n (0) \, \delta \left(Q - \bar n \cdot \cal P \right)
    \delta^{(2)} ( \vec{\cal P}_{\perp} ) \delta \left(m_J^2 - \hat m_J^2\right)
    \bar \chi_n (0)
    \vert
    0 \bigr\rangle_\mu \,,
\end{align}
In this expression, $\chi_n(0)$ the gauge invariant SCET quark ``jet'' field, given by $\chi_n \equiv W_n^\dagger \xi_n$ to leading power in SCET. Here $\xi_n$ is the SCET $n$-collinear quark field and $W_n\big[\bn \cdot A_n\big]$ is a Wilson built out of $\bn \cdot A_n$ collinear gluon fields. Here $\bn$ is an auxiliary vector such that $\bn \cdot n \sim 1$. The measurement operator $\hat m_J^2$ measures the invariant squared mass of the jet, and acts on a state $\vert X_n \rangle$ of $n$-collinear particles as
\eqh{
    \hat m_J^2 \vert X_n \rangle = \Big( \sum_{i \in X_{n}} p_i \Big)^2\vert X_{n} \rangle \,.
}
Similarly, the gluon jet function is written in terms of the SCET gluon, gauge invariant building block $\mathcal{B}_{\perp \nu} = g^{-1}\big[W_n^\dagger i D_{n\perp}^\mu W_n\big]$ as
\eqh{
    J_g (m_J^2) = \frac{\left(2\pi\right)^3}{N_C}
    {\rm tr} \bigl \langle 0 \big|
    \mathcal{B}_\perp ^{\nu} (0)  \delta \left(Q - \bar n \cdot \cal{P} \right)
    \delta^{(2)} ( \vec{\cal P}_{\perp} ) \delta \left(m_J^2 - \hat m_J^2 \right)
    \mathcal{B}_{\perp \nu} (0) 
    \vert
    0 \bigr\rangle_\mu \,,
}
The one-loop expressions are obtained by simply starting from \eq{GInteg} without the jet clustering constraint as the collinear partons lie at angles parametrically smaller than jet radius. The results are given by
\begin{align}\label{eq:Jmom}
	J_q(m_J^2 ,\mu ) &=  \delta(m_J^2) + \frac{\alpha_s(\mu) C_F}{\pi} \bigg( {\cal L}_1(m_J^2, \mu^2)- \frac{6}{8}{\cal L}_0(m_J^2, \mu^2) 
	-\frac{\delta (m_J^2)}{4} \Big(\pi^2 - 7\Big)
	\bigg)
	\, ,  \\ 
	J_g(m_J^2, \mu)  &= \delta (m_J^2)   + \frac{\alpha_s(\mu)}{\pi} \bigg(\frac{C_A}{\pi}  {\cal L}_1(m_J^2, \mu^2)- \frac{\beta_0}{4}{\cal L}_0(m_J^2, \mu^2)  \nn \\ 
	&\qquad \qquad\qquad\qquad+ 
	\frac{\delta(m_J^2) }{36}\Big(C_A\big(67 - 9\pi^2\big) - 20 n_F T_R \Big)
	\bigg) \nn \, , 
\end{align}
where we follow the convention
\begin{align}
    {\cal L}_n (\ell^+, \mu ) \equiv \frac{1}{\mu} {\cal L}_n\bigg(\frac{\ell^+}{\mu} \bigg)   \, .
\end{align}

After Laplace transforming using
\begin{align}
    \tilde J_\kappa(s,\mu) &= \int e^{-s \, m_J^2} J_\kappa(m_J^2,\mu) \, ,
\end{align}
the quark and gluon jet functions at one loop become
\begin{align}
    J_q (s,\mu) & = 1 + \frac{\alpha_s C_F}{\pi} \left[ \frac{1}{2}
    \log^2 \left(   s \, \mu^2 e^{\gamma_E} \right)
    + \frac{3}{4}
    \log \left(  s \, \mu^2 e^{\gamma_E} \right)
    + \frac{7}{4}
    -
    \frac{\pi^2}{6}
    \right] \,, \\
    J_g (s,\mu) & = 1 + \frac{\alpha_s}{\pi} \left[ \frac{C_A}{2}
    \log^2 \left( s \, \mu^2 e^{\gamma_E} \right)
    + \frac{\beta_0}{4}
    \log \left(   s \mu^2 e^{\gamma_E} \right)
    + C_A
    \left(\frac{67}{36} - \frac{\pi^2}{6} \right) -
    n_F T_R \, \frac{5}{9}
    \right] \,.  \nn 
\end{align}
Here the cusp anomalous dimension satisfies Casmir scaling to the 3-loop order
we need here
\eqh{
    \Gamma_{J}^\kappa = 2 \, C_\kappa \Gamma^{\text{cusp}}[\alpha_s] \,,
}
while the one and two-loop non-cusp anomalous dimensions needed for NNLL resummation are given by
\begin{align}
	\gamma_0^{J_q} &= 6 C_F \, ,\qquad
	\gamma_0^{J_g}  = 2 \beta_0 \, ,
\end{align}
\begin{align}
    \gamma_1^{J_q} & = C_F \left[ C_F \left(3-4 \pi^2+48 \zeta_3 \right) + C_A \left( \frac{1769}{27} + \frac{22 \pi^2}{9} - 80 \zeta_3 \right) + T_R n_f \left(-\frac{484}{27} - \frac{8 \pi^2}{9} \right) \right] \,,\nn 
    \\
    \gamma_1^{J_g} & = C_A^2 \left( \frac{2192}{27} - \frac{22 \pi^2}{9} - 32 \zeta_3 \right) + C_A T_R n_f \left(-\frac{736}{27}+ \frac{8 \pi^2}{9} \right) - 8 C_F T_R n_f \,.
\end{align}
for quarks and gluons respectively.
%

%=======================================================================
\subsection{Ungroomed-soft function}
%=======================================================================
The ungroomed soft function measures $p^+$ with a jet radius constraint. The operator definition  is given by
\eqh{
    S_{c_m}^{\kappa } (\ell^+,\mu) =
    \frac{1}{N_c}
    {\rm tr} \bigl\langle 0 \big|  {\rm T} \{ Y_{n\kappa}^\dagger Y_{\bar{n} \kappa} \} \delta \bigl(\ell^+- \overline  \Theta_{R}^{(0)} \, n\cdot \hat p_{us}\bigr) \bar {\rm T} \{ Y_{n \kappa}^\dagger Y_{\bar{n}  \kappa} \}
    \big| 0 \bigr\rangle_\mu \, .
    \label{eq:ScmDef}
}
The operator $\hat p_{us}^\mu$ acts on the ultrasoft fields passing the cut and satisfying the jet radius constraint $\overline \Theta_R^{(0)}$ in the soft limit. In absence of any real soft radiation the $\overline \Theta_R^{(0)}$ is trivially satisfied (as there is always a collinear parton in the background that sources the ultrasoft gluons). Here $Y_{n,\kappa}\big[n\cdot A_{us}\big]$ and $Y_{\bn,\kappa}\big[\bn \cdot A_{us}\big]$ are Wilson lines built out of ultrasoft gluons resulting from a field redefinitions of the SCET collinear quark and gluon fields that result in decoupling of collinear-ultrasoft interactions at leading power.  For $\kappa = g$ these correspond to adjoint Wilson lines. The subscript $\mu$ on the matrix element denotes $\MS$ renormalization. 

In terms of the standard definition of light cone coordinates, the jet radius constraint is given by
\begin{align}
    \overline \Theta_R^{(0)} \equiv \Theta \Big(\zeta - \sqrt{\frac{p^+}{p^-}}\Big) \, , 
\end{align}
where $\zeta$ was defined above in \eq{zetaDef}. Upon expressing in terms of $n,\bn$ reference vectors defined in \eq{LCDef}, this constraint simply becomes that of hemisphere jets, $\overline \Theta_R^{(0)} = \Theta(p^- - p^+)$, such that
\begin{align}\label{eq:ScmNLO}
    S_{c_m}^{\kappa} (\ell^+ ,\epsilon) &= 	\frac{\alpha_s C_\kappa}{\pi}
    \frac{(\mu^2 e^{\gamma_E})^{\eps}}{\Gamma(1-\eps)} 
    \int \frac{dp^+ dp^-}{(p^+p^-)^{1+\eps}} 
    \Theta(p^- - p^+)
    \big[
    \delta(\ell^+ - p^+ ) - \delta(\ell^+)
    \big]
    \nn 
    \\
    &\qquad =\frac{1}{\eps_{\rm UV}}\frac{\alpha_s C_\kappa}{\pi}
    \frac{( e^{\gamma_E})^{\eps}}{\Gamma(1-\eps)} \frac{1}{\mu}
    \bigg[
    \Big(
    \frac{\mu}{\ell^+}
    \Big)^{1+ 2\eps_{\rm IR}}
    - \delta\big(\ell^+ \big)\Big(\frac{1}{2\eps_{\rm UV}} - \frac{1}{2\eps_{\rm IR}}\Big)
    \bigg]
    \, .
\end{align}
We can expand the $(\mu/\ell^+)^{1+2\eps_{\rm IR}}$ using the identity:
\begin{align}
    \lim\limits_{\eps \ra 0}  \frac{\Theta(x)}{x^{a+b\eps}} = \frac{\delta (x)}{1-a-b\eps} + \Big[\frac{\Theta(x)}{x^a}\Big]_+ - b\eps \Big[\Theta(x)\frac{\ln\, x}{x^{a}}\Big]_+  + \ldots \,, 
\end{align}
such that the $1/\eps_{\rm IR}$ poles cancel. The renormalized expression reads
\begin{align}\label{eq:Scm1L}
    S_{c_m}^\kappa  \big(\ell^+,  \mu\big)  &=
    \delta (\ell^+) + \frac{\alpha_s C_\kappa}{2\pi} \bigg[-4 {\cal L}_1 \big(\ell^+, \mu\big) + \frac{\pi^2}{12} \delta (\ell^+)  \bigg] \, .
\end{align}
We note that when expressed in terms of the dimensionless variables $x$ and $y$ defined in \eqs{xyDef}{yDef2}, we find precisely the same one-loop integral leading to the expression in \eq{G0nosd}:
\begin{align}
    S_{c_m}^\kappa (\ell^+, \mu)   = \frac{1}{Q} {\cal G}_{\kappa,\rm No\,sd}^{(0)} \bigg(\frac{\ell^+}{Q}, Q, \alpha_s(\mu)\bigg) \, .
\end{align}
The cusp anomalous dimension is given by
\begin{align}
    \Gamma_{S_{c_m}}^\kappa[\alpha_s] = -2 C_\kappa \Gamma^{\rm cusp} [\alpha_s]
\end{align}
The non-cusp anomalous dimensions are given by
\begin{align}
    \gamma_0^{S_{c_m}^\kappa} &= 0 \, ,\\ 
    \gamma_1^{S_{c_m}^\kappa} &= C_\kappa \bigg( C_A \Big(-\frac{808}{27} + \frac{11 \pi^2}{9} + 28 \zeta_3\Big) 
    + n_f T_F \Big(\frac{224}{27} - \frac{4\pi^2}{9}\Big)
    \bigg) 
    \nn \, .
\end{align}
%=======================================================================

\subsection{Collinear-soft function}
%=======================================================================
\label{app:csoft}
The collinear-soft function is defined by
\begin{align}
    S_c^\kappa (\tilde k, \beta, \mu) \equiv \frac{1}{N_c} \tr \langle 0 | {\rm T} \big\{ X_{n}^\dagger V_n \big \} \delta \Big(\tilde k  - \overline \Theta_{\rm sd}^{cs}(\hat p_{cs}^\mu) \,(\qcut)^{\frac{1}{1+\beta}} n\cdot \hat p_{cs}\Big ) \overline {\rm T} \big\{V_n^\dagger X_n \big \} | 0 \rangle_{\mu} 
\end{align}
Here $X_{n}\big[n\cdot A_{n}^{cs}\big]$ denotes collinear-soft Wilson line resulting from the same ultrasoft decoupling field redefinition of collinear quarks and gluons as in the case of ungroomed soft function above. The Wilson line $V_n\big[\bn \cdot A_{cs}^n\big]$, on the other hand, arises from integrating out off-shell modes generated through collinear-soft gluons emitted by partons in directions other than $n$. The measurement function involves soft drop constraint expanded in the collinear-soft limit. In general, the groomer involves clustering of collinear soft particles along with the energetic collinear partons (at parametrically smaller angles), with subsequent grooming tests. Thus, it is not straightforward to write down the expression for arbitrary number of partons. At one-loop accuracy for a single collinear-soft gluon carrying momentum $p^\mu$ we have
\begin{align}
    \overline \Theta_{\rm sd}^{cs}  (p^\mu)   \equiv \Theta \bigg(\bn \cdot p - (\qcut)^{\frac{1}{2+\beta}} (n\cdot p)^{\frac{\beta}{2+\beta}}\bigg) \, .
\end{align}
Comparing with the earlier expression for the soft limit in \eq{Constr0}, we have additionally made a small-angle approximation which leads to dropping the $y\zeta^2$ term in \eq{Constr0}. Note that unlike \eq{ScmDef} we do not include the jet radius constraint as the collinear soft particles  lie deep within the jet.  

Despite various appearances of $\qcut$ factors, the function in fact only depends on $\beta$. This can be seen by rewriting the above constraint as
\begin{align}
    \delta \big( \tilde k -  \overline \Theta_{\rm sd}^{cs}  (p^\mu)  (\qcut)^\frac{1}{1+\beta} n\cdot p\big)  = \delta \bigg( \tilde k -  \Theta \big( \bn \cdot \tilde p - \omega^{\frac{2}{2+\beta}}  (n\cdot  \tilde p  )^{\frac{\beta}{2+\beta}}\big) n\cdot \tilde{ p} \:  \omega^{\frac{1}{1+\beta}}\bigg) \, , 
\end{align} 
where $\tilde p$ is related to $p$ as
\begin{align}\label{eq:pprimedef}
        \tilde p  = \Big(p^+ (\qcut/\omega)^\frac{1}{1+\beta} ,  \frac{p^-}{(\qcut/\omega)^\frac{1}{1+\beta} }, p_\perp\Big) \, ,
\end{align}
which being a reparametrization symmetry transformation of the $n$ and $\bn$ reference vectors  results in the matrix element being independent of $\qcut$. Here $\omega$ is an arbitrary dimension-1 parameter.

To evaluate the ${\cal O}(\alpha_s)$ expression we can proceed by making the same change of variables under the integral as in \eq{pprimedef}, setting $\omega = 1$, which yields  
\begin{align}
    S_c^\kappa \big(\tilde k, \beta, \mu\big) &= 	\frac{\alpha_s C_\kappa}{\pi}
    \frac{(\mu^2 e^{\gamma_E})^{\eps}}{\Gamma(1-\eps)} 
    \int \frac{d\tilde p^+ d\tilde p^-}{(\tilde p^+\tilde p^-)^{1+\eps}} 
    \Theta\big(\tilde p^- - (\tilde p^+)^{\frac{\beta}{2+\beta}}\big) 
    \big[
    \delta(\tilde k - \tilde p^+ ) - \delta(\tilde k)
    \big]
    \\
    &=\frac{1}{\eps_{\rm UV}}\frac{\alpha_s C_\kappa}{\pi}
    \frac{( e^{\gamma_E})^{\eps}}{\Gamma(1-\eps)} \frac{1}{\mu^{\frac{2+\beta}{1+\beta}}}
    \Bigg[
    \bigg(
    \frac{\mu^{\frac{2+\beta}{1+\beta}}}{\tilde k}
    \bigg)^{1+ \frac{2(1+\beta)}{2+\beta}\eps_{\rm IR}}   
    - \delta\big(\ell^+ \big)\frac{2+\beta}{2(1+\beta)}\Big(\frac{1}{\eps_{\rm UV}} - \frac{1}{\eps_{\rm IR}}\Big)
    \Bigg]
    \, .    \nn 
\end{align}
Thus, we find 
\begin{align}	\label{eq:sc1loop}
    S_c^\kappa \big(\tilde k,  \mu   ,\beta\big)
    &= \delta(\tilde k)+ \frac{\alpha_s C_\kappa}{\pi} \Biggl[  
    \frac{-2(1+\beta)}{2 + \beta} 
    \,{\cal L}_1 \left(\tilde k , \mu^{\frac{2+\beta}{1+\beta}}  \right)
    + \frac{\pi^2}{24} \frac{2 + \beta}{1 + \beta}  
    \delta(\tilde k)
    \Biggr]  \, .
\end{align}
The Laplace-space transform of $S_c^\kappa(\tilde k ,\beta,\mu)$ is defined as
\eqh{
    \tilde S_c^\kappa (s,\beta,\mu) =  \int \df \tilde k \:  e^{-s \tilde k } S_c^{\kappa} (\tilde k , \beta ,\mu ),
}
which gives
\begin{align}\label{eq:LapSc}
    \tilde S_c^\kappa (s)
    &=
    1 + \frac{\alpha_s C_\kappa}{\pi}
    \bigg[
    - \Big(
    \frac{1+\beta}{2+\beta}
    \Big)
    \log^2 \big( s e^{\gamma_E}\mu^{\frac{2+\beta}{1+\beta}} \big)
    -\frac{\pi^2}{24}
    \frac{\beta(3\beta+ 4)}{(1+\beta)(2+\beta)}
    \bigg]
    \, .
\end{align}

The cusp anomalous dimension is given by
\begin{align}
    \Gamma_{S_{c}}^\kappa[\alpha_s] = -2 C_\kappa \Gamma^{\rm cusp} [\alpha_s] \, .
\end{align}
The one-loop non-cusp anomalous dimension is again zero,
\begin{align}
    \gamma_0^{S_c^\kappa} = 0 \, .
\end{align}
We determine the two loop non-cusp anomalous dimension via RG consistency:
\begin{align}
    \gamma_1^{S_c^\kappa}(\beta) = - \gamma_1^{N^\kappa} - \gamma_1^{J_\kappa} - \gamma_1^{S_G^\kappa}(\beta) \, ,
\end{align}
where $\gamma_1^{S_G^\kappa}(\beta)$ is stated below in \eq{gamma1SG}.
%=======================================================================
%=======================================================================
\subsection{Global soft function}
%=======================================================================
%=======================================================================
The operator definition of the global soft function is given by
\begin{align}
    S_G^\kappa \big(\qcut, \beta,\zeta, \mu\big) 
    \equiv \frac{1}{N_c} \tr \langle 0 | {\rm T} \big\{Y^\dagger_{n\kappa} Y_{\bn \kappa} \big \}   \Theta_{\rm sd}^{(gs)} (\hat p_{gs}^\mu) \overline {\rm T} \big\{Y_{n\kappa}^\dagger Y_{\bn \kappa}\big \} | 0 \rangle_\mu  \, .
\end{align}
The function involves precisely the same Wilson lines as in the case of ungroomed soft function (with the fields labeled as $A^\mu_{gs}$), with the following constraint\footnote{Note that in absence of any global soft radiation  at tree level, the constraint is vacuously true. For one-loop virtual term one can appropriately convert the $\eps_{\rm IR}$ poles to $\eps_{\rm UV}$ poles by imposing a constraint $V^{(gs)} = \overline \Theta_R^{(0)} - \overline \Theta_{\rm sd}^{(0)}$ in the virtual matrix element.}
\begin{align}
     \Theta_{\rm sd}^{(gs)} (\hat p_{gs}^\mu) \equiv \overline \Theta^{(0)}_R \Theta^{(0)}_{\rm sd}  \, ,
\end{align}
where the jet radius constraint is the same as in the case of ungroomed soft function, but here we have the soft drop failing constraint $\Theta_{\rm sd}^{(0)}(\hat p^\mu)$ with momenta expanded in the soft limit. For the case of a single gluon, this was stated above in \eq{Constr0} which in terms of $p^+ = y Q$ and $p^- = xQ$ reads
\begin{align}
    \Theta_{\rm sd}^{(0)}(p^\mu) \equiv \Theta \bigg((\qcut)^{\frac{\beta}{2+\beta}} (p^+)^\frac{\beta}{2+\beta} - p^-  - p^+ \zeta^2 \bigg)\,.
\end{align}
The one-loop result is obtained by evaluating
\begin{align}\label{eq:SGbare1}
    & S_G^{\kappa } \big(\qcut, \beta, \mu\big) =
    \frac{\alpha_s C_\kappa}{\pi}
    \frac{(\mu^2 e^{\gamma_E})^{\eps}}{\Gamma(1-\eps)} 
    \int \frac{dp^+ dp^-}{(p^+p^-)^{1+\eps}}  \overline \Theta_{R, \rm sd}^{(gs)}   
    \, ,
\end{align}
with the renormalized result being
\begin{align}\label{eq:SG1loop}
    S_G^\kappa\big(\qcut, \beta, \mu\big)&= 
    1 + \frac{\alpha_s(\mu)C_\kappa}{\pi} 
    \bigg[
    \frac{1}{(1+\beta)} \log^2\Big(\frac{\mu}{\qcut}\Big)
    -\frac{\pi^2}{24}
    \Big(
    \frac{1}{1+\beta}
    \Big)
    \\
    &\qquad \qquad \qquad \qquad 
    -\frac{(2+\beta)}{4}
    \Big(
    2{\rm Li}_2\Big[ \frac{\zeta^2}{1 + \zeta^2} \Big]
    + \log^2\big [1 + \zeta^2\big]
    \Big)
    \bigg] \nn \, .
\end{align}
The cusp anomalous dimension is given by
\begin{align}
    \Gamma_{S_G^\kappa} [\alpha_s] = + 2 C_\kappa \Gamma^{\rm cusp} [\alpha_s] \, ,
\end{align}
The one-loop non-cusp anomalous dimension $\gamma_0^{S_G^\kappa} = 0$. We use the following fit to the numerical determination of the two-loop non-cusp anomalous dimension for arbitrary $\beta$ computed in \Ref{Bell:2018vaa}:
\begin{align}
    \gamma_1^{S_G^\kappa}(\beta) = \frac{C_\kappa}{1 +\beta} \bigg( \gamma_{C_F}^{S_G^\kappa} (\beta)+n_f \gamma_{T_F}^{S_G^\kappa}(\beta) + \gamma_{C_A}^{S_G^\kappa}(\beta)
    \bigg) \, ,
\end{align}
where
\begin{align}\label{eq:gamma1SG}
    \gamma_{C_F}^{S_G^\kappa} (\beta)
    &= C_F \big(  0.00563338 \beta^3 - 0.621462  \beta^2 - 1.11337 \beta + 16.9974\big) \, , 
    \\
    \gamma_{T_F}^{S_G^\kappa}(\beta)
    &= T_F \big(- 0.26041 \beta^3 + 2.01765 \beta^2 + 3.48117 \beta - 10.9341\big) \, , \nn 
    \\ 
    \gamma_{C_A}^{S_G^\kappa}(\beta)
    &= C_A \big(
    +0.640703 \beta^3 +3.37308 \beta^2 +3.68876 \beta - 20.4351
    \big) \, . \nn 
\end{align}

%=====================

%=======================================================================
%=======================================================================
 
\section{Resummation in Laplace space}

The resummation kernels are defined as described in App.~A of \Ref{Pathak:2020iue}. For expressions for $K(\Gamma, \mu,\mu_0)$ and $\eta(\Gamma, \mu,\mu_0)$ to N$^3$LL order see \Ref{Bachu:2020nqn}.
To derive the form given in \eq{FactPlainResum} we first include resummation in the Laplace space for the differential cross section without the NGL piece:
\begin{align}\label{eq:DSigDxi}
    \frac{d \Sigma (\xi)}{d\xi}  &=  Q^2e^{K_J + K_s}  \big(Q \mu_s\big)^{\omega_s} \big(\mu_J^2\big)^{\omega_J} 
    \int \frac{dx }{2\pi i} \: e^{x Q^2 \xi} \big(e^{\gamma_E} x\big)^{\omega_J + \omega_s} \tilde J(x) \tilde S_{\rm plain}^{\kappa} (Q x)  \, , 
\end{align}
where the jet and soft functions are evolved from scales $\mu_J$ and $\mu_s$ respectively to final scale $\mu$, such that
\begin{align}
    K_J \equiv K_J (\mu, \mu_J) \, , \qquad K_{s} \equiv K_{cs_m} (\mu, \mu_s) \, , 
\end{align}
and likewise for the $\omega_J$ and $\omega_s$.  The Laplace transform of $S_{\rm plain}^\kappa$ is defined as
\begin{align}\label{eq:ScmLapDef}
    \tilde S_{\rm plain}^\kappa\bigl[  u, \mu\bigr]  
    &=
    \int_0^\infty  d \ell^+
    e^{-u \ell^+}
    S_{\rm plain}^\kappa\big[\ell^+, \mu\big]   
    \nn \\
    &=\tilde  S_{c_m} \big[u,\mu] +   \Delta \tilde S_{\rm sd}^{\kappa} \big[ u , \qcut, \beta, R , \mu \big] 
    \, ,
\end{align}
At one-loop, the ungroomed soft function in Laplace space, renormalized in $\overline{{\rm MS}}$, is given by
\begin{align}\label{eq:ScmLap}
    \tilde S_{c_m}^\kappa\bigl[u,\mu \bigr]
    &=
    1 + \frac{\alpha_s C_\kappa}{\pi}
    \bigg[
    -
    \log^2 \big( u e^{\gamma_E}\mu  \big)
    -\frac{\pi^2}{8}
    \bigg]
    \, ,
\end{align}
and
\begin{align}
     \Delta \tilde S_{\rm sd}^{\kappa} \big[ u , Q\xi_0' , \beta, \zeta , \mu  \big] 
    &= \int_0^\infty d \ell^+  \:      e^{-u \ell^+} \Delta S^{\kappa}_{\rm sd} \big[\ell^+  , Q\xi_0' , \beta, \zeta , \mu \big]  \, ,
\end{align}
Following the arguments of Ref.~\cite{Pathak:2020iue} we can show that
\begin{align}
     &\int \frac{dx }{2\pi i} \: e^{x Q^2 \xi}  \big(e^{\gamma_E}x)^{\Omega} \tilde J(x) \tilde S_{c_m}^{\kappa} (Q x)   \\
     &\qquad=\Big(\frac{1}{Q^2\xi}  \Big)^{\Omega + 1}  \int \frac{\df \varrho  }{2\pi i} \: e^{\varrho}  \big(e^{\gamma_E}\varrho)^{\Omega} \tilde J\Big(\frac{\varrho}{Q^2 \xi}\Big) \tilde S_{c_m}^{\kappa} \Big(\frac{\varrho}{Q \xi}\Big)  \nn \\
     &\qquad =
      \Big(\frac{1}{Q^2\xi}  \Big)^{\Omega + 1}  \tilde J_\kappa\Big [ \partial_{\Omega} + \log\Big(\frac{\mu_J^2}{Q^2 \xi}\Big),\, \alpha_s(\mu_J) \Big] \, 
     \tilde S_{c_m}^\kappa\Big [\partial_{\Omega} + \log\Big(\frac{\mu_{s} }{Q \xi}\Big) , \alpha_s(\mu_{cs_m})\Big ] \frac{e^{\gamma_E\Omega}}{\Gamma(-\Omega)}
     \nn \, .
\end{align}
Here the derivatives of $\Omega$ correctly insert the inverse Laplace transforms logarithms of $\varrho $ by using the result
\begin{align}
    \int \frac{\df \varrho}{2\pi i} e^{\varrho } \big(e^{\gamma_E} \varrho\big)^{\Omega} \log^{n}(e^{\gamma_E}\varrho) = \frac{\df^n}{\df \Omega^n}   \int \frac{\df \varrho}{2\pi i} e^{\varrho } \big(e^{\gamma_E} \varrho\big)^{\Omega}= \frac{\df^n}{\df \Omega^n} \frac{e^{\gamma_E \Omega} }{\Gamma(-\Omega)} \, .
\end{align}
Thus, we have
\begin{align}
    \frac{d \Sigma (\xi)}{d\xi}  &= {\cal J}^\kappa_{\rm plain} [\partial_\Omega; \xi, Q, \mu]\frac{e^{\gamma_E \Omega} }{\Gamma(-\Omega)} \bigg|_{\Omega = \omega_J + \omega_s}
\end{align}
where
\begin{align}\label{eq:PlainOperator}
    {\cal J}^\kappa_{\rm plain} [\partial_\Omega; \xi, Q, \mu]& \equiv \frac{1}{\xi} e^{K_J + K_s}  \frac{ \big(Q \mu_s\big)^{\omega_s} \big(\mu_J^2\big)^{\omega_J} }{(\xi Q^2)^{\Omega}}  
    \:  
    \nn
    \\
    &\times 
    \,
    \tilde J_\kappa\Big [ \partial_{\Omega} + \log\Big(\frac{\mu_J^2}{Q^2 \xi}\Big),\, \alpha_s(\mu_J) \Big] \, 
    \tilde S_{c_m}^\kappa\Big [\partial_{\Omega} + \log\Big(\frac{\mu_{s} }{Q \xi}\Big) , \alpha_s(\mu_{s})\Big ]
    \,, 
\end{align}
which captures the first term in \eq{xsecResummedPlain}. 
In \eq{PlainOperator} we have rewritten the Laplace transforms of the jet function and the $S_{c_m}^\kappa$ function in an alternative notation~\cite{Pathak:2020iue}:
\begin{align}\label{eq:LapAlternative}
	\tilde J_\kappa \big [ \log (e^{\gamma_E} x \mu_J^2), \alpha_s (\mu_J)\big] &\equiv \tilde J_\kappa (x, \mu_J)  \, ,
	\\
	\tilde S_{c_m}^\kappa \Big[\log \big( u e^{\gamma_E}\mu_s  \big) ,  \alpha(\mu_{s})\Big]   &\equiv
	\tilde S_{c_m}^\kappa\bigl(u,\mu_{s} \bigr)\,,
	\nn 
\end{align}
such that the first argument involving the derivative $\partial_{\Omega}$ in the Laplace transforms of the jet and collinear-soft functions is understood to replace the logarithms that appear in the corresponding Laplace space expressions.

To obtain the cumulative cross section we simply integrate over $\xi$ from $-\infty$ to $\xi_c$, such that
\begin{align}\label{eq:DSigDxi2}
    \Sigma (\xi^c) &=  e^{\gamma_E} e^{K_J + K_s}  \big(Q \mu_s\big)^{\omega_s} \big(\mu_J^2\big)^{\omega_J} 
    \int \frac{dx }{2\pi i} \: e^{x Q^2 \xi} \big(e^{\gamma_E} x\big)^{\omega_J + \omega_s - 1} \tilde J(x) \tilde S_{\rm plain}^{\kappa} (Q x)  \, , 
    \\
    &= \xi {\cal J}^\kappa_{\rm plain} [\partial_\Omega; \xi, Q, \mu] \frac{e^{\gamma_E \Omega} }{\Gamma(-\Omega + 1)} \bigg|_{\Omega = \omega_J + \omega_s} \, .
\end{align}
 
Next, the soft drop piece in \eq{ScmLapDef} involves the kernel
\begin{align}\label{eq:Qsd}
    {\cal Q}_{\rm sd}^{\kappa} \bigg[\Omega,  \frac{\xi}{\xi_0}, \beta,R,  \mu \bigg] &\equiv  \int \frac{\df \varrho }{2\pi i} \: e^{\varrho } \big(e^{\gamma_E}\varrho)^{\Omega}
        \Delta \tilde S_{\rm sd}^{\kappa} \bigg[ \frac{\varrho }{Q \xi} , \qcut, \beta, R , \mu \bigg]  \\
    &=\int_0^\infty \df \ell^+  \Bigg[ \int \frac{\df \varrho }{2\pi i} \: e^{\varrho\big(1 - \frac{\ell^+}{Q \xi}\big) } \big(e^{\gamma_E}\varrho)^{\Omega}\Bigg]\Delta S^{\kappa}_{\rm sd} \big[\ell^+  , \qcut , \beta, R , \mu \big] \nn \\
    &= \frac{e^{\gamma_E \Omega} }{\Gamma(-\Omega)} \int_0^\infty \df \ell^+ \: 
    {\cal L}_0^{-\Omega}\bigg(1 - \frac{\ell^+}{Q \xi}\bigg)
   \Delta S^{\kappa}_{\rm sd} \big[\ell^+  , \qcut , \beta, R , \mu \big] 
    \nn \, ,
\end{align}
where
\begin{align}
    {\cal L}_0^{a} (x) \equiv {\cal L}^a (x)+ \frac{1}{a} \delta (x) %= \frac{1}{x^{1-a}} 
    \, , \qquad {\cal L}^{a}(x) \equiv \Big [\frac{\Theta(x)}{x^{1-a}}\Big]_+ \, , \qquad a\neq 0 \, ,
\end{align}
where the plus function ${\cal L}^a(x)$ is defined as usual with the boundary condition at $x = 1$.
The plus function ${\cal L}_0^{a} (x)$ for $x > 0$ simply equates to $1/x^{1-a}$, but the prescription above makes the singularity integrable. 
Using the result in \eq{Splain1} this simplifies to the following expression:
\begin{align}
    &{\cal Q}_{\rm sd}^{\kappa} \bigg[\Omega, z =  \frac{\xi_0'}{\xi}  , \beta,R,  \mu \bigg] = 
    \\
    &\qquad 
    - \frac{\alpha_s C_\kappa}{\pi} \frac{e^{\gamma_E \Omega}}{\Gamma(-\Omega)} \int_0^1 \df x\: {\cal L}_0^{-\Omega}(1-zx ) \: 
    \Bigg[\frac{\Theta(1-x)\Theta(x)}{x} \ln \bigg( \frac{1 + \zeta^2}{x^{2+\beta}}- \zeta^2\bigg)
    \Bigg]_+
    \nn \, .
\end{align}
We see that unlike the fixed order expression in \eq{GkappaExplicit} the expression above as a result of resummation (for $\Omega < 0$) is non-zero for $\xi > \xi_0'$. 

To evaluate this integral we note that for  a function $f(x)$ that has integrable singularity at $x\ra 0$,
\begin{align}\label{eq:fcalLInteg}
    &\int_0^1 \df x \:{\cal L}_0^{-\Omega}(1-zx) \big[ \Theta(x)\Theta(1-x) f(x) \big]_+
	= \int_0^1 \df x\: f(x) \big[{\cal L}_0^{-\Omega}(1-zx) - 1\big]
	\\
	&\qquad = \Theta(z - 1)\bigg[  \int_0^{1/z} \df x \: f(x) \big[{\cal L}_0^{-\Omega}(1-zx) - 1\big] 
	- \int_{1/z}^1 \df x \: f(x)
	\bigg]
		\nn
	\\
	&\qquad \qquad+ \Theta(1-z)  \int_0^1 \df x \: f(x) \bigg(\frac{1}{(1-zx)^{1+\Omega}} - 1\bigg)
	\, ,
	\nn
\end{align}
where the plus prescription is only needed to regulate the integrable singularity at $x = 1/z$. We simplify the first term further (assuming $z > 1$):
\begin{align}
	& \int_0^{1/z} \df x \: f(x) \big[{\cal L}_0^{-\Omega}(1-zx) - 1\big]
		 = \int_0^{z^\inv} \df x \bigg(
	\frac{1}{(1-zx)^{1+\Omega}} \big[f(x) - f (z^\inv)\big] -f(x)
	-\frac{1}{\Omega}f (z^\inv)
	\bigg)
	\, ,
\end{align}
such that this form will improve the convergence at both $x \ra 0$ and $x\ra z^\inv$. 

By following the same logic, we can define the corresponding operator for resummation in the soft drop region:
\begin{align}\label{eq:SDOperator}
   {\cal J}^\kappa_{\rm sd} [\partial_\Omega; \xi, Q, \mu]
   & \equiv \frac{1}{\xi} 
    \:  
    e^{\big [  K_{cs} (\mu, \mu_{cs} ) + K_{J} (\mu, \mu_{J} )\big] }
    \Big(\frac{\mu_J^2}{Q^2 \xi}\Big)^{ \omega_J(\mu, \mu_J)}
    \Bigg(\frac{ \mu_{cs}}{Q \xi}\Big(\frac{\mu_{cs}}{\qcut}\Big)^\frac{1}{1+\beta}\Bigg)^{\omega_{cs} (\mu, \mu_{s})}
    \\
    &\times 
    \,
    \tilde J_\kappa\Big [ \partial_{\Omega} + \log\Big(\frac{\mu_J^2}{Q^2 \xi}\Big),\, \alpha_s(\mu_J) \Big] \, 
    \tilde S_{c}^\kappa\Bigg [\partial_{\Omega} + \log\Bigg(\frac{ \mu_{cs}}{Q \xi}\Big(\frac{\mu_{cs}}{\qcut}\Big)^\frac{1}{1+\beta}\Bigg) , \alpha_s(\mu_{cs})\Bigg ]    \nn
    \,.
\end{align}
%

 %=======================================================================
 %=======================================================================
 \section{Jet-mass power corrections in the soft drop resummation region}
 \label{app:MassSub}
 %=======================================================================
 %=======================================================================
 
 To obtain the non-singular fixed order correction we calculate 
 \begin{align}
     \xi \Delta {\cal G}^{\rm p.c.}
     &= e^{\big [ K_N + K_{gs} + K_{\cal C} + K_{cs_g}\big] }\Big(\frac{\mu_N}{Q}\Big)^{\omega_N} 
     \Big(\frac{\mu_{gs}}{\qcut}\Big)^{\omega_{gs}}  \Big(\frac{\mu_{cs}}{\qcut (r_g^{\rm max})^{1+\beta}}\Big)^{\omega_{cs_g}}
     \Big(\frac{\mu_{\cal C}}{Q r_g^{\rm max} } \Big)^{\omega_{\cal C}} \\
     &\quad \times\xi  \Bigg( 
     {\cal C}^{\kappa[1]} \big(\xi, Q ,\mu_{\cal C}) - \bigg[J^{[1]}_\kappa \Big(Q^2\xi, \mu_{\cal C})\Big) + 
     S_{c_m}^{\kappa[1]} \Big(Q\xi, \mu_{\cal C}r_g^{\rm max}\Big)
     \bigg]
     \Bigg) \nn \, .
 \end{align}
 The superscript $[1]$ corresponds to one-loop pieces. The factor of $\xi$ on the left hand side corresponds to power correction to the cross section, differential in $\ln \xi$. Note that the $\xi \delta(\xi)$ terms are set to zero, and thus only the logarithmic pieces contribute to the power correction.
 This  expression is obtained by taking the $r_g \ra r_g^{\rm max}(\xi)$ limit of the small-$R_g$ cross section where the jet-mass measurement is treated in fixed order. The scale $\mu_{gs}$ is given by \eq{muNmuGS}, the scale $\mu_{cs}$ with variations was defined in \eq{rhodef}, and finally the collinear function scale $\mu_{\cal C}$ is defined as
 \begin{align}
     \mu_{\cal C}(\xi) = \mu_N \Big(\frac{\mu_{cs}(\xi)}{\mu_{gs}}\Big)^{\frac{1}{1+\beta}} \, .
 \end{align}
 Thus $\mu_{\cal C}$ inherits its variations from $\mu_{cs}$ and $\mu_{gs}$.
 The evolution factors involve kernels corresponding to the collinear function ${\cal C}^\kappa$ and the groomed jet radius collinear soft function $S_{c_g}$ defined in \Ref{Pathak:2020iue}. Their anomalous dimensions are exactly negative of $N_\kappa$ and $S_G^\kappa$ functions, respectively:
 \begin{align}
     &\Gamma_{\cal C^{\kappa}}[\alpha_s] = - \Gamma_{N^\kappa}[\alpha_s] \, ,&
     &\gamma_{\cal C^{\kappa}}[\alpha_s] = - \gamma_{N^\kappa}[\alpha_s] \, ,& \\
     &\Gamma_{S_{c_g}^\kappa}[\alpha_s]  = - \Gamma_{S_G^\kappa}[\alpha_s] \, ,& 
     &\gamma_{S_{c_g}^\kappa}[\alpha_s]  = - \gamma_{S_G^\kappa}[\alpha_s] \, .&\nn 
 \end{align}
 As a result, in the ungroomed region with $\mu_{cs_g} = \mu_{gs}$ and in the fixed order region where $\mu_{\cal C} = \mu_{N}$, the evolution factors for $S_G^\kappa$ and $S_{c_g}^\kappa$, and $N_\kappa$ and ${\cal C}^\kappa$ respectively cancel each other. 
 
 Finally, since the jet and soft functions are evaluated at the same scale $\mu_{\cal C}$, one need not take the Laplace transform. We directly evaluate the power correction using the momentum space expressions for jet and the soft functions given in \eqs{Jmom}{Scm1L}.

\section{Individual profile variations for different values of \texorpdfstring{$\bm{\beta}$}{beta}}
\label{app:profvar}

\begin{figure}[t!]
    \centering
    \includegraphics[width=0.85\textwidth]{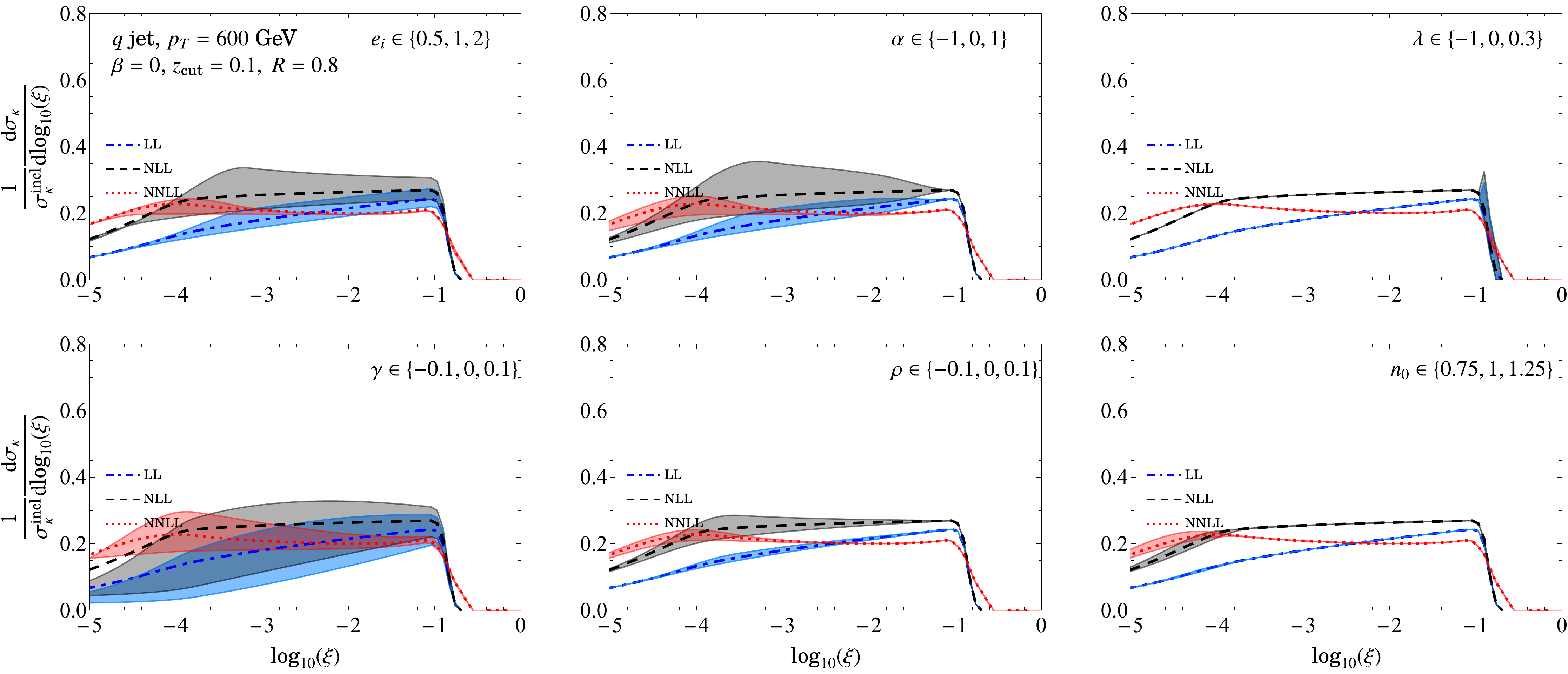}
    \\[10pt]
    \includegraphics[width=0.85\textwidth]{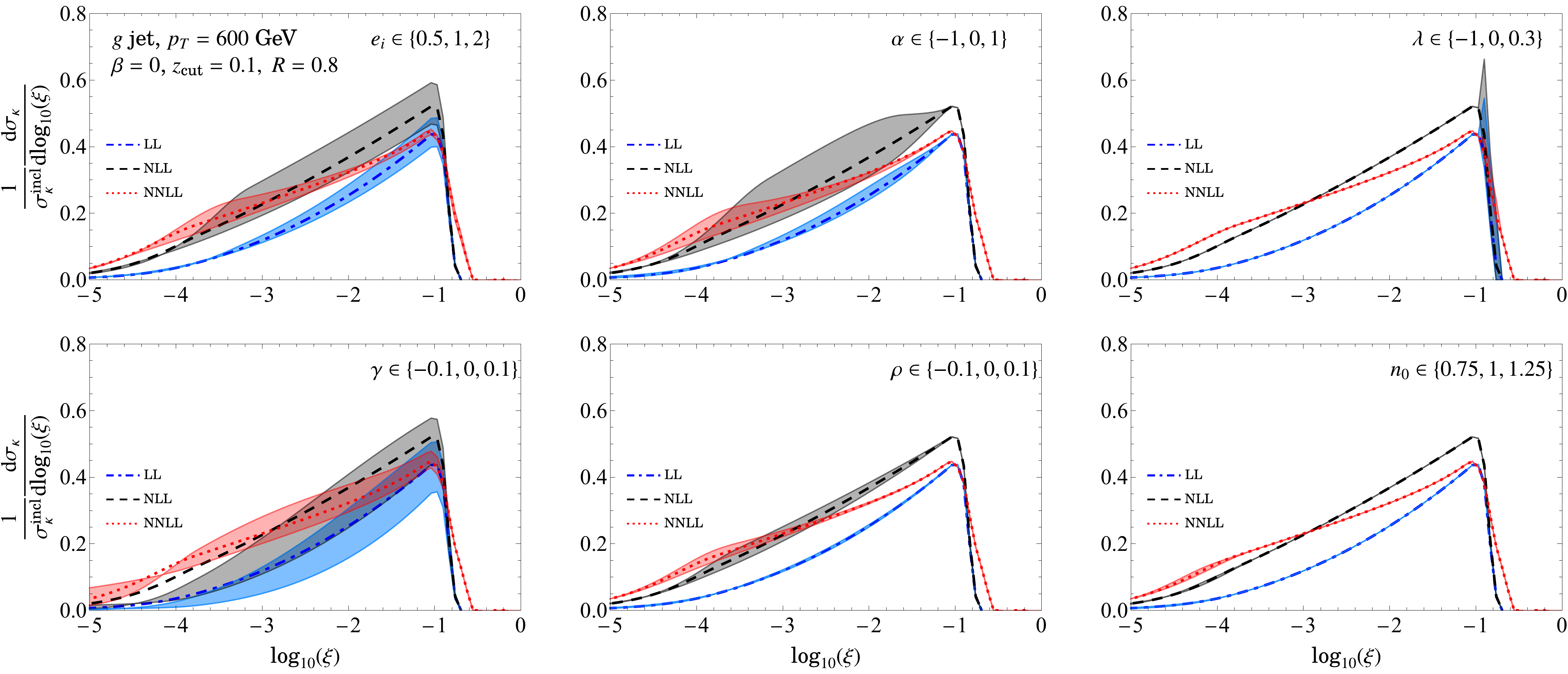}
    \caption{Variations for each of the six different profile-variation parameters individually, for quark and gluon jets at $p_T=600$ GeV, with $\beta=0$.}
    \label{fig:profvarb0}
\end{figure}

\begin{figure}[H]
    \centering
    \includegraphics[width=0.35\textwidth]{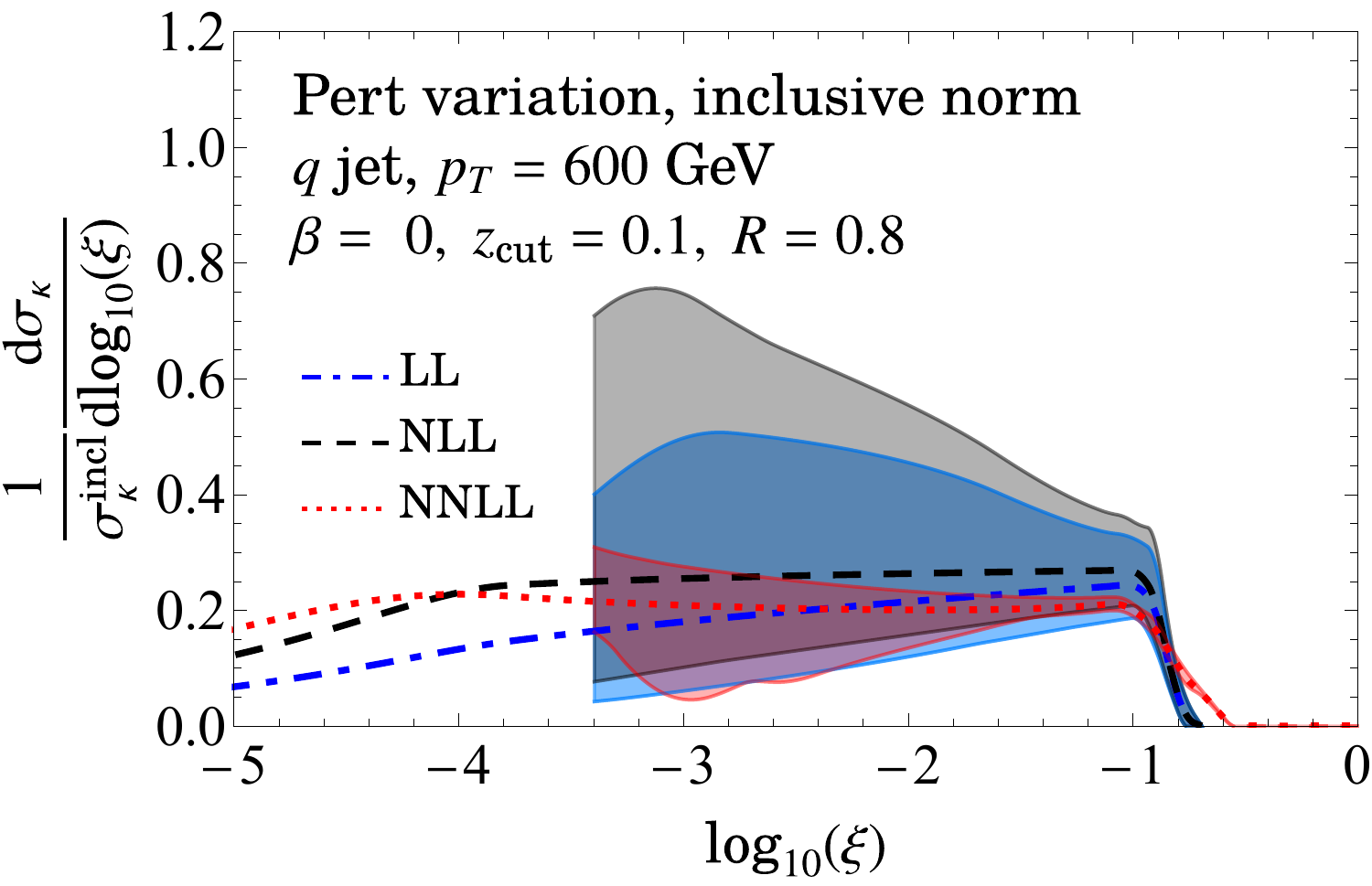}
    \includegraphics[width=0.35\textwidth]{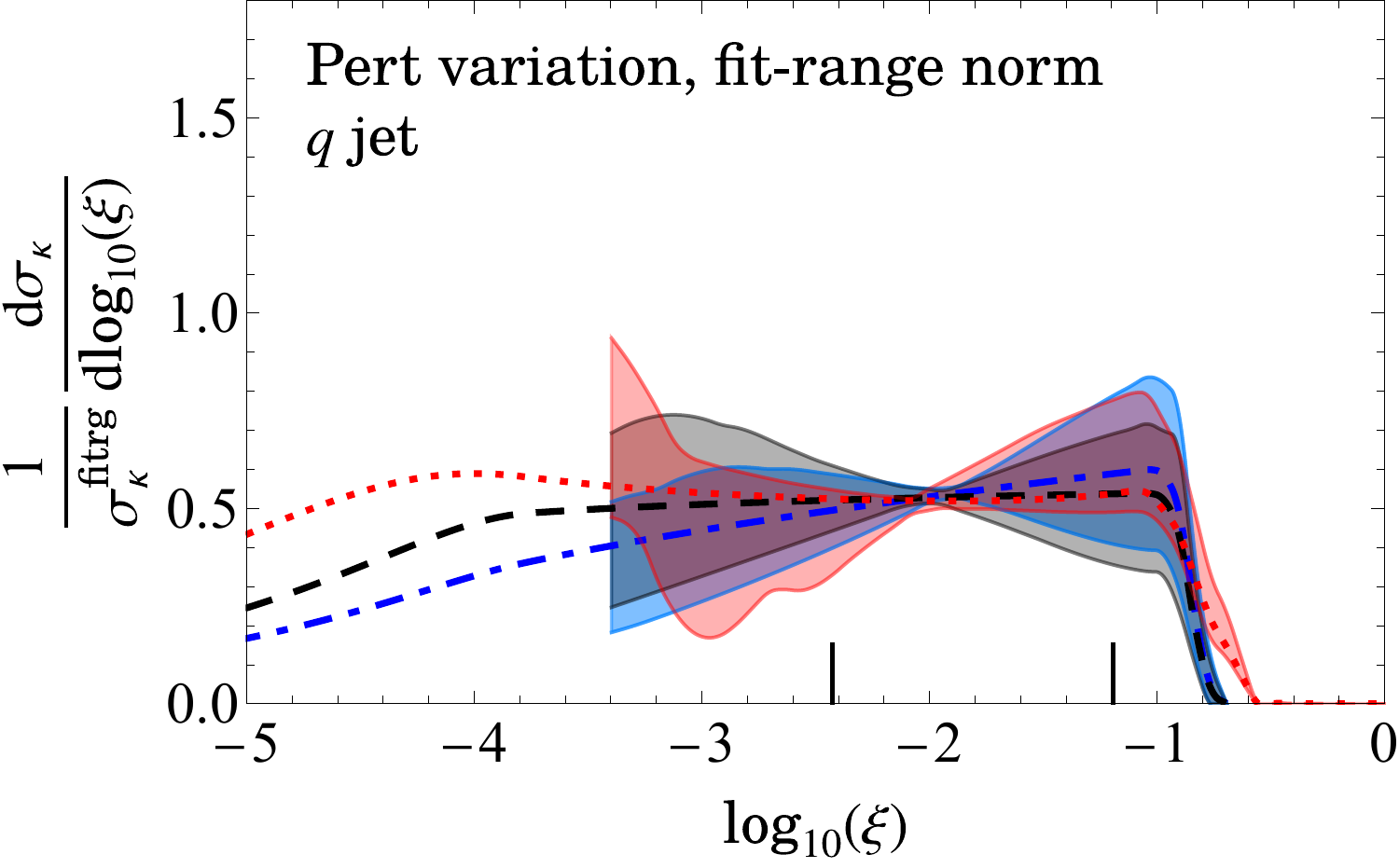}\\
    \includegraphics[width=0.35\textwidth]{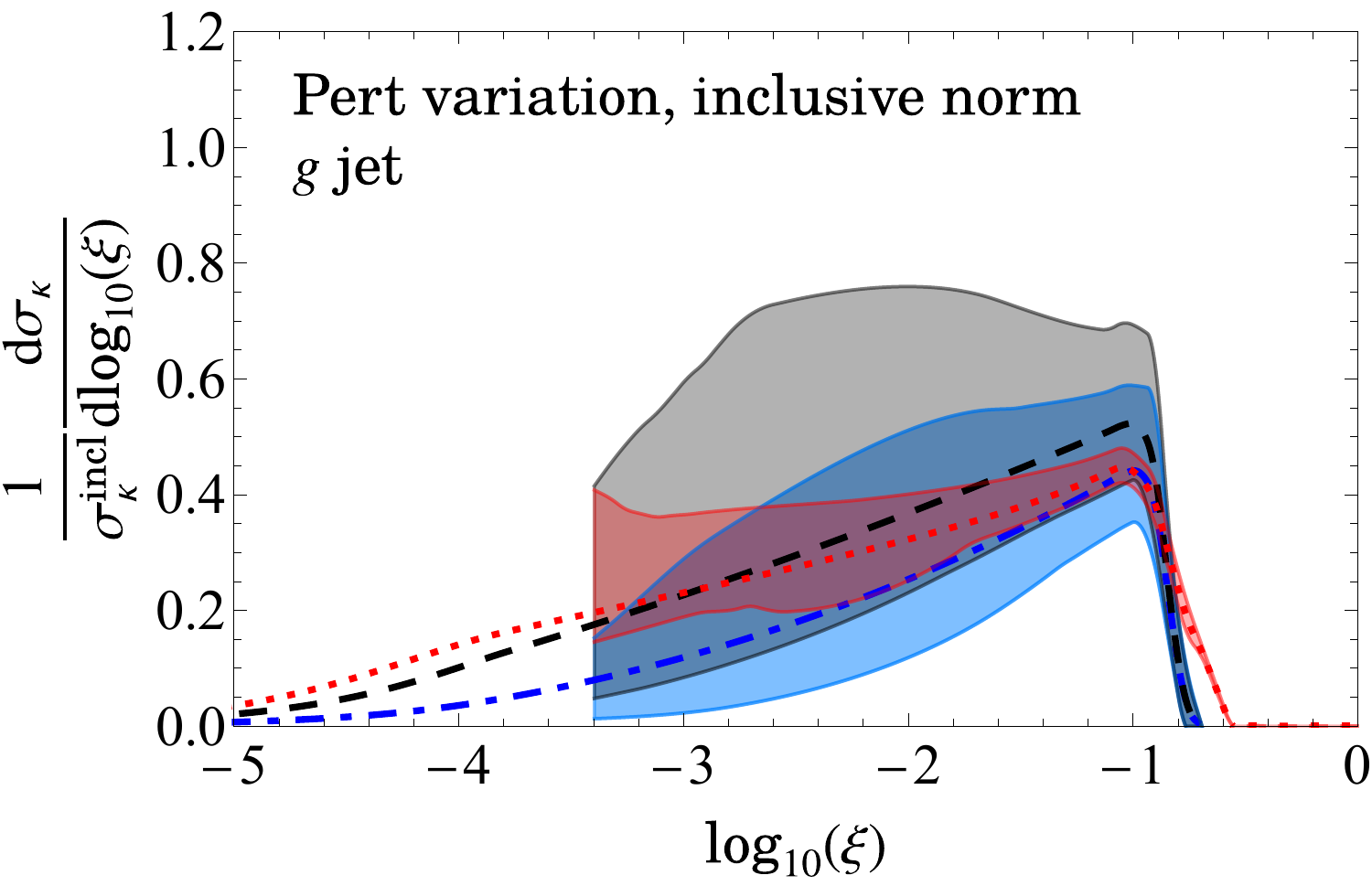}
    \includegraphics[width=0.35\textwidth]{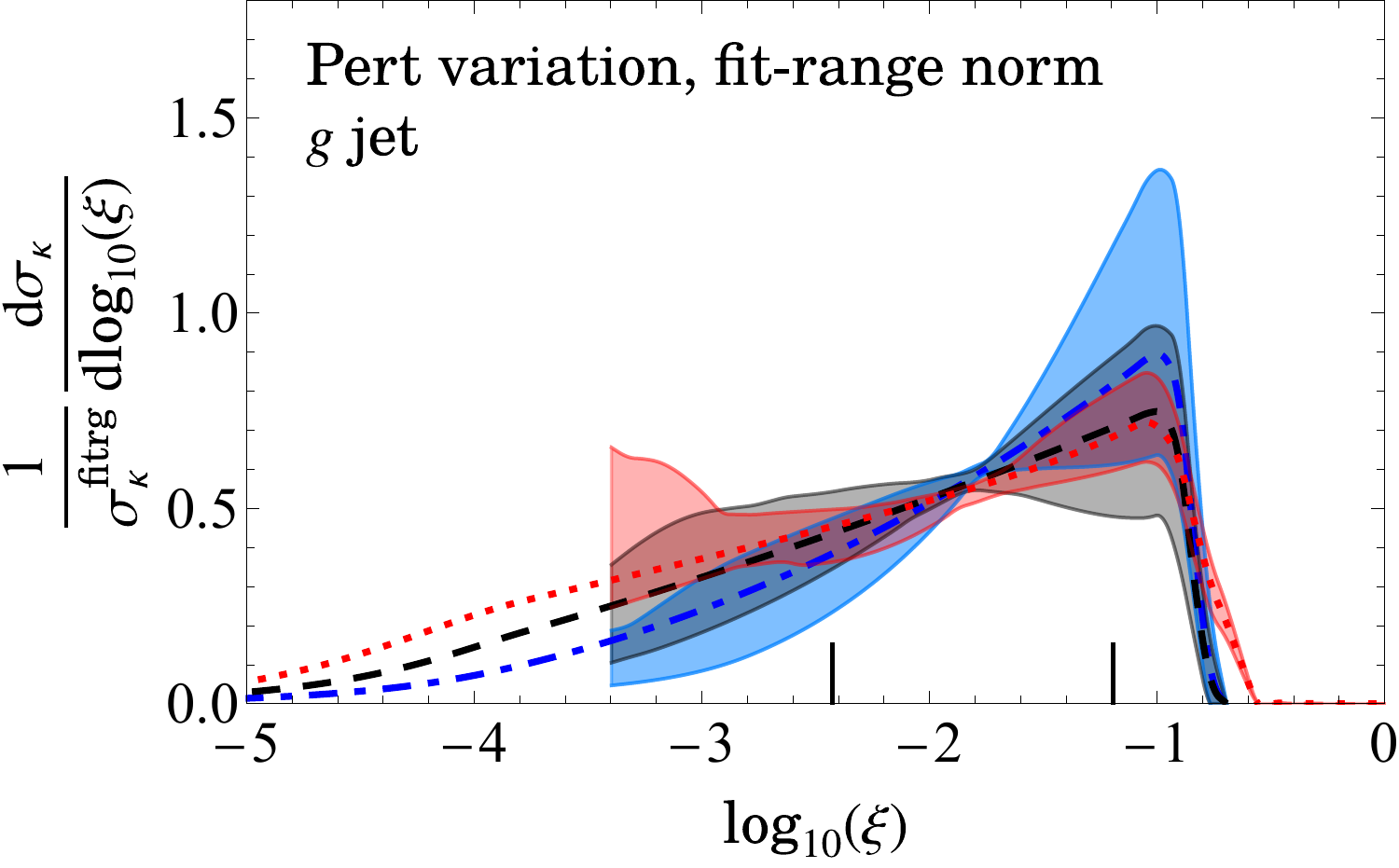}
    \caption{An envelope of the profile variations for quark and gluon jets respectively, at LL, NLL and NNLL.}
    \label{fig:profile_vars_allb0}
\end{figure}

\begin{figure}[t]
    \centering
    \includegraphics[width=0.85\textwidth]{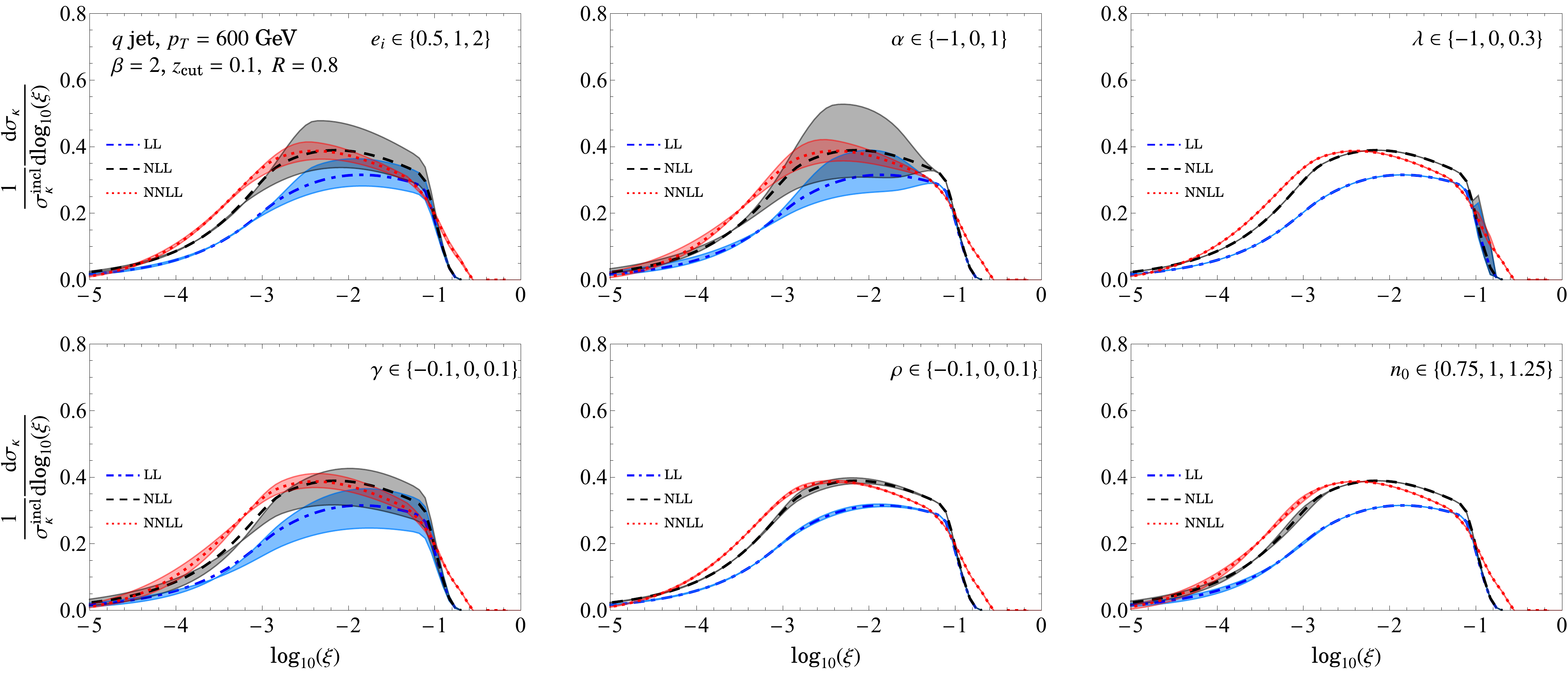}
    \\[10pt]
    \includegraphics[width=0.85\textwidth]{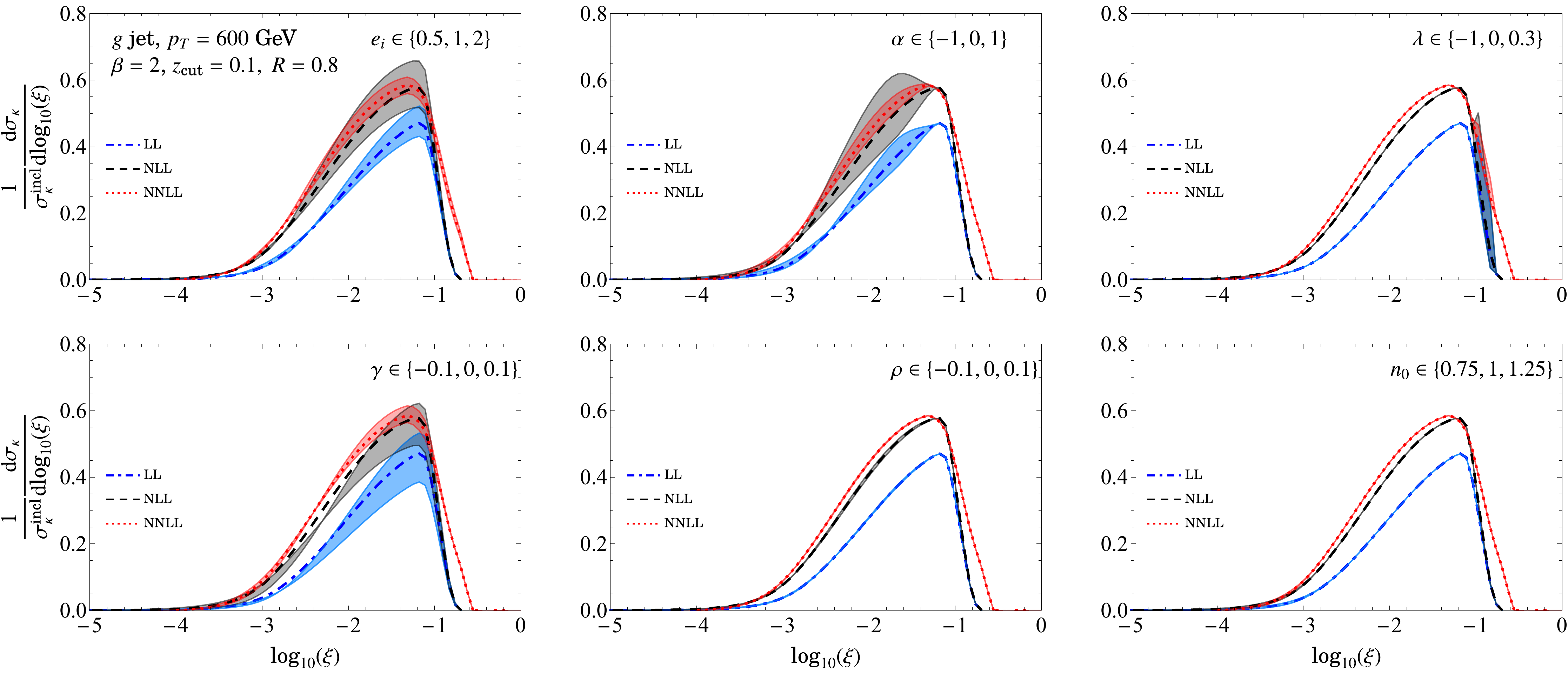}
    \caption{Variations for each of the six different profile-variation parameters individually, for quark and gluon jets at $p_T=600$ GeV, with $\beta=2$.}
    \label{fig:profvarb2}
\end{figure}

\begin{figure}[H]
    \centering
    \includegraphics[width=0.35\textwidth]{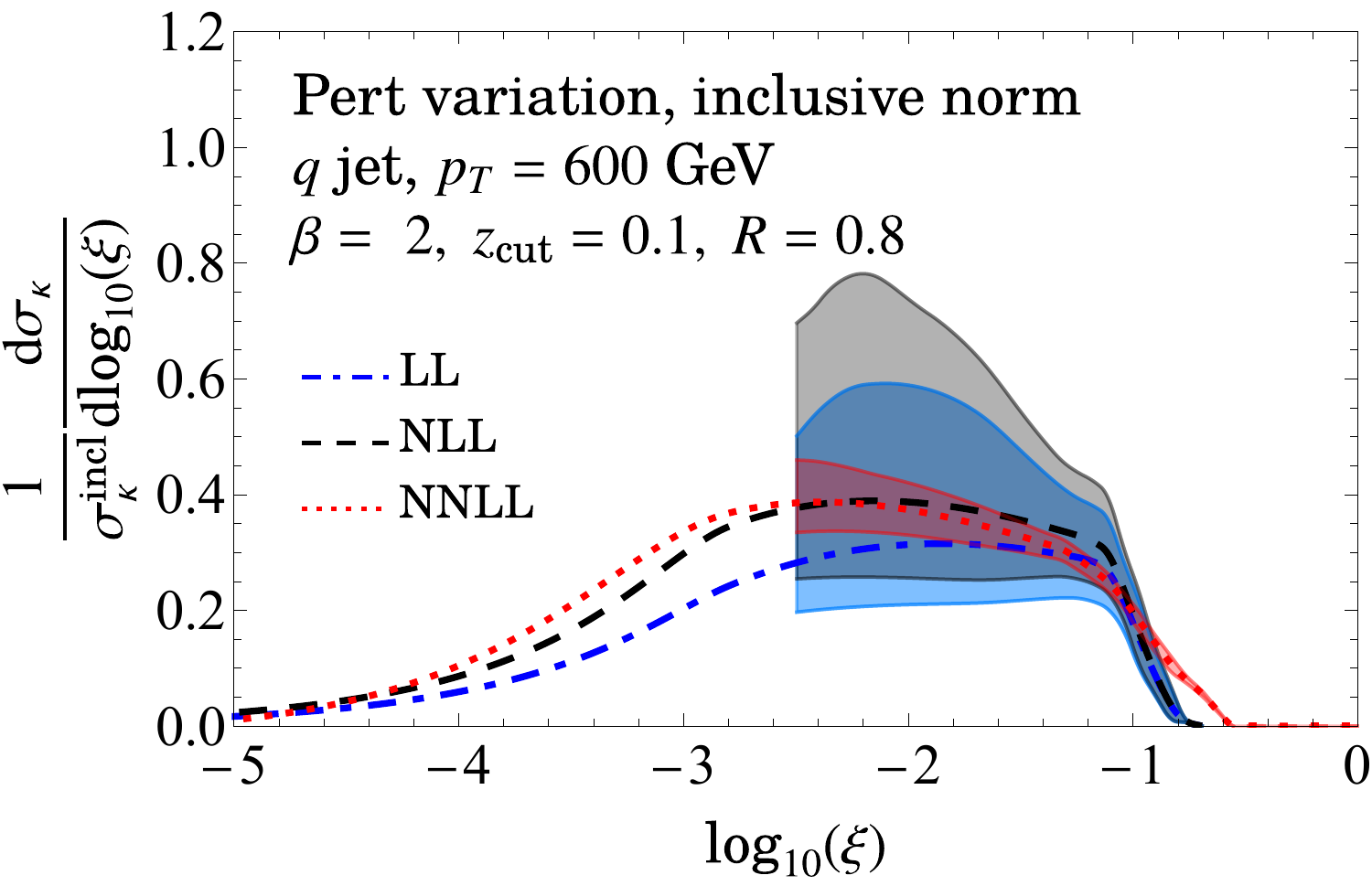}
    \includegraphics[width=0.35\textwidth]{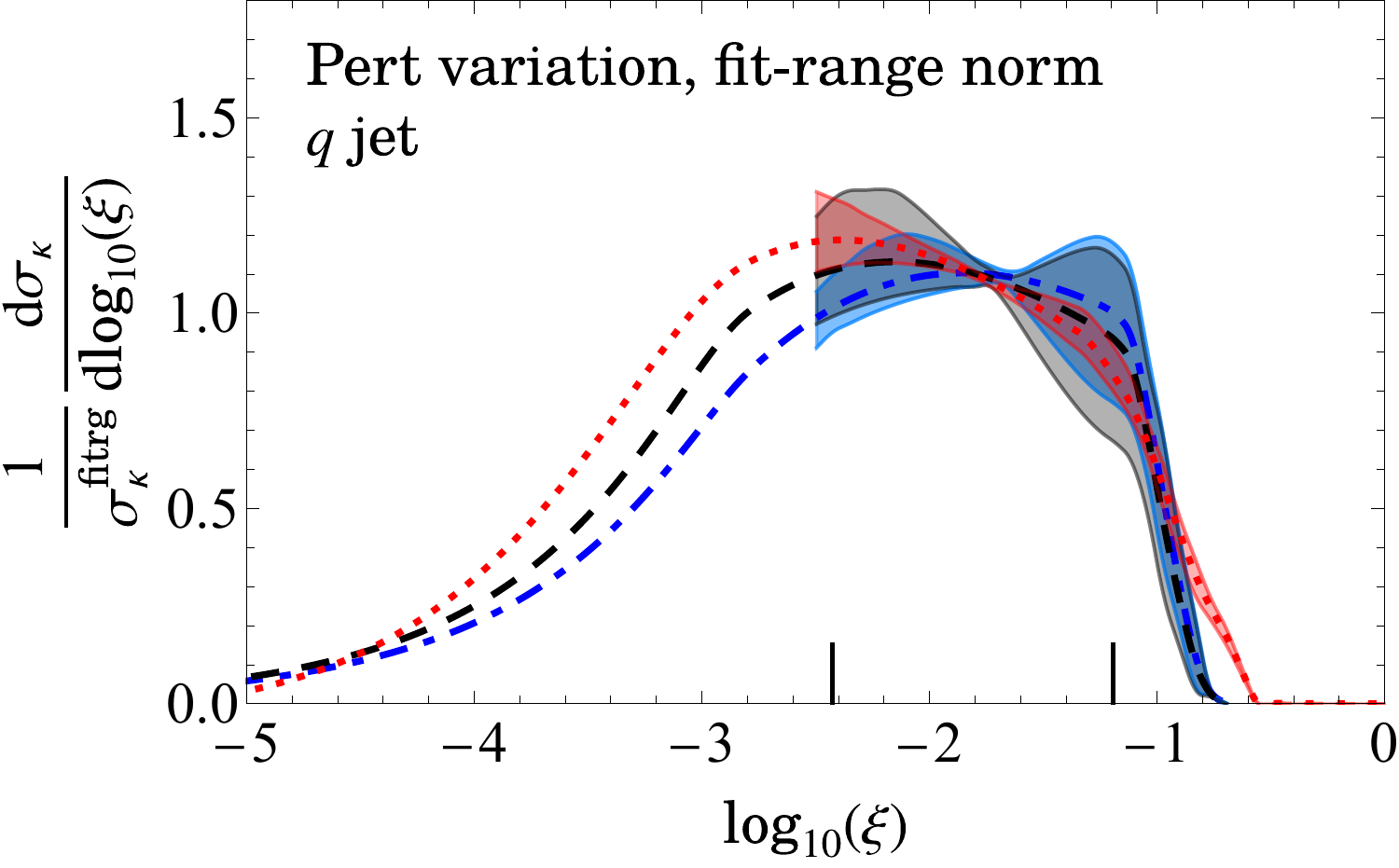}\\
    \includegraphics[width=0.35\textwidth]{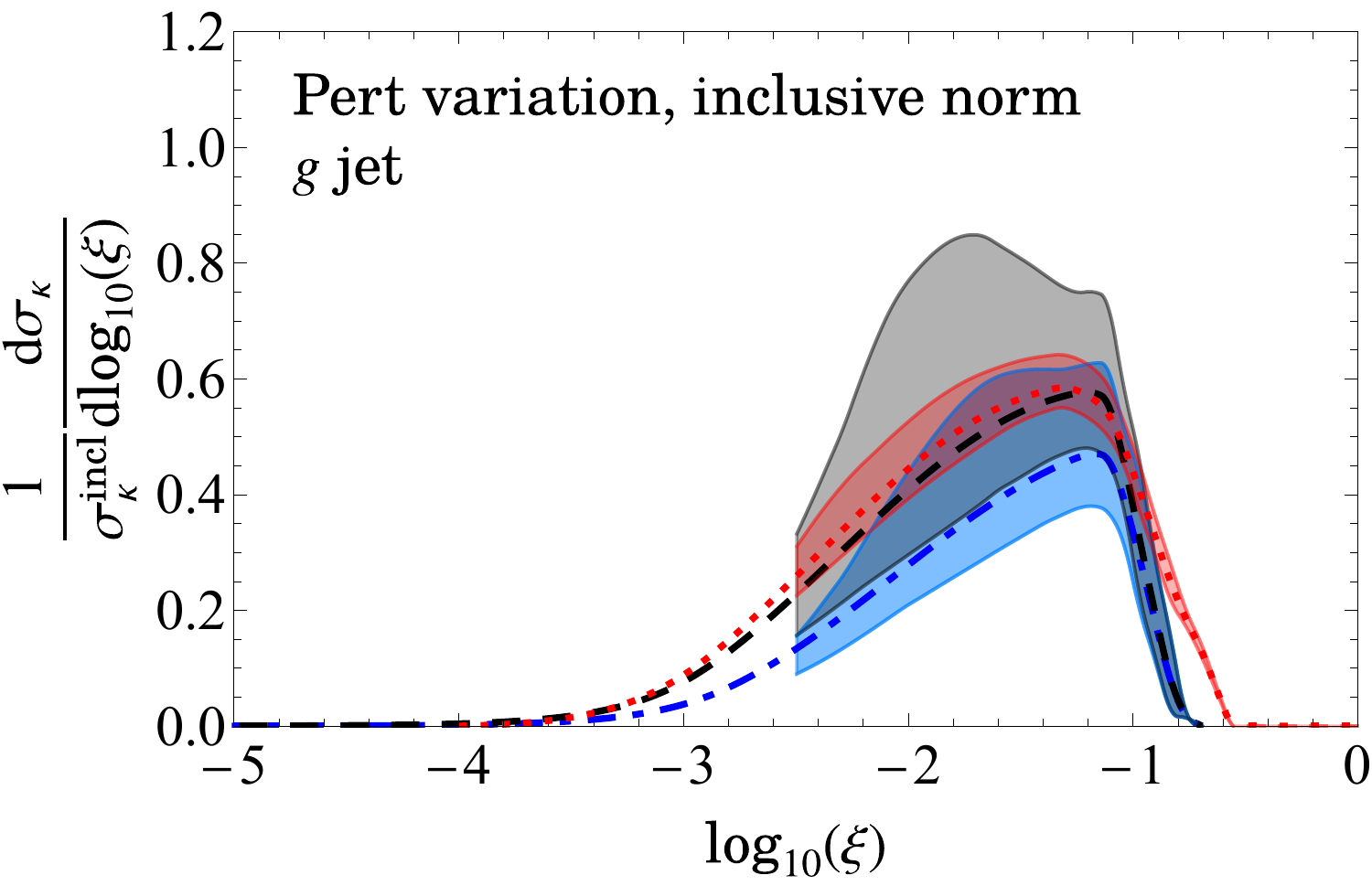}
    \includegraphics[width=0.35\textwidth]{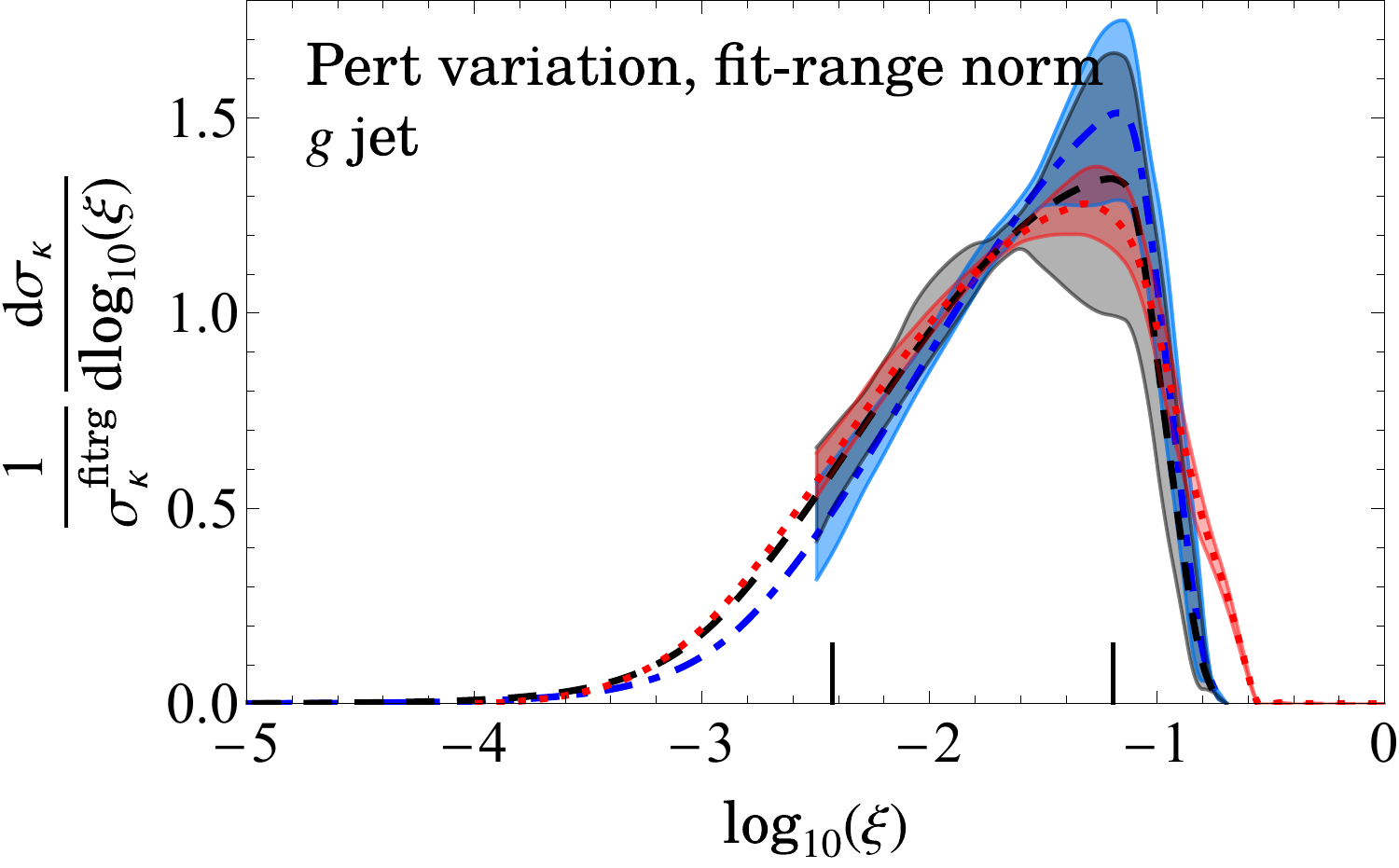}
    \caption{An envelope of the profile variations for quark and gluon jets respectively, at LL, NLL and NNLL.}
    \label{fig:profile_vars_allb2}
\end{figure}

In \fig{profvar} in the text we showed for $\beta=1$ the results of individual variations of profile parameters, which are used to assess the perturbative uncertainties order by order. In this appendix we give the analogous results for $\beta=0$ in \fig{profvarb0} and for $\beta=2$ in \fig{profvarb2}.  Once again, the dominant contributions to the uncertainty come from the parameters $e_i$, $\alpha$, and $\gamma$. In \fig{profile_vars_allb0} and \fig{profile_vars_allb2}, we show envelopes of the profile variations, for $\beta=0$ and $\beta=2$, respectively.

\newpage

\clearpage

\bibliography{sd}

\providecommand{\href}[2]{#2}\begingroup\raggedright\begin{thebibliography}{10}

\bibitem{L3:1992nwf}
{\scshape L3} collaboration, B.~Adeva et~al., \emph{{Studies of hadronic event
  structure and comparisons with QCD models at the Z0 resonance}},
  \href{https://doi.org/10.1007/BF01558288}{\emph{Z. Phys. C} {\bfseries 55}
  (1992) 39}.

\bibitem{SLD:1994idb}
{\scshape SLD} collaboration, K.~Abe et~al., \emph{{Measurement of alpha-s
  (M(Z)**2) from hadronic event observables at the Z0 resonance}},
  \href{https://doi.org/10.1103/PhysRevD.51.962}{\emph{Phys. Rev. D} {\bfseries
  51} (1995) 962} [\href{https://arxiv.org/abs/hep-ex/9501003}{{\ttfamily
  hep-ex/9501003}}].

\bibitem{DELPHI:1996oqw}
{\scshape DELPHI} collaboration, P.~Abreu et~al., \emph{{Measurement of event
  shape and inclusive distributions at S**(1/2) = 130-GeV and 136-GeV}},
  \href{https://doi.org/10.1007/s002880050312}{\emph{Z. Phys. C} {\bfseries 73}
  (1997) 229}.

\bibitem{ALEPH:2003obs}
{\scshape ALEPH} collaboration, A.~Heister et~al., \emph{{Studies of QCD at e+
  e- centre-of-mass energies between 91-GeV and 209-GeV}},
  \href{https://doi.org/10.1140/epjc/s2004-01891-4}{\emph{Eur. Phys. J. C}
  {\bfseries 35} (2004) 457}.

\bibitem{DELPHI:2004omy}
{\scshape DELPHI} collaboration, J.~Abdallah et~al., \emph{{The Measurement of
  alpha(s) from event shapes with the DELPHI detector at the highest LEP
  energies}}, \href{https://doi.org/10.1140/epjc/s2004-01889-x}{\emph{Eur.
  Phys. J. C} {\bfseries 37} (2004) 1}
  [\href{https://arxiv.org/abs/hep-ex/0406011}{{\ttfamily hep-ex/0406011}}].

\bibitem{OPAL:2004wof}
{\scshape OPAL} collaboration, G.~Abbiendi et~al., \emph{{Measurement of event
  shape distributions and moments in e+ e- ---\ensuremath{>} hadrons at 91-GeV
  - 209-GeV and a determination of alpha(s)}},
  \href{https://doi.org/10.1140/epjc/s2005-02120-6}{\emph{Eur. Phys. J. C}
  {\bfseries 40} (2005) 287}
  [\href{https://arxiv.org/abs/hep-ex/0503051}{{\ttfamily hep-ex/0503051}}].

\bibitem{Dissertori:2007xa}
G.~Dissertori, A.~Gehrmann-De~Ridder, T.~Gehrmann, E.~W.~N. Glover, G.~Heinrich
  and H.~Stenzel, \emph{{First determination of the strong coupling constant
  using NNLO predictions for hadronic event shapes in e+ e- annihilations}},
  \href{https://doi.org/10.1088/1126-6708/2008/02/040}{\emph{JHEP} {\bfseries
  02} (2008) 040} [\href{https://arxiv.org/abs/0712.0327}{{\ttfamily
  0712.0327}}].

\bibitem{Becher:2008cf}
T.~Becher and M.~D. Schwartz, \emph{{A Precise determination of $\alpha_s$ from
  LEP thrust data using effective field theory}},
  \href{https://doi.org/10.1088/1126-6708}{\emph{JHEP} {\bfseries 07} (2008)
  034} [\href{https://arxiv.org/abs/0803.0342}{{\ttfamily 0803.0342}}].

\bibitem{Davison:2009wzs}
R.~A. Davison and B.~R. Webber, \emph{{Non-Perturbative Contribution to the
  Thrust Distribution in e+ e- Annihilation}},
  \href{https://doi.org/10.1140/epjc/s10052-008-0836-7}{\emph{Eur. Phys. J. C}
  {\bfseries 59} (2009) 13} [\href{https://arxiv.org/abs/0809.3326}{{\ttfamily
  0809.3326}}].

\bibitem{Bethke:2009ehn}
{\scshape JADE} collaboration, S.~Bethke, S.~Kluth, C.~Pahl and J.~Schieck,
  \emph{{Determination of the Strong Coupling alpha(s) from hadronic Event
  Shapes with O(alpha**3(s)) and resummed QCD predictions using JADE Data}},
  \href{https://doi.org/10.1140/epjc/s10052-009-1149-1}{\emph{Eur. Phys. J. C}
  {\bfseries 64} (2009) 351} [\href{https://arxiv.org/abs/0810.1389}{{\ttfamily
  0810.1389}}].

\bibitem{Dissertori:2009ik}
G.~Dissertori, A.~Gehrmann-De~Ridder, T.~Gehrmann, E.~W.~N. Glover,
  G.~Heinrich, G.~Luisoni et~al., \emph{{Determination of the strong coupling
  constant using matched NNLO+NLLA predictions for hadronic event shapes in
  e+e- annihilations}},
  \href{https://doi.org/10.1088/1126-6708/2009/08/036}{\emph{JHEP} {\bfseries
  08} (2009) 036} [\href{https://arxiv.org/abs/0906.3436}{{\ttfamily
  0906.3436}}].

\bibitem{Abbate:2010xh}
R.~Abbate, M.~Fickinger, A.~H. Hoang, V.~Mateu and I.~W. Stewart, \emph{{Thrust
  at N${}^3$LL with Power Corrections and a Precision Global Fit for
  $\alpha_s(m_Z)$}},
  \href{https://doi.org/10.1103/PhysRevD.83.074021}{\emph{Phys. Rev.}
  {\bfseries D83} (2011) 074021}
  [\href{https://arxiv.org/abs/1006.3080}{{\ttfamily 1006.3080}}].

\bibitem{Abbate:2012jh}
R.~Abbate, M.~Fickinger, A.~H. Hoang, V.~Mateu and I.~W. Stewart,
  \emph{{Precision Thrust Cumulant Moments at $N^3$LL}},
  \href{https://doi.org/10.1103/PhysRevD.86.094002}{\emph{Phys. Rev. D}
  {\bfseries 86} (2012) 094002}
  [\href{https://arxiv.org/abs/1204.5746}{{\ttfamily 1204.5746}}].

\bibitem{Hoang:2015hka}
A.~H. Hoang, D.~W. Kolodrubetz, V.~Mateu and I.~W. Stewart, \emph{{Precise
  determination of $\alpha_s$ from the $C$-parameter distribution}},
  \href{https://doi.org/10.1103/PhysRevD.91.094018}{\emph{Phys. Rev.}
  {\bfseries D91} (2015) 094018}
  [\href{https://arxiv.org/abs/1501.04111}{{\ttfamily 1501.04111}}].

\bibitem{Hoang:2014wka}
A.~H. Hoang, D.~W. Kolodrubetz, V.~Mateu and I.~W. Stewart,
  \emph{{$C$-parameter distribution at N$^3$LL' including power corrections}},
  \href{https://doi.org/10.1103/PhysRevD.91.094017}{\emph{Phys. Rev. D}
  {\bfseries 91} (2015) 094017}
  [\href{https://arxiv.org/abs/1411.6633}{{\ttfamily 1411.6633}}].

\bibitem{CMS:2019oeb}
{\scshape CMS} collaboration, A.~M. Sirunyan et~al., \emph{{Determination of
  the strong coupling constant $\alpha_{S}(m_\mathrm{Z})$ from measurements of
  inclusive W$^\pm$ and Z boson production cross sections in proton-proton
  collisions at $ \sqrt{\mathrm{s}} $ = 7 and 8 TeV}}, {\emph{JHEP} {\bfseries
  06} (2020) 018} [\href{https://arxiv.org/abs/1912.04387}{{\ttfamily
  1912.04387}}].

\bibitem{ATLAS:2020mee}
{ATLAS collaboration}, \emph{{Determination of the strong coupling constant and
  test of asymptotic freedom from Transverse Energy-Energy Correlations in
  multijet events at $\sqrt{s} = 13$ TeV with the ATLAS detector}},  2020.

\bibitem{Aoki:2021kgd}
Y.~Aoki et~al., \emph{{FLAG Review 2021}},
  \href{https://arxiv.org/abs/2111.09849}{{\ttfamily 2111.09849}}.

\bibitem{ParticleDataGroup:2018ovx}
{\scshape Particle Data Group} collaboration, M.~Tanabashi et~al.,
  \emph{{Review of Particle Physics}},
  \href{https://doi.org/10.1103/PhysRevD.98.030001}{\emph{Phys. Rev. D}
  {\bfseries 98} (2018) 030001}.

\bibitem{ParticleDataGroup:2020ssz}
{\scshape Particle Data Group} collaboration, P.~A. Zyla et~al., \emph{{Review
  of Particle Physics}},
  \href{https://doi.org/10.1093/ptep/ptaa104}{\emph{PTEP} {\bfseries 2020}
  (2020) 083C01}.

\bibitem{Larkoski:2014wba}
A.~J. Larkoski, S.~Marzani, G.~Soyez and J.~Thaler, \emph{{Soft Drop}},
  \href{https://doi.org/10.1007/JHEP05(2014)146}{\emph{JHEP} {\bfseries 05}
  (2014) 146} [\href{https://arxiv.org/abs/1402.2657}{{\ttfamily 1402.2657}}].

\bibitem{Larkoski:2017cqq}
A.~J. Larkoski, I.~Moult and D.~Neill, \emph{{Factorization and Resummation for
  Groomed Multi-Prong Jet Shapes}},
  \href{https://doi.org/10.1007/JHEP02(2018)144}{\emph{JHEP} {\bfseries 02}
  (2018) 144} [\href{https://arxiv.org/abs/1710.00014}{{\ttfamily
  1710.00014}}].

\bibitem{Larkoski:2017iuy}
A.~J. Larkoski, I.~Moult and D.~Neill, \emph{{Analytic Boosted Boson
  Discrimination at the Large Hadron Collider}},
  \href{https://arxiv.org/abs/1708.06760}{{\ttfamily 1708.06760}}.

\bibitem{Baron:2018nfz}
J.~Baron, S.~Marzani and V.~Theeuwes, \emph{{Soft-Drop Thrust}},
  \href{https://doi.org/10.1007/JHEP08(2018)105}{\emph{JHEP} {\bfseries 08}
  (2018) 105} [\href{https://arxiv.org/abs/1803.04719}{{\ttfamily
  1803.04719}}].

\bibitem{Kang:2018vgn}
Z.-B. Kang, K.~Lee, X.~Liu and F.~Ringer, \emph{{Soft drop groomed jet
  angularities at the LHC}},
  \href{https://doi.org/10.1016/j.physletb.2019.04.018}{\emph{Phys. Lett.}
  {\bfseries B793} (2019) 41}
  [\href{https://arxiv.org/abs/1811.06983}{{\ttfamily 1811.06983}}].

\bibitem{Makris:2018npl}
Y.~Makris and V.~Vaidya, \emph{{Transverse Momentum Spectra at Threshold for
  Groomed Heavy Quark Jets}},
  \href{https://doi.org/10.1007/JHEP10(2018)019}{\emph{JHEP} {\bfseries 10}
  (2018) 019} [\href{https://arxiv.org/abs/1807.09805}{{\ttfamily
  1807.09805}}].

\bibitem{Kardos:2018kth}
A.~Kardos, G.~Somogyi and Z.~Tr{\'o}cs{\'a}nyi, \emph{{Soft-drop event shapes
  in electron--positron annihilation at next-to-next-to-leading order
  accuracy}}, \href{https://doi.org/10.1016/j.physletb.2018.10.014}{\emph{Phys.
  Lett.} {\bfseries B786} (2018) 313}
  [\href{https://arxiv.org/abs/1807.11472}{{\ttfamily 1807.11472}}].

\bibitem{Chay:2018pvp}
J.~Chay and C.~Kim, \emph{{Factorized groomed jet mass distribution in
  inclusive jet processes}},
  \href{https://doi.org/10.3938/jkps.74.439}{\emph{J. Korean Phys. Soc.}
  {\bfseries 74} (2019) 439}
  [\href{https://arxiv.org/abs/1806.01712}{{\ttfamily 1806.01712}}].

\bibitem{Napoletano:2018ohv}
D.~Napoletano and G.~Soyez, \emph{{Computing $N$-subjettiness for boosted
  jets}}, \href{https://doi.org/10.1007/JHEP12(2018)031}{\emph{JHEP} {\bfseries
  12} (2018) 031} [\href{https://arxiv.org/abs/1809.04602}{{\ttfamily
  1809.04602}}].

\bibitem{Lee:2019lge}
C.~Lee, P.~Shrivastava and V.~Vaidya, \emph{{Predictions for energy correlators
  probing substructure of groomed heavy quark jets}},
  \href{https://doi.org/10.1007/JHEP09(2019)045}{\emph{JHEP} {\bfseries 09}
  (2019) 045} [\href{https://arxiv.org/abs/1901.09095}{{\ttfamily
  1901.09095}}].

\bibitem{Hoang:2019ceu}
A.~H. Hoang, S.~Mantry, A.~Pathak and I.~W. Stewart, \emph{{Nonperturbative
  Corrections to Soft Drop Jet Mass}},
  \href{https://arxiv.org/abs/1906.11843}{{\ttfamily 1906.11843}}.

\bibitem{Gutierrez-Reyes:2019msa}
D.~Gutierrez-Reyes, Y.~Makris, V.~Vaidya, I.~Scimemi and L.~Zoppi,
  \emph{{Probing Transverse-Momentum Distributions With Groomed Jets}},
  \href{https://doi.org/10.1007/JHEP08(2019)161}{\emph{JHEP} {\bfseries 08}
  (2019) 161} [\href{https://arxiv.org/abs/1907.05896}{{\ttfamily
  1907.05896}}].

\bibitem{Kardos:2019iwa}
A.~Kardos, A.~Larkoski and Z.~Tr{\'o}cs{\'a}nyi, \emph{{Soft-Dropped
  Observables with CoLoRFuLNNLO}},
  \href{https://doi.org/10.5506/APhysPolB.50.1891}{\emph{Acta Phys. Polon. B}
  {\bfseries 50} (2019) 1891}.

\bibitem{Marzani:2019evv}
S.~Marzani, D.~Reichelt, S.~Schumann, G.~Soyez and V.~Theeuwes, \emph{{Fitting
  the Strong Coupling Constant with Soft-Drop Thrust}},
  \href{https://arxiv.org/abs/1906.10504}{{\ttfamily 1906.10504}}.

\bibitem{Mehtar-Tani:2019rrk}
Y.~Mehtar-Tani, A.~Soto-Ontoso and K.~Tywoniuk, \emph{{Dynamical grooming of
  QCD jets}},  \href{https://arxiv.org/abs/1911.00375}{{\ttfamily 1911.00375}}.

\bibitem{Kardos:2020ppl}
A.~Kardos, A.~J. Larkoski and Z.~Tr\'ocs\'anyi, \emph{{Two- and three-loop data
  for the groomed jet mass}},
  \href{https://doi.org/10.1103/PhysRevD.101.114034}{\emph{Phys. Rev. D}
  {\bfseries 101} (2020) 114034}
  [\href{https://arxiv.org/abs/2002.05730}{{\ttfamily 2002.05730}}].

\bibitem{Larkoski:2020wgx}
A.~J. Larkoski, \emph{{Improving the understanding of jet grooming in
  perturbation theory}},
  \href{https://doi.org/10.1007/JHEP09(2020)072}{\emph{JHEP} {\bfseries 09}
  (2020) 072} [\href{https://arxiv.org/abs/2006.14680}{{\ttfamily
  2006.14680}}].

\bibitem{Lifson:2020gua}
A.~Lifson, G.~P. Salam and G.~Soyez, \emph{{Calculating the primary Lund Jet
  Plane density}},  \href{https://arxiv.org/abs/2007.06578}{{\ttfamily
  2007.06578}}.

\bibitem{Caucal:2021cfb}
P.~Caucal, A.~Soto-Ontoso and A.~Takacs, \emph{{Dynamically groomed jet radius
  in heavy-ion collisions}},
  \href{https://arxiv.org/abs/2111.14768}{{\ttfamily 2111.14768}}.

\bibitem{Frye:2016aiz}
C.~Frye, A.~J. Larkoski, M.~D. Schwartz and K.~Yan, \emph{{Factorization for
  groomed jet substructure beyond the next-to-leading logarithm}},
  \href{https://doi.org/10.1007/JHEP07(2016)064}{\emph{JHEP} {\bfseries 07}
  (2016) 064} [\href{https://arxiv.org/abs/1603.09338}{{\ttfamily
  1603.09338}}].

\bibitem{Kardos:2020gty}
A.~Kardos, A.~J. Larkoski and Z.~Tr\'ocs\'anyi, \emph{{Groomed jet mass at high
  precision}},
  \href{https://doi.org/10.1016/j.physletb.2020.135704}{\emph{Phys. Lett. B}
  {\bfseries 809} (2020) 135704}
  [\href{https://arxiv.org/abs/2002.00942}{{\ttfamily 2002.00942}}].

\bibitem{Benkendorfer:2021unv}
K.~Benkendorfer and A.~J. Larkoski, \emph{{Grooming at the Cusp: All-Orders
  Predictions for the Transition Region of Jet Groomers}},
  \href{https://arxiv.org/abs/2108.02779}{{\ttfamily 2108.02779}}.

\bibitem{Pathak:2020iue}
A.~Pathak, I.~W. Stewart, V.~Vaidya and L.~Zoppi, \emph{{EFT for Soft Drop
  Double Differential Cross Section}},
  \href{https://arxiv.org/abs/2012.15568}{{\ttfamily 2012.15568}}.

\bibitem{AP}
A.~Pathak, ``{The catchment area of groomed jets}.'' Publication forthcoming,
  2022.

\bibitem{Ferdinand:2022xxx}
A.~E. Ferdinand, K.~Lee and A.~Pathak, \emph{{Dissecting groomed soft radiation
  with factorization}},  2022.

\bibitem{ATLAS:2017zda}
{\scshape ATLAS} collaboration, M.~Aaboud et~al., \emph{{Measurement of the
  Soft-Drop Jet Mass in pp Collisions at $\sqrt{s} = 13$ TeV with the ATLAS
  Detector}}, \href{https://doi.org/10.1103/PhysRevLett.121.092001}{\emph{Phys.
  Rev. Lett.} {\bfseries 121} (2018) 092001}
  [\href{https://arxiv.org/abs/1711.08341}{{\ttfamily 1711.08341}}].

\bibitem{ATLAS:2019mgf}
{\scshape ATLAS} collaboration, G.~Aad et~al., \emph{{Measurement of soft-drop
  jet observables in $pp$ collisions with the ATLAS detector at $\sqrt {s}$ =13
  TeV}}, \href{https://doi.org/10.1103/PhysRevD.101.052007}{\emph{Phys. Rev. D}
  {\bfseries 101} (2020) 052007}
  [\href{https://arxiv.org/abs/1912.09837}{{\ttfamily 1912.09837}}].

\bibitem{ATLAS:2021urs}
{ATLAS collaboration}, \emph{{A precise interpretation for the top quark mass
  parameter in ATLAS Monte Carlo simulation}}, {\emph{ATL-PHYS-PUB-2021-034}
  (2021) }.

\bibitem{Bauer:2000ew}
C.~W. Bauer, S.~Fleming and M.~E. Luke, \emph{{Summing Sudakov logarithms in B
  ---> X(s gamma) in effective field theory}},
  \href{https://doi.org/10.1103/PhysRevD.63.014006}{\emph{Phys.\ Rev.\ D}
  {\bfseries 63} (2000) 014006}
  [\href{https://arxiv.org/abs/hep-ph/0005275}{{\ttfamily hep-ph/0005275}}].

\bibitem{Bauer:2000yr}
C.~W. Bauer, S.~Fleming, D.~Pirjol and I.~W. Stewart, \emph{{An Effective field
  theory for collinear and soft gluons: Heavy to light decays}},
  \href{https://doi.org/10.1103/PhysRevD.63.114020}{\emph{Phys.\ Rev.\ D}
  {\bfseries 63} (2001) 114020}
  [\href{https://arxiv.org/abs/hep-ph/0011336}{{\ttfamily hep-ph/0011336}}].

\bibitem{Bauer:2001ct}
C.~W. Bauer and I.~W. Stewart, \emph{{Invariant operators in collinear
  effective theory}},
  \href{https://doi.org/10.1016/S0370-2693(01)00902-9}{\emph{Phys.\ Lett.\ B}
  {\bfseries 516} (2001) 134}
  [\href{https://arxiv.org/abs/hep-ph/0107001}{{\ttfamily hep-ph/0107001}}].

\bibitem{Bauer:2001yt}
C.~W. Bauer, D.~Pirjol and I.~W. Stewart, \emph{{Soft collinear factorization
  in effective field theory}},
  \href{https://doi.org/10.1103/PhysRevD.65.054022}{\emph{Phys.\ Rev.\ D}
  {\bfseries 65} (2002) 054022}
  [\href{https://arxiv.org/abs/hep-ph/0109045}{{\ttfamily hep-ph/0109045}}].

\bibitem{Marzani:2017kqd}
S.~Marzani, L.~Schunk and G.~Soyez, \emph{{The jet mass distribution after Soft
  Drop}}, \href{https://doi.org/10.1140/epjc/s10052-018-5579-5}{\emph{Eur.
  Phys. J. C} {\bfseries 78} (2018) 96}
  [\href{https://arxiv.org/abs/1712.05105}{{\ttfamily 1712.05105}}].

\bibitem{Forte:2020pyp}
S.~Forte and Z.~Kassabov, \emph{{Why $\alpha _s$ cannot be determined from
  hadronic processes without simultaneously determining the parton
  distributions}},
  \href{https://doi.org/10.1140/epjc/s10052-020-7748-6}{\emph{Eur. Phys. J. C}
  {\bfseries 80} (2020) 182}
  [\href{https://arxiv.org/abs/2001.04986}{{\ttfamily 2001.04986}}].

\bibitem{Alwall:2014hca}
J.~Alwall, R.~Frederix, S.~Frixione, V.~Hirschi, F.~Maltoni, O.~Mattelaer
  et~al., \emph{{The automated computation of tree-level and next-to-leading
  order differential cross sections, and their matching to parton shower
  simulations}}, \href{https://doi.org/10.1007/JHEP07(2014)079}{\emph{JHEP}
  {\bfseries 07} (2014) 079} [\href{https://arxiv.org/abs/1405.0301}{{\ttfamily
  1405.0301}}].

\bibitem{Broggio:2014hoa}
A.~Broggio, A.~Ferroglia, B.~D. Pecjak and Z.~Zhang, \emph{{NNLO hard functions
  in massless QCD}}, \href{https://doi.org/10.1007/JHEP12(2014)005}{\emph{JHEP}
  {\bfseries 12} (2014) 005} [\href{https://arxiv.org/abs/1409.5294}{{\ttfamily
  1409.5294}}].

\bibitem{Kelley:2010fn}
R.~Kelley and M.~D. Schwartz, \emph{{1-loop matching and NNLL resummation for
  all partonic 2 to 2 processes in QCD}},
  \href{https://doi.org/10.1103/PhysRevD.83.045022}{\emph{Phys. Rev. D}
  {\bfseries 83} (2011) 045022}
  [\href{https://arxiv.org/abs/1008.2759}{{\ttfamily 1008.2759}}].

\bibitem{Ellis:1996mzs}
R.~K. Ellis, W.~J. Stirling and B.~R. Webber, \emph{{QCD and collider
  physics}}, vol.~8. Cambridge University Press, 2, 2011.

\bibitem{Kang:2018jwa}
Z.-B. Kang, K.~Lee, X.~Liu and F.~Ringer, \emph{{The groomed and ungroomed jet
  mass distribution for inclusive jet production at the LHC}},
  \href{https://doi.org/10.1007/JHEP10(2018)137}{\emph{JHEP} {\bfseries 10}
  (2018) 137} [\href{https://arxiv.org/abs/1803.03645}{{\ttfamily
  1803.03645}}].

\bibitem{Kang:2016mcy}
Z.-B. Kang, F.~Ringer and I.~Vitev, \emph{{The semi-inclusive jet function in
  SCET and small radius resummation for inclusive jet production}},
  \href{https://doi.org/10.1007/JHEP10(2016)125}{\emph{JHEP} {\bfseries 10}
  (2016) 125} [\href{https://arxiv.org/abs/1606.06732}{{\ttfamily
  1606.06732}}].

\bibitem{Dasgupta:2014yra}
M.~Dasgupta, F.~Dreyer, G.~P. Salam and G.~Soyez, \emph{{Small-radius jets to
  all orders in QCD}},
  \href{https://doi.org/10.1007/JHEP04(2015)039}{\emph{JHEP} {\bfseries 04}
  (2015) 039} [\href{https://arxiv.org/abs/1411.5182}{{\ttfamily 1411.5182}}].

\bibitem{Becher:2015hka}
T.~Becher, M.~Neubert, L.~Rothen and D.~Y. Shao, \emph{{Effective Field Theory
  for Jet Processes}}, {\emph{Phys. Rev. Lett.} {\bfseries 116} (2016) 192001}
  [\href{https://arxiv.org/abs/1508.06645}{{\ttfamily 1508.06645}}].

\bibitem{Chien:2015cka}
Y.-T. Chien, A.~Hornig and C.~Lee, \emph{{Soft-collinear mode for jet cross
  sections in soft collinear effective theory}}, {\emph{Phys. Rev. D}
  {\bfseries 93} (2016) 014033}
  [\href{https://arxiv.org/abs/1509.04287}{{\ttfamily 1509.04287}}].

\bibitem{Hornig:2016ahz}
A.~Hornig, Y.~Makris and T.~Mehen, \emph{{Jet Shapes in Dijet Events at the LHC
  in SCET}}, {\emph{JHEP} {\bfseries 04} (2016) 097}
  [\href{https://arxiv.org/abs/1601.01319}{{\ttfamily 1601.01319}}].

\bibitem{Dasgupta:2016bnd}
M.~Dasgupta, F.~A. Dreyer, G.~P. Salam and G.~Soyez, \emph{{Inclusive jet
  spectrum for small-radius jets}}, {\emph{JHEP} {\bfseries 06} (2016) 057}
  [\href{https://arxiv.org/abs/1602.01110}{{\ttfamily 1602.01110}}].

\bibitem{Kolodrubetz:2016dzb}
D.~W. Kolodrubetz, P.~Pietrulewicz, I.~W. Stewart, F.~J. Tackmann and W.~J.
  Waalewijn, \emph{{Factorization for Jet Radius Logarithms in Jet Mass Spectra
  at the LHC}}, {\emph{JHEP} {\bfseries 12} (2016) 054}
  [\href{https://arxiv.org/abs/1605.08038}{{\ttfamily 1605.08038}}].

\bibitem{Neill:2021std}
D.~Neill, F.~Ringer and N.~Sato, \emph{{Leading jets and energy loss}},
  \href{https://doi.org/10.1007/JHEP07(2021)041}{\emph{JHEP} {\bfseries 07}
  (2021) 041} [\href{https://arxiv.org/abs/2103.16573}{{\ttfamily
  2103.16573}}].

\bibitem{Cal:2019hjc}
P.~Cal, F.~Ringer and W.~J. Waalewijn, \emph{{The jet shape at NLL'}},
  \href{https://doi.org/10.1007/JHEP05(2019)143}{\emph{JHEP} {\bfseries 05}
  (2019) 143} [\href{https://arxiv.org/abs/1901.06389}{{\ttfamily
  1901.06389}}].

\bibitem{Cal:2020flh}
P.~Cal, K.~Lee, F.~Ringer and W.~J. Waalewijn, \emph{{Jet energy drop}},
  \href{https://doi.org/10.1007/JHEP11(2020)012}{\emph{JHEP} {\bfseries 11}
  (2020) 012} [\href{https://arxiv.org/abs/2007.12187}{{\ttfamily
  2007.12187}}].

\bibitem{Kang:2019prh}
Z.-B. Kang, K.~Lee, X.~Liu, D.~Neill and F.~Ringer, \emph{{The soft drop
  groomed jet radius at NLL}},
  \href{https://doi.org/10.1007/JHEP02(2020)054}{\emph{JHEP} {\bfseries 02}
  (2020) 054} [\href{https://arxiv.org/abs/1908.01783}{{\ttfamily
  1908.01783}}].

\bibitem{Stewart:2010tn}
I.~W. Stewart, F.~J. Tackmann and W.~J. Waalewijn, \emph{{N-Jettiness: An
  Inclusive Event Shape to Veto Jets}},
  \href{https://doi.org/10.1103/PhysRevLett.105.092002}{\emph{Phys. Rev. Lett.}
  {\bfseries 105} (2010) 092002}
  [\href{https://arxiv.org/abs/1004.2489}{{\ttfamily 1004.2489}}].

\bibitem{Berger:2010xi}
C.~F. Berger, C.~Marcantonini, I.~W. Stewart, F.~J. Tackmann and W.~J.
  Waalewijn, \emph{{Higgs Production with a Central Jet Veto at NNLL+NNLO}},
  \href{https://doi.org/10.1007/JHEP04(2011)092}{\emph{JHEP} {\bfseries 04}
  (2011) 092} [\href{https://arxiv.org/abs/1012.4480}{{\ttfamily 1012.4480}}].

\bibitem{Banfi:2012yh}
A.~Banfi, G.~P. Salam and G.~Zanderighi, \emph{{NLL+NNLO predictions for
  jet-veto efficiencies in Higgs-boson and Drell-Yan production}},
  \href{https://doi.org/10.1007/JHEP06(2012)159}{\emph{JHEP} {\bfseries 06}
  (2012) 159} [\href{https://arxiv.org/abs/1203.5773}{{\ttfamily 1203.5773}}].

\bibitem{Becher:2012qa}
T.~Becher and M.~Neubert, \emph{{Factorization and NNLL Resummation for Higgs
  Production with a Jet Veto}}, {\emph{JHEP} {\bfseries 07} (2012) 108}
  [\href{https://arxiv.org/abs/1205.3806}{{\ttfamily 1205.3806}}].

\bibitem{Dasgupta:2012hg}
M.~Dasgupta, K.~Khelifa-Kerfa, S.~Marzani and M.~Spannowsky, \emph{{On jet mass
  distributions in Z+jet and dijet processes at the LHC}},
  \href{https://doi.org/10.1007/JHEP10(2012)126}{\emph{JHEP} {\bfseries 10}
  (2012) 126} [\href{https://arxiv.org/abs/1207.1640}{{\ttfamily 1207.1640}}].

\bibitem{Banfi:2012jm}
A.~Banfi, P.~F. Monni, G.~P. Salam and G.~Zanderighi, \emph{{Higgs and Z-boson
  production with a jet veto}},
  \href{https://doi.org/10.1103/PhysRevLett.109.202001}{\emph{Phys. Rev. Lett.}
  {\bfseries 109} (2012) 202001}
  [\href{https://arxiv.org/abs/1206.4998}{{\ttfamily 1206.4998}}].

\bibitem{Chien:2012ur}
Y.-T. Chien, R.~Kelley, M.~D. Schwartz and H.~X. Zhu, \emph{{Resummation of Jet
  Mass at Hadron Colliders}},
  \href{https://doi.org/10.1103/PhysRevD.87.014010}{\emph{Phys. Rev. D}
  {\bfseries 87} (2013) 014010}
  [\href{https://arxiv.org/abs/1208.0010}{{\ttfamily 1208.0010}}].

\bibitem{Jouttenus:2013hs}
T.~T. Jouttenus, I.~W. Stewart, F.~J. Tackmann and W.~J. Waalewijn, \emph{{Jet
  mass spectra in Higgs boson plus one jet at next-to-next-to-leading
  logarithmic order}},
  \href{https://doi.org/10.1103/PhysRevD.88.054031}{\emph{Phys. Rev. D}
  {\bfseries 88} (2013) 054031}
  [\href{https://arxiv.org/abs/1302.0846}{{\ttfamily 1302.0846}}].

\bibitem{Becher:2013xia}
T.~Becher, M.~Neubert and L.~Rothen, \emph{{Factorization and
  $N^{3}LL_{p}$+NNLO predictions for the Higgs cross section with a jet veto}},
  {\emph{JHEP} {\bfseries 10} (2013) 125}
  [\href{https://arxiv.org/abs/1307.0025}{{\ttfamily 1307.0025}}].

\bibitem{Stewart:2013faa}
I.~W. Stewart, F.~J. Tackmann, J.~R. Walsh and S.~Zuberi, \emph{{Jet $p_T$
  resummation in Higgs production at $NNLL'+NNLO$}}, {\emph{Phys. Rev. D}
  {\bfseries 89} (2014) 054001}
  [\href{https://arxiv.org/abs/1307.1808}{{\ttfamily 1307.1808}}].

\bibitem{Stewart:2014nna}
I.~W. Stewart, F.~J. Tackmann and W.~J. Waalewijn, \emph{{Dissecting Soft
  Radiation with Factorization}},
  \href{https://doi.org/10.1103/PhysRevLett.114.092001}{\emph{Phys. Rev. Lett.}
  {\bfseries 114} (2015) 092001}
  [\href{https://arxiv.org/abs/1405.6722}{{\ttfamily 1405.6722}}].

\bibitem{Stewart:2015waa}
I.~W. Stewart, F.~J. Tackmann, J.~Thaler, C.~K. Vermilion and T.~F. Wilkason,
  \emph{{XCone: N-jettiness as an Exclusive Cone Jet Algorithm}},
  \href{https://doi.org/10.1007/JHEP11(2015)072}{\emph{JHEP} {\bfseries 11}
  (2015) 072} [\href{https://arxiv.org/abs/1508.01516}{{\ttfamily
  1508.01516}}].

\bibitem{Banfi:2015pju}
A.~Banfi, F.~Caola, F.~A. Dreyer, P.~F. Monni, G.~P. Salam, G.~Zanderighi
  et~al., \emph{{Jet-vetoed Higgs cross section in gluon fusion at
  N$^{3}$LO+NNLL with small-$R$ resummation}}, {\emph{JHEP} {\bfseries 04}
  (2016) 049} [\href{https://arxiv.org/abs/1511.02886}{{\ttfamily
  1511.02886}}].

\bibitem{Chien:2019osu}
Y.-T. Chien and I.~W. Stewart, \emph{{Collinear Drop}},
  \href{https://doi.org/10.1007/JHEP06(2020)064}{\emph{JHEP} {\bfseries 06}
  (2020) 064} [\href{https://arxiv.org/abs/1907.11107}{{\ttfamily
  1907.11107}}].

\bibitem{Dasgupta:2001sh}
M.~Dasgupta and G.~P. Salam, \emph{{Resummation of non-global QCD
  observables}}, {\emph{Phys. Lett.} {\bfseries B512} (2001) 323}
  [\href{https://arxiv.org/abs/hep-ph/0104277}{{\ttfamily hep-ph/0104277}}].

\bibitem{Ligeti:2008ac}
Z.~Ligeti, I.~W. Stewart and F.~J. Tackmann, \emph{{Treating the b quark
  distribution function with reliable uncertainties}},
  \href{https://doi.org/10.1103/PhysRevD.78.114014}{\emph{Phys. Rev.}
  {\bfseries D78} (2008) 114014}
  [\href{https://arxiv.org/abs/0807.1926}{{\ttfamily 0807.1926}}].

\bibitem{Lustermans:2019plv}
G.~Lustermans, J.~K.~L. Michel, F.~J. Tackmann and W.~J. Waalewijn,
  \emph{{Joint two-dimensional resummation in $q_{T}$ and $0$-jettiness at
  NNLL}}, \href{https://doi.org/10.1007/JHEP03(2019)124}{\emph{JHEP} {\bfseries
  03} (2019) 124} [\href{https://arxiv.org/abs/1901.03331}{{\ttfamily
  1901.03331}}].

\bibitem{Marzani:2017mva}
S.~Marzani, L.~Schunk and G.~Soyez, \emph{{A study of jet mass distributions
  with grooming}}, \href{https://doi.org/10.1007/JHEP07(2017)132}{\emph{JHEP}
  {\bfseries 07} (2017) 132}
  [\href{https://arxiv.org/abs/1704.02210}{{\ttfamily 1704.02210}}].

\bibitem{Anderle:2020mxj}
D.~Anderle, M.~Dasgupta, B.~K. El-Menoufi, J.~Helliwell and M.~Guzzi,
  \emph{{Groomed jet mass as a direct probe of collinear parton dynamics}},
  \href{https://doi.org/10.1140/epjc/s10052-020-8411-y}{\emph{Eur. Phys. J. C}
  {\bfseries 80} (2020) 827}
  [\href{https://arxiv.org/abs/2007.10355}{{\ttfamily 2007.10355}}].

\bibitem{Larkoski:2015zka}
A.~J. Larkoski, I.~Moult and D.~Neill, \emph{{Non-Global Logarithms,
  Factorization, and the Soft Substructure of Jets}},
  \href{https://doi.org/10.1007/JHEP09(2015)143}{\emph{JHEP} {\bfseries 09}
  (2015) 143} [\href{https://arxiv.org/abs/1501.04596}{{\ttfamily
  1501.04596}}].

\bibitem{Dasgupta:2007wa}
M.~Dasgupta, L.~Magnea and G.~P. Salam, \emph{{Non-perturbative QCD effects in
  jets at hadron colliders}},
  \href{https://doi.org/10.1088/1126-6708/2008/02/055}{\emph{JHEP} {\bfseries
  02} (2008) 055} [\href{https://arxiv.org/abs/0712.3014}{{\ttfamily
  0712.3014}}].

\bibitem{Mateu:2012nk}
V.~Mateu, I.~W. Stewart and J.~Thaler, \emph{{Power Corrections to Event Shapes
  with Mass-Dependent Operators}},
  \href{https://doi.org/10.1103/PhysRevD.87.014025}{\emph{Phys. Rev.}
  {\bfseries D87} (2013) 014025}
  [\href{https://arxiv.org/abs/1209.3781}{{\ttfamily 1209.3781}}].

\bibitem{Proceedings:2018jsb}
\emph{{Les Houches 2017: Physics at TeV Colliders Standard Model Working Group
  Report}}, 3, 2018.

\bibitem{Gallicchio:2011xc}
J.~Gallicchio and M.~D. Schwartz, \emph{{Pure Samples of Quark and Gluon Jets
  at the LHC}}, \href{https://doi.org/10.1007/JHEP10(2011)103}{\emph{JHEP}
  {\bfseries 10} (2011) 103} [\href{https://arxiv.org/abs/1104.1175}{{\ttfamily
  1104.1175}}].

\bibitem{scetlib}
M.~A. Ebert, J.~K.~L. Michel, F.~J. Tackmann et~al., \emph{{SCETlib: A C++
  Package for Numerical Calculations in QCD and Soft-Collinear Effective
  Theory}}, {\emph{DESY-17-099} (2018) }.

\bibitem{scetlibSD}
A.~Pathak, ``\texttt{scetlib::sd}: A {C}++ library for soft drop resummed
  observables in \texttt{SCETlib}.'' Forthcoming, 2022.

\bibitem{Bell:2018vaa}
G.~Bell, R.~Rahn and J.~Talbert, \emph{{Two-loop anomalous dimensions of
  generic dijet soft functions}}, {\emph{Nucl. Phys. B} {\bfseries 936} (2018)
  520} [\href{https://arxiv.org/abs/1805.12414}{{\ttfamily 1805.12414}}].

\bibitem{Bachu:2020nqn}
B.~Bachu, A.~H. Hoang, V.~Mateu, A.~Pathak and I.~W. Stewart, \emph{{Boosted
  top quarks in the peak region with N3LL resummation}},
  \href{https://doi.org/10.1103/PhysRevD.104.014026}{\emph{Phys. Rev. D}
  {\bfseries 104} (2021) 014026}
  [\href{https://arxiv.org/abs/2012.12304}{{\ttfamily 2012.12304}}].

\end{thebibliography}\endgroup

\end{document}